\documentclass{emulateapj}
\usepackage{amsmath}

\def\kmsec{\mbox{km~s$^{\rm -1}$}}
\def\logg{\mbox{log~{\it g}}}

\def\teff{\mbox{$T_{\rm eff}$}}
\def\vt{\mbox{$v_{\rm t}$}}
\def\logrw{\mbox{$\log$(EW/$\lambda$)}}
\def\rpro{\mbox{$r$-process}}
\def\spro{\mbox{$s$-process}}

\def\loggf{$\log gf$}

\shorttitle{Detailed Abundances of 313 Metal-Poor Stars}
\shortauthors{Roederer et al.}
\slugcomment{Accepted for publication in the Astronomical Journal}

\begin{document}

\title{A Search for Stars of Very Low Metal Abundance.\ VI.\ \\
Detailed Abundances of 313 Metal-Poor Stars\footnotemark[1]}

\footnotetext[1]{
This paper includes data gathered with the 6.5 meter 
Magellan Telescopes located at Las Campanas Observatory, Chile,
and The McDonald Observatory of The University of Texas at Austin.
The Hobby-Eberly Telescope is a 
joint project of the University of Texas at Austin, the Pennsylvania State 
University, Stanford University, Ludwig-Maximilians-Universit\"{a}t 
M\"{u}nchen, and Georg-August-Universit\"{a}t G\"{o}ttingen.
}

\author{Ian U.\ Roederer,\altaffilmark{2,4}
George W.\ Preston,\altaffilmark{2}
Ian B.\ Thompson,\altaffilmark{2}
Stephen A. Shectman,\altaffilmark{2}
Christopher Sneden,\altaffilmark{3}
Gregory S.\ Burley,\altaffilmark{2}
Daniel D.\ Kelson\altaffilmark{2}
}

\altaffiltext{2}{Carnegie Observatories, 
813 Santa Barbara St., Pasadena, CA 91101, USA}

\altaffiltext{3}{Department of Astronomy, University of Texas at Austin,
1 University Station, C1400, Austin, TX 78712, USA}

\altaffiltext{4}{Present address:\
Department of Astronomy, University of Michigan,
500 Church Street, Ann Arbor, MI 48109, USA;
iur@umich.edu}


\addtocounter{footnote}{4}

\begin{abstract}

We present radial velocities, equivalent widths, 
model atmosphere parameters, and 
abundances or upper limits for 53~species of 48~elements 
derived from high resolution optical spectroscopy
of 313~metal-poor stars.
A majority of these stars 
were selected from the metal-poor
candidates of the HK~Survey
of Beers, Preston, and Shectman.
We derive detailed abundances for 61\% of these stars 
for the first time.
Spectra were obtained during a 10-year observing
campaign using the
Magellan Inamori Kyocera Echelle 
spectrograph
on the Magellan Telescopes at Las Campanas Observatory,
the Robert G.\ Tull Coud\'{e} Spectrograph
on the Harlan J.\ Smith Telescope at McDonald Observatory,
and the High Resolution Spectrograph 
on the Hobby-Eberly Telescope 
at McDonald Observatory.
We perform a standard LTE abundance analysis
using MARCS model atmospheres,
and we apply line-by-line statistical corrections
to minimize systematic abundance differences 
arising when different sets of lines are available for analysis.
We identify several abundance correlations with 
effective temperature.
A comparison with previous abundance analyses 
reveals significant differences in stellar parameters,
which we investigate in detail.
Our metallicities are, on average,
lower by $\approx$~0.25~dex for red giants and 
$\approx$~0.04~dex for subgiants.
Our sample contains 19~stars with [Fe/H]~$\leq -$3.5, 
84~stars with [Fe/H]~$\leq -$3.0, and 
210~stars with [Fe/H]~$\leq -$2.5.
Detailed abundances are presented
here or elsewhere
for 91\% of the 209~stars with
[Fe/H]~$\leq -$2.5 as estimated from medium resolution
spectroscopy by Beers, Preston, and Shectman.
We will discuss the interpretation of these abundances
in subsequent papers.

\end{abstract}


\section{Introduction}
\label{introduction}

Observations of the high redshift Universe
reveal carbon, magnesium, and
other metals in the clouds of hydrogen that fueled
the rising star formation rate density during the
first several Gyr after the Big Bang
(e.g., \citealt{sargent88,cooke11,matejek12}).
Surely some of those gas clouds
evolve into galaxies like today's Milky Way,
whose stellar halos retain a chemical memory of 
those early epochs of star formation
and metal production.
If, however, one is interested in studying the early nucleosynthesis
of less abundant metals, like
holmium \citep{sneden96}
or uranium \citep{cayrel01},
stars in the Milky Way
are the only practical targets.
For this and many other reasons,
the importance of expanding the inventory of 
halo stars
whose heavy metal abundances are known in great detail
has long been recognized.

\subsection{Previous Surveys of Metal-Poor Stars}

Noteworthy in regard to the storyline of the present
study are the surveys of \citet{bond70,bond80} and
\citet{bidelman73}.
The photographic plates for their objective-prism surveys were 
taken with the University of Michigan's 0.61~m 
Curtis Schmidt Telescope.
This telescope, initially located
near Ann Arbor, 
was relocated in 1966 to
Cerro Tololo Inter-American Observatory (CTIO).
Most of the observations for \citet{bond70}
were made in Michigan, while 
\citeauthor{bidelman73} and \citet{bond80}
made their observations at CTIO.
Bond also used Str\"{o}mgren photometry to assign spectral types
and luminosity classes to his candidates,
and he measured radial velocities from followup 
coud\'{e} spectroscopy when possible.

Many of the
well-known and bright metal-poor stars in the Henry Draper (HD),
Bonner Durchmusterung (BD), or 
C\'{o}rdoba Durchmusterung (CD) Catalogs were
identified during this period by these surveys.
Other metal-poor stars were found among 
the high proper motion stars in the
Lowell Proper Motion Survey 
(\citealt{giclas71,giclas78};
these stars are identified with a ``G'' prefix
before their catalog designation)
and the New Luyten Two Tenths Catalog
(NLTT, \citealt{luyten79}).
\citet{ryan89} discusses the
methods used to identify metal-poor stars in these surveys.

\subsection{The HK Objective-Prism Survey}

Our own work on the subject began with 
an objective-prism survey at the Curtis Schmidt (CS) Telescope
at CTIO, initiated by G.\ Preston and S.\ Shectman,
in 1978--1979.
The key advance was the use of an interference filter to expose
only the region around the stellar Ca~\textsc{ii}~H and K 
absorption lines
at a spectral resolution of $\approx$~5~\AA\
(R~$\equiv \lambda/\Delta\lambda \sim$~800).
This interference filter reduced crowding and sky fog and 
allowed longer exposure times
(90~minutes) than had been practical previously.
The ``HK'' Survey plates reach $B \approx$~15,
several magnitudes fainter than
the surveys of
\citet{bond70,bond80} and \citet{bidelman73}.
Visual inspection of the plates using a 
low-power binocular microscope 
yielded about 1800 metal-poor candidate stars
on 72~plates.

Broadband $UBV$ photometry and
followup medium resolution (1~\AA; R~$\sim$~4,000)
spectroscopy 
covering the 3700--4500~\AA\ wavelength range
were obtained for 450~candidates.
\citet{beers85} presented metallicity estimates for
134~metal-poor candidates with [Fe/H]~$< -$2.0.
Using a revised metallicity calibration 
\citep{beers90},
Beers, Preston, \& Shectman 
(1992; hereafter \citeauthor{beers92})
published metallicity estimates, radial velocity measurements,
and distances for 1044 dwarfs and giants
with subsolar [Fe/H] from 135~unique fields
covering 3375 square degrees (8\% of the sky).

The relocation of Case Western Reserve University's 
Burrell Schmidt (BS) Telescope 
from Cleveland to Kitt Peak National Observatory (KNPO) in 1979
enabled a similar survey in the Northern Hemisphere.
Followup medium resolution spectroscopy for these candidates
was not yet available in 1992.
\citet{beers13} describes the impressive worldwide network of
2--4~m class telescopes involved in the subsequent medium resolution
spectroscopic followup of metal-poor candidates from
both the Northern and Southern portions of the HK~Survey.
Detailed abundances of metal-poor candidates 
from the Northern portion of the HK~Survey have been published 
elsewhere (e.g., \citealt{honda04a,honda04b};
\citealt{lai04,lai08}) and will not be considered in 
the sample presented here.

An analysis of digital scans of the original HK~Survey plates
was made by \citet{rhee01} and in subsequent
unpublished work by J.\ Rhee, T.\ Beers, and coworkers
(see also Section~3.3.1 of \citealt{beers05}).
The goal of this ``HK-II'' Survey was to
identify metal-poor red giant stars that
may have been overlooked in the original 
visual scans of the plates due to the
unavoidable temperature bias against
cool stars.
These candidates will not be considered in 
the sample presented here.

\subsection{Subsequent Surveys}

The Hamburg/ESO (HE) Survey
\citep{wisotzki00}
introduced quantitative methods to identify
metal-poor candidates from digitized spectra
\citep{christlieb08}.
These techniques increase the 
effective yields of genuine metal-poor stars,
especially among giants, when color information
is included as part of the selection criteria.
This survey also reaches several magnitudes deeper 
than the HK~Survey.

Recent surveys have built on these
quantitative techniques to identify even greater numbers
of candidate metal-poor stars from low resolution spectroscopy,
including the Radial Velocity Experiment 
(RAVE; see \citealt{fulbright10}),
the Sloan Digital Sky Survey (SDSS), and the 
Sloan Extensions for Galactic Understanding and Exploration
(SEGUE-1 and SEGUE-2; \citealt{yanny09,rockosi12}).
Many metal-poor candidates are
also expected to be found among ongoing surveys by the
Large sky Area Multi-Object fiber Spectroscopic Telescope
(LAMOST; e.g., \citealt{deng12}); 
and the SkyMapper Telescope
\citep{keller12}.

\subsection{Detailed Abundance Followup of Metal-Poor Candidates}

Bright stars identified by the
surveys of \citet{bond70,bond80} and \citet{bidelman73}
have been analyzed in great chemical detail by numerous
investigators, including 
\citet{luck81,luck85}, \citet{hartmann88}, 
\citet{gilroy88}, 
\citet{gratton88,gratton91,gratton94}, 
\citet{magain89},
\citet{peterson90}, \citet{zhao90},
and \citet{johnson02}.
\citet{ryan91b}, 
\citet{fulbright00,fulbright02}, \citet{stephens02}, and
\citet{ishigaki10} studied the compositions of 
stars selected from the early objective prism and proper motion surveys.
Detailed abundance studies of large samples of stars 
from the HE Survey have been conducted by
\citet{carretta02},
\citet{cohen02,cohen04,cohen08},
\citet{barklem05}, 
\citet{aoki07},
\citet{hollek11},
\citet{norris13}, and
\citet{yong13}.
Candidates from the HK-II Survey have been observed as part of the
Chemical Abundances of Stars in the Halo (CASH)
project at the University of Texas
\citep{frebel08b,roederer08b}.
Detailed chemical followup of large numbers of stars
from the SDSS and SEGUE have
been performed by
\citet{aoki08,aoki13}, \citet{caffau11}, and \citet{bonifacio12}.

Over the last several decades,
high resolution optical spectroscopic followup 
of candidates from \citeauthor{beers92} 
has confirmed hundreds of
them as genuine metal-poor stars.
Chemical abundances of handfuls of stars from 
Beers et al.\ (\citeyear{beers85})
were presented by \citet{molaro90a}, \citet{molaro90b},
\citet{norris93}, and \citet{primas94}.
\citet{mcwilliam95a,mcwilliam95b},
\citet{norris96}, and \citet{ryan96} 
were the first to analyze larger samples of stars (34~stars between them)
from \citeauthor{beers92}.~
Since then, the number of detailed abundance studies 
conducted on candidates from \citeauthor{beers92}
has grown tremendously, and there are far too many 
excellent ones to list here individually.
There have been several dedicated observing campaigns 
to obtain high resolution spectroscopy of substantial numbers of stars
(typically $\approx$~10--30), including analyses by 
\citet{aoki05,aoki07},
\citet{honda04a,honda04b}, 
\citet{lai08},
the ``First Stars'' team
\citep{cayrel04,spite05,francois07,bonifacio09},
and a reanalysis of the published values from many of these studies by
\citet{yong13}.

The detailed chemical analysis 
performed by \citet{mcwilliam95a,mcwilliam95b}
launched our efforts to use these
stars as probes of the earliest epoch of metal
enrichment in the Galaxy.
Our subsequent abundance studies based on
high resolution spectroscopy of metal-poor candidates
from the HK~Survey have examined
carbon rich metal-poor stars
\citep{preston01,sneden03b,roederer14},
individual stars of interest 
\citep{sneden94,ivans05,preston06b,thompson08}, 
stars on the horizontal branch \citep{preston06}, and
stars with kinematics indicative of a cold stellar stream \citep{roederer10}.

In this paper we present abundance results for
313~stars, including 217~stars from the HK~Survey,
using high resolution spectroscopy obtained from 2003--2013
at the Magellan Telescopes at Las Campanas Observatory.
As of July, 2013,
detailed abundances for
91\% (191/209)
of the stars with estimated [Fe/H]~$\leq -$2.5
in \citeauthor{beers92} are presented
here or have been published elsewhere previously.
Sixty-one percent (132/217) of the stars from the HK~Survey
presented in this work are analyzed 
in such a manner for the first time.
Abundances in the other 85~stars have been examined previously
by the studies above or others named below.
A limited selection of 
stars from the BD, CD, G, HD, and HE catalogs are also
(re)analyzed in the present study.

\section{Abundance Notation}
\label{notation}

We adopt standard definitions of elemental abundances and ratios.
For element X, the logarithmic absolute abundance is defined
as the number of atoms of element X per 10$^{12}$ hydrogen atoms,
$\log\epsilon$(X)~$\equiv \log_{10}(N_{\rm X}/N_{\rm H}) +$~12.0.
For elements X and Y, the logarithmic abundance ratio relative to the
solar ratio is defined as
[X/Y]~$\equiv \log_{10} (N_{\rm X}/N_{\rm Y}) -
\log_{10} (N_{\rm X}/N_{\rm Y})_{\odot}$.

Abundances or ratios denoted with the ionization state
are defined to be 
the total elemental abundance as derived from transitions of
that particular ionization state 
after ionization corrections,
assuming \citet{saha21} equilibrium,
have been applied.
For example,
$\log\epsilon$(Fe~\textsc{ii}) denotes the number density of all
iron atoms as derived from Fe~\textsc{ii} lines.

When reporting relative abundance ratios for a specific element X
(e.g., [X/Fe]),
these ratios are constructed
by comparing total abundances derived 
from species in the same ionization state.
For example, if X is a neutral species,
the ratio [X/Fe] is calculated using
the total abundance of element X derived from the neutral species
with the total iron abundance derived from Fe~\textsc{i}.
Similarly, if X is an ionized species,
the ratio [X/Fe] is calculated using
the total abundance of element X derived from the ionized species 
with the total iron abundance derived from Fe~\textsc{ii}.

\section{Observations}
\label{obs}

\subsection{Target Selection}

Over the course of this program we have
observed 88\% (184/209) of the stars with
estimated [Fe/H]~$\leq -$2.5 in \citeauthor{beers92},
excluding three duplicate identifications
in Table~5 of \citeauthor{beers92}.
Abundances derived from previous moderate
or high resolution spectroscopy 
of seven of the remaining 25~stars have been presented
elsewhere by 
\citet{norris93}, 
\citet{spite00}, 
\citet{preston01}, 
\citet{lai04}, 
and T.\ Masseron et al.\ (unpublished;
see \citealt{masseron10}).  
The 18 remaining unobserved candidates are faint by our standards,
$V >$~14.5, and most are fainter than $V >$~15.0.

We also observed other metal-poor stars
to expand the sample to higher metallicity
and into the Northern hemisphere.
These additional targets
include higher metallicity candidates from \citeauthor{beers92}
and bright stars from the catalogs discussed in Section~\ref{introduction}.
Some of these stars have been previously analyzed elsewhere.
Our final sample includes observations of 217~stars from 
the HK~Survey and 96~stars from other sources.

\begin{deluxetable}{ll}
\tablecaption{Repeat Star Identifications
\label{multipleidtab}}
\tablewidth{0pt}
\tabletypesize{\scriptsize}
\tablehead{
\colhead{Primary Name} &
\colhead{Additional Designations} }
\startdata 
CS~22169--035 & HE~0409$-$1212 \\
CS~22172--029 & HE~0328$-$1047 \\
CS~22185--007 & HE~0315$-$1528 \\
CS~22189--009 & HE~0239$-$1340 \\
CS~22873--128 & HE~2002$-$5843 \\
CS~22886--003 & CS~29512--030  \\
CS~22886--012 & CS~29512--015  \\
CS~22886--013 & CS~29512--013  \\
CS~22888--014 & CS~30493--023  \\
CS~22890--064 & CS~30306--117  \\
CS~22892--052 & HE~2214$-$1654 \\
CS~22894--023 & CS~22952--011  \\
CS~22937--072 & CS~29501--051  \\
              & CS~30492--102  \\
CS~22942--002 & HE~0044$-$2459 \\
CS~22948--066 & CS~30343--064  \\
CS~22949--037 & HE~2323$-$0256 \\
CS~22952--015 & HE~2334$-$0604 \\
CS~22954--015 & HE~0236$-$0242 \\
CS~22957--027 & HE~2356$-$0410 \\
CS~22968--014 & HE~0305$-$5442 \\
CS~29517--042 & CS~31060--052  \\
CS~30339--069 & HE~0027$-$3613 \\
\enddata
\end{deluxetable}

All stars in our sample are present in the SIMBAD database,
and their coordinates may be found there.
During the course of the objective-prism survey, 
portions of some fields were observed multiple times 
with different plates. 
This led to multiple identification numbers
for a few candidates.
In such cases we adopt the number associated with the
earliest observation.
Table~\ref{multipleidtab} lists the 
stars in our survey that received
multiple identification numbers.
Several of these stars were also rediscovered
during the course of the HE Survey, and the 
HE designations are listed in Table~\ref{multipleidtab}.
The HK~Survey itself rediscovered stars with previous
catalog names, and these are listed in Table~6 and on page~2033 of
\citeauthor{beers92}.

\subsection{High Resolution Spectroscopy}

Observations conducted at Las Campanas Observatory
were made with the Magellan Inamori Kyocera Echelle (MIKE)
spectrograph \citep{bernstein03}.
This instrument is currently mounted on the f/11 Nasmyth platform at the
6.5~m Landon Clay Telescope (Magellan~II).
Early observations for our program were taken while MIKE
was mounted on the 6.5~m Walter Baade Telescope (Magellan~I) in 2003.
The MIKE spectra were taken with the 0\farcs7\,$\times$\,5\farcs0 slit, 
yielding
a resolving power of $R \sim$~41,000 in the blue 
and $R \sim$~35,000 in the red as measured from isolated ThAr lines.
A dichroic splits the two arms around 4950~\AA, although the wavelength
at the split is bluer in earlier observations.
This setup gives approximately 2.4~pixels per resolution element (RE)
in the blue and 2.1~pixels~RE$^{-1}$ in the red.
Many of the observations taken in 2003 and 2004 were made 
using the double aperture 0\farcs7\,$\times$\,2\farcs0 slit (``A-B mode'').
In this mode, exposures are made with the star in 
aperture A and the sky in aperture B simultaneously;
the pattern is then reversed for the subsequent exposure.
This observing procedure was adopted to optimize sky subtraction
for faint targets.
This strategy proved unnecessary for our targets, which are not
sky-noise limited, and all subsequent
MIKE observations were made using the single 0\farcs7\,$\times$\,5\farcs0 slit.
This setup achieves complete wavelength coverage from 
3350--9150~\AA, although some of the early exposures only extend to
7250~\AA\ in the red. 

For the exposures made in the A-B mode, bias subtraction, flat-fielding,
sky and scattered light subtraction, cosmic ray removal, and
correction of the slit tilt were 
accomplished with software written by S.A.S.
Aperture extraction, wavelength calibration (derived from 
ThAr frames taken before or after each stellar integration), co-addition
of separate observations, and continuum normalization were 
performed within the IRAF environment.
For the exposures made using the single 0\farcs7\,$\times$\,5\farcs0 slit, 
data reduction,
extraction, and wavelength calibration were performed using 
the MIKE data reduction pipeline
written by D.\ Kelson
(see also \citealt{kelson03}).
Observations were often broken into several 
sub-exposures (not longer than 2400~s per sub-exposure).
We refer to the addition of these sub-exposures as one observation.
Coaddition of repeat observations 
and continuum normalization were performed within the 
IRAF environment.

Other observations were made with the 
Robert G.\ Tull Coud\'{e} Spectrograph
\citep{tull95} on the 2.7~m Harlan J.\ Smith
Telescope at McDonald Observatory.
These spectra were taken with the 2\farcs4\,$\times$\,8\farcs0 slit, yielding 
a resolving power $R \sim$~33,000 and approximately 4.0~pixels~RE$^{-1}$.
This setup delivers complete wavelength coverage from 3700--5700~\AA,
with small gaps between the echelle orders further to the red.
For our abundance analysis we only use the spectra blueward of 8000~\AA.

Additional observations were made with the High Resolution Spectrograph
(HRS; \citealt{tull98}) on the 9.2~m Hobby-Eberly Telescope
\citep{ramsey98} at McDonald Observatory.
These exposures were taken during the standard queue observing mode
\citep{shetrone07}.
Using the 1\farcs0 slit yields a resolving power $R \sim$~30,000
and approximately 3.1~pixels~RE$^{-1}$.
In the bluemost cross-dispersion setting
this instrument delivers complete wavelength coverage from 3900--6800~\AA. 

For the data obtained with the Tull Spectrograph and HRS, 
reduction, extraction, sky and scattered light removal, 
and wavelength calibration (derived
from ThAr exposures taken before or after each stellar exposure)
of the spectra were accomplished using the REDUCE software package
\citep{piskunov02}.
These observations were also broken into several sub-exposures
with exposure times typically not longer than 1800~s.
Coaddition and continuum normalization were performed within the 
IRAF environment.

\begin{deluxetable*}{lccccccc}
\tablecaption{Log of Observations
\label{obstab}}
\tablewidth{0pt}
\tabletypesize{\scriptsize}
\tablehead{
\colhead{Star} &
\colhead{Telescope/} &
\colhead{Observer} &
\colhead{Exp.} & 
\colhead{Date} &
\colhead{UT mid-} &
\colhead{Heliocentric} &
\colhead{Heliocentric} \\
\colhead{} &
\colhead{Instrument} &
\colhead{} &
\colhead{time (s)} &
\colhead{} &
\colhead{exposure} &
\colhead{JD} &
\colhead{$V_{\rm r}$ (\kmsec)} }
\startdata
CS 22166--016   & Magellan-Clay/MIKE      & GWP  & 1600   & 2004 Aug 12 & 10:19 & 2453229.934   & $-$210.0      \\
CS 22169--008   & Magellan-Clay/MIKE      & IBT  & 4000   & 2004 Sep 23 & 08:55 & 2453271.874   & $+$184.2      \\
CS 22169--035   & Magellan-Baade/MIKE     & GWP  & 1400   & 2003 Jan 18 & 04:22 & 2452657.684   & $+$15.4       \\
CS 22169--035   & Magellan-Baade/MIKE     & GWP  & 2400   & 2003 Jan 20 & 02:05 & 2452659.589   & $+$14.1       \\
CS 22169--035   & Magellan-Clay/MIKE      & GWP  & 1700   & 2007 Aug 24 & 09:56 & 2454336.914   & $+$14.1       \\
CS 22171--031   & Magellan-Clay/MIKE      & GWP  & 2800   & 2008 Sep 10 & 07:59 & 2454719.837   & $+$38.8       \\
CS 22171--037   & Magellan-Clay/MIKE      & IBT  & 7200   & 2003 Nov 01 & 05:53 & 2452944.750   & $-$261.4      \\
CS 22171--037   & Magellan-Clay/MIKE      & IUR  & 7200   & 2012 Aug 26 & 06:12 & 2456165.762   & $-$261.1      \\
\enddata                                                                                                        
\tablecomments{
The complete version of Table~\ref{obstab} is available online only.
A short version is shown here to
illustrate its form and content.
}
\end{deluxetable*}

We have observed 250~stars with Magellan$+$MIKE,
52~stars with Smith$+$Tull, and
19~stars with HET$+$HRS.
After accounting for duplicate observations of the same star
and three double-lined spectroscopic binaries whose abundances
are not examined in this paper,
we are left with a total of 313 stars.
These 532 individual observations account for 
nearly 495~hours of integration time.
In Table~\ref{obstab} we present a record of all observations.
A full version of Table~\ref{obstab} is available in the online 
edition of the journal.

\begin{figure}
\begin{center}
\includegraphics[angle=0,width=3.35in]{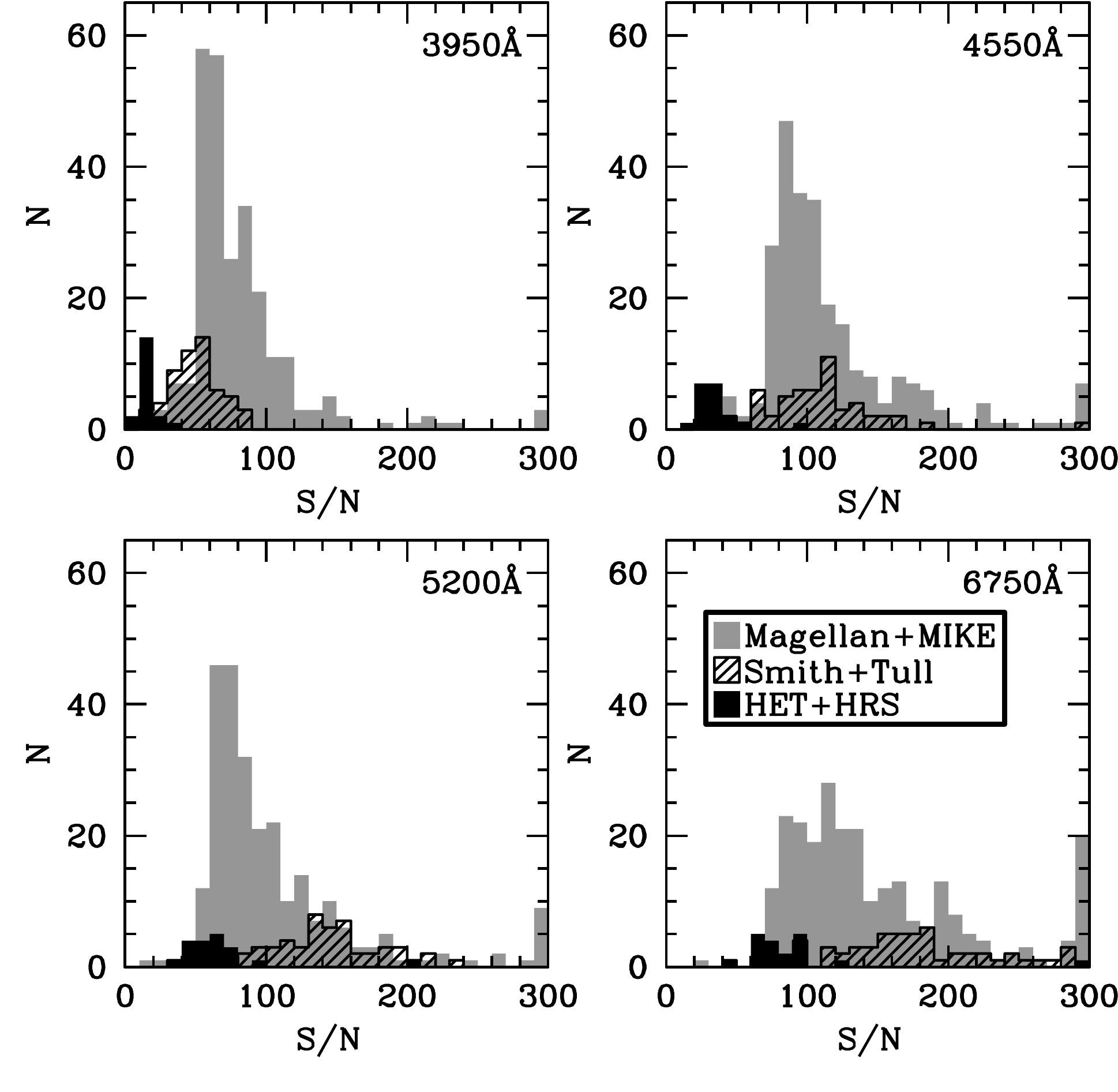}
\end{center}
\caption{
\label{snfig}
S/N estimates at four wavelengths.
The gray shaded histogram 
represents spectra obtained with the MIKE spectrograph,
the hatched histogram represents spectra obtained with the Tull spectrograph,
and the black shaded histogram 
represents spectra obtained with the HRS.
The last bar to the right in each panel indicates the number of stars with
S/N~$>$~300/1.
}
\end{figure}

\begin{deluxetable*}{lccccccccc}
\tablecaption{Observational Stellar Data
\label{rvtab}}
\tablewidth{0pt}
\tabletypesize{\scriptsize}
\tablehead{
\colhead{Star} &
\colhead{$\langle$RV$\rangle$} & 
\colhead{Binary} &  
\colhead{Literature} &
\colhead{Total exp.} &
\colhead{No.} & 
\colhead{S/N} & 
\colhead{S/N} & 
\colhead{S/N} & 
\colhead{S/N} \\  
\colhead{} &
\colhead{(\kmsec)} &
\colhead{flag\tablenotemark{1}} &
\colhead{RV Ref.} &
\colhead{time (s)} &
\colhead{obs.} &
\colhead{3950\AA} &
\colhead{4550\AA} &
\colhead{5200\AA} &
\colhead{6750\AA} }
\startdata
CS 22166--016   & $-$210.0  & 1  & 1     & 1600  & 1     & 115   & 175   & 95    & 140    \\
CS 22169--008   & $+$184.2  & 0  &\nodata& 4000  & 1     & 100   & 135   & 70    & 95     \\
CS 22169--035   & $+$14.5   & 1  & 2     & 5500  & 3     & 75    & 135   & 170   & 285    \\
CS 22171--031   & $+$38.8   & 0  &\nodata& 2800  & 1     & 60    & 80    & 70    & 115    \\
CS 22171--037   & $-$261.3  & 1  &\nodata& 14400 & 2     & 95    & 125   & 125   & 180    \\
CS 22172--029   & \nodata   & 2  & 3     & 7600  & 2     & 80    & 110   & 105   & 180    \\
\enddata 
\tablerefs{
 (1) \citealt{giridhar01};
 (2) \citealt{bonifacio09};
 (3) \citealt{barklem05};
 (4) \citealt{honda04a};
 (5) \citealt{mcwilliam95a};
 (6) \citealt{lai08};
 (7) \citealt{aoki02a};
 (8) \citealt{tsangarides04};
 (9) \citealt{cohen02};
(10) \citealt{lai04};
(11) \citealt{preston09};
(12) \citealt{preston06};
(13) \citealt{hollek11};
(14) \citealt{preston01};
(15) \citealt{primas94};
(16) \citealt{norris01};
(17) \citealt{depagne02};
(18) \citealt{preston00};
(19) \citealt{norris97};
(20) \citealt{bonifacio98};
(21) \citealt{roederer14};
(22) \citealt{sivarani06};
(23) \citealt{aoki07};
(24) \citealt{sneden03b};
(25) \citealt{aoki02b};
(26) \citealt{sbordone10};
(27) \citealt{aoki09};
(28) \citealt{hill02};
(29) \citealt{carney03};
(30) \citealt{zhang09};
(31) \citealt{latham02};
(32) \citealt{latham91};
(33) \citealt{ito13};
(34) \citealt{carney08};
(35) \citealt{pourbaix04};
(36) \citealt{bonifacio99};
(37) \citealt{norris86};
(38) \citealt{nordstrom04};
(39) \citealt{gratton94};
(40) \citealt{roederer08};
(41) \citealt{ryan91};
(42) \citealt{aoki02c};
(43) \citealt{vaneck03};
(44) \citealt{lucatello05};
(45) \citealt{cohen08};
(46) \citealt{fulbright02}
}
\tablenotetext{1}{Binary flags:
(0) unknown binary status;
(1) no RV variations detected in multiple epochs;
(2) RV variations detected, suspected binary, no systemic velocity listed;
(3) spectroscopic binary confirmed by other studies, systemic velocity listed.
}
\tablecomments{
The complete version of Table~\ref{rvtab} is available online only.
A short version is shown here to
illustrate its form and content.
}
\end{deluxetable*}

Signal to noise (S/N) estimates, listed in Table~\ref{rvtab}, are based on
Poisson statistics for the number of photons collected in the continuum
at several reference wavelengths
once all observations of a given target have been coadded together.
Spectra obtained with different instruments were not coadded.
These S/N estimates are illustrated in Figure~\ref{snfig}.
We have succeeded in achieving a relatively high S/N 
at 3950~\AA, between the Ca~\textsc{ii}~H and K lines,
for most stars observed with the MIKE and Tull spectrographs
(S/N~$\gtrsim$~50 and 40, respectively).
The HRS is not optimized for blue response, and
similar S/N levels were not practical in stars observed with this setup.
The S/N continues to increase when moving toward the red for stars
observed with all three instruments.
High S/N levels in the blue region of the spectrum
are essential for deriving abundances of many species
whose most promising transitions are found
in this spectral region.

\section{Radial Velocities}
\label{rv}

To measure the radial velocity (RV) of each of our target stars,
we cross correlate our spectra against standard template stars
using the \textit{fxcor} task in IRAF.
The RV with respect to the ThAr lamp
is found by cross correlating the echelle
order containing the Mg~\textsc{i}~b lines.
For spectra taken with the MIKE and Tull spectrographs,
we also cross correlate the echelle
order containing the telluric O$_{2}$ B band near 6900~\AA\
with a template to remove any velocity shifts resulting from 
thermal and mechanical motions in the spectrographs.
We use empirical O$_{2}$ wavelengths from \citet{griffin73}
to create this zero-velocity template from one spectrum
with high S/N ratios obtained with each instrument.
The HRS spectra do not contain this band, and we cross-correlate
against the telluric O$_{2}$ $\alpha$ band near 6300~\AA\ 
using laboratory wavelengths from \citet{babcock48}.
These measurements are consistent with no shift.
Heliocentric corrections are computed using the IRAF \textit{rvcorrect} task.
Heliocentric RV measurements for each observation are
listed in Table~\ref{obstab}.

\begin{figure}
\begin{center}
\includegraphics[angle=0,width=3.35in]{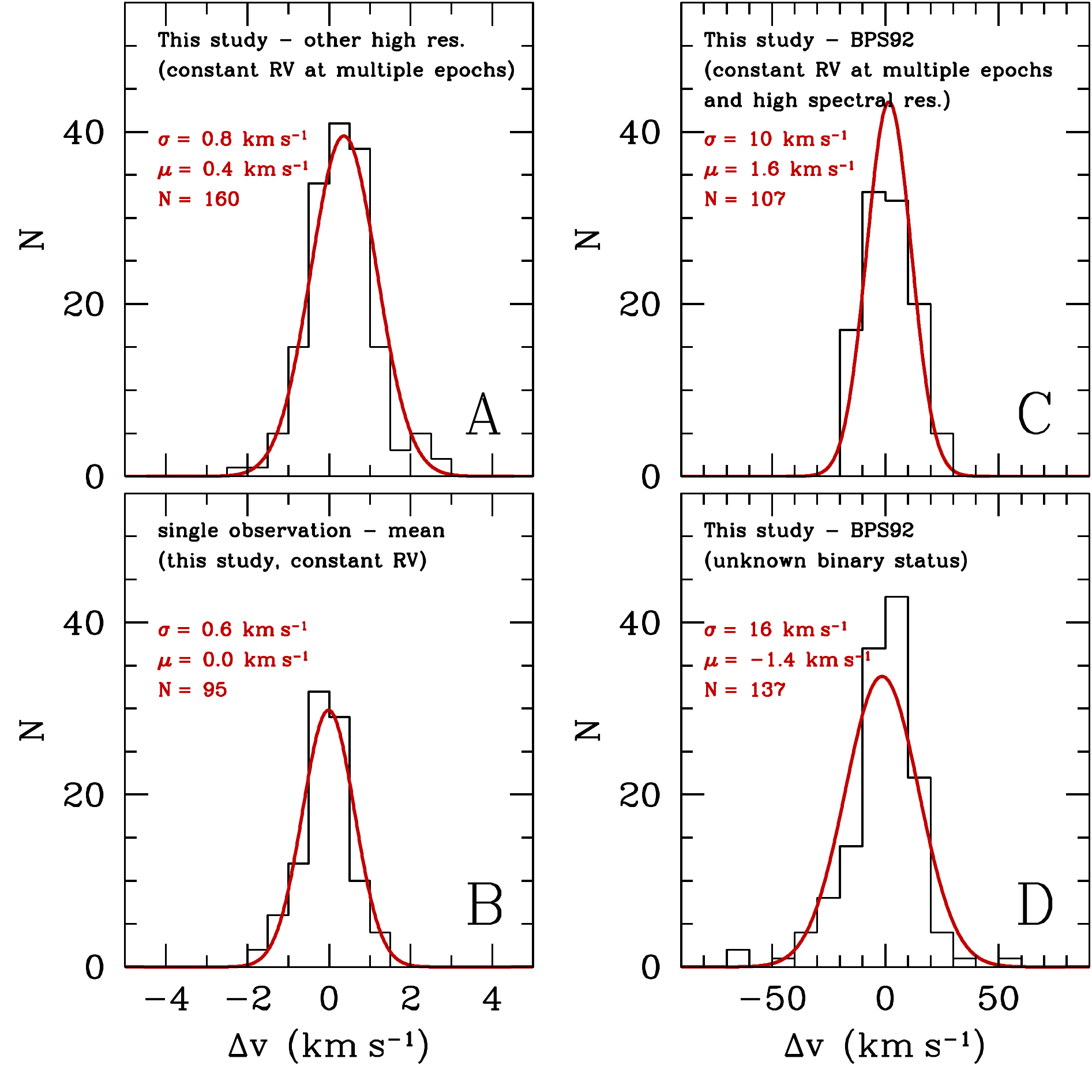}
\end{center}
\caption{
\label{rvfig}
Differences in our measured RV displayed as histograms when compared to
RV constant stars observed by other investigators (Panel~A),
repeated measurements of RV constant stars made by us (Panel~B),
measurements made from moderate resolution spectroscopy 
by \citeauthor{beers92} 
of RV constant stars (Panel~C), and
measurements made from moderate resolution spectroscopy 
by \citeauthor{beers92}
for stars with an unknown binary status (Panel~D).
The mean ($\mu$), standard deviation ($\sigma$), and
number of points included ($N$)
are listed in each panel.
}
\end{figure}

Table~\ref{rvtab} lists the mean RV derived for each of our targets.
We have searched the literature for previous RV measurements
to identify stars that are members of binary or multiple star systems.
We compare our heliocentric RV measurements to those measured by
previous investigators.
These references are listed in Table~\ref{rvtab}.
Our measurements for known RV-constant stars are all in good agreement
with previous investigations.
Typical RV uncertainties for previous studies are
$\approx$~1~\kmsec\ when the previous measurement was made
from high resolution spectra, and 
RV uncertainties are often several \kmsec\ for lower resolution data.
Panel~A of Figure~\ref{rvfig} displays the RV differences 
that arise from comparison of
our mean measurements to those obtained by previous investigations.
We only make this comparison when the previous study reported 
an error less than 1~\kmsec; many of these stars also have been
observed at many epochs (i.e., $\gg$~3), so it is probable that they 
are indeed RV constant.
For the 160~measurements presented in Panel~A, $\sigma =$~0.81~\kmsec.
Panel~B compares the differences when a single measurement is compared
to the mean of other measurements made by us of the same star
for stars that show no RV variations in excess of 2~\kmsec.
This subset has $\sigma =$~0.64~\kmsec.
For a sample of 95 observations, the total shown in Panel~B, and a
normal distribution with a dispersion of 0.64~\kmsec, we would 
expect less than one observation to deviate by more than 2~\kmsec.
This suggests that 2~\kmsec\ is a reasonable
discriminant for identifying RV constant stars at the $\approx$~3$\sigma$
level.
Based on these tests, we estimate a total uncertainty in each RV
measurement of $\approx$~0.6--0.8~\kmsec.

Column~3 of Table~\ref{rvtab} indicates the 
binary status of each star.
A flag of ``0'' indicates that the star has only been observed during
a single epoch (e.g., one observation, or several observations
separated in time by less than one week).  
For stars lacking RV variations during this 
limited time, we assume that we
cannot distinguish between a single star or 
a multiple star system with a relatively long orbital period, so we
classify the binary nature of these stars as unknown.
A flag of ``1'' indicates that the star has been observed in multiple
epochs, either by us or in combination with previous investigators,
and does not exhibit any RV variations beyond the mutual
uncertainties of the measurements. 
A flag of ``2'' indicates that the star exhibits RV variations,
either in our measurements or in combination with previous investigators.
These stars are likely in binary or multiple star systems, but we lack
sufficient information to determine an orbital solution or systemic RV.
A flag of ``3'' indicates that the star has previously been identified 
as a member of a binary or multiple star system.
For these stars, we list the systemic velocity as determined by 
\citet{preston00,preston01}, 
\citet{latham02}, \citet{carney03}, \citet{sneden03b}, 
\citet{pourbaix04}, or \citet{roederer14}.

Many of our targets from the HK~Survey have not been observed previously 
at high spectral resolution, but \citeauthor{beers92} made RV estimates 
from moderate resolution ($\sim$~1\AA) spectroscopy.
Figure~\ref{rvfig} also displays the differences in RV measured from the 
moderate resolution and high resolution spectroscopy.
Panel~C shows the differences for stars that have been observed at 
high spectral resolution at multiple epochs (either by us or in 
combination with other investigators) and do not exhibit any RV 
variations 
as measured from the high resolution spectroscopy.
The standard deviation, 10~\kmsec, is identical to the uncertainty estimate
reported by \citeauthor{beers92}, and the mean offset is only 1.6~\kmsec.
This indicates that their estimates are generally reliable.
Panel~D shows the differences for stars that have only been observed
at high spectral resolution at a single epoch.
The larger $\sigma$ of this sample, 18~\kmsec, leads us to speculate
that at least some of these stars may be in binary or multiple star systems;
however, caution is warranted regarding this conclusion.
Several of the RV estimates reported in \citeauthor{beers92} deviate
significantly from the RV measurements derived from 
high resolution spectroscopy.
Stars whose RV estimate in \citeauthor{beers92} differs by more than 30~\kmsec\
from our high resolution RV measurements 
are listed in Table~\ref{deviantrv} and are not
shown in Figure~\ref{rvfig}.
Barring any misidentifications in the present survey, we attribute
these discrepancies to irregularities of unknown origin that occurred
during the course of the \citeauthor{beers92} survey.
In light of this, we refrain from assigning a binary status classification
to any star with only a single high resolution RV measurement, even if 
the star has a RV estimate from \citeauthor{beers92}.

\begin{deluxetable}{lccc}
\tablecaption{Stars with Deviant RV Measurements in BPS92
\label{deviantrv}}
\tablewidth{0pt}
\tabletypesize{\scriptsize}
\tablehead{
\colhead{Star} &
\colhead{BPS92 RV} &
\colhead{High res.\ RV\tablenotemark{a}} &
\colhead{Other Ref.\tablenotemark{a}} \\
\colhead{} &
\colhead{(\kmsec)} &
\colhead{(\kmsec)} &
\colhead{} }
\startdata
 CS~22892--052 & $-$75  & $+$13.0  & 1, 2 \\
 CS~22945--058 & $-$36  & $+$23.4  & \nodata \\
 CS~22949--037 & $-$79  & $-$125.4 & 1, 3 \\
 CS~22949--048 & $-$95  & $-$160.8 & 1 \\
 CS~22954--004 & $+$28  & $-$10.5  & \nodata \\
 CS~22956--081 & $+$210 & $+$244.6 & \nodata \\
 CS~22957--022 & $+$25  & $-$31.6  & 4 \\
 CS~22957--024 & $-$37  & $-$67.4  & \nodata \\
 CS~22957--026 & $+$14  & $-$18.9  & \nodata \\
 CS~22957--036 & $-$88  & $-$154.5 & \nodata \\
\enddata
\tablenotetext{a}{
The RV listed in Column~3 is derived from the present study.
In all cases this value agrees with previous investigations,
referenced in Column~4, within 2~\kmsec.
}
\tablerefs{
(1) \citealt{mcwilliam95a};
(2) \citealt{bonifacio09};
(3) \citealt{depagne02};
(4) \citealt{lai08} }
\end{deluxetable}

Finally, our observations reveal two new double-lined spectroscopic
binary stars, 
\object[BPS CS 22884-033]{CS~22884--033}
and 
\object[HE 2047-5612]{HE~2047$-$5612}.
We also reaffirm earlier work by \citet{masseron12} showing that
the spectrum of 
\object[BPS CS 22949-008]{CS~22949--008}
exhibits two sets of lines.
We present the observational data for these three stars in
Table~\ref{obstab}, but we 
omit them from the subsequent analysis.

\section{Equivalent Widths}
\label{ew}

We measure equivalent widths (EWs) from our spectra using a
semi-automatic routine that fits Voigt absorption line profiles
to continuum-normalized spectra.
This routine presents a plot of every fit to the user
for final approval or modification.
The local continuum is identified automatically by an iterative 
clipping procedure using a region of 3.5~\AA\ on either side of the 
line of interest, but the user can identify a different continuum level
for each line when necessary.
Typically 7--11 points surrounding the line center are used for the fit,
covering $\sim$~3--4 times the Gaussian
full width at half-maximum depth of the line for weak lines.
For stronger lines with clearly visible wings and no obvious blending
features, the number of fitting points is increased.
We choose to fit Voigt profiles to the absorption lines 
to simultaneously account for the Gaussian core 
and the dispersion wings present in some stronger lines;
for weak lines, the Voigt profile effectively resembles a Gaussian profile.

\begin{deluxetable*}{lccccc}
\tablecaption{Sample Table of EW Measurements
\label{ewtab}}
\tablewidth{0pt}
\tabletypesize{\scriptsize}
\tablehead{
\colhead{Star} &
\colhead{Wavelength (\AA)} &
\colhead{Species} &
\colhead{E.P.\ (eV)} &
\colhead{$\log gf$} &
\colhead{E.W.\ (m\AA)\tablenotemark{a}}}
\startdata
 CS 22166-016 &   6707.80 & Li I & 0.00 &  0.17 &   limit \\
 CS 22166-016 &   6300.30 & [O I]& 0.00 & -9.78 & \nodata \\
 CS 22166-016 &   7771.94 & O I  & 9.14 &  0.37 & \nodata \\
 CS 22166-016 &   7774.17 & O I  & 9.14 &  0.22 & \nodata \\
 CS 22166-016 &   7775.39 & O I  & 9.14 &  0.00 & \nodata \\
 CS 22166-016 &   4057.51 & Mg I & 4.35 & -0.89 &    24.5 \\
 CS 22166-016 &   4167.27 & Mg I & 4.35 & -0.71 &    39.4 \\
 CS 22166-016 &   4702.99 & Mg I & 4.33 & -0.38 & \nodata \\
 CS 22166-016 &   5172.68 & Mg I & 2.71 & -0.45 &   180.1 \\
 CS 22166-016 &   5183.60 & Mg I & 2.72 & -0.24 &   195.4 \\
 CS 22166-016 &   5528.40 & Mg I & 4.34 & -0.50 &    58.8 \\
 CS 22166-016 &   5711.09 & Mg I & 4.34 & -1.72 & \nodata \\
 CS 22166-016 &   3943.99 & Al I & 0.00 & -0.64 & \nodata \\
 CS 22166-016 &   3961.52 & Al I & 0.01 & -0.34 &   synth \\
\enddata
\tablecomments{
The complete version of Table~\ref{ewtab} is available online only.
A short version is shown here to
illustrate its form and content.
}
\tablenotetext{a}{
``Synth'' indicates an abundance was derived from spectrum synthesis;
``limit'' indicates that an upper limit on the abundance was derived from 
the line.
}
\end{deluxetable*}

The complete list of 47,744 EW measurements for all stars analyzed
is given in Table~\ref{ewtab}.
Other lines fit by spectrum synthesis (9,268~lines) 
or used to derive upper limits (12,279~lines)
are indicated using ``synth'' or ``limit,'' respectively.
The complete version of Table~\ref{ewtab} is available 
in the online edition of the journal, and only
a sample is shown in the printed edition
to demonstrate its form and content.

\subsection{Comparison of Equivalent Widths Measured from Spectra
Obtained with Different Spectrographs}
\label{diffinstr}

\begin{figure*}
\begin{center}
\includegraphics[angle=270,width=6.5in]{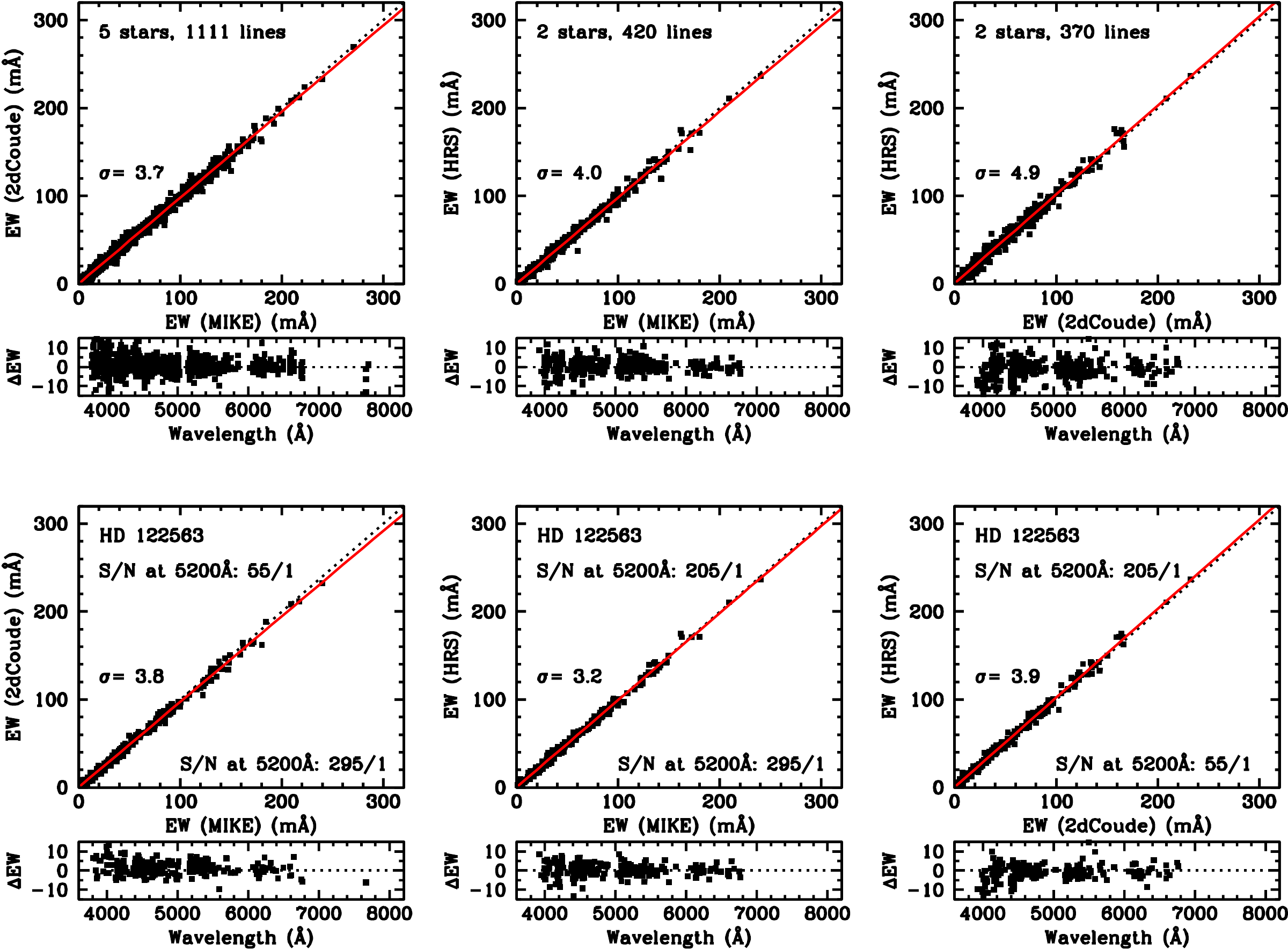}
\end{center}
\caption{
\label{diffinstrfig}
Comparison of EWs measured from spectra of the same
stars obtained with different spectrographs.
The 1:1 correlation is indicated as a dotted line, and the least-squares
fit is indicated by the solid red line.
The standard deviation ($\sigma$) 
is computed relative to the 1:1 correlation, not the 
least-squares fit.
The top row of panels examines the combined measurement for stars 
observed with both the MIKE and Tull (``2dCoude'') spectrographs 
(\hbox{HD~106373}, \hbox{HD~108317},
\hbox{HD~122563}, \hbox{HD~132475}, and \hbox{HE~1320$-$1339}),
MIKE and HRS (\hbox{G015-010} and \hbox{HD~122563}), and
the Tull and HRS (\hbox{G025-024} and \hbox{HD~122563}).
The bottom row of panels examines the measurements for one star
observed with all three instruments, \hbox{HD~122563}.
}
\end{figure*}

We have deliberately observed a small number of bright stars with more than 
one instrument to check the reliability of the reduction processes at 
handling cosmic rays, scattered light, flat-fielding, etc.
We have five stars in common to Magellan$+$MIKE and Smith$+$Tull,
two stars in common to Magellan$+$MIKE and HET$+$HRS, and
two stars in common to HET$+$HRS and Smith$+$Tull;
one star, 
\object[HD 122563]{HD~122563},
was observed with all three setups.
In Figure~\ref{diffinstrfig} we compare the EWs measured
from each of these spectra.
The standard deviation
of each set of comparisons is driven strongly by the
S/N of the individual spectra.
The largest deviations regularly arise from strong lines that suffer
from blending and lower S/N in the blue.
We find no evidence of differences in the spectra beyond 
the EW measurement uncertainties, typically 3--4~m\AA.
It is reassuring that
the spectra obtained with the different instruments
are effectively interchangeable in the wavelength regions where they
overlap with adequate S/N.

\subsection{Comparison of Equivalent Widths with Those
Measured by Other Investigators}
\label{otherew}

\begin{figure}
\begin{center}
\includegraphics[angle=0,width=3.35in]{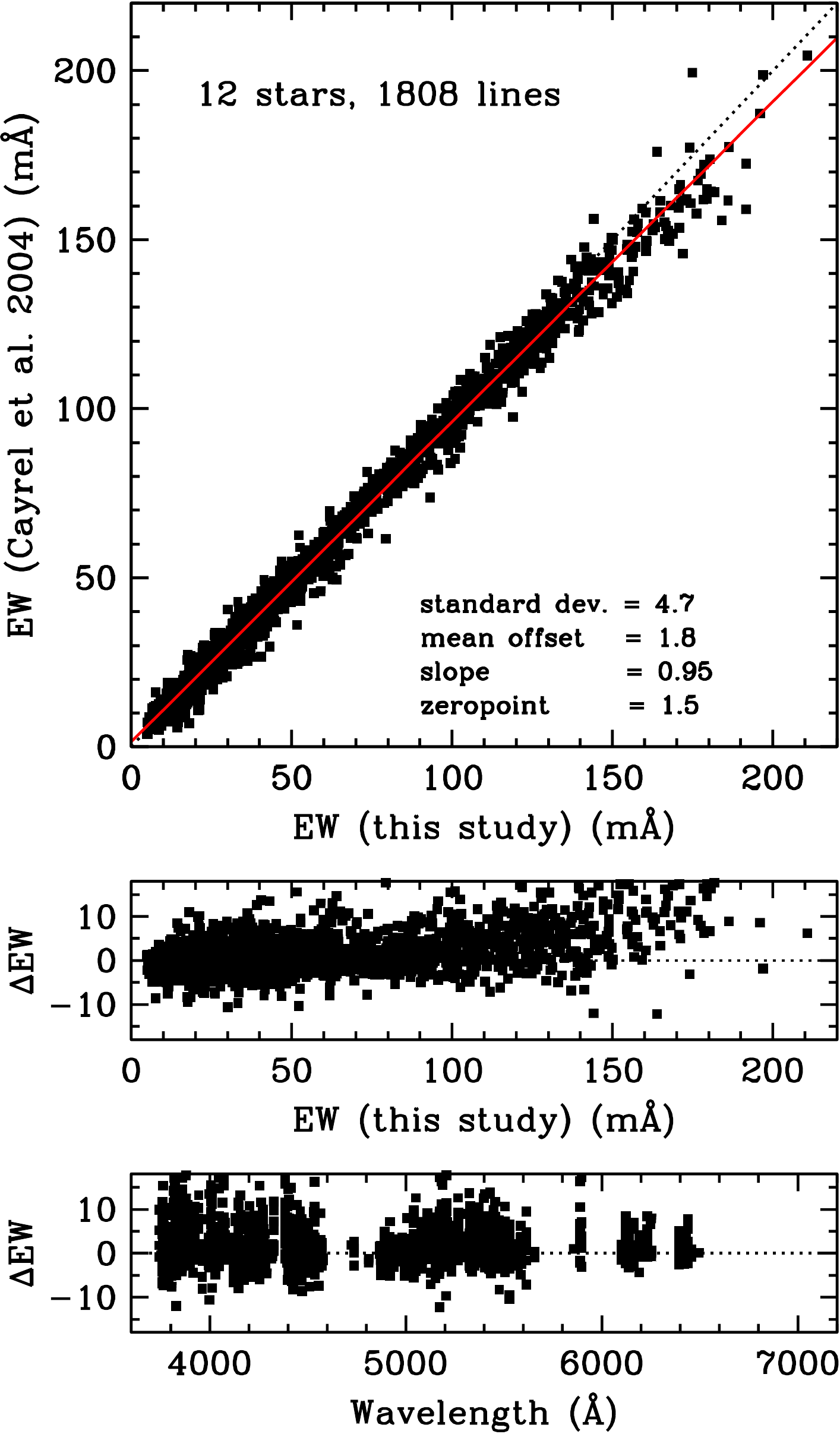}
\end{center}
\caption{
\label{ewcayrel}
Comparison of EW measurements with those of
\citet{cayrel04}.
The 1:1 correlation is indicated as a dotted line, and the least-squares
fit is indicated by the solid red line in the top panel.
The standard deviation is computed relative to the 1:1 correlation, not the 
least-squares fit.
The differences are in the sense of 
this study \textit{minus} \citeauthor{cayrel04}~
}
\end{figure}

\begin{deluxetable*}{llcccccc}
\tablecaption{Comparison of Equivalent Width Measurements
\label{ewcomptab}}
\tablewidth{0pt}
\tabletypesize{\scriptsize}
\tablehead{
\colhead{Study} &
\colhead{Line subset} &
\colhead{$N_{\rm stars}$} &
\colhead{$N_{\rm lines}$} &
\colhead{slope} &
\colhead{zeropoint} &
\colhead{mean offset} &
\colhead{$\sigma$ (m\AA)} }
\startdata
\citet{carretta02}           & all lines        &  2 &  155 & 0.88 & 5.2 & $+$2.3 &  7.0 \\
                             & EW $<$ 100 m\AA\ &  2 &  123 & 0.88 & 5.2 & $+$0.3 &  5.3 \\
\citet{cayrel04}             & all lines        & 12 & 1808 & 0.95 & 1.5 & $+$1.8 &  4.7 \\
                             & EW $<$ 100 m\AA\ & 12 & 1415 & 0.96 & 0.8 & $+$0.7 &  3.6 \\
\citet{honda04a}  &&&&&&& \\ 
\hspace*{4ex} bright targets & all lines        &  2 &  276 & 0.95 & 1.2 & $+$1.7 &  3.7 \\
                             & EW $<$ 100 m\AA\ &  2 &  231 & 0.95 & 1.2 & $+$1.0 &  3.0 \\
\hspace*{4ex} faint targets  & all lines        &  2 &  176 & 0.85 & 7.0 & $+$1.9 &  9.6 \\
                             & EW $<$ 100 m\AA\ &  2 &  137 & 0.86 & 6.6 & $-$0.9 &  7.2 \\
\citet{ivans03}              & all lines        &  2 &  262 & 0.96 & 1.5 & $+$0.3 &  4.2 \\
                             & EW $<$ 100 m\AA\ &  2 &  241 & 0.94 & 1.8 & $+$0.1 &  3.8 \\
\citet{johnson02}            & all lines        &  3 &  763 & 0.93 & 1.3 & $+$1.9 &  5.0 \\
                             & EW $<$ 100 m\AA\ &  3 &  693 & 0.95 & 0.6 & $+$1.3 &  3.4 \\
\citet{lai08}                & all lines        &  3 &  397 & 0.97 & 0.5 & $+$1.3 &  4.2 \\
                             & EW $<$ 100 m\AA\ &  3 &  355 & 0.97 & 0.4 & $+$1.0 &  3.6 \\
\citet{mcwilliam95a}         & all lines        &  8 & 1350 & 0.93 & 5.1 & $-$0.3 & 11.3 \\
                             & EW $<$ 100 m\AA\ &  8 & 1033 & 0.94 & 4.7 & $-$1.6 &  9.6 \\
\enddata
\end{deluxetable*}

Our sample has many stars in common with earlier
studies by other investigators.
Figure~\ref{ewcayrel} compares EWs for 12 of the
stars in common with the First Stars sample \citep{cayrel04}.  
For 1808 lines in common with 3700~$< \lambda <$~6500~\AA\ and
0~$<$~EW~$<$~220~m\AA, we find a standard deviation ($\sigma$)
of 4.7~m\AA\ and a very slight trend for the strongest lines
(EW~$\gtrsim$~150~m\AA), where our EW measurements 
are larger by $\approx$~4--5\% on average.
The weakest lines (EW~$<$~100~m\AA) show very good agreement and have a 
smaller $\sigma$, 3.6~m\AA.
This standard deviation is the same as found in comparisons of
EWs measured by us using spectra from different instruments.
No significant variations from these values are found when EWs
from these 12~stars are considered on a star-by-star basis.
Similar results are obtained when we compare our EWs 
with those of other investigators.
Table~\ref{ewcomptab} lists the standard deviation found when our
EWs are compared with those of
stars in common for six other studies.
Table~\ref{ewcomptab} also lists
the slope and zeropoint of the linear regression (where 1.0 and 0.0
represent the slope and zeropoint for perfect agreement) and the 
mean offset.
The slopes range from 0.88 to 0.97 with positive zeropoints;
our EWs are systematically larger for the strongest lines.
The mean offsets are negligible.
The large standard deviation 
with respect to \citet{mcwilliam95a} is a result of 
the relatively low S/N of the spectra in that study (typical
S/N of 30--45~pix$^{-1}$).
For most other studies, $\sigma$ for all lines is $\approx$~4.0--5.0~m\AA;
when only weak lines are considered (EW~$<$~100~m\AA), $\sigma$
drops to $\approx$~3.5--4.0~m\AA.

\begin{figure}
\begin{center}
\includegraphics[angle=0,width=3.35in]{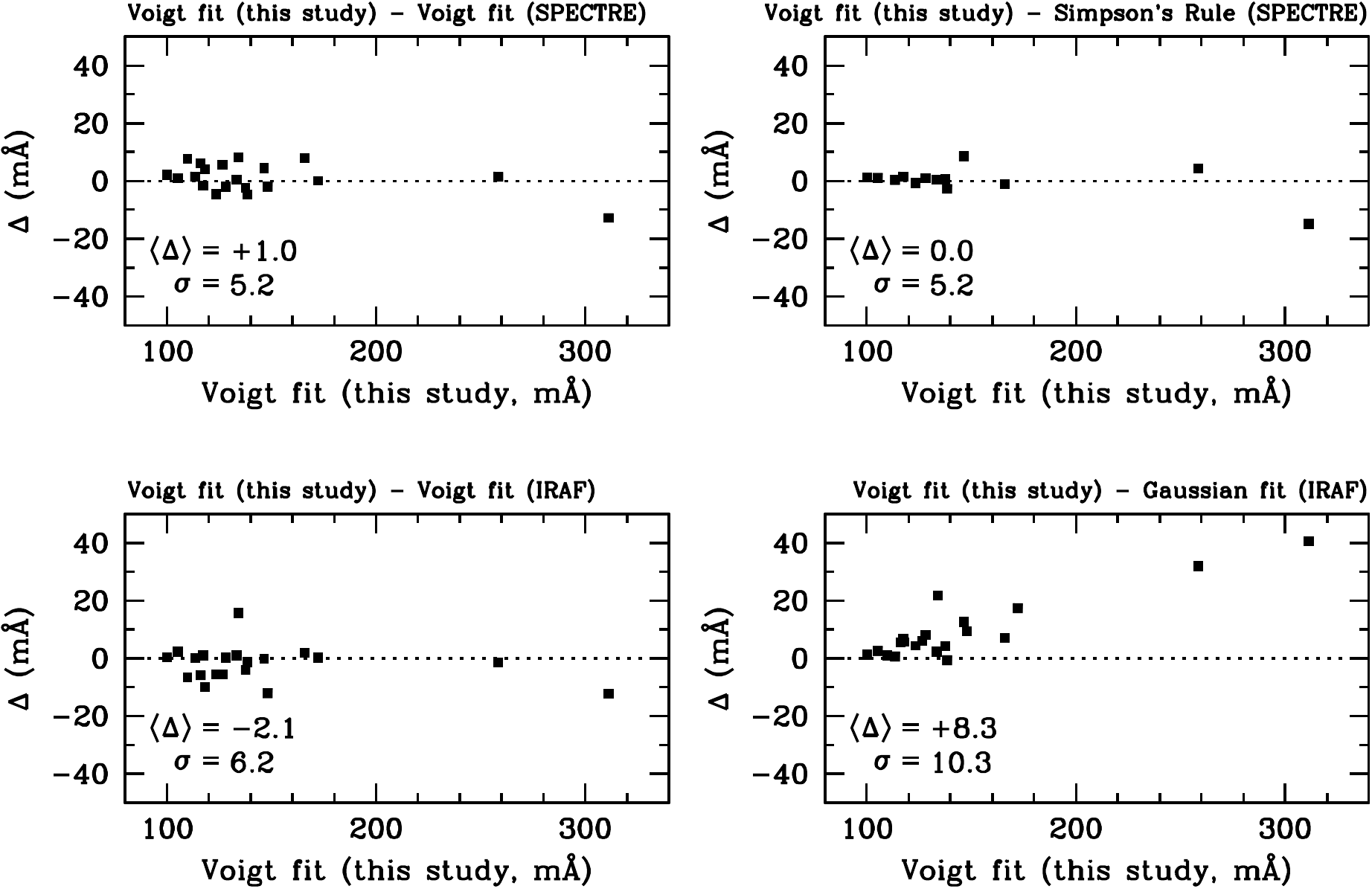}
\end{center}
\caption{
\label{strongew}
Comparison of EWs measured using various techniques and
software packages in the 
high S/N spectrum of \mbox{HD~128279}.
Only lines with EW~$>$~100~m\AA\ are considered here.
Note that only 13 of the 20 lines have been measured by Simpson's rule
due to weak blends in the wings of these lines.
The dotted line indicates a difference of zero.
The mean offset and standard deviation of the residuals are shown.
}
\end{figure}

To further investigate the nature of the discrepancy for the 
strong lines, we examine in detail 20 of the strongest lines commonly
used for abundance analyses in the bright, metal-poor giant 
\object[HD 128279]{HD~128279}
from a high S/N MIKE spectrum.
Figure~\ref{strongew} compares our EW measurements for these 
lines with measurements made by fitting Voigt and Gaussian profiles
in IRAF, by fitting a Voigt profile in SPECTRE \citep{fitzpatrick87},
and by direct integration in SPECTRE via Simpson's rule.
We find good agreement between the EW measurements
from the Voigt profiles and by direct integration, even for the 
strongest lines with EW~$>$~250~m\AA.
Gaussian profiles clearly underestimate 
absorption in the wings of the strongest lines. 
Other investigations do account for extra absorption in the line wings,
even if the formal Voigt profile is not explicitly invoked.
We conclude that the small systematic difference for the strongest lines
result from the different methods used 
to account for this extra absorption.

\section{Photometry}
\label{photometry}

Optical and infrared broadband photometry is 
available for most stars in our sample, but
these data are only used to calculate
initial guesses for the stellar temperatures.
Reddening values are taken from the \citet{schlegel98} maps,
altered in the case of
high reddening according to the prescription of \citet{bonifacio00}.
We adopt the extinction coefficients of \citet{cardelli89}.
We employ the \citet{ramirez05} color transformations to place the
Two Micron All Sky Survey (2MASS; \citealt{skrutskie06}) infrared photometry
on the Telescopio Carlos S\'{a}nchez (TCS) 
scale of \citet{alonso99b}.
The optical photometry originates from many different sources, including
\citet{harris64},
\citet{sandage69},
\citet{nicolet78},
\citet{carney79a,carney83},
\citet{carney79b},
\citet{bond80},
\citet{norris85,norris99},
\citet{carney86},
\citet{sandage86}, 
\citet{lazauskaite90},
\citet{preston91,preston94,preston06},
\citeauthor{beers92}, 
\citet{carney94},
\citet{beers95},
\citet{beers00,beers07},
\citet{schuster04},
\citet{rossi05}, and
\citet{christlieb08}.
The $V$~magnitudes, adopted reddenings, and dereddened $B-V$ and $V-K$ colors
are listed in Table~\ref{atmtab}.

\section{Atomic Data}

\begin{deluxetable}{ccccc}
\tablecaption{Atomic Data
\label{atomictab}}
\tablewidth{0pt}
\tabletypesize{\scriptsize}
\tablehead{
\colhead{$\lambda$ (\AA)} &
\colhead{Species} &
\colhead{E.P. (eV)} &
\colhead{$\log gf$} &
\colhead{Ref.} }
\startdata
   6707.80 & Li~\textsc{i}  & 0.00 & $+$0.17 &  1 \\
   \multicolumn{4}{c}{CH $A^2\Delta - X^2\Pi$ G band} &  2 \\
   \multicolumn{4}{c}{NH $A^3\Pi    - X^3\Sigma$ band}    &  3 \\
   \multicolumn{4}{c}{CN $B^2\Sigma - X^2\Sigma$ band}    &  3 \\
   6300.30 & [O~\textsc{i}] & 0.00 & $-$9.78 &  4 \\ 
   7771.94 & O~\textsc{i}   & 9.15 & $+$0.37 &  4 \\
   7774.17 & O~\textsc{i}   & 9.15 & $+$0.22 &  4 \\
   7775.39 & O~\textsc{i}   & 9.15 & $+$0.00 &  4 \\
\enddata
\tablecomments{
The complete version of Table~\ref{atomictab} is available online only.
A short version is shown here to
illustrate its form and content.
}
\tablerefs{
 (1) \citealt{smith98}; 
 (2) B.\ Plez (2007, private communication);
 (3) \citealt{kurucz95};
 (4) \citealt{fuhr09};
 (5) \citealt{chang90};
 (6) \citealt{aldenius07};
 (7) \citealt{aldenius09};
 (8) \citealt{lawler89}, using hfs from \citealt{kurucz95};
 (9) \citealt{lawler13};
(10) \citealt{wood13};
(11) \citealt{pickering01}, with corrections given in \citealt{pickering02};
(12) \citealt{doerr85a}, using hfs from \citealt{kurucz95};
(13) \citealt{biemont89};
(14) \citealt{sobeck07};
(15) \citealt{nilsson06};
(16) \citealt{denhartog11} for both $\log(gf)$ value and hfs;
(17) \citealt{obrian91};
(18) \citealt{melendez09};
(19) \citealt{cardon82}, using hfs from \citealt{kurucz95};
(20) \citealt{nitz99}, using hfs from \citealt{kurucz95};
(21) \citealt{fuhr09}, using hfs from \citealt{kurucz95};
(22) \citealt{roederer12a};
(23) \citealt{biemont11};
(24) \citealt{ljung06};
(25) \citealt{nilsson10};
(26) \citealt{whaling88}; 
(27) \citealt{palmeri05};
(28) \citealt{wickliffe94};
(29) \citealt{fuhr09}, using hfs/IS from \citealt{mcwilliam98} when available;
(30) \citealt{lawler01a}, using hfs from \citealt{ivans06};
(31) \citealt{lawler09}; 
(32) \citealt{li07}, using hfs from \citealt{sneden09};
(33) \citealt{denhartog03}, using hfs/IS from \citealt{roederer08} when available;
(34) \citealt{lawler06}, using hfs/IS from \citealt{roederer08} when available;
(35) \citealt{lawler01b}, using hfs/IS from \citealt{ivans06};
(36) \citealt{denhartog06};
(37) \citealt{lawler01c}, using hfs from \citealt{lawler01d};
(38) \citealt{wickliffe00}; 
(39) \citealt{lawler04} for both $\log(gf)$ value and hfs;
(40) \citealt{lawler08};
(41) \citealt{wickliffe97};
(42) \citealt{sneden09} for both $\log(gf)$ value and hfs/IS;
(43) \citealt{lawler07};
(44) \citealt{ivarsson03}, using hfs/IS from \citealt{cowan05};
(45) \citealt{biemont00}, using hfs/IS from \citealt{roederer12b}; 
(46) \citealt{nilsson02}. 
}
\end{deluxetable}

We start with a list of 474~lines to be considered
for each star, 
but all lines are not observed, detected, and
unblended in each star.
This list is found in Table~\ref{atomictab}.
The full version of Table~\ref{atomictab} is available 
only in the online edition of the journal.
This table includes the line wavelength, species identification,
excitation potential (E.P.) of the lower level of the transition, 
\loggf\ value, and references to the
source for the \loggf\ value.
We perform spectral synthesis for lines broadened by
hyperfine splitting (hfs) or in cases where
a significant isotope shift (IS) may be present
(see Section~\ref{abundances}).
The references for these data are also included 
in Table~\ref{atomictab}.
When available, we use damping constants from \citet{barklem00}
and \citet{barklem05b}.
In all other cases we resort to the standard \citet{unsold55} 
approximation.

\section{Model Atmospheres}
\label{modelatm}

Numerous challenges in modeling stellar atmospheres
persist, and they limit the accuracy of the derived abundances.
These challenges include
departures from 
local thermodynamic equilibrium (LTE) in the
stellar atmosphere,
time-dependent three-dimensional (3D) versus 
static one-dimensional (1D) representations
of the atmosphere, inadequate treatment of convection,
inclusion and proper calculation 
of all relevant sources of continuous opacity, and
inaccurate or incomplete atomic and molecular data.
Significant progress has been made 
to improve the situation in recent years, but it is still
impractical to analyze large datasets except by means of the
general procedures in common use for decades: namely,
EW analysis or spectral synthesis of lines assuming 1D, plane-parallel,
static model atmospheres in LTE throughout the line-forming layers.
We make use of a recent set of MARCS models
\citep{gustafsson08}, and we demonstrate in 
Section~\ref{modeluncertainties} that very similar results
are obtained from the ATLAS9 grid \citep{castelli04}.

\subsection{Model Atmosphere Parameters}
\label{modelmethod}

The inhomogeneous nature of the optical photometry (Section~\ref{photometry})
leads us to adopt model atmosphere parameters 
that are primarily derived from our spectra.
Effective temperatures (\teff) are derived by requiring that
abundances derived from Fe~\textsc{i} lines show no trend with
the E.P.\ of the lower level of the transition.
In many stars the Fe~\textsc{i} lines span a range of 0.0 to 4.5~eV,
a far broader range than is available for any other species.

Microturbulence velocities (\vt) are derived by requiring that 
abundances derived from Fe~\textsc{i} lines show no trend with
line strength, expressed as the unitless quantity \logrw.
Lines with \logrw~$\gtrsim -$5.0
(e.g., EW~$\gtrsim$~40~m\AA\ at $\lambda =$~4000~\AA) are sensitive to \vt,
and the sensitivity to \vt\ diminishes on the damping portion of the curve
of growth, with \logrw~$\gtrsim -$4.4
(e.g., EW~$>$~160~m\AA\ near 4000~\AA).
In many stars the strongest optical Fe~\textsc{i} lines used have 
\logrw~$> -$4.5 (EW~$>$~130~m\AA\ near 4000~\AA),
which is on the saturated portion of the
curve of growth in both dwarfs and giants.
The weakest Fe~\textsc{i} lines used often have 
\logrw~$< -$6.0 (EW~$<$~4~m\AA\ near 4000~\AA), but the weakest lines
detected are a strong function of S/N.

Overionization is the name given to the effect that occurs
when the local mean radiation intensity exceeds that
predicted by the Planck source function, and
overionization leads to a potentially significant
underestimate of the number density of minority species.
This non-LTE effect 
can lead to systematic underestimates
of the Fe~\textsc{i} number density when using the LTE approximation for
the source function in metal-poor stars
(e.g., \citealt{thevenin99}, \citealt{asplund05}).
To reduce the impact of this effect on
the ionization balance between Fe~\textsc{i} and Fe~\textsc{ii}, 
in most stars we derive
surface gravities (\logg, in cgs units) from theoretical
isochrones in the Y$^{2}$ grid \citep{demarque04}.
For stars between the main sequence turnoff (MSTO) and the tip of the 
red giant branch,
we interpolate the Y$^{2}$ isochrones
in \teff\ for the appropriate metallicity of each star.
We assume an input age of 12~Gyr for all stars. 
Stars with ages 12.0~$\pm$~1.5~Gyr 
were formed at redshifts $z >$~2, so this range includes 
the ages of most metal-poor Milky Way globular clusters
(e.g., \citealt{dotter10}),
the ages of individual halo stars as computed from radioactive
decay of heavy elements (e.g., \citealt{roederer09}), 
and the redshifts of the
rising inferred star formation rate density in galaxies
that may grow to the size of the Milky Way
(e.g., \citealt{bouwens07}).
For stars on the horizontal branch (HB), below the
isochrone inflection point (i.e., the main sequence; MS), 
or hotter than the MSTO, we derive 
surface gravities from the usual method of requiring 
the ionization balance between Fe~\textsc{i} and Fe~\textsc{ii}.
Uncertainties inherent to each of these methods are discussed below.

Fe~\textsc{ii} is the dominant ionization state,
so overionization has little impact on the iron abundance
derived from Fe~\textsc{ii} lines
in stars of types F-G-K.
We assume that the overall metallicity, [M/H], is
represented by the iron abundance 
derived from Fe~\textsc{ii} lines.
The trade off, of course, is that the iron abundance
derived from Fe~\textsc{ii} lines is more sensitive to the 
surface gravity of the model atmosphere than the 
iron abundance derived from Fe~\textsc{i} lines.

We use the following procedure to converge to a final set of
model parameters.
Broadband photometry is used only to produce an initial estimate
of \teff\ from dereddened $V-K$ or $J-K$ colors
and the \citet{alonso99b} color--\teff\ relations.
We have had to extrapolate the \citeauthor{alonso99b}\ calibrations
to stars with [Fe/H]~$< -$3 in our sample, 
but the metallicity sensitivity of such relations 
decreases with decreasing metallicity 
and these extrapolations are used for initial guesses only.
Other relations, e.g., those of \citet{gonzalezhernandez09}, are
calibrated directly to the 2MASS photometric system
and are valid at [Fe/H]~$< -$3.
These relations are used less frequently in prior abundance surveys,
so we adopt the \citeauthor{alonso99b}\ relations 
for the sake of comparison.
We estimate the initial \logg\ by interpolating (in $V-K$ or $J-K$) 
the appropriate Y$^{2}$ isochrones.
We estimate the initial \vt\ from \logg\ using
the relation of \citet{gratton96}.
We compile metallicity estimates from many of the literature
sources listed above, but they have relatively little impact on the 
initial estimates of the other parameters.

We use a recent version (c.\ 2010) of the
spectral line analysis code MOOG\footnote{
\texttt{http://www.as.utexas.edu/{\raise.17ex\hbox{$\scriptstyle\sim$}}chris/moog.html}}
 \citep{sneden73} to
derive abundances of iron from
Fe~\textsc{i} and Fe~\textsc{ii} lines using our EW measurements
and a MARCS model atmosphere interpolated for our initial 
model atmosphere parameters.
We refine our initial \teff, \vt, and [Fe/H] estimates 
to enforce no trend of derived iron abundance 
(from Fe~\textsc{i}) with each line's E.P.\
or \logrw\ value, and we require that the input
model metallicity agrees with the iron abundance
derived from Fe~\textsc{ii} lines.
We maintain \logg\ fixed through these iterations.
Then, for stars along the subgiant branch and red giant branch,
the revised \teff\ and [Fe/H] estimates are
used to refine \logg\ from the isochrones.
For stars on the HB, MS, or warmer than the MSTO, iron ionization
equilibrium is enforced to derive \logg.
Lines whose inferred abundance deviates by more than 2$\sigma$
from the mean are discarded.
We repeat this process until all four parameters converge,
and this usually occurs within two to four iterations.
We consider convergence to mean that the model atmosphere parameters
are stable from one iteration to the next to the nearest
10~K in \teff, 0.05~dex in \logg, 0.05~\kmsec\ in \vt,
and 0.01~dex in metallicity.
We perform this iteration scheme using a version of the
batch mode capabilities of MOOG, and all steps are supervised
by the user.

\begin{figure}
\begin{center}
\includegraphics[angle=0,width=3.35in]{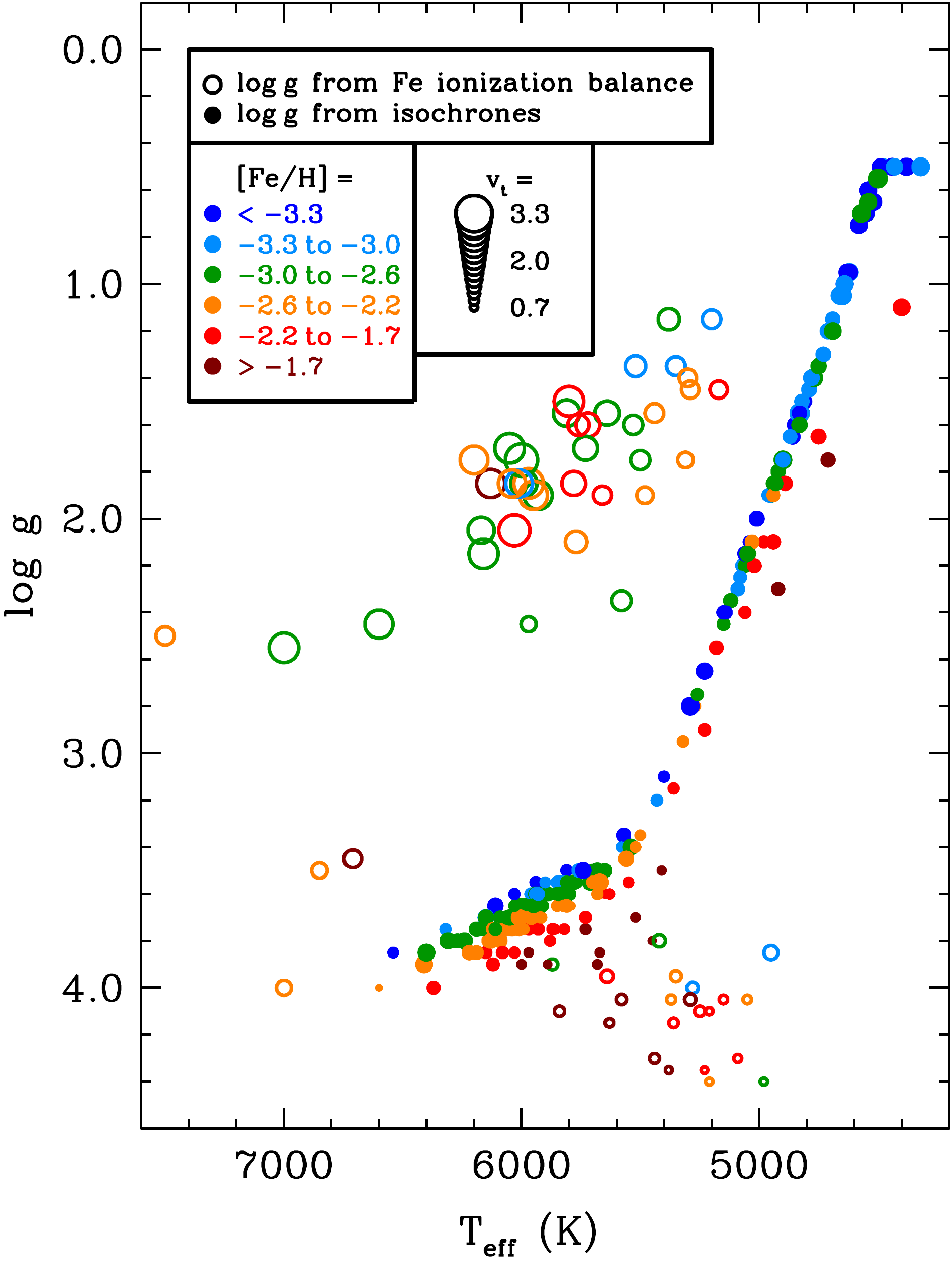}
\end{center}
\caption{
\label{teffloggplot}
\teff\ versus \logg\ diagram for all 313~stars in our sample.
Color-coding indicates the metallicity (from Fe~\textsc{ii}),
and the point size indicates \vt. 
Open circles represent stars whose \logg\ was derived from Fe
ionization balance, and closed circles represent stars whose
\logg\ was calculated from isochrones.
The ``bend'' observed for few stars at the top of the 
red giant branch is a result of 
encountering the edge of the grid of model atmospheres at 
\logg~$=$~0.5 at low metallicity.
}
\end{figure}

Table~\ref{atmtab} lists the final \teff, \logg,
\vt, and [Fe/H] for all 313 stars in the survey. 
Figure~\ref{teffloggplot} illustrates these values.
Different colors represent different metallicity ranges.
The size of each point corresponds to the size of the 
microturbulence velocity parameter.
Open circles represent stars whose gravity is derived from Fe
ionization balance, while filled circles represent stars
whose gravity is derived from the Y$^{2}$ grid of isochrones.
This approach leads to the narrow width of
the subgiant and red giant branches and the non-physical
gap between the red giant branch and the HB.
The latter effect is reminiscent of the difference
between absolute magnitudes of field red giants
and the fiducial of globular cluster M92
found in Figure~4 of \citet{luck85}.
The bulk of the stars in our sample are subgiants (151~stars, 
denoted ``SG'' in Table~\ref{atmtab}) 
or red giants (98~stars, denoted ``RG'').
Smaller numbers of stars are on the HB (39~stars, denoted ``HB''), 
MS (22~stars, denoted ``MS''), 
or are warmer than the MSTO (3~stars, denoted ``BS'' in analogy
with the blue straggler stars found in globular clusters).

\subsection{Statistical Uncertainties}
\label{modeluncertainties}

We use the statistical (internal) uncertainty in the derived
Fe~\textsc{i} and Fe~\textsc{ii} abundances 
($\sigma_{\rm Fe\,I}$ and $\sigma_{\rm Fe\,II}$, respectively)
to assess the statistical uncertainties in the model parameters.
We relate the sensitivity of \teff\ and \vt\
to their correlation with E.P.\ and \logrw\ through
the statistical uncertainty in $\sigma_{\rm Fe\,I}$.
Our tests for several stars spanning the range of the SG and RG classes
suggest that simple linear relations with \teff\ and \vt\ 
are appropriate and lead to the relations
$\sigma_{\teff}^{2} = (-0.17\teff + 540)^{2}\times\sigma_{\rm Fe\,I}^{2}$
and
$\sigma_{\vt}^{2} = (0.32\vt - 0.90)^{2}\times\sigma_{\rm Fe\,I}^{2}$.
For other stars in our sample we adopt relationships
that are independent of \teff\ and \vt.
For stars on the MS, we find
$\sigma_{\teff}^{2} = (400 \sigma_{\rm Fe\,I})^{2}$
and
$\sigma_{\vt}^{2} = (1.5 \sigma_{\rm Fe\,I})^{2}$.
For stars on the HB, we find 
$\sigma_{\teff}^{2} = (520 \sigma_{\rm Fe\,I})^{2}$
and
$\sigma_{\vt}^{2} = (0.36 \sigma_{\rm Fe\,I})^{2}$.
For stars warmer than the MSTO (BS), we find
$\sigma_{\teff}^{2} = (400 \sigma_{\rm Fe\,I})^{2}$
and
$\sigma_{\vt}^{2} = (0.48 \sigma_{\rm Fe\,I})^{2}$.

\begin{figure}
\begin{center}
\includegraphics[angle=0,width=3.35in]{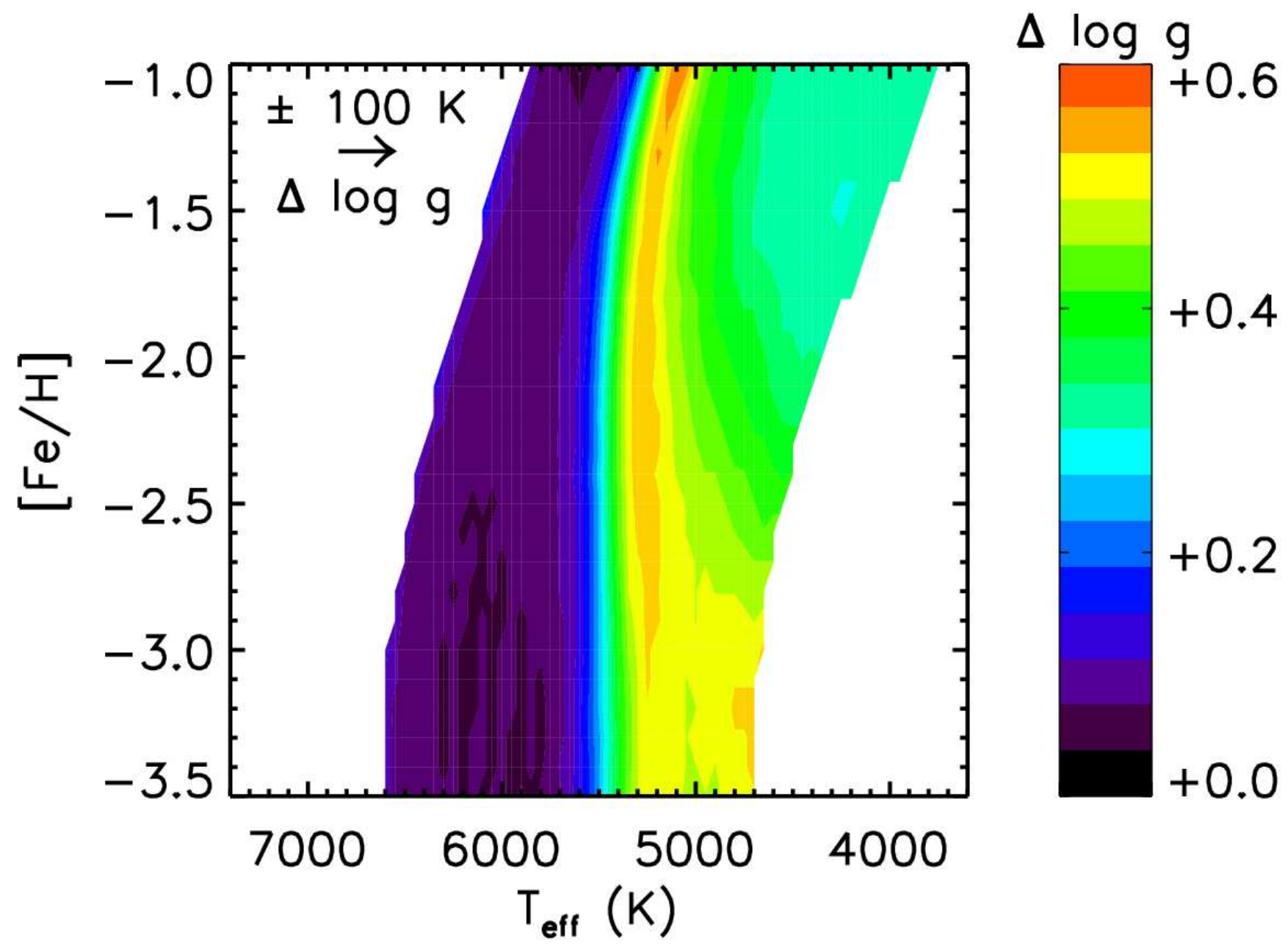}
\end{center}
\caption{
\label{errorplot1}
Variations in \logg\ from the Y$^{2}$ grid of
isochrones corresponding to uncertainties in \teff.
At given values of \teff\ and [Fe/H], the corresponding
uncertainty in \logg\ can be read off from the plot.
The color bar on the right indicates the magnitude of the
uncertainty in \logg\ for a change in \teff\ of 100~K.
}
\end{figure}

\begin{figure}
\begin{center}
\includegraphics[angle=0,width=3.35in]{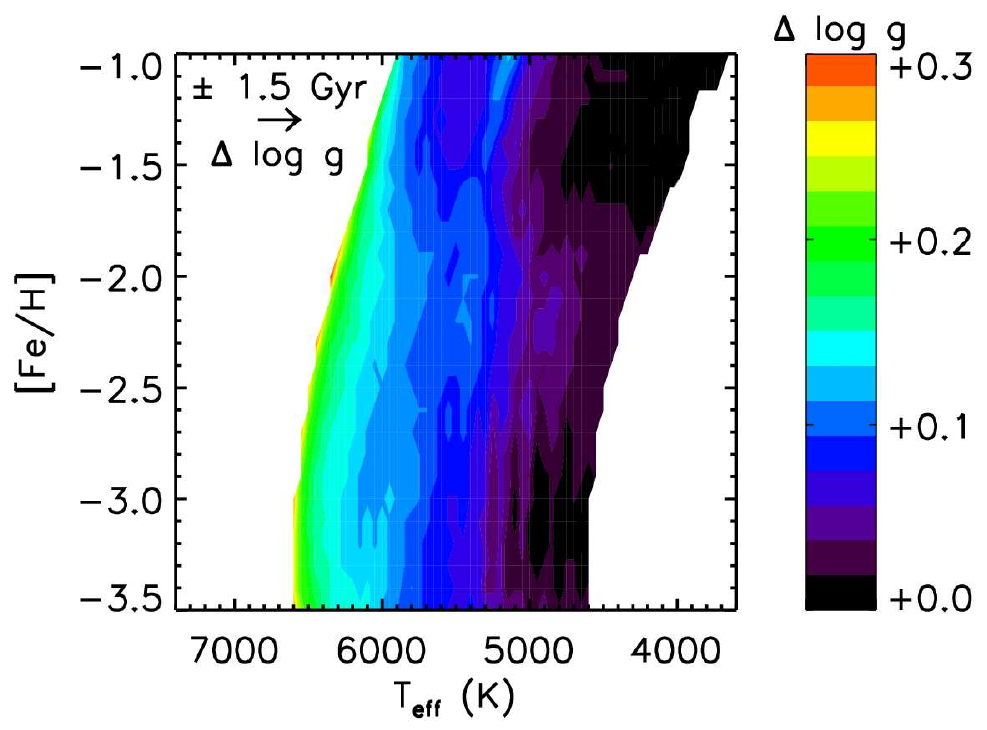}
\end{center}
\caption{
\label{errorplot2}
Variations in \logg\ from the Y$^{2}$ grid of
isochrones corresponding to uncertainties in age.
At given values of \teff\ and [Fe/H], the corresponding
uncertainty in \logg\ can be read off from the plot.
The color bar on the right indicates the magnitude of the
uncertainty in \logg\ if the isochrone age is varied by 1.5~Gyr
(relative to the 12~Gyr isochrones).
}
\end{figure}

\begin{figure}
\begin{center}
\includegraphics[angle=0,width=3.35in]{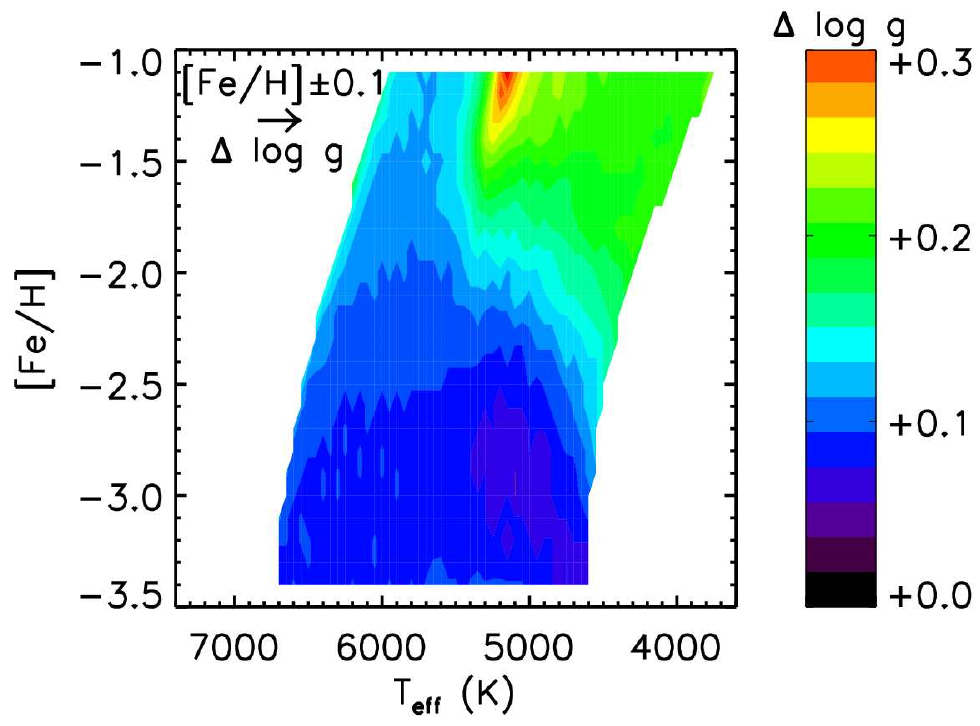}
\end{center}
\caption{
\label{errorplot3}
Variations in \logg\ from the Y$^{2}$ grid of
isochrones corresponding to uncertainties in metallicity.
At given values of \teff\ and [Fe/H], the corresponding
uncertainty in \logg\ can be read off from the plot.
The color bar on the right indicates the magnitude of the
uncertainty in \logg\ for a change in [Fe/H] of 0.1~dex.
}
\end{figure}

For stars in the SG and RG classes, 
we assess statistical uncertainties in \logg\ by 
varying \teff\ and [Fe/H] as input parameters to the
grid of Y$^{2}$ isochrones.
In Figure~\ref{errorplot1} we illustrate the change in 
\logg\ when \teff\ is varied by $\pm$~100~K.
Stars along the giant branch are most sensitive because of its
steep slope, with 
$\Delta \logg$ approaching 0.50 per 100~K, whereas
stars in the SG class and near the MSTO show
$\Delta \logg$ ranging from only 0.05 to 0.20 per 100~K.
Figures~\ref{errorplot2} and \ref{errorplot3} illustrate the change in
\logg\ when the isochrone age is varied by $\pm$~1.5~Gyr,
our assumed age uncertainty,
or the input metallicity is varied by $\pm$~0.10~dex.
(Recall that we assume an age of 12~Gyr in the isochrones.)
The gravity is most sensitive to age near the MSTO, whereas
stars on the red giant branch have almost no age sensitivity in old stellar 
populations.
The gravity is most sensitive to the isochrone metallicity in more
metal-rich stars where line blanketing has a significant impact
on the temperature and color.
For most of the stars in our sample, the uncertainty in \logg\ 
resulting from the uncertainty in metallicity is very small
relative to other sources of uncertainty.
For stars on the MS, HB, and warmer than the MSTO, we assess the 
statistical uncertainty in \logg\ by 
varying the gravity such that the iron abundances derived from
Fe~\textsc{i} and Fe~\textsc{ii} remain in agreement 
within $\pm (\sigma_{\rm Fe~I}^{2}+\sigma_{\rm Fe~II}^{2})^{1/2}$.
We consider statistical uncertainties in [M/H] equal to the standard deviation
of Fe~\textsc{i},
since that species is used in deriving \vt, its uncertainties, and the 
cross terms discussed in Section~\ref{abundsigma}.
The statistical uncertainties in \logg\ listed in Table~\ref{atmtab}
include the uncertainty from each of these factors.
All statistical uncertainties include a ``softening'' factor to guard
against unreasonably small uncertainties
(10~K in \teff; 
0.05~dex in \logg;
0.05~\kmsec\ in \vt; and
0.02~dex in [M/H]).

To empirically test the sensitivity 
of the derived model parameters to the
choice of model grid, we select two subsets of 
stars from the full sample and rederive model parameters using
the ATLAS9 grid of $\alpha$-enhanced models with
convective overshooting turned off.
We begin the derivation from the original (unculled) set of 
EW measurements for each of these stars and follow the
same set of procedures.
One subset of stars is comprised of 
32~subgiants with 5800~$\,\leq \teff \leq$~6000~K
(as derived from the MARCS grid) and
$-$3.5~$\leq$~[Fe/H]~$\leq -$2.5.
The other subset is comprised of 22~giants with 
4790~$\,\leq \teff \leq$~5090~K
(as derived from the MARCS grid) and
$-$3.5~$\leq$~[Fe/H]~$\leq -$2.5.
The model parameters derived for the subset of subgiants 
are nearly identical, with 
$\Delta \teff = -$8~K ($\sigma =$~29~K),
$\Delta \logg = -$0.005 ($\sigma =$~0.027),
$\Delta \vt = +$0.002~\kmsec\ ($\sigma =$~0.039~\kmsec), and
$\Delta$~[Fe/H]~$= -$0.006~dex ($\sigma =$~0.023~dex).
Differences are in the sense of MARCS minus ATLAS9.
The model parameters derived for the subset of giants are
similar but not identical, with
$\Delta \teff = +$39~K ($\sigma =$~26~K),
$\Delta \logg = +$0.11~ ($\sigma =$~0.08),
$\Delta \vt = -$0.07~\kmsec\ ($\sigma =$~0.03~\kmsec), and
$\Delta$~[Fe/H]~$= +$0.04~dex ($\sigma =$~0.03~dex).
Thus, the ATLAS9 grid tends to move stars slightly up the 
red giant branch to 
cooler temperatures, lower gravities, higher microturbulence velocities,
and lower metallicities.
Fortunately, the magnitude of the effect is rather small.
The standard deviations derived in each of these tests also
provide an estimate of the dispersion in model parameters that
could be expected in the convergence routine.
These uncertainties are 
comparable to the size of the statistical uncertainties.
We include a model convergence uncertainty of 30~K in \teff,
0.08~dex in \logg, and 0.04~\kmsec\ in \vt\ in the 
statistical uncertainties listed in Table~\ref{atmtab}.

\subsection{Comparison with Previous Surveys}
\label{compareprevious}

\begin{deluxetable*}{lccccc}
\tablecaption{Comparison of Atmospheric Parameters with Previous Surveys
\label{comparetab}}
\tablewidth{0pt}
\tabletypesize{\scriptsize}
\tablehead{
\colhead{Comparison} &
\colhead{$N$} &
\colhead{$\Delta\teff\ (\sigma)$} &
\colhead{$\Delta\logg\ (\sigma)$} &
\colhead{$\Delta\vt\   (\sigma)$} &
\colhead{$\Delta$Fe~\textsc{ii} ($\sigma$)} \\
\colhead{} &
\colhead{} &
\colhead{(K)} &
\colhead{} &
\colhead{(\kmsec)} &
\colhead{} }
\startdata      
\citet{luck85}       \tablenotemark{b}   &  11 & $-$169 (140) & $-$0.17 (0.44) & $-$0.57 (0.46) & $-$0.49 (0.18) \\
\citet{mcwilliam95b} \tablenotemark{a}   &  26 & $-$174 (117) & $-$0.44 (0.50) & $-$0.56 (0.26) & $-$0.25 (0.15) \\
\citet{ryan96}       \tablenotemark{a}   &  10 & $-$233 (203) & $-$0.82 (0.41) & $-$0.22 (0.41) & $-$0.38 (0.20) \\
\citet{fulbright00}  \tablenotemark{b}   &  17 & $-$43  (140) & $-$0.33 (0.33) & $-$0.08 (0.28) & $-$0.14 (0.12) \\
\citet{johnson02}    \tablenotemark{b}   &   6 & $+$12   (45) & $-$0.12 (0.26) & $-$0.27 (0.28) & $-$0.16 (0.06) \\
\citet{stephens02}   \tablenotemark{b}   &   6 & $-$44  (218) & $-$0.17 (0.16) & $+$0.18 (0.37) & $-$0.01 (0.19) \\
\citet{gratton03}    \tablenotemark{a}   &   6 & $-$210  (98) & $-$0.41 (0.24) & $+$0.00 (0.22) & $-$0.10 (0.15) \\
\citet{cayrel04}     \tablenotemark{a}   &  18 & $-$236 (104) & $-$0.68 (0.41) & $-$0.22 (0.14) & $-$0.20 (0.13) \\
\citet{honda04b}     \tablenotemark{a}   &  10 & $-$192 (130) & $-$0.76 (0.29) & $-$0.20 (0.31) & $-$0.31 (0.23) \\
\citet{simmerer04}   \tablenotemark{a}   &  19 & $-$82  (155) & $-$0.47 (0.35) & $-$0.29 (0.43) & $-$0.22 (0.17) \\
\citet{preston06}    \tablenotemark{b}   &  19 & $-$58  (124) & $-$0.59 (0.28) & $-$0.02 (0.27) & $-$0.15 (0.12) \\
\citet{lai08}        \tablenotemark{a}   &  12 & $-$233 (204) & $-$0.50 (0.39) & $-$0.10 (0.08) & $-$0.08 (0.18) \\
\citet{bonifacio09}  \tablenotemark{a}   &   9 & $-$184 (198) & $-$0.56 (0.43) & $+$0.07 (0.30) & $-$0.09 (0.13) \\
\citet{ishigaki10}   \tablenotemark{b}   &  10 & $+$86  (164) & $-$0.13 (0.31) & $+$0.00 (0.35) & $+$0.00 (0.20) \\
\citet{roederer10}   \tablenotemark{b}   &  11 & $+$75  (163) & $+$0.24 (0.35) & $-$0.04 (0.14) & $+$0.10 (0.16) \\
\hline
\multicolumn{6}{c}{All Stars} \\
\hline
All                                      & 190 & $-$118 (175) & $-$0.43 (0.44) & $-$0.20 (0.36) & $-$0.18 (0.20) \\
Group (a)                                & 110 & $-$185 (154) & $-$0.57 (0.42) & $-$0.26 (0.35) & $-$0.21 (0.18) \\
Group (b)                                &  80 & $-$28  (161) & $-$0.24 (0.41) & $-$0.11 (0.36) & $-$0.13 (0.22) \\
\hline
\multicolumn{6}{c}{Stars in class ``RG''} \\
\hline
All                                      & 108 & $-$154 (160) & $-$0.48 (0.49) & $-$0.30 (0.35) & $-$0.25 (0.20) \\
Group (a)                                &  73 & $-$215 (125) & $-$0.65 (0.43) & $-$0.33 (0.31) & $-$0.27 (0.17) \\
Group (b)                                &  35 & $-$27  (151) & $-$0.13 (0.40) & $-$0.23 (0.41) & $-$0.21 (0.24) \\
\hline
\multicolumn{6}{c}{Stars in class ``SG''} \\
\hline
All                                      &  40 & $-$94  (211) & $-$0.30 (0.37) & $-$0.04 (0.38) & $-$0.04 (0.18) \\
Group (a)                                &  24 & $-$144 (199) & $-$0.41 (0.36) & $-$0.08 (0.41) & $-$0.06 (0.15) \\
Group (b)                                &  16 & $-$18  (211) & $-$0.14 (0.34) & $+$0.01 (0.33) & $+$0.01 (0.22) \\
\hline
\multicolumn{6}{c}{Stars in class ``HB''} \\
\hline
All                                      &  28 & $-$7   (160) & $-$0.47 (0.42) & $-$0.11 (0.31) & $-$0.12 (0.17) \\
Group (a)                                &   5 & $+$44  (189) & $-$0.52 (0.47) & $-$0.38 (0.37) & $-$0.17 (0.18) \\
Group (b)                                &  23 & $-$19  (156) & $-$0.46 (0.42) & $-$0.05 (0.26) & $-$0.10 (0.16) \\
\hline
\multicolumn{6}{c}{Stars in class ``MS''} \\
\hline
All                                      &  14 & $-$138  (99) & $-$0.30 (0.19) & $-$0.05 (0.25) & $-$0.15 (0.08) \\
Group (a)                                &   8 & $-$172 (108) & $-$0.29 (0.21) & $-$0.09 (0.22) & $-$0.14 (0.10) \\
Group (b)                                &   6 & $-$92   (71) & $-$0.31 (0.16) & $+$0.01 (0.30) & $-$0.15 (0.07) \\
\enddata
\tablecomments{
Differences are in the sense of this study minus other study.
Quantities in parenthesis refer to the standard deviation.
}
\tablenotetext{a}{
Studies that do not use Fe~\textsc{i} abundance versus E.P.\ as the primary means of determining \teff\
}
\tablenotetext{b}{
Studies that use Fe~\textsc{i} abundance versus E.P.\ as the primary means of determining \teff\
}
\end{deluxetable*}

We compare our model atmosphere parameters with those determined by
studies from the last 30~years that analyzed 
at least six stars in common with us:
\citet{luck85},
\citet{mcwilliam95b},
\citet{ryan96},
\citet{fulbright00},
\citet{johnson02},
\citet{stephens02},
\citet{gratton03},
\citet{cayrel04},
\citet{honda04b},
\citet{preston06},
\citet{lai08},
\citet{bonifacio09},
\citet{ishigaki10}, and
\citet{roederer10}.
These studies can be divided into two general categories based on 
the primary means by which \teff\ has been determined.
One group (group ``b'' in Table~\ref{comparetab})
uses the usual spectroscopic technique of determining \teff\
by minimizing the dependence of Fe~\textsc{i} abundance with E.P.,
as we have done.
The other group (group ``a'' in Table~\ref{comparetab})
relies on color--\teff\ relations or
fits to the wings of Balmer series lines to calculate \teff.
Some of these studies use
a hybrid of these techniques, and we have made our best
attempt to divide them into one of the two categories for purposes of 
this comparison.

\begin{figure}
\begin{center}
\includegraphics[angle=0,width=3.35in]{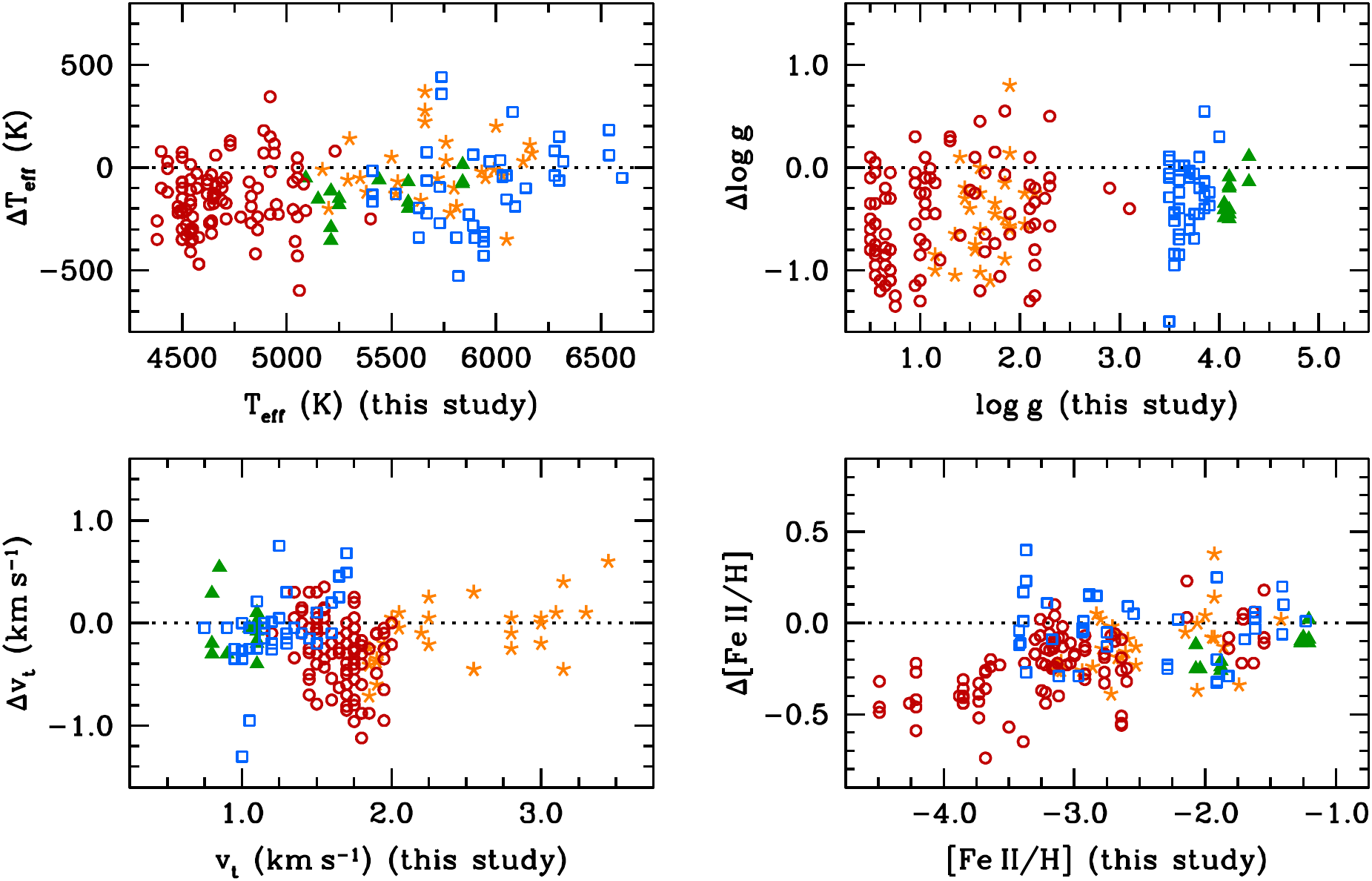}
\end{center}
\caption{
\label{compareplot}
Comparison of model atmosphere parameters and derived metallicities
between our study and previous work.
All differences are in the sense of ``this study'' minus ``previous.''
Different evolutionary states are illustrated by different point symbols:
open red circles mark stars in class RG,
open blue squares mark stars in class SG,
orange starred symbols mark stars in class HB, and
filled green triangles mark stars in class MS.
The dotted lines indicate a difference of zero.
}
\end{figure}

The differences and standard deviations 
in \teff, \logg, \vt, and metallicity 
(expressed as [Fe~\textsc{ii}/H];
i.e., the iron abundance as derived from Fe~\textsc{ii} lines) are
listed in Table~\ref{comparetab}
and illustrated in Figure~\ref{compareplot}.
As has long been known 
(e.g., \citealt{luck85}),
photometric \teff\ determinations consistently predict warmer 
temperatures and higher gravities than purely spectroscopic 
\teff\ predictions.
The values listed in Table~\ref{comparetab} reaffirm this situation.

There are no significant 
temperature differences between our study and previous ones that used
the abundance versus E.P.\ approach, with a mean 
$\Delta\teff = -$28~K ($\sigma =$~161~K) from 80 stars.
Differences in the gravities
($\Delta\logg = -$0.24, $\sigma =$~0.41), 
microturbulence parameters
($\Delta\vt = -$0.11~\kmsec, $\sigma =$~0.36~\kmsec), and
derived metallicities 
($\Delta$~[Fe/H]~$= -$0.13, $\sigma =$~0.22~dex)
are negative but not statistically significant.
These offsets show only slight variations if the stellar
evolutionary status is considered.

We find significant differences in the derived stellar parameters
when we compare with studies that compute \teff\ by other methods.
For the 110~stars in common, 
our temperatures are cooler
($\Delta\teff = -$185~K, $\sigma =$~154~K),
our gravities are lower
($\Delta\logg = -$0.57, $\sigma =$~0.42), and our
microturbulence parameters are smaller
($\Delta\vt = -$0.26~\kmsec, $\sigma =$~0.35~\kmsec).
These offsets conspire to lower our derived mean metallicities
by 0.21~dex ($\sigma =$~0.18~dex).
These offsets show some dependence on evolutionary state.
The most pronounced offsets are in red giants, where
our mean metallicities are lower by 0.27~dex ($\sigma =$~0.17~dex).
Except for \vt, 
these differences are moderately significant.

Given the present state of modeling of the line-forming regions
of stellar atmospheres, it is unclear whether one approach is 
preferable to another.
Both are likely inadequate at some level.
Abundance studies often adopt a technique based on the 
information available about the stellar sample or the
quality of the spectra themselves
(reliable photometry, knowledge of stellar distance, 
interstellar reddening, spectral coverage, etc.).~
Most studies of stars with [Fe/H]~$< -$2.5 
selected from \citeauthor{beers92}
have adopted temperatures calculated from color--\teff\ relations
or Balmer line profiles
(e.g., \citealt{mcwilliam95b}, \citealt{ryan96}, 
\citealt{cayrel04}, \citealt{lai08}, \citealt{bonifacio09}),
and so on average our derived metallicities will always be 
lower than theirs.
Nevertheless, despite these differences,
Figure~\ref{compareplot} demonstrates that
the \textit{relative} metallicities are in good agreement
for all but the most metal-poor giants.
We discuss this point in more detail in Section~\ref{atmsystematic}.

\citet{lind12} have considered how
line-by-line departures from LTE on Fe~\textsc{i} and Fe~\textsc{ii}
lines can act to influence the derivation of stellar parameters.
For all parameter ranges included in our sample,
and for the choice of the hydrogen collision parameter
adopted by \citeauthor{lind12}, their study finds that
LTE calculations of
the Fe~\textsc{ii} abundance reflect that of non-LTE
calculations to within 0.02~dex.
This principle underlies our decision to adopt the iron
abundance derived from Fe~\textsc{ii} lines as our metallicity indicator.
For the metallicity range of our sample,
\citeauthor{lind12}\ also suggest that temperatures derived 
by our method 
will underestimate \teff\ by less than 30~K for red giants,
overestimate \teff\ by less than 30~K for stars on the 
subgiant branch and main sequence turn-off,
overestimate \teff\ by 40--120~K for stars on the horizontal branch,
and
overestimate \teff\ by less than 20~K for stars on the main sequence.
In summary, it seems unlikely that departures from LTE
alone can account for the differences between the
photometric and spectroscopic temperatures estimated
for our stellar sample.

\begin{figure}
\begin{center}
\includegraphics[angle=0,width=3.35in]{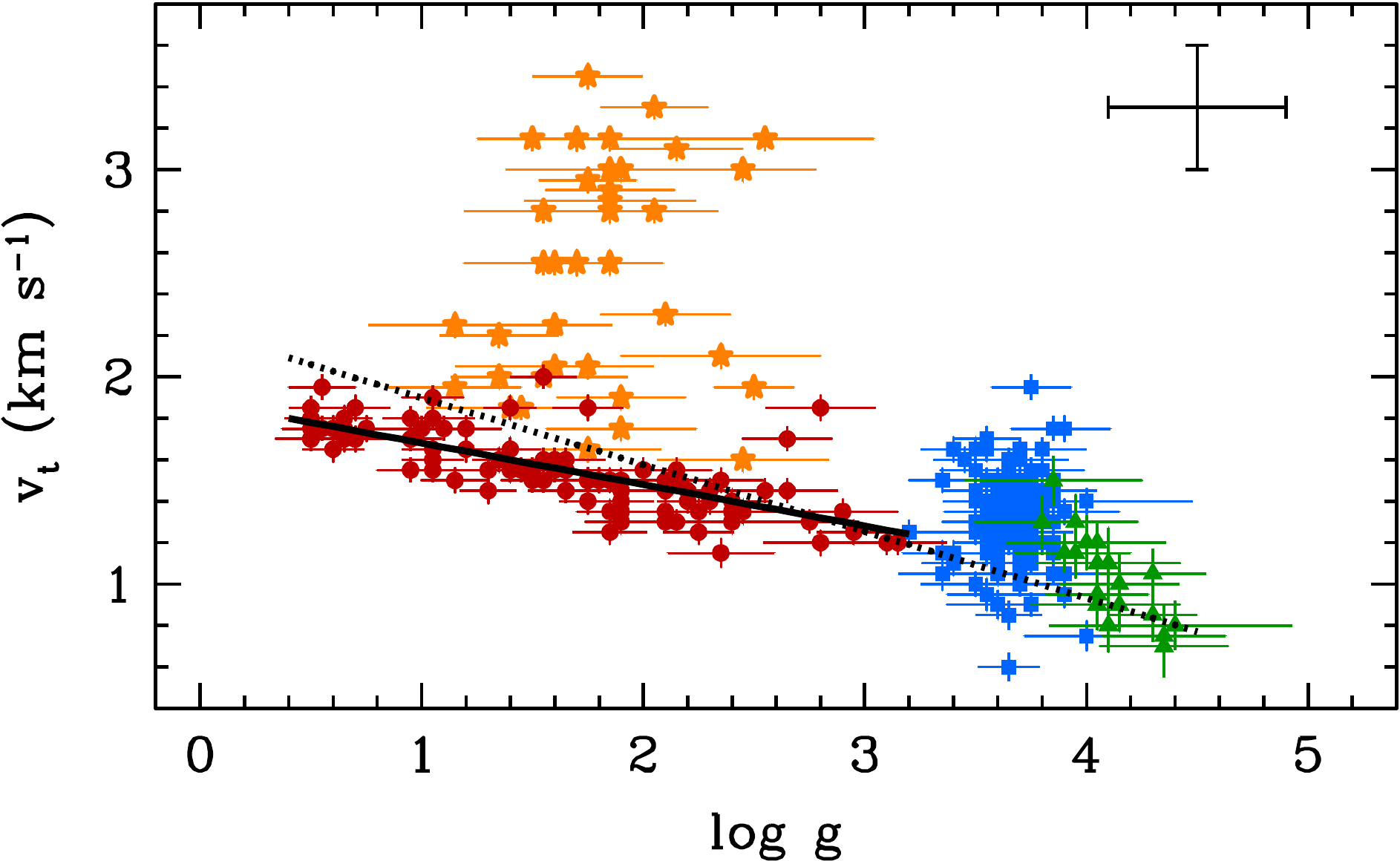}
\end{center}
\caption{
\label{vtplot}
Relationship between \logg\ and \vt.
Different evolutionary states are illustrated by different point symbols:
red circles mark stars in class RG,
blue squares mark stars in class SG,
orange starred symbols mark stars in class HB, and
green triangles mark stars in class MS.
The solid line represents the relationship 
defined by the RG stars in our sample.
The dotted line represents the
relationship found by \citet{gratton96}.
Statistical uncertainties are shown on the points,
and the black cross in the upper right corner
illustrates a typical systematic uncertainty.
}
\end{figure}

As illustrated in Figure~\ref{vtplot}, 
our derived \vt\ values 
for the most luminous giants are slightly lower
that our initial guesses
calculated from the relation given by \citet{gratton96}.
The \citeauthor{gratton96}\ relation is shown
by the dotted line.
The solid line shows a linear least-squares fit
to the giants in our sample, given by
\vt\ $=$ $-$0.20~\logg\ $+$ 1.88
($\sigma =$~0.13).
For less luminous giants, the
two relations agree well.
Despite the shallower slope, \logg\ and \vt\ 
show a tight correlation for the giants in our sample.
The difference for the most luminous giants
can be attributed mostly to the lower \logg\
values derived by our methods.
Shifting the \logg\ values for our RG stars
an average of 0.65~dex to the right
(Table~\ref{comparetab}, group ``a'')
would bring the \vt\ values for the two relations into better agreement
overall.
No significant relationship appears between \logg\ and \vt\
for the stars in our SG, MS, or HB classes.

The standard deviations reported in Table~\ref{comparetab}
are considerably larger than the
statistical uncertainties reported in Table~\ref{atmtab},
and the difference is likely due to 
systematic effects.
By definition, these are not included in the statistical error budget.
Zeropoint differences in [Fe/H] are generally insignificant
considering the magnitude of the dispersions.
We address the source of the remaining zeropoint differences
in Section~\ref{ironcomparison}, and
systematic uncertainties are discussed in Section~\ref{atmsystematic}.

\subsection{The Impact of EW Measurements, Line Lists,
Model Grids, and Analysis Codes}
\label{ironcomparison}

We assess the impact that the EW measurements, line list, \loggf\ values,
and general machinery (i.e., grid of model atmospheres and
line analysis software)
have on the derived metallicities
when compared with previous work.
To do this, we rederive iron abundances 
for stars in common with other studies
using published EW measurements, line lists, 
and model atmosphere parameters.
Five recent sets of studies have several stars
in common with us:\
\citet{johnson02},
the ``OZ'' project \citep{carretta02,cohen06,cohen08},
the First Stars project (including
\citealt{cayrel04}, \citealt{spite05}, \citealt{francois07}, and others),
\citet{honda04a,honda04b}, and
\citet{lai08}.

\begin{deluxetable*}{lccccc}
\tablecaption{Comparison of Derived Iron Abundances Using Different 
Equivalent Widths, Line Lists, and Analysis Tools
\label{compare2tab}}
\tablewidth{0pt}
\tabletypesize{\scriptsize}
\tablehead{
\colhead{Abundance} &
\colhead{No.\ stars} &
\colhead{$\Delta\ (\sigma)$} &
\colhead{$\Delta\ (\sigma)$} &
\colhead{$\Delta\ (\sigma)$} &
\colhead{$\Delta\ (\sigma)$} \\
\colhead{ratio} &
\colhead{in common} &
\colhead{``A''} &
\colhead{``B''} &
\colhead{``C''} &
\colhead{``C''$-$``A''} }
\startdata      
\multicolumn{6}{c}{\citet{johnson02}} \\
\hline
~[Fe~\textsc{i}/H]  &  7 & $-$0.114 (0.021) & $-$0.054 (0.039) & $-$0.067 (0.032) & $+$0.047 (0.031) \\
~[Fe~\textsc{ii}/H] &  7 & $-$0.064 (0.028) & $-$0.063 (0.032) & $-$0.010 (0.034) & $+$0.054 (0.034) \\
\hline
\multicolumn{6}{c}{OZ:\ \citet{carretta02}; \citet{cohen06,cohen08}} \\
\hline
~[Fe~\textsc{i}/H]  &  5 & $-$0.154 (0.055) & $+$0.066 (0.046) & $-$0.214 (0.051) & $-$0.060 (0.070) \\
~[Fe~\textsc{ii}/H] &  5 & $-$0.066 (0.055) & $-$0.142 (0.207) & $+$0.042 (0.156) & $+$0.108 (0.172) \\
\hline
\multicolumn{6}{c}{First Stars:\ \citet{cayrel04}} \\
\hline
~[Fe~\textsc{i}/H]  & 18 & $-$0.027 (0.008) & $+$0.016 (0.039) & $-$0.032 (0.046) & $-$0.005 (0.046) \\
~[Fe~\textsc{ii}/H] & 18 & $-$0.004 (0.033) & $-$0.058 (0.068) & $+$0.039 (0.064) & $+$0.043 (0.017) \\
\hline
\multicolumn{6}{c}{\citet{honda04a,honda04b}} \\
\hline
~[Fe~\textsc{i}/H]  & 10 & $-$0.260 (0.021) & $-$0.037 (0.078) & $-$0.189 (0.090) & $+$0.071 (0.082) \\
~[Fe~\textsc{ii}/H] & \nodata & \nodata     & \nodata          & \nodata          & \nodata          \\
\hline
\multicolumn{6}{c}{\citet{lai08}} \\
\hline
~[Fe~\textsc{i}/H]  & 12 & $-$0.007 (0.016) & $+$0.022 (0.025) & $-$0.029 (0.025) & $-$0.023 (0.023) \\
~[Fe~\textsc{ii}/H] & 12 & $-$0.008 (0.017) & $-$0.139 (0.075) & $+$0.125 (0.091) & $+$0.133 (0.083) \\
\enddata
\tablecomments{
Comparison ``A'' derives the iron abundances using published EWs and model parameters 
from the other study and our machinery.
Comparison ``B'' derives the iron abundances using published EWs from the other study,
our derived model parameters, and our machinery.
Comparison ``C'' derives the iron abundances using EWs measured by us, stellar parameters
derived by the other study, and our machinery.
Comparison ``C''$-$``A'' is the difference in iron abundances when only the EWs and linelists
change.
Differences are in the sense of ``as derived here'' minus ``published.''
Quantities in parenthesis refer to the standard deviation.
}
\end{deluxetable*}

The results of several comparisons are listed in Table~\ref{compare2tab}.
In comparison ``A,'' we rederive the iron abundances using
published EW measurements, line lists, and model atmosphere parameters 
from another study, but we use the ATLAS9 grid of model atmospheres
and MOOG to perform the calculations.
We may think of comparison ``A'' as 
representing the differences in [Fe/H] we would have
derived if we had used the EW measurements, 
line lists, and temperature/gravity scales
of other studies.
In comparison ``B,'' we rederive the iron abundances using
the published EW measurements and line lists from other studies,
our grid of MARCS models, and MOOG.
We may think of comparison ``B'' as
representing the differences in [Fe/H] we would have
derived if we had adopted other EW measurements and line lists 
for our analysis.
In comparison ``C,'' we rederive the iron abundances using
our EW measurements and line list, 
the model atmosphere parameters from other studies,
the ATLAS9 model grid, and MOOG.
We may think of comparison ``C'' as
representing the differences in [Fe/H] that would have
been derived by other studies using our EW measurements and line list.

The final column in Table~\ref{compare2tab}
(``C''$-$``A'') lists the
differences in derived iron abundances when 
using stellar parameters from other studies if
only the EW measurements and line list are changed.
We calculate that the \loggf\ values for Fe~\textsc{i}
and Fe~\textsc{ii} lines
in common are different by $<$~0.03~dex on average.
Thus any residual differences in the final column in 
Table~\ref{compare2tab} are the result of
which lines are actually used in the analysis.

These comparisons are imperfect.
For example, some of these other studies 
used Turbospectrum, SPTOOL, or employed earlier versions
of MOOG that did not include Rayleigh scattering
in the source function.
When appropriate, we have reverted to the earlier version of MOOG
when making these calculations for comparison.
The OSMARCS and earlier ATLAS grids of model atmospheres
were used in some other studies, 
and the interpolation codes for the ATLAS9 grid are also different.
The adopted solar iron abundance varies from
$\log \epsilon$~(Fe)~$=$~7.50 to 7.52 among these studies,
and the scaled solar compositions adopted
vary from one model grid to another.
Finally, 
we simply adopt published model atmosphere
parameters without rederiving them from scratch 
before determining the iron abundances.

Nevertheless, Table~\ref{compare2tab} indicates
that the differences in the iron abundances
are often small or not significant.
It is beyond the scope of the present study
to identify the source of the remaining discrepancies.
Using the comparison with the First Stars sample as an example,
we conclude based on comparison ``A'' that our analysis
would have derived Fe~\textsc{i} abundances lower by 0.027~dex
using their methods.
The differences in the EW measurements and lines used for
analysis are negligible, as demonstrated by 
comparisons ``B,'' ``C,'' ``C''$-$``A,''
and Figure~\ref{ewcayrel}.
These comparisons should give some guidance
to investigators attempting to place our study
in the context of others.

\subsection{Systematic Uncertainties}
\label{atmsystematic}

The true magnitude of
systematic uncertainties is more difficult to quantify.
The scatter observed when comparing our model parameters to
those in previous studies of stars in common 
may give some guidance here.
For stars in the evolutionary classes RG/SG/HB/MS, 
when comparing with studies where model parameters were derived
by similar techniques,
these comparisons yield standard deviations of 
151/211/156/71~K in \teff,
0.40/0.34/0.42/0.16 in \logg,
0.41/0.33/0.26/0.30~\kmsec\ in \vt, and
0.24/0.22/0.16/0.07~dex in [Fe~\textsc{ii}/H].
In most cases these uncertainties dominate the statistical uncertainties.

\begin{figure}
\begin{center}
\includegraphics[angle=0,width=3.35in]{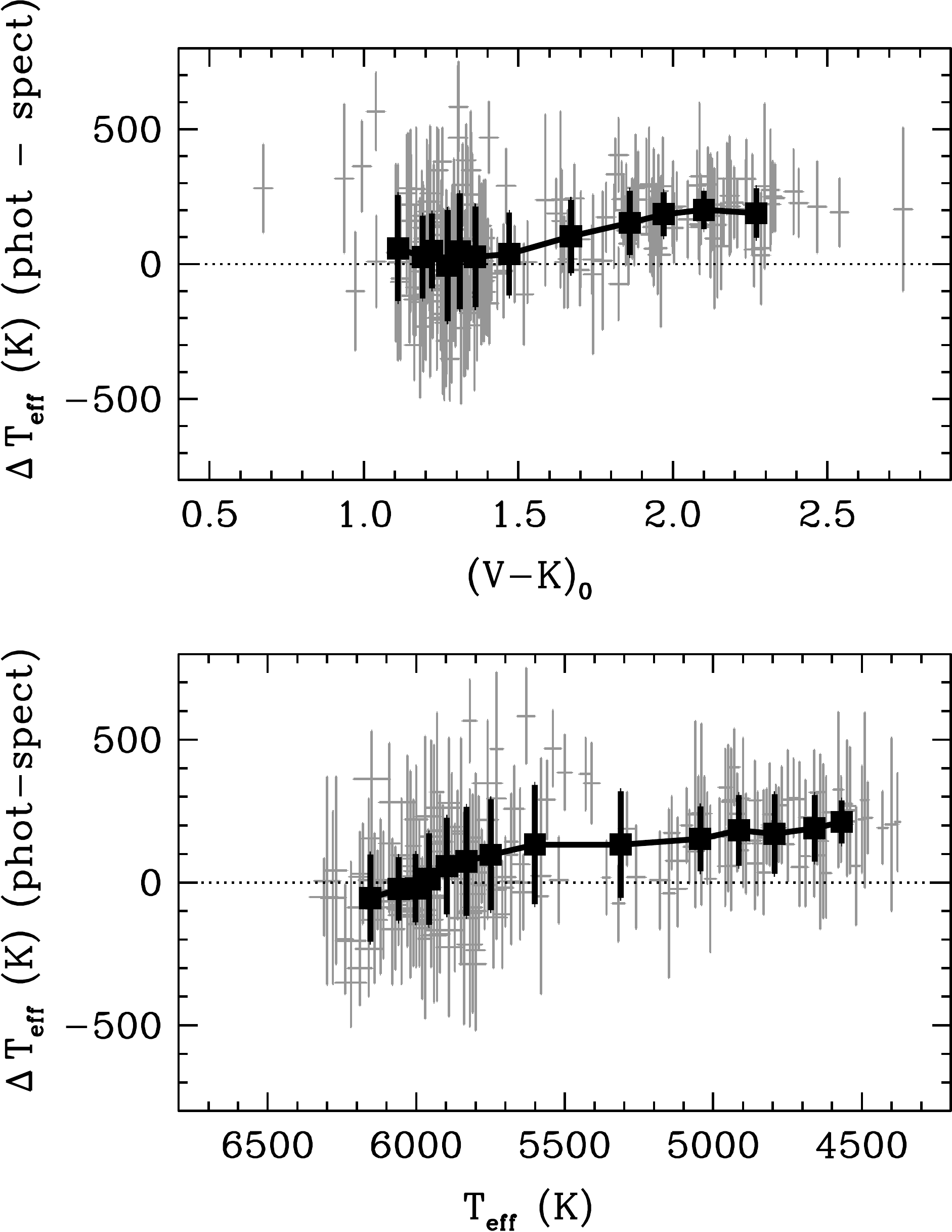}
\end{center}
\caption{
\label{delteffplot}
Difference in photometric and spectroscopic \teff\ as a 
function of $(V-K)_{0}$ (top) and spectroscopic \teff\ (bottom).
The photometric \teff\ values are predicted from the
\citet{alonso99b} ($J-H$) and ($J-K$) relations, using
the methods described in Section~\ref{photometry}.
Stars with discrepant photometry are ignored.
Only stars evolved beyond the MSTO but not yet on the HB are shown.
Gray crosses indicate individual stars, and the large black points
mark the weighted average and standard deviation
computed by passing a box of 30~stars
with an overlap of 15~stars through the data.
The dotted line indicates a difference of zero.
}
\end{figure}

We may further investigate the scale of systematic uncertainties by
comparing our model parameters with those derived by alternative techniques.
We compare our spectroscopically-derived temperatures with
ones derived from color--\teff\ relations (hereafter ``photometric''
temperatures).
We compute photometric temperatures from the
$J-H$ and $J-K$ relations of \citet{alonso99b}.
This is illustrated in Figure~\ref{delteffplot}, where we show
the difference in photometric and spectroscopic temperature
for 219~stars in our sample.
There is a large amount of scatter at any given temperature, but
there is a significant offset (approaching 200~K on average) for 
stars cooler than $\sim$~5100~K
and $(V-K)_{0} >$~1.8.
This finding echoes that of many previous studies, including
\citet{johnson02},
\citet{cayrel04},
\citet{aoki07},
\citet{cohen08}, and
\citet{lai08}. 
The underlying physical cause of this discrepancy
is not fully understood, but it likely stems at least in part
from the inability of one-dimensional hydrostatic 
LTE model atmospheres to 
capture the essential physics of convection,  
radiative transfer, and so on.
This naturally affects the predicted colors and line formation.
\citet{frebel13} discuss this issue at length and 
propose an empirical calibration based on
seven well-studied nearby stars to bring the 
two scales into agreement.
The relationship between the \teff\ differences and \teff\
shown in their Figure~2 agrees nearly perfectly with the
mean offsets shown in our Figure~10.  
\citeauthor{frebel13}\ also note that the discrepancy
between the two scales becomes even more exaggerated
for stars with [Fe/H]~$< -$3.5.
We see a similar outcome:\ for the 27~stars (counting repeat analyses)
with [Fe/H]~$< -$3.5 in common between our sample and previous ones
(Figure~\ref{compareplot}),
our \teff\ values are lower by 323~K ($\sigma =$~90~K) on average.
This demonstrates that the cooler temperature scale in our
study relative to previous studies that used color-\teff\
relations is a consequence to the method used to derive \teff,
as discussed previously in Section~\ref{compareprevious}.

\begin{figure}
\begin{center}
\includegraphics[angle=0,width=3.35in]{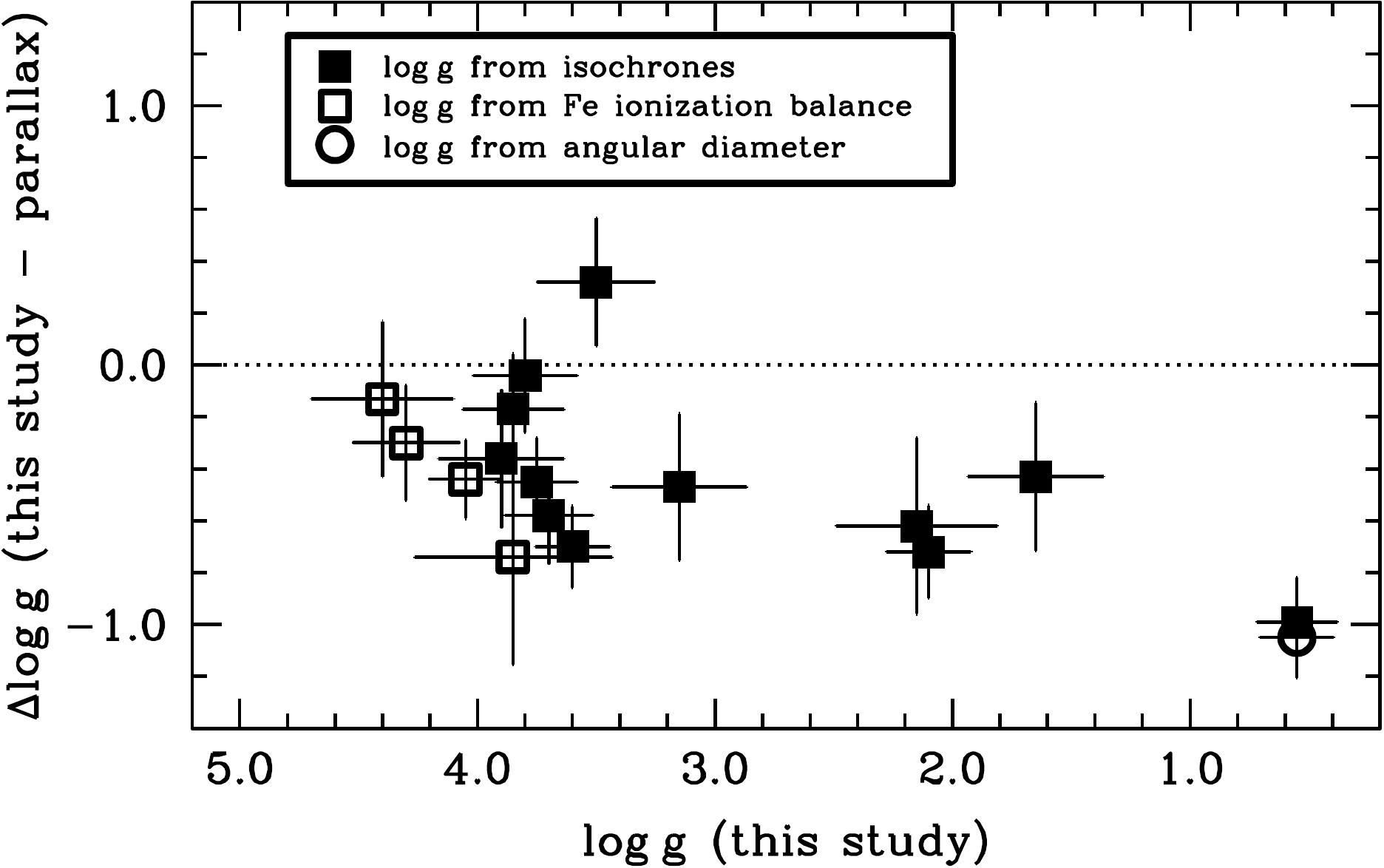}
\end{center}
\caption{
\label{parallax}
Comparison of \logg\ derived in this study with those computed from
Hipparcos parallaxes.
Only the 
16~stars with $E(B-V) <$~0.12 and Hipparcos uncertainties smaller than
20\% are considered.
The fill and shape of the points 
indicates the method by which the \logg\ was obtained.
For display purposes, the difference in the
``angular diameter'' measurement
is plotted alongside the ``parallax'' measurements,
but these are really two separate methods to derive \logg.
The dotted line indicates a difference of zero.
}
\end{figure}

For a few nearby stars, 
we compare our gravities with those calculated from 
parallaxes measured by the Hipparcos mission \citep{perryman97},
using the data from the reduction by \citet{vanleeuwen07}.
We restrict this comparison to stars with Hipparcos parallax
precisions better than 20\% and $E(B-V) <$~0.12;
only 16~stars in our sample meet these criteria.
Using the parallax, apparent magnitude, reddening,
bolometric corrections (BC; \citealt{alonso99a}), temperature, metallicity,
mass (assumed to be 0.8~$M_{\odot}$),
and the solar constants
$M_{\rm bol, \odot} =$~4.74, 
\teff$_{\odot} =$~5780~K, and
\logg$_{\odot} =$~4.44, 
we can calculate \logg\ by means of the relation
\begin{align}
& \log g = 0.4(M_{K,\star} + BC_{K} - M_{\rm bol,\odot}) 
+ \log g_{\odot} \nonumber \\
& + 4\log(T_{\rm eff,\star}/T_{\rm eff,\odot})
+ \log(m_{\star}/m_{\odot}).
\end{align}
Figure~\ref{parallax} compares these ``physical'' \logg\ values 
with our derived \logg\ values for these 16~stars.
We also compare our gravity for 
\object[HD 122563]{HD~122563}
with that
derived from the recent measurement of its angular diameter
by \citet{creevey12}.
Our \logg\ values are systematically lower 
than the physical ones by 0.49~dex ($\sigma =$~0.32~dex),
independent of the method we have used to derive \logg.
This offset is similar to that found 
($-$0.57, $\sigma =$~0.42) when comparing our
\logg\ values with those calculated from
color-\teff\ relations by other investigators 
(Section~\ref{compareprevious}).

We use the \teff\ predicted by
$(V-K)_{0}$ instead of a spectroscopically-derived \teff\
as an alternative method to calculate \logg\ from isochrones.
For the 11~stars to which we can apply
both the parallax method and the isochrone method, 
the photometric temperatures give \logg\ values
greater by only 0.15~dex than the physical ones.
This would be encouraging if not for the fact that 
the dispersion in \logg\ for the photometric \teff\ input is
substantially larger ($\sigma =$~0.63~dex) than that
using the spectroscopic \teff\ input
($\sigma =$~0.36~dex).

\begin{figure}
\begin{center}
\includegraphics[angle=0,width=3.35in]{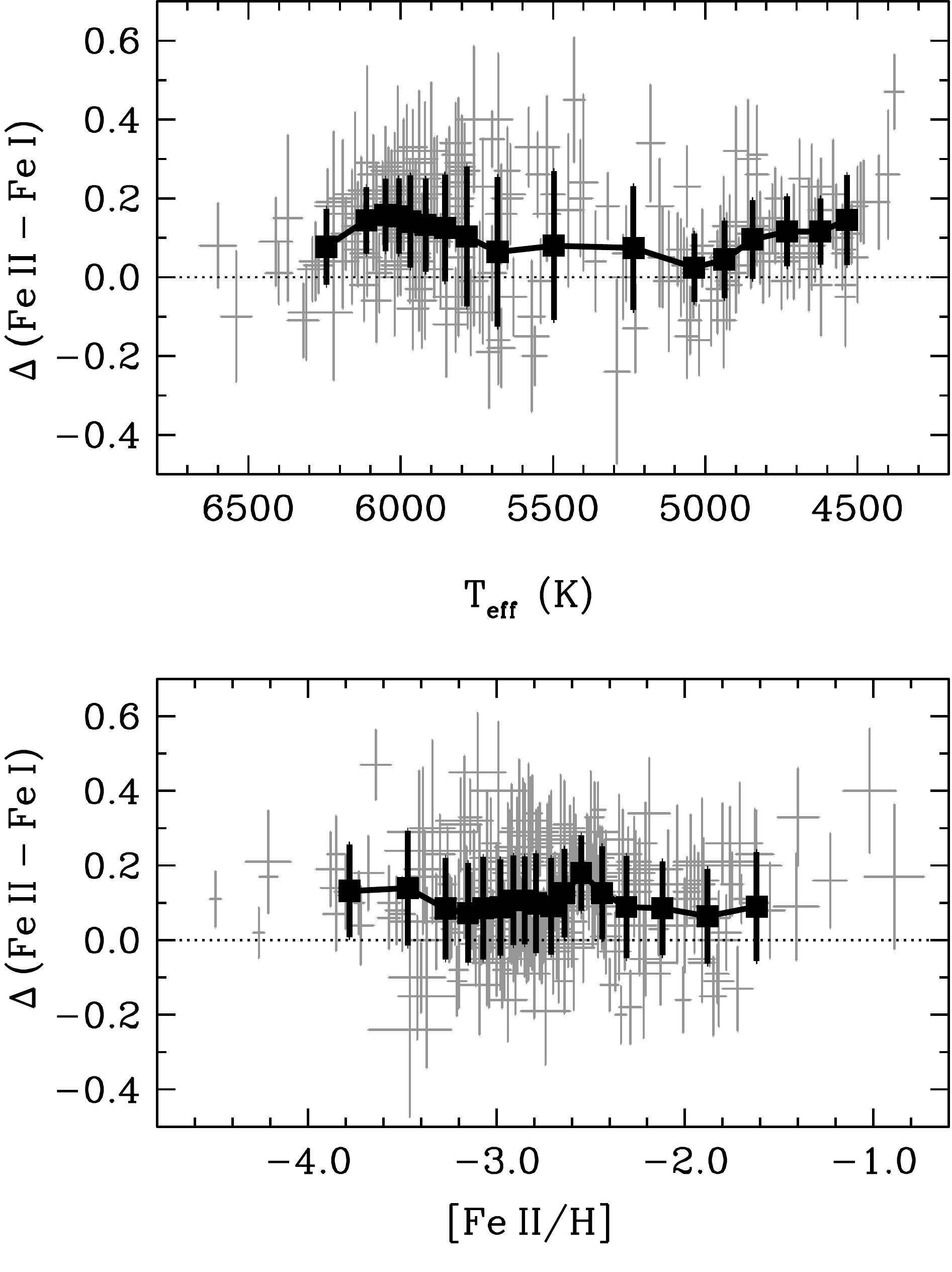}
\end{center}
\caption{
\label{delfeplot}
Difference in iron abundance derived from 
Fe~\textsc{ii} and Fe~\textsc{i} as a function of 
\teff\ and [Fe~\textsc{ii}/H].
Only stars in the SG and RG classes are shown.
Gray crosses indicate individual stars, and the large black points
mark the weighted average and standard deviation
computed by passing a box of 30~stars
with an overlap of 15~stars through the data.
The dotted line indicates a difference of zero.
}
\end{figure}

Figure~\ref{delfeplot} shows the difference in the derived
iron abundances using Fe~\textsc{ii} or Fe~\textsc{i} lines.
Differences are mostly found in the range 
$-$0.1~$<$~[Fe~\textsc{ii}/Fe~\textsc{i}]~$< +$0.3~dex,
with a mean difference of 0.10~dex ($\sigma =$~0.13~dex).
These differences do not show a strong dependence on 
metallicity, but they are 
slightly larger for the warmest ($\approx$~6100~K) 
and coolest ($\approx$~4600~K) stars in the sample.
The differences are comparable to what is expected if
overionization of Fe~\textsc{i} is responsible 
(e.g., \citealt{thevenin99}).

We suggest that the statistical uncertainties, listed in Table~\ref{atmtab},
should be considered when comparing
abundance ratios of stars with similar parameters.
The total uncertainties, which are more difficult to assess,
should be used when absolute abundances are considered.
The values listed at the beginning of this section may
be considered representative of the total uncertainties.

\subsection{The Impact of log $g$ on Metallicity}

Our \logg\ values are systematically lower than 
those implied by several other derivation methods.
As demonstrated in Section~\ref{abundsigma},
ionized species are more sensitive to \logg\ than neutral species are.
Our metallicities are based on the iron abundance
derived from Fe~\textsc{ii} lines, 
so this difference does have an impact.

Tables~\ref{comparetab} and
\ref{compare2tab} can be used to quantify this impact
on the iron abundances.
Two of the sets of comparisons listed,
the 0Z studies and the \citet{lai08} study,
calculated photometric \teff\ values and used these
to calculate \logg\ from isochrones.
Their values of \teff\ and \logg\ are equivalent to 
what we would have calculated using this approach.
Note that the 0Z set of studies listed in 
Table~\ref{compare2tab} did not have enough stars
in common to merit inclusion in Table~\ref{comparetab},
so we perform an identical 
comparison separately and list the results here.
For the five stars in common, the mean differences are
$-$227~K ($\sigma =$~169~K) in \teff,
$-$0.48~dex ($\sigma =$~0.42~dex) in \logg,
$-$0.55~\kmsec\ ($\sigma =$~0.22~\kmsec) in \vt, and
$-$0.15~dex ($\sigma =$~0.22~dex) in [Fe~\textsc{ii}/H].

These results suggest that our metallicities
are lower by 0.15~dex (0Z) or
0.08~dex \citep{lai08}
in a straight-up comparison.
Comparison ``C'' in Table~\ref{compare2tab}
lists the [Fe~\textsc{i}/H] and 
[Fe~\textsc{ii}/H] abundances that we would have derived for our sample
using others' photometric \teff\ and \logg\ values
with our EW measurements, linelist, model grid, and MOOG.
These corrections must be applied to the values above.
Thus, we conclude that our metallicities are lower
by 0.19~dex (0Z) or 0.20~dex (\citeauthor{lai08})
because we have used spectroscopic \teff\ values 
to predict \logg\ using the Y$^{2}$ isochrones.

For the stars in common with the 
0Z and \citet{lai08} studies, using our approach the mean
[Fe~\textsc{ii}/Fe~\textsc{i}] ratio is 
$+$0.16~$\pm$~0.04~dex ($\sigma =$~0.15~dex).
Using the alternate approach the mean is
$+$0.12~$\pm$~0.02~dex ($\sigma =$~0.08~dex).
We also divide these stars into classes of giants or subgiants.
For the giants, 
using the alternate approach would reduce the
mean difference from 
$+$0.20~$\pm$~0.04 ($\sigma =$~0.14) to
$+$0.14~$\pm$~0.02 ($\sigma =$~0.07).
For the subgiants, the differences are even smaller,
$+$0.10~$\pm$~0.07 ($\sigma =$~0.17) using our approach and
$+$0.09~$\pm$~0.04 ($\sigma =$~0.10) using the alternate approach.
These results suggest that our approach may
slightly overestimate the amount of overionization 
occurring for Fe~\textsc{i} in the giants.

\subsection{Iron Abundance Trends with Wavelength}

\citet{roederer12b} identified a relationship between 
wavelength and iron abundance derived from Fe~\textsc{i} lines
in four metal-poor giants,
and \citet{lawler13} found a similar effect in a metal-poor turnoff star.
This effect is characterized as a decrease in the average abundance by
$\approx$~0.05 to 0.20~dex 
at short wavelengths
(mainly 2280~$< \lambda <$~4000~\AA) compared to 
long wavelengths ($\lambda >$~4400~\AA).~
\citeauthor{roederer12b}\ investigated several explanations for this effect and 
favored an unidentified extra source of continuous opacity
at short wavelengths that was not accounted for.
Ultimately that study only adopted an empirical correction to the abundances
derived from lines at short wavelengths so as to match the abundances
derived from longer wavelengths.  
The four stars in that study are also in our study
(\object[HD 108317]{HD~108317},
\object[HD 122563]{HD~122563},
\object[HD 126238]{HD~126238},
and 
\object[HD 128279]{HD~128279}).
Our derived temperatures are different by 30 to 70~K
because of the different techniques used to derive the model atmosphere
parameters.
We do not find a similar effect for these four stars, although
the number of Fe~\textsc{i} lines studied at short wavelengths
is considerably smaller than the sample examined by \citeauthor{roederer12b} 

\begin{figure}
\begin{center}
\includegraphics[angle=0,width=3.35in]{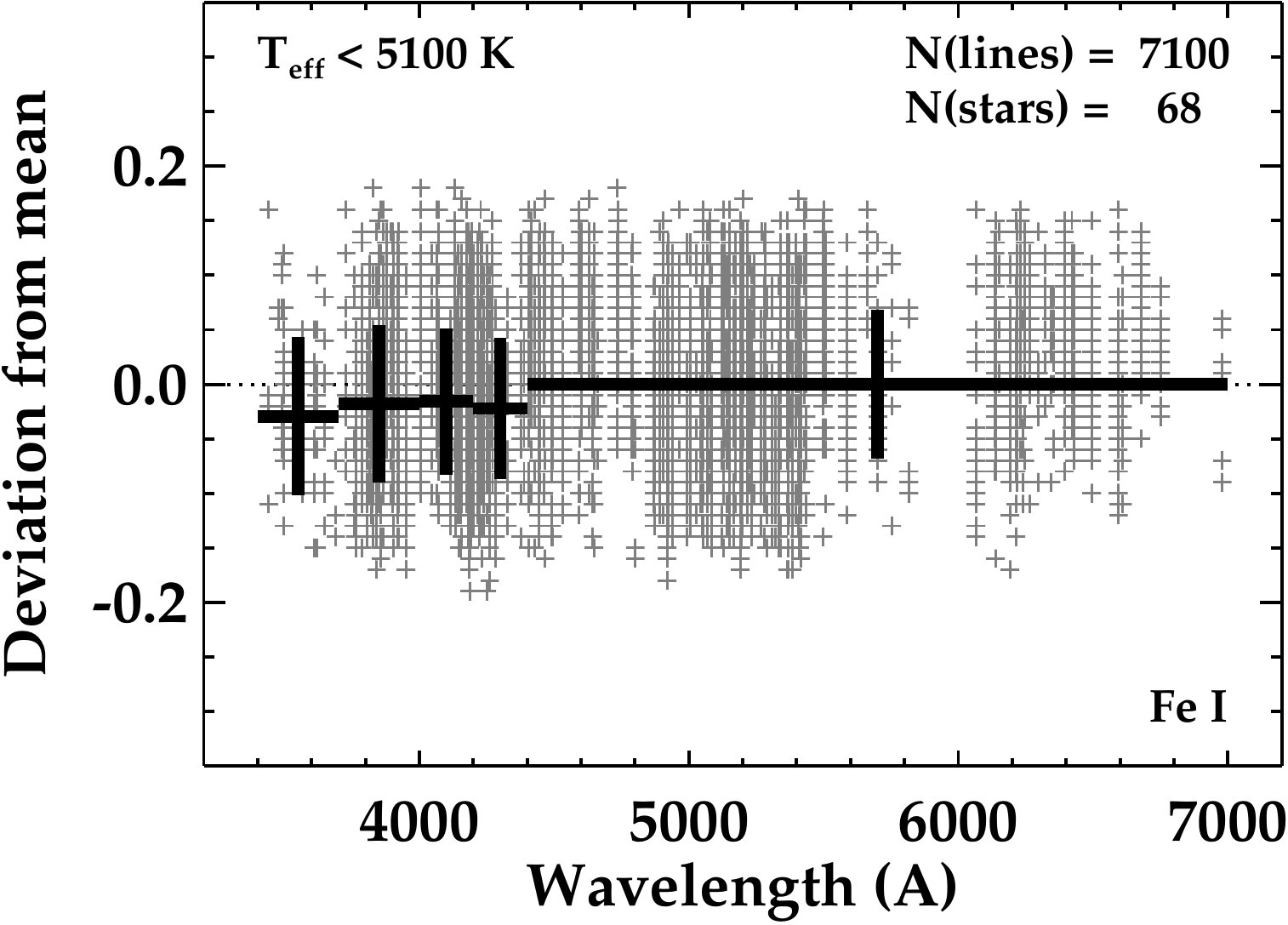} \\
\vspace*{0.2in}
\includegraphics[angle=0,width=3.35in]{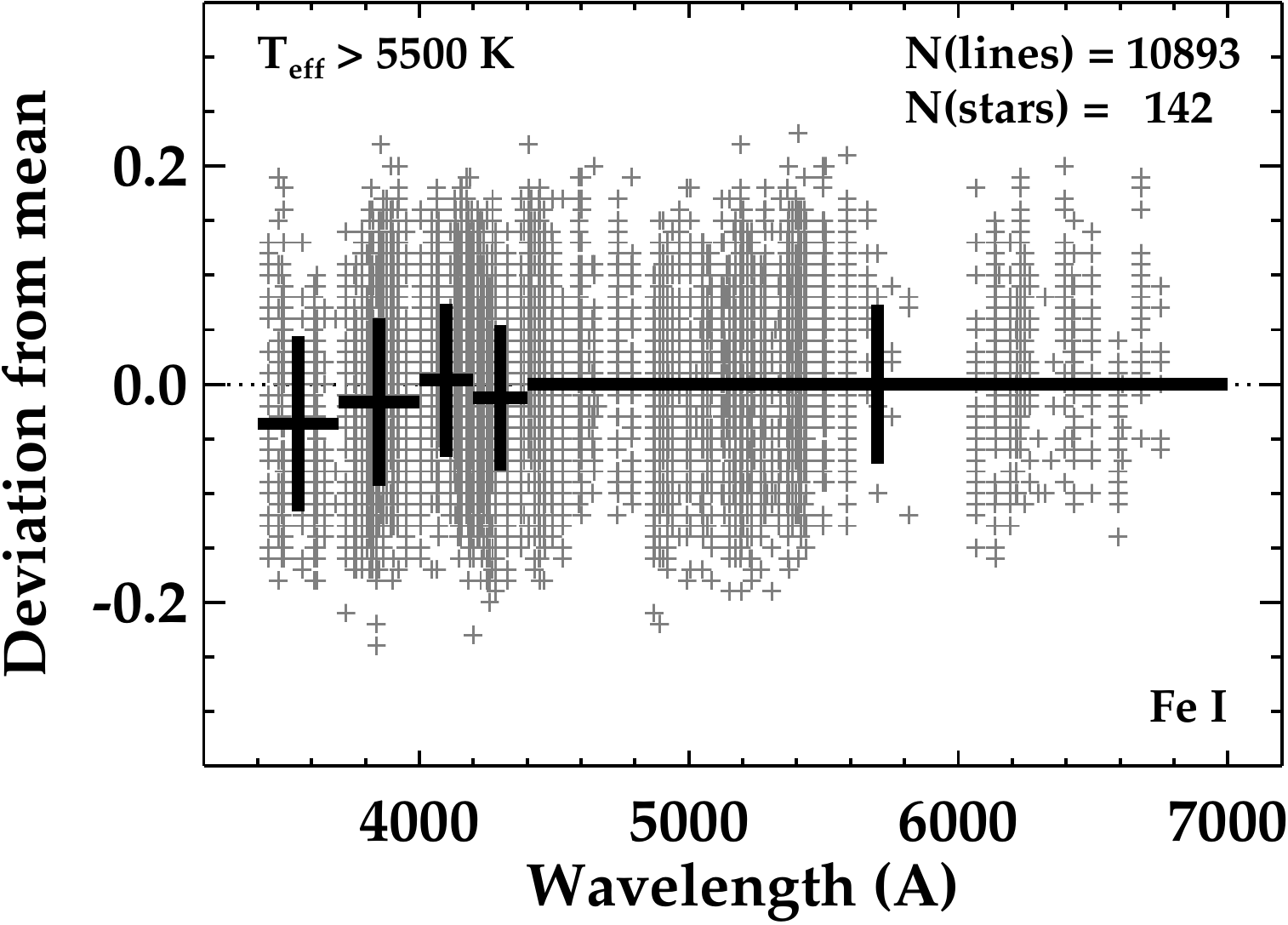}
\end{center}
\caption{
\label{fe1waveplot}
Line-by-line abundances derived from Fe~\textsc{i} lines
as a function of wavelength.
The top panel shows lines in stars on the upper giant branch with 
\teff~$< 5100$~K, and the 
bottom panel shows lines in stars with \teff~$> 5500$~K
and cooler than the MSTO.
Small gray crosses mark individual lines, and bold black crosses
mark the mean and standard deviation in each of several wavelength bins.
All abundances are normalized to the mean abundance for each star
derived from lines with $\lambda >$~4400~\AA,
which is marked by the dotted black line.
}
\end{figure}

We take advantage of our large sample of stars
to improve the statistics at short wavelengths.
Figure~\ref{fe1waveplot} illustrates the wavelength dependence of
abundances derived from Fe~\textsc{i} lines in our study for
68~giants and 142~subgiants.
Abundances derived from Fe~\textsc{i} 
lines at short wavelengths are, on average,
lower by 0.04~dex or less than lines at long wavelengths.
The results do not change when comparing smaller subsets of stars with
similar \teff\ and metallicity.
This is a very small difference compared to what \citet{roederer12b} found,
and the difference may be related to the methods used to derive
model parameters.
\citeauthor{roederer12b}\ derived model parameters mainly using 
lines at long wavelengths, only considering the abundances 
from lines at short wavelengths once the model parameters had been set.
In the present study, we have included all Fe~\textsc{i} lines 
when deriving the model parameters.
This choice was necessary because a substantial part of our sample
is warmer and more metal-poor than the stars considered by 
\citeauthor{roederer12b},
hence fewer lines at long wavelengths are available to us.
If lines at short wavelength yield systematically lower abundances,
they will preferentially be culled by our 2$\sigma$-clipping 
algorithm described in Section~\ref{modelmethod}.
Our experience suggests that this is a possibility, so the
results shown in Figure~\ref{fe1waveplot} should be
taken as lower limits to the differences at short wavelengths.

Lines of other species are not affected by the 2$\sigma$-clipping
algorithm but would be affected if there is a physical
origin of this effect, like missing continuous opacity.
We devote our attention in Section~\ref{offsets}
to finding other evidence of such an effect.

\subsection{Other Considerations}

\citet{mcwilliam95b} noted that the first (uppermost) levels
of the grids of model atmospheres 
available at the time, those of \citet{bell76} and Kurucz\footnote{
Cited by 
\citet{mcwilliam95b} as Kurucz, R.~L., 1992, private communication},
began at different atmosphere layers.
For a typical metal-poor giant 
(\object[BD-18 5550]{BD$-$18~5550}),
these models began 
at column masses (``RHOX'') near 13 and 0.2 g~cm$^{-2}$, respectively.
In the \citeauthor{bell76}\ model, the optical depth at line center
of lines stronger than
\logrw~$> -$4.7 was nonzero in the first layer, indicating that
a significant portion of the line was formed above this layer.  
While the Kurucz models covered the entire line-forming 
region, they too failed to include the temperature inversion 
expected in the low density layers of the chromosphere.
This introduced a different set of problems since the cores of 
strong lines may be formed in these layers.
Models of 
\object[BD-18 5550]{BD$-$18~5550}
from the current 
MARCS and ATLAS9 grids extend to 1.3 and 0.02~g~cm$^{-2}$, respectively.
This covers the line-forming region for lines up to 
\logrw~$\gg -$4.0 in the ATLAS9 models and 
\logrw~$= -$4.45 in the MARCS models.
Fortuitously, \logrw~$= -$4.45 
is also the typical upper bound for the EWs we have measured in giants,
so both model grids treat the lines in our dataset comparably in this regard.
Tests by \citet{mcwilliam95b} to account for the presence of
a chromosphere indicate a zero-point uncertainty of 
approximately 0.1~dex for Fe~\textsc{i} lines with 
\logrw~$> -$4.7 in metal-poor giants.
This issue is not resolved by the updated models and will 
affect our results for the giants.  
For typical metal-poor dwarf or subgiant models we obtain similar results,
in that the MARCS models do not encompass the entire line-forming region
for \logrw~$> -$4.50.
Fe~\textsc{i} lines with \logrw~$> -$4.50 ($-$4.45)
comprise less than 2.0\% (0.7\%) of the total Fe~\textsc{i} lines 
in our sample
and are unlikely to skew the derived atmospheric parameters significantly.
This could, however, impact
species whose abundance can only be deduced from strong lines.
In Section~\ref{offsets}
we consider systematic abundance offsets associated with
different lines of the same species.

Using spherically-symmetric model atmospheres with an analysis 
code that treats the line-forming layers as parallel slabs 
can lead to systematic errors in the derived abundances.
The spherical models have a slightly lower temperature structure
in the uppermost layers due to dilution of the radiation
field from lower layers \citep{gustafsson08}.
We find that the magnitude of this effect is small, less than a few percent
in the most extreme cases (i.e., stars with very low surface gravity),
where the main difference lies in 
optically-thin layers with $\log \tau < -$1.
We therefore make no corrections for this effect.

The current version of MOOG includes Rayleigh scattering from neutral H atoms
in the source function \citep{sobeck11}.
Our calculations indicate that opacity from Rayleigh scattering
contributes 14\% of the total continuous opacity at 3700~\AA\ in
typical metal-poor red giants at $\tau =$~1. 
To the extent that the Rayleigh scattering contribution
is being properly computed, it should be included.
(The other dominant contribution to the continuous opacity in these stars
comes from bound-free absorption by H$^{-}$.)
In typical metal-poor subgiants, the contribution from Rayleigh
scattering is much less than 1\% at all wavelengths considered.

The classical technique of deriving the microturbulence velocity
from Fe~\textsc{i} lines, which we have used, may systematically
overestimate \vt\ because of correlated errors in the
measured EWs and derived abundances \citep{magain84}.
Theoretical EWs computed from the stellar model could
eliminate this particular bias.
\citet{mucciarelli11} has reexamined the situation,
finding that the two methods produce equivalent results when 
moderately high S/N data are used
(S/N~$\gtrsim$~70/1).  
Our data generally fall in this regime
since we are not typically measuring weak lines 
(i.e., those most prone to the bias) at blue wavelengths 
where the S/N is lower.
Even in our HET spectra, where the S/N is lowest, no more than
4\% of our Fe~\textsc{i} lines would be susceptible to this bias. 
We do not pursue the matter further.

Distances calculated from absolute magnitudes computed from
spectroscopically-derived 
\teff\ or \logg\ will be systematically overestimated
relative to distances calculated from photometric determinations 
of \teff\ or \logg.
Any analysis that makes use of these distances to
examine the kinematic properties of our sample will be affected.
A proper analysis of the stellar kinematics 
is deferred for future work.

\section{Abundance Analysis}
\label{abundances}

On average, we derive abundances or upper limits
from 216 ($\sigma =$~48) of the 474~lines considered
in each star.
For all lines with an EW measurement listed in 
Table~\ref{ewtab}, we use
MOOG to compute theoretical EWs that 
are forced to match the measured EWs
by adjusting the input abundance.
For lines denoted ``synth'' in Table~\ref{ewtab},
we use MOOG to generate a series of synthetic spectra
where the abundance of the element producing the
line of interest is varied to match the observed spectrum.
Each synthesis spans
$\pm$~3~\AA\ on either side of this line.
We generate the line lists used in the synthesis
using the \citet{kurucz95} line lists
as a starting point.
We then replace the \loggf\ values of 
the line of interest and any other lines
with data from laboratory or more detailed
theoretical studies, where available.
Occasionally, especially for lines in the blue regions
of the spectrum, we are forced to 
alter the theoretical \loggf\ values
to produce a reasonable fit to the observed spectrum.
These lists are then employed,
unchanged, for syntheses of all stars in the sample.
We do allow for abundance variations
of elements other than the one of interest
based on our EW analysis for each star.
Abundances derived from lines of
Li~\textsc{i}, CH, CN, NH, Al~\textsc{i},
Sc~\textsc{ii}, V~\textsc{i}, V~\textsc{ii},
Mn~\textsc{i}, Mn~\textsc{ii},
Co~\textsc{i}, Cu~\textsc{i}, and
elements with $Z >$~30 are determined
via the synthesis method.
All others are derived using EWs.

We estimate an upper limit on the abundance
for lines not detected in our spectra.
The 3$\sigma$ upper limits are calculated 
from a version of the formula presented on p.\ 590 of
\citet{frebel08}, which itself is derived from
equation A8 of \citet{bohlin83}.
When multiple lines of the same species are not detected,
we adopt the upper limit that provides the
strongest constraint on the abundance.

The presence of multiple isotopes
of some lines of interest may lead to 
small energy shifts in the transition
wavelength, and our syntheses account 
for the IS for
Ba~\textsc{ii}, Nd~\textsc{ii}, Sm~\textsc{ii},
Eu~\textsc{ii}, Yb~\textsc{ii}, Ir~\textsc{i}, 
and Pb~\textsc{i}.
In all cases we adopt the \rpro\ isotopic ratios
presented in \citet{sneden08} unless
our analysis reveals a substantial
contribution of \spro\ material.
In that case, we adopt an appropriate mix
of $r$- and \spro\ isotopes based on
our derived abundances.

\begin{deluxetable}{lcccc}
\tablecaption{Abundances Derived from Individual Lines 
\label{lineabundtab}}
\tablewidth{0pt}
\tabletypesize{\scriptsize}
\tablehead{
\colhead{Star} & 
\colhead{Species} &
\colhead{Wavelength} &
\colhead{$\log\epsilon$} &
\colhead{$\sigma$} \\
\colhead{} &
\colhead{} &
\colhead{(\AA)} &
\colhead{} &
\colhead{} }
\startdata      
 CS 22166-016        & Li I & 6707.80 & $<$ 0.44 & \nodata \\
 CS 22166-016        & Mg I & 4057.51 &     5.02 &    0.17 \\
 CS 22166-016        & Mg I & 4167.27 &     5.07 &    0.18 \\
 CS 22166-016        & Mg I & 5172.68 &     5.15 &    0.39 \\
 CS 22166-016        & Mg I & 5183.60 &     5.02 &    0.37 \\
 CS 22166-016        & Mg I & 5528.40 &     5.13 &    0.19 \\
\enddata
\tablecomments{
The complete version of Table~\ref{lineabundtab} is available online only.  
A short version is shown here to
illustrate its form and content.
}
\end{deluxetable}

Abundances and uncertainties for 
each line in each star
are reported in Table~\ref{lineabundtab}.
The mean abundances and uncertainties are presented
in Table~\ref{finalabundtab}.
The full versions of Tables~\ref{lineabundtab} 
and \ref{finalabundtab}
are available in the online edition of the journal.
The meanings of the different uncertainty estimates in
Table~\ref{finalabundtab} are discussed in 
Section~\ref{abundsigma}.
The ratios relative to the Solar values, [X/H] or [X/Fe], where
X stands for any metal, are computed relative to the 
\citet{asplund09} Solar photospheric abundances,
listed in Table~\ref{solartab}.
In cases where the photospheric value is poorly known, 
we adopt the abundance measured in CI-type
carbonaceous meteorites instead.
These cases are noted in Table~\ref{solartab}.
We remind readers that [X/Fe] ratios are 
constructed using the abundances derived from 
species in the same ionization state;
i.e., neutrals to neutrals and ions to ions.
Only $\log \epsilon$ abundances or upper limits,
not [X/Fe] ratios, are presented
for lithium and technetium.

\begin{deluxetable*}{ccccccccc}
\tablecaption{Mean Abundances 
\label{finalabundtab}}
\tablewidth{0pt}
\tabletypesize{\scriptsize}
\tablehead{
\colhead{Star} &
\colhead{Species} &
\colhead{$N_{\rm lines}$} &
\colhead{$\log\epsilon$} &
\colhead{[X/Fe]\tablenotemark{a}} &
\colhead{$\sigma_{\rm statistical}$} &
\colhead{$\sigma_{\rm total}$} &
\colhead{$\sigma_{\rm neutrals}$} &
\colhead{$\sigma_{\rm ions}$} }
\startdata      
CS~22166--016     & Fe~\textsc{i}  &  96 &     4.28 & -3.22 &  0.06 &  0.15 &  0.00 &  0.00  \\
CS~22166--016     & Fe~\textsc{ii} &  11 &     4.41 & -3.09 &  0.07 &  0.15 &  0.00 &  0.00  \\
CS~22166--016     & Li~\textsc{i}  &   1 & $<$ 0.52 &\nodata&\nodata&\nodata&\nodata&\nodata \\
CS~22166--016     & C~(CH)         &   1 &     5.35 &  0.01 &  0.15 &  0.25 &  0.19 &  0.19  \\
CS~22166--016     & N~(NH)         &   1 &     5.34 &  0.60 &  0.30 &  0.36 &  0.32 &  0.32  \\
CS~22166--016     & O~\textsc{i}   &   0 &  \nodata &\nodata&\nodata&\nodata&\nodata&\nodata \\
CS~22166--016     & Na~\textsc{i}  &   0 &  \nodata &\nodata&\nodata&\nodata&\nodata&\nodata \\
CS~22166--016     & Mg~\textsc{i}  &   5 &     5.07 &  0.69 &  0.05 &  0.15 &  0.08 &  0.18  \\
CS~22166--016     & Al~\textsc{i}  &   1 &     2.56 & -0.67 &  0.09 &  0.29 &  0.24 &  0.30  \\
CS~22166--016     & Si~\textsc{i}  &   0 &  \nodata &\nodata&\nodata&\nodata&\nodata&\nodata \\
CS~22166--016     & K~\textsc{i}   &   0 &  \nodata &\nodata&\nodata&\nodata&\nodata&\nodata \\
CS~22166--016     & Ca~\textsc{i}  &  11 &     3.73 &  0.62 &  0.11 &  0.17 &  0.12 &  0.20  \\
CS~22166--016     & Sc~\textsc{ii} &   6 &     0.01 & -0.05 &  0.06 &  0.15 &  0.17 &  0.10  \\
CS~22166--016     & Ti~\textsc{i}  &  15 &     2.01 &  0.29 &  0.07 &  0.15 &  0.09 &  0.18  \\
CS~22166--016     & Ti~\textsc{ii} &  22 &     2.08 &  0.22 &  0.06 &  0.15 &  0.17 &  0.09  \\
CS~22166--016     & V~\textsc{i}   &   1 &     0.68 & -0.03 &  0.14 &  0.19 &  0.15 &  0.22  \\
CS~22166--016     & V~\textsc{ii}  &   2 &     0.99 &  0.15 &  0.20 &  0.24 &  0.26 &  0.21  \\
CS~22166--016     & Cr~\textsc{i}  &   6 &     2.27 & -0.15 &  0.06 &  0.14 &  0.08 &  0.18  \\
CS~22166--016     & Cr~\textsc{ii} &   2 &     2.63 &  0.08 &  0.11 &  0.18 &  0.20 &  0.13  \\
CS~22166--016     & Mn~\textsc{i}  &   4 &     1.78 & -0.42 &  0.08 &  0.16 &  0.10 &  0.20  \\
CS~22166--016     & Mn~\textsc{ii} &   4 &     1.74 & -0.60 &  0.15 &  0.21 &  0.23 &  0.17  \\
CS~22166--016     & Co~\textsc{i}  &   5 &     1.77 &  0.00 &  0.14 &  0.20 &  0.15 &  0.23  \\
CS~22166--016     & Ni~\textsc{i}  &   6 &     3.07 &  0.07 &  0.11 &  0.21 &  0.16 &  0.23  \\
CS~22166--016     & Cu~\textsc{i}  &   0 &  \nodata &\nodata&\nodata&\nodata&\nodata&\nodata \\
CS~22166--016     & Zn~\textsc{i}  &   0 &  \nodata &\nodata&\nodata&\nodata&\nodata&\nodata \\
CS~22166--016     & Ga~\textsc{i}  &   0 &  \nodata &\nodata&\nodata&\nodata&\nodata&\nodata \\
CS~22166--016     & Rb~\textsc{i}  &   0 &  \nodata &\nodata&\nodata&\nodata&\nodata&\nodata \\
CS~22166--016     & Sr~\textsc{ii} &   2 &    -0.50 & -0.28 &  0.04 &  0.25 &  0.25 &  0.25  \\
CS~22166--016     & Y~\textsc{ii}  &   1 &    -1.41 & -0.53 &  0.10 &  0.17 &  0.20 &  0.12  \\
CS~22166--016     & Zr~\textsc{ii} &   3 &    -0.57 & -0.06 &  0.12 &  0.18 &  0.21 &  0.14  \\
CS~22166--016     & Nb~\textsc{ii} &   1 & $<$ 0.12 &  1.75 &\nodata&\nodata&\nodata&\nodata \\
CS~22166--016     & Mo~\textsc{i}  &   1 & $<$-0.41 &  0.93 &\nodata&\nodata&\nodata&\nodata \\
CS~22166--016     & Tc~\textsc{i}  &   1 & $<$-0.11 &\nodata&\nodata&\nodata&\nodata&\nodata \\
CS~22166--016     & Ru~\textsc{i}  &   0 &  \nodata &\nodata&\nodata&\nodata&\nodata&\nodata \\
CS~22166--016     & Sn~\textsc{i}  &   1 & $<$ 0.95 &  2.10 &\nodata&\nodata&\nodata&\nodata \\
CS~22166--016     & Ba~\textsc{ii} &   3 &    -1.35 & -0.44 &  0.05 &  0.15 &  0.17 &  0.09  \\
CS~22166--016     & La~\textsc{ii} &   1 &    -2.07 & -0.08 &  0.18 &  0.22 &  0.24 &  0.19  \\
CS~22166--016     & Ce~\textsc{ii} &   5 & $<$-1.62 & -0.11 &\nodata&\nodata&\nodata&\nodata \\
CS~22166--016     & Pr~\textsc{ii} &   1 &    -1.84 &  0.53 &  0.17 &  0.22 &  0.24 &  0.18  \\
CS~22166--016     & Nd~\textsc{ii} &   1 &    -1.85 & -0.18 &  0.11 &  0.18 &  0.20 &  0.13  \\
CS~22166--016     & Sm~\textsc{ii} &   0 &  \nodata &\nodata&\nodata&\nodata&\nodata&\nodata \\
CS~22166--016     & Eu~\textsc{ii} &   3 &    -2.37 &  0.19 &  0.13 &  0.19 &  0.21 &  0.15  \\
CS~22166--016     & Gd~\textsc{ii} &   2 & $<$-1.21 &  0.81 &\nodata&\nodata&\nodata&\nodata \\
CS~22166--016     & Tb~\textsc{ii} &   0 &  \nodata &\nodata&\nodata&\nodata&\nodata&\nodata \\
CS~22166--016     & Dy~\textsc{ii} &   1 &    -1.70 &  0.29 &  0.14 &  0.20 &  0.22 &  0.15  \\
CS~22166--016     & Ho~\textsc{ii} &   0 &  \nodata &\nodata&\nodata&\nodata&\nodata&\nodata \\
CS~22166--016     & Er~\textsc{ii} &   1 &    -1.70 &  0.47 &  0.40 &  0.43 &  0.44 &  0.41  \\
CS~22166--016     & Tm~\textsc{ii} &   0 &  \nodata &\nodata&\nodata&\nodata&\nodata&\nodata \\
CS~22166--016     & Yb~\textsc{ii} &   1 &    -2.10 &  0.07 &  0.12 &  0.19 &  0.21 &  0.14  \\
CS~22166--016     & Hf~\textsc{ii} &   2 & $<$-0.94 &  1.30 &\nodata&\nodata&\nodata&\nodata \\
CS~22166--016     & Ir~\textsc{i}  &   0 &  \nodata &\nodata&\nodata&\nodata&\nodata&\nodata \\
CS~22166--016     & Pb~\textsc{i}  &   1 & $<$ 0.53 &  1.71 &\nodata&\nodata&\nodata&\nodata \\
CS~22166--016     & Th~\textsc{ii} &   3 & $<$-1.95 &  1.08 &\nodata&\nodata&\nodata&\nodata \\
\enddata
\tablecomments{
The complete version of Table~\ref{finalabundtab} is available online only.  
A short version illustrating the results for one star is shown here to
illustrate its form and content.
}
\tablenotetext{a}{[Fe/H] is indicated for Fe~\textsc{i} and Fe~\textsc{ii}}
\end{deluxetable*}

\begin{deluxetable}{ccc}
\tablecaption{Adopted Solar Abundances
\label{solartab}}
\tablewidth{0pt}
\tabletypesize{\scriptsize}
\tablehead{
\colhead{Element} &
\colhead{$Z$} &
\colhead{$\log\epsilon$} }
\startdata      
Li &  3 &\nodata \\
C  &  6 &  8.43  \\
N  &  7 &  7.83  \\
O  &  8 &  8.69  \\
Na & 11 &  6.24  \\
Mg & 12 &  7.60  \\
Al & 13 &  6.45  \\
Si & 14 &  7.51  \\
K  & 19 &  5.03  \\
Ca & 20 &  6.34  \\
Sc & 21 &  3.15  \\
Ti & 22 &  4.95  \\
V  & 23 &  3.93  \\
Cr & 24 &  5.64  \\
Mn & 25 &  5.43  \\
Fe & 26 &  7.50  \\
Co & 27 &  4.99  \\
Ni & 28 &  6.22  \\
Cu & 29 &  4.19  \\
Zn & 30 &  4.56  \\
Ga & 31 &  3.04  \\
Rb & 37 &  2.52  \\
Sr & 38 &  2.87  \\
Y  & 39 &  2.21  \\
Zr & 40 &  2.58  \\
Nb & 41 &  1.46  \\
Mo & 42 &  1.88  \\
Tc & 43 &\nodata \\
Ru & 44 &  1.75  \\
Sn & 50 &  2.07\tablenotemark{a} \\
Ba & 56 &  2.18  \\
La & 57 &  1.10  \\
Ce & 58 &  1.58  \\
Pr & 59 &  0.72  \\
Nd & 60 &  1.42  \\
Sm & 62 &  0.96  \\
Eu & 63 &  0.52  \\
Gd & 64 &  1.07  \\
Tb & 65 &  0.30  \\
Dy & 66 &  1.10  \\
Ho & 67 &  0.48  \\
Er & 68 &  0.92  \\
Tm & 69 &  0.10  \\
Yb & 70 &  0.92\tablenotemark{a} \\
Hf & 72 &  0.85  \\
Ir & 77 &  1.38  \\
Pb & 82 &  2.04\tablenotemark{a} \\
Th & 90 &  0.06\tablenotemark{a} \\
\enddata
\tablenotetext{a}{Meteoritic abundance}
\end{deluxetable}

Molecular abundances are derived by spectrum synthesis.
For the NH lines near 3360~\AA,
we adopt the \citet{kurucz95} line list
with \loggf\ values reduced by a factor of two
and a dissociation potential of 3.45~eV,
as recommended by \citet{johnson07}.
For the CN lines near 3880~\AA,
we adopt the \citeauthor{kurucz95} line lists
without change and a dissociation potential
of 7.65~eV.
For the CH lines near 4310~\AA,
we adopt the line list of B.\ Plez (2007, private communication)
and a dissociation potential of 3.47~eV.
We adopt a default fitting uncertainty of 0.15~dex
for the CH lines,
0.20~dex for the CN lines, and
0.30~dex for the NH lines.
The final abundances listed in Table~\ref{finalabundtab}
reflect the CH and NH abundances, when possible, 
otherwise the nitrogen abundance is derived from 
the CN abundance after carbon has been set using the CH lines.
Molecular formation in cool stellar atmospheres
is sensitive to the temperature and density,
especially the presence and degree of granulation 
found in three-dimensional hydrodynamical
models (e.g., \citealt{collet07}).
These effects are difficult to quantify.
Our one-dimensional LTE results should be 
treated with due caution
when using them for anything besides gross 
discriminants of carbon and nitrogen enrichment.

For all lines with $\lambda >$~5670~\AA,
we examine the stellar spectrum simultaneously
with a telluric spectrum of earth's atmosphere
\citep{hinkle00}.
In general, we did not observe hot telluric standards
during our observing program, so
any lines that appear to be compromised by 
telluric absorption are discarded from further consideration.
Furthermore, the Na~\textsc{i} D lines
are sometimes blended with interstellar sodium
absorption.
We do not attempt to
derive abundances from these lines
when they appear asymmetric, broadened beyond the typical
stellar line widths, or
when the telluric spectrum suggests that they may be compromised.

\begin{deluxetable*}{cccccccc}
\tablecaption{Abundance Differences Derived from $\alpha$-enhanced and $\alpha$-normal Models
\label{alphatab}}
\tablewidth{0pt}
\tabletypesize{\scriptsize}
\tablehead{
\colhead{} &
\multicolumn{3}{c}{G004-036} &
\colhead{} &
\multicolumn{3}{c}{BD$+$80~245} \\
\cline{2-4}\cline{6-8} 
\colhead{Species or Ratio} &
\colhead{$\Delta$} &
\colhead{$\sigma$} &
\colhead{N$_{\rm lines}$} &
\colhead{} &
\colhead{$\Delta$} &
\colhead{$\sigma$} &
\colhead{N$_{\rm lines}$} }
\startdata      
$\log \epsilon$(Mg~\textsc{i})   & $+$0.004 & 0.005 &   7    & & $+$0.010 & 0.012 &   4    \\
$\log \epsilon$(Ca~\textsc{i})   & $+$0.004 & 0.005 &  10    & & $+$0.001 & 0.005 &  11    \\
$\log \epsilon$(Ti~\textsc{i})   & $+$0.005 & 0.005 &  15    & & $-$0.002 & 0.004 &  13    \\
$\log \epsilon$(Ti~\textsc{ii})  & $+$0.007 & 0.005 &  19    & & $+$0.019 & 0.003 &  17    \\
$\log \epsilon$(Cr~\textsc{i})   & $+$0.003 & 0.005 &   9    & & $+$0.000 & 0.004 &  11    \\
$\log \epsilon$(Cr~\textsc{ii})  & $+$0.007 & 0.006 &   3    & & $+$0.013 & 0.006 &   3    \\
$\log \epsilon$(Fe~\textsc{i})   & $+$0.003 & 0.005 &  87    & & $+$0.002 & 0.006 &  66    \\
$\log \epsilon$(Fe~\textsc{ii})  & $+$0.006 & 0.005 &  10    & & $+$0.017 & 0.005 &  10    \\
$\log \epsilon$(Ni~\textsc{i})   & $+$0.006 & 0.005 &   7    & & $+$0.001 & 0.004 &   8    \\
~[Mg~\textsc{i}/Fe~\textsc{i}]   & $+$0.001 & 0.007 &\nodata & & $+$0.008 & 0.013 &\nodata \\
~[Ca~\textsc{i}/Fe~\textsc{i}]   & $+$0.001 & 0.007 &\nodata & & $-$0.002 & 0.008 &\nodata \\
~[Ti~\textsc{i}/Fe~\textsc{i}]   & $+$0.001 & 0.007 &\nodata & & $-$0.005 & 0.007 &\nodata \\
~[Ti~\textsc{ii}/Fe~\textsc{ii}] & $+$0.001 & 0.007 &\nodata & & $+$0.002 & 0.006 &\nodata \\
~[Cr~\textsc{i}/Fe~\textsc{i}]   & $+$0.000 & 0.007 &\nodata & & $-$0.002 & 0.007 &\nodata \\
~[Cr~\textsc{ii}/Fe~\textsc{ii}] & $+$0.001 & 0.008 &\nodata & & $-$0.004 & 0.008 &\nodata \\
~[Ni~\textsc{i}/Fe~\textsc{i}]   & $+$0.002 & 0.007 &\nodata & & $-$0.001 & 0.007 &\nodata \\
~[Ti~\textsc{i}/Ti~\textsc{ii}]  & $-$0.003 & 0.007 &\nodata & & $-$0.021 & 0.006 &\nodata \\   
~[Cr~\textsc{i}/Cr~\textsc{ii}]  & $-$0.003 & 0.008 &\nodata & & $-$0.013 & 0.007 &\nodata \\   
~[Fe~\textsc{i}/Fe~\textsc{ii}]  & $-$0.003 & 0.007 &\nodata & & $-$0.015 & 0.008 &\nodata \\     
\enddata
\end{deluxetable*}

Several stars in our sample have low [$\alpha$/Fe] ratios,
so using the $\alpha$-enhanced grid of models may not,
in principle, be appropriate.
To test how much of an effect this may have on our
analysis, we have performed the abundance analysis
for several key species using an $\alpha$-enhanced model
and an $\alpha$-normal model for the
two most $\alpha$-poor stars in our sample,
\object[G 4-36]{G004-036}
and
\object[BD+80 245]{BD$+$80~245}.
The results of this test are listed in Table~\ref{alphatab}.
For 
\object[G 4-36]{G004-036},
a subgiant,
the differences in derived $\log \epsilon$
are all smaller than 0.007~dex, and 
ratios constructed among abundances derived from
like ionization states differ by 0.002~dex or less.
For 
\object[BD+80 245]{BD$+$80~245},
the differences in derived
$\log \epsilon$ are larger than 0.002~dex only for
ionized species, for which the differences are as large as 0.019~dex.
Ratios constructed among abundances derived from
like ionization states differ by 0.005~dex or less.
Differences among abundances derived from 
unlike ionization states, e.g., [Fe~\textsc{i}/Fe~\textsc{ii}], 
are 0.021~dex or smaller, and we
(again) advise against constructing abundance ratios this way.
For stars with intermediate [$\alpha$/Fe] ratios,
these differences will be smaller.
On the basis of this test, we conclude that 
using the $\alpha$-enhanced models for all stars in the sample
will have, at most, a minimal effect on the derived abundance
ratios.

\subsection{Uncertainties}
\label{abundsigma}

We estimate abundance uncertainties following the
formalism presented in \citet{mcwilliam95b}.
The standard deviation of a single line is calculated 
according to equation A5 of \citeauthor{mcwilliam95b}~
To evaluate the partial derivatives, we have selected
model atmospheres representing stars in our sample on
the red giant branch, subgiant branch, horizontal branch, and
main sequence.  
Then, following \citeauthor{mcwilliam95b}, we alter the 
model parameters one by one to estimate the change in the 
abundance that results in fictitious iron lines whose
strength spans the full range of line strengths 
considered in our sample.
We repeat this exercise for both neutral lines and
singly-ionized lines.

\begin{figure}
\begin{center}
\includegraphics[angle=0,width=3.35in]{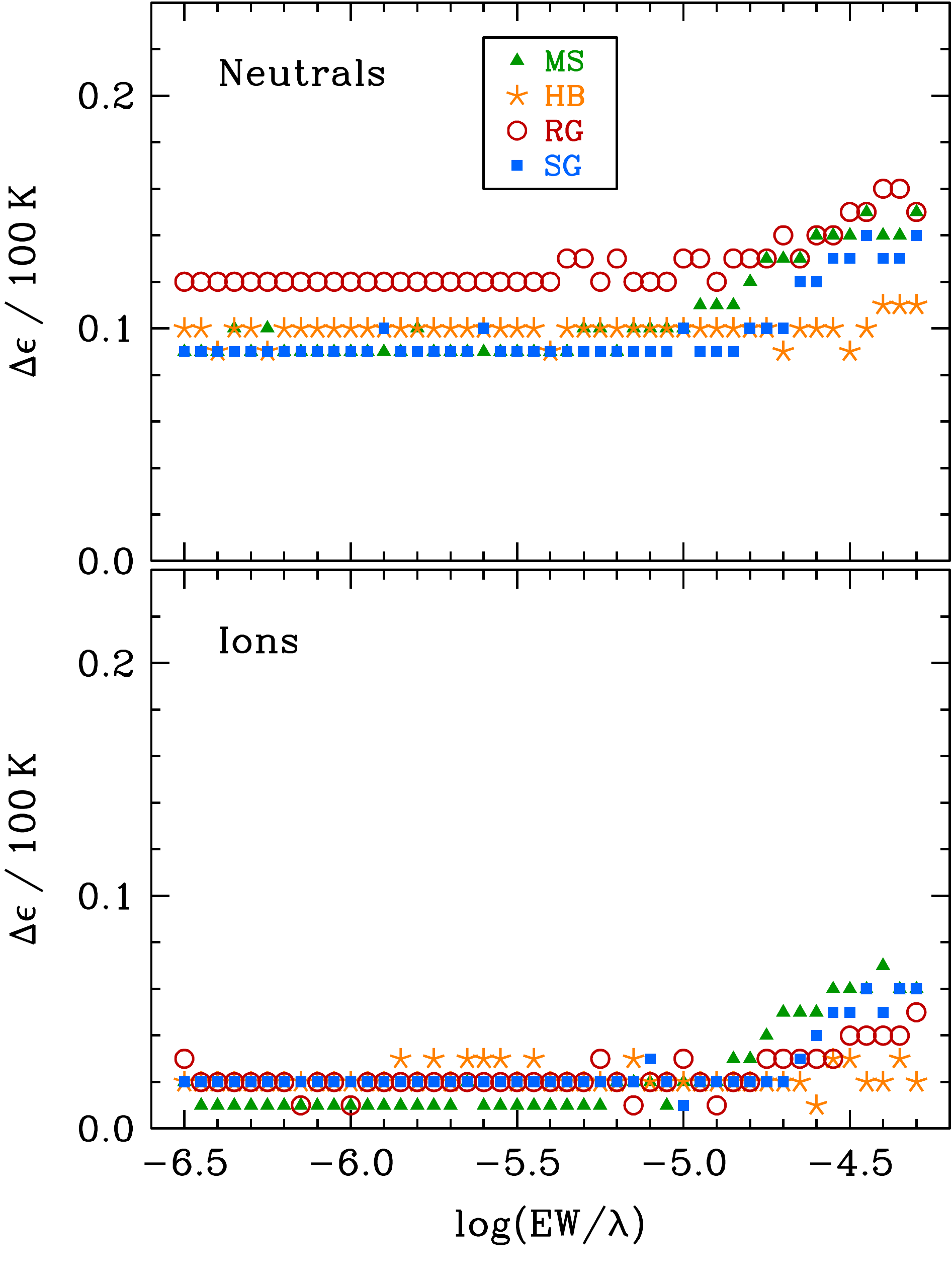}
\end{center}
\caption{
\label{deltaabundteff}
Change in derived abundances of neutral lines (top)
and singly-ionized lines (bottom)
resulting from a change in \teff\ of 100~K.
}
\end{figure}

\begin{figure}
\begin{center}
\includegraphics[angle=0,width=3.35in]{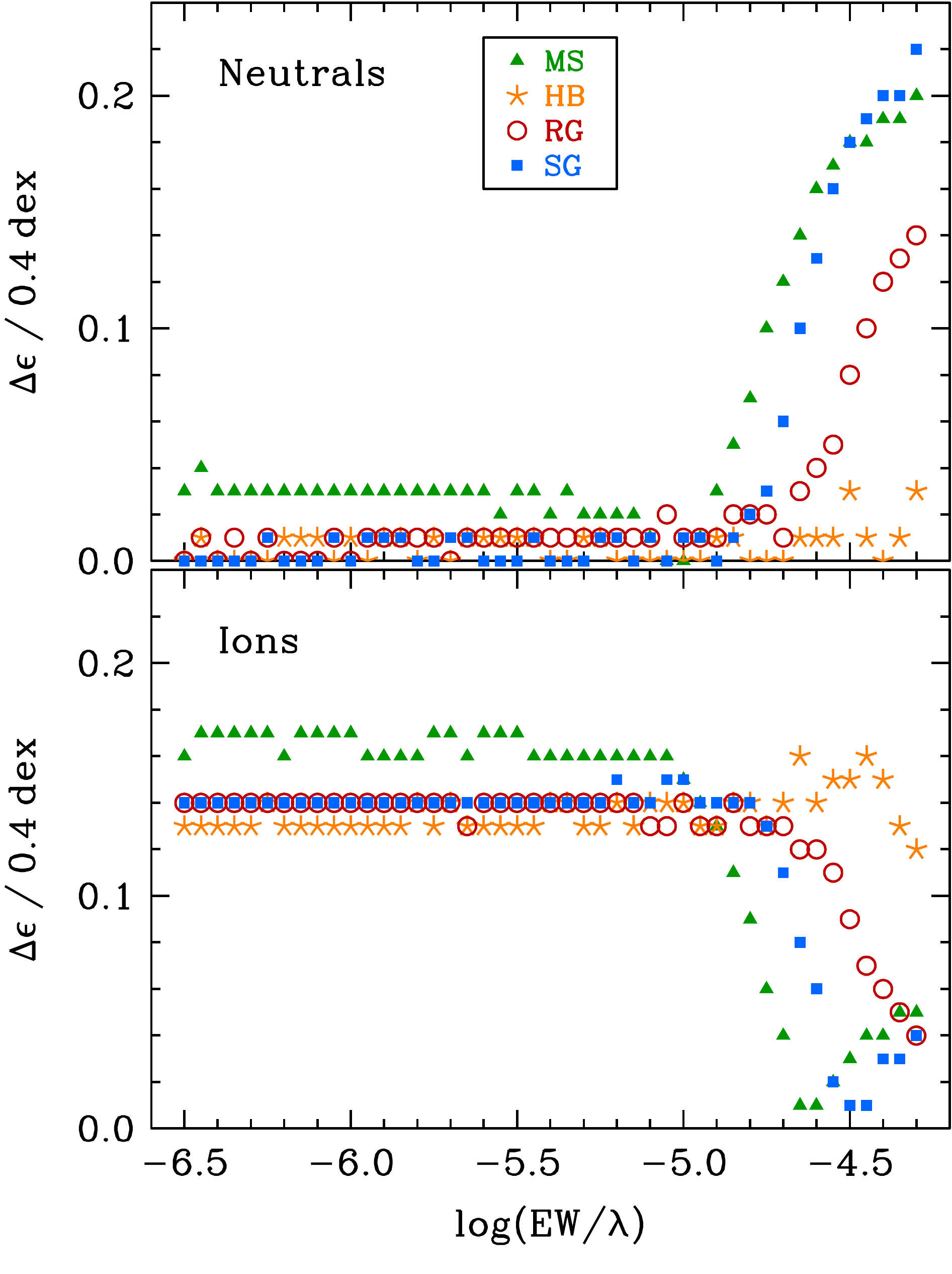}
\end{center}
\caption{
\label{deltaabundlogg}
Change in derived abundances of neutral lines (top)
and singly-ionized lines (bottom)
resulting from a change in \logg\ of 0.4~dex.
Symbols are the same as in Figure~\ref{deltaabundteff}.
}
\end{figure}

\begin{figure}
\begin{center}
\includegraphics[angle=0,width=3.35in]{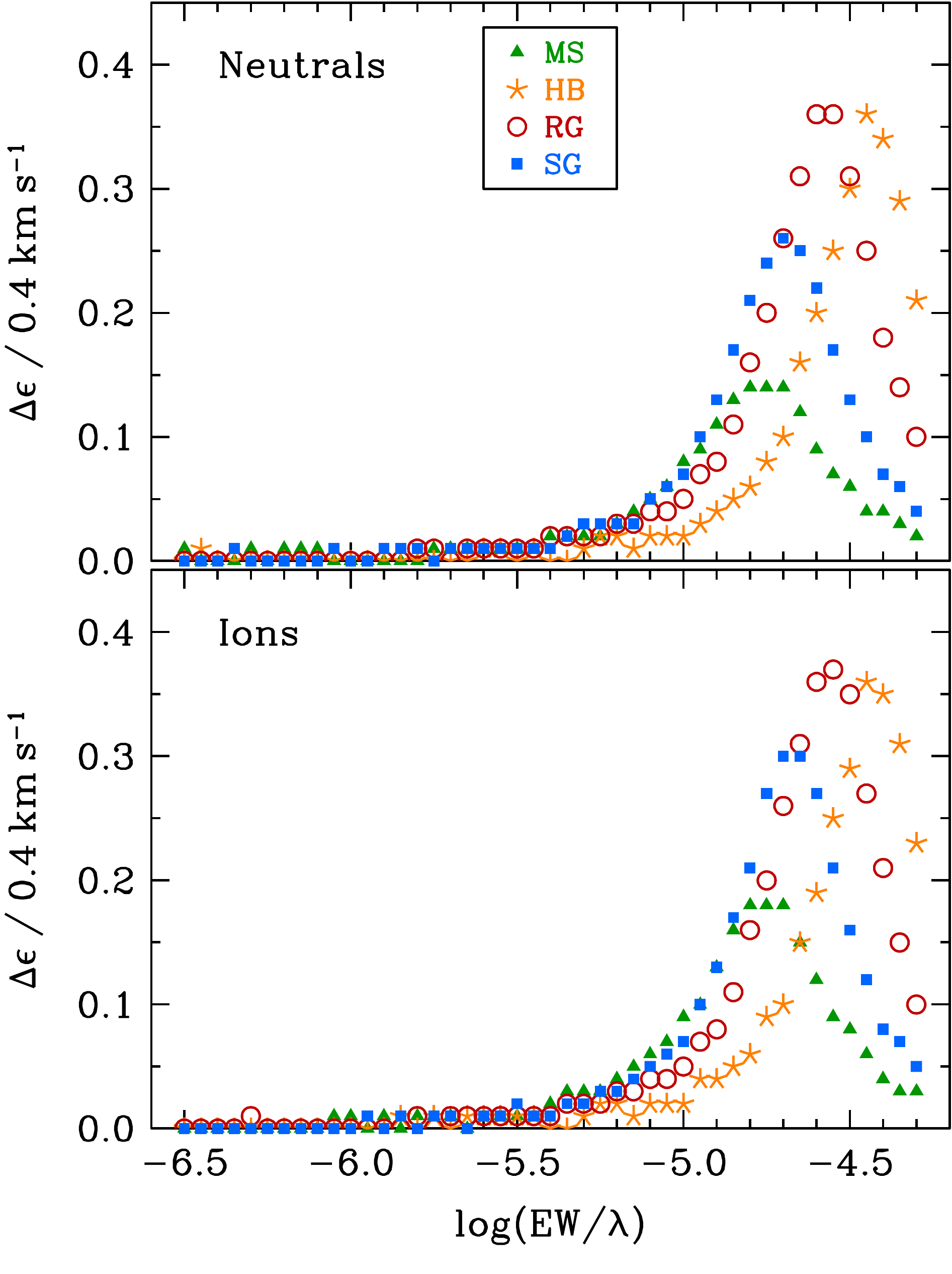}
\end{center}
\caption{
\label{deltaabundvt}
Change in derived abundances of neutral lines (top)
and singly-ionized lines (bottom)
resulting from a change in \vt\ of 0.4~\kmsec.
Symbols are the same as in Figure~\ref{deltaabundteff}.
}
\end{figure}

\begin{figure}
\begin{center}
\includegraphics[angle=0,width=3.35in]{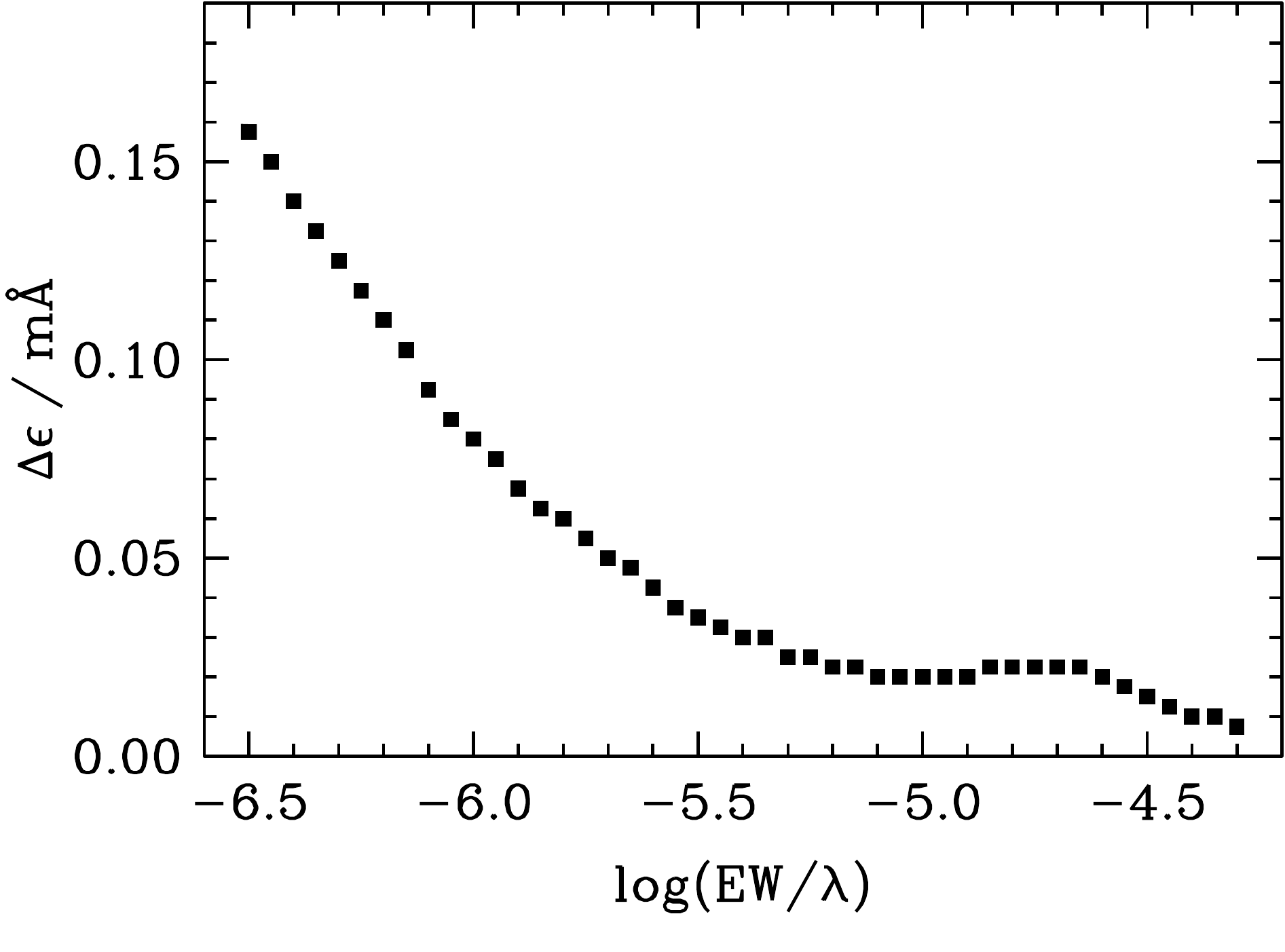}
\end{center}
\caption{
\label{deltaabundew}
Change in derived abundances of lines of various strength
resulting from a change in EW of 1~m\AA.
}
\end{figure}

The results of this exercise are illustrated in
Figures~\ref{deltaabundteff} through \ref{deltaabundvt}.
As expected, lines of neutral atoms are more sensitive to \teff\ than 
lines of ionized atoms are, and the opposite is
true for pressure sensitivity (\logg).
For strong lines, the microturbulence velocity parameter
dominates the uncertainties.
In practice, we fit the relationships shown in 
Figures~\ref{deltaabundteff} through \ref{deltaabundvt}
by polynomial functions (often just constants over much of
the range of line strength) and use these in our calculations.
The uncertainty in abundance resulting from 
uncertainty in the EW measurement
is estimated in a similar manner by altering the line strength
by 1~m\AA, as shown in Figure~\ref{deltaabundew}.
As expected, this corresponds to a proportionally 
larger uncertainty for weaker lines.
In practice, we adopt a wavelength-dependent uncertainty
for the EW based on 
the median S/N ratios. 

The cross term for \teff\ and \logg\ 
in equation A5 of \citet{mcwilliam95b} is
evaluated using the procedure outlined by \citet{johnson02}.
For each of our model atmospheres of representative stars,
we compare the \logg\ parameters derived when altering 
the input \teff\ by an amount corresponding
to a random draw of \teff\ from a normal distribution with 
$\sigma = \sigma_{\teff}$, 
where $\sigma_{\teff}$ is typical for stars in each evolutionary state.
We repeat this exercise 10~times for each covariance
for a representative case for each evolutionary state.
The covariance, $\sigma_{T \log\,g}$, is then estimated 
according to equation~3 of \citet{johnson02}.
For stars in the RG, SG, HB, and MS classes,
$\sigma_{T~\log\,g}$ is 5.1, 1.2, 2.8, and 2.2; 
$\sigma_{\log\,g~vt}$ is $-$0.04, $-$0.02, $-$0.03, and $-$0.01; and
$\sigma_{vt~T}$ is 44, 38, 45, and 16. 
The 1$\sigma$
uncertainties are listed for each line in each 
star in Table~\ref{lineabundtab}.

We compute mean abundances weighted by the 
uncertainty given by equation~A5 of \citet{mcwilliam95b}.
These abundances are reported in Table~\ref{finalabundtab}.
Several sets of uncertainties are listed in this table.
The statistical uncertainty, $\sigma_{\rm statistical}$, 
is that given by equation~A17 of \citeauthor{mcwilliam95b},
which includes uncertainties in the EW measurement and 
\loggf\ values.
This uncertainty generally decreases as the number of
lines examined increases, although we have forced
an artificial minimum uncertainty of 0.02~dex to
guard against unreasonably small values.
The total uncertainty, $\sigma_{\rm total}$, 
is that given by equation~A16 of \citeauthor{mcwilliam95b}~
This includes the statistical uncertainty and 
uncertainties in the model atmosphere parameters
and does not decrease appreciably as the number of lines
increases.
For this calculation,
we adopt the estimates of the systematic uncertainties
in the model atmosphere parameters given in Section~\ref{atmsystematic}.
The remaining two uncertainties listed in Table~\ref{finalabundtab}
are approximations to the abundance ratio uncertainties
given by equations~A19 and A20 of \citeauthor{mcwilliam95b}~
Rather than calculating and presenting 
this uncertainty for every possible element pair,
we have computed these uncertainties using Fe~\textsc{i} and 
Fe~\textsc{ii} as representative cases.
We suggest that $\sigma_{\rm neutrals}$ for element A 
should be added in quadrature with $\sigma_{\rm statistical}$ for
element B when computing the ratio [A/B] when B is 
derived from neutral lines.
Similarly, we suggest that $\sigma_{\rm ions}$ for element A 
should be added in quadrature with $\sigma_{\rm statistical}$ for
element B when element B is derived from ionized lines.
(The uncertainty in the ratio [B/A] may not 
necessarily equal that for [A/B], but 
in general they are comparable.)

\subsection{Corrections for Departures from LTE 
in the Line Formation Calculations}
\label{nlte}

Great effort
in recent years has been dedicated
to identifying transitions that are not
well represented by the assumptions of LTE
in late-type stellar atmospheres.
We adopt non-LTE corrections for the abundances derived
from lines of Li~\textsc{i}, O~\textsc{i},
Na~\textsc{i}, and K~\textsc{i}.
Grids of non-LTE calculations 
spanning a range of stellar parameters and metallicities
have been presented for various lines of these species,
as discussed below.
The abundances presented in Tables~\ref{lineabundtab} and 
\ref{finalabundtab} reflect
these corrections on Li~\textsc{i}, O~\textsc{i},
Na~\textsc{i}, and K~\textsc{i}.
For investigators who wish to make use of our uncorrected LTE
abundances of these species, 
we list the corrections in Table~\ref{nltecorr}.
The complete version of this table is available only in the online edition
of the journal.
Lines of other species may not be formed in LTE,
but grids of non-LTE abundance calculations 
for these species in late-type stars are not readily available.

\begin{deluxetable*}{cccccccccccccc}
\tablecaption{Non-LTE Abundance Corrections
\label{nltecorr}}
\tablewidth{0pt}
\tabletypesize{\scriptsize}
\tablehead{
\colhead{} &
\colhead{Li~\textsc{i}} &
\colhead{} &
\multicolumn{3}{c}{O~\textsc{i}} &
\colhead{} &
\multicolumn{4}{c}{Na~\textsc{i}} &
\colhead{} &
\multicolumn{2}{c}{K~\textsc{i}} \\
\cline{2-2} \cline{4-6} \cline{8-11} \cline{13-14}
\colhead{Star} &
\colhead{6707} &
\colhead{} &
\colhead{7771} &
\colhead{7774} &
\colhead{7775} &
\colhead{} &
\colhead{5682} &
\colhead{5688} &
\colhead{5889} &
\colhead{5895} &
\colhead{} &
\colhead{7664} &
\colhead{7698} }
\startdata      
 CS~22166--016    & $+$0.08 & & \nodata & \nodata & \nodata & & \nodata & \nodata & \nodata & \nodata & & \nodata & \nodata \\
 CS~22169--008    & $-$0.03 & & $-$0.22 & $-$0.22 & \nodata & & \nodata & \nodata & \nodata & \nodata & & \nodata & \nodata \\
 CS~22169--035    & $+$0.14 & & $-$0.05 & $-$0.05 & $-$0.05 & & \nodata & \nodata & \nodata & \nodata & & \nodata & $-$0.27 \\
 CS~22171--031    & $-$0.04 & & $-$0.26 & $-$0.26 & $-$0.25 & & \nodata & \nodata & \nodata & \nodata & & \nodata & \nodata \\
 CS~22171--037    & $-$0.03 & & $-$0.46 & $-$0.45 & $-$0.45 & & \nodata & \nodata & $-$0.05 & $-$0.05 & & \nodata & \nodata \\
\enddata
\tablecomments{
The complete version of Table~\ref{nltecorr} is available online only.  
A short version is shown here to illustrate its form and content.
 }
\end{deluxetable*}

For the Li~\textsc{i} 6707~\AA\ line, we use the corrections
computed by \citet{lind09}.
For the few stars with parameters outside the grid of their calculations 
(4000~$\leq$~\teff~$\leq$~8000~K, 
1.0~$\leq$~\logg~$\leq$~5.0,
1.0~$\leq$~\vt~$\leq$~5.0 \kmsec,
$-$5.0~$\leq$~[Fe/H]~$\leq$~0),
we use the value at the nearest point on the grid.
The corrections are generally positive for cool stars and 
negative for warm stars.
The corrections for upper limits 
are calculated based on the 3$\sigma$ EW estimated 
from the S/N measurements, 
so they should be treated with caution.
Most corrections are small, $< \pm$~0.05~dex, but a few
are as large as $\pm$~0.13~dex for lines detected.
We include an additional 0.05~dex statistical uncertainty
in our error estimates to account for uncertainties
in the corrections.

The O~\textsc{i} triplet at 7771, 7774, and 7775~\AA\ 
is not formed in LTE, and we adopt the corrections
presented by \citet{fabbian09}.
For the few stars with parameters outside the grid of their calculations 
(4500~$\leq$~\teff~$\leq$~6500~K, 
2.0~$\leq$~\logg~$\leq$~5.0,
$-$3.0~$\leq$~[Fe/H]~$\leq$~0),
we use the value at the nearest point on the grid.
We include an additional 0.1~dex statistical uncertainty
in our error estimates to account for uncertainties
in the corrections.
After correcting these abundances, however, the
resulting oxygen abundances are, on average,
0.50~$\pm$~0.06~dex ($\sigma =$~0.25~dex)
higher than those derived from the 
[O~\textsc{i}]~6300~\AA\ line
in the seven stars where both
abundance indicators could be reliably measured.
The [O~\textsc{i}]~6300~\AA\ line
is generally considered to be a reliable abundance
indicator formed under conditions of LTE
\citep{kiselman01}.
The offset between abundances derived
from [O~\textsc{i}] and O~\textsc{i} is 
reminiscent of the result found
by \citet{garciaperez06} in metal-poor giants.
We apply a correction for this offset as discussed
in Section~\ref{offsets}.

The lines of the Na~\textsc{i} D resonance doublet 
at 5898 and 5895~\AA\ are not formed under conditions of LTE.
We adopt corrections to our LTE abundances 
using the grid presented by \citet{lind11}.
The corrections are always negative, in the sense that 
LTE underestimates the line strength and overestimates the abundance.
\citeauthor{lind11}\ also present corrections for 
the higher excitation Na~\textsc{i} 5682 and 5688~\AA\ lines.
We also include these corrections, which are generally small
($<$~0.1~dex) for consistency.
For the few stars with parameters outside the grid of their calculations 
(4000~$\leq$~\teff~$\leq$~8000~K, 
1.0~$\leq$~\logg~$\leq$~5.0,
1.0~$\leq$~\vt~$\leq$~5.0 \kmsec,
$-$5.0~$\leq$~[Fe/H]~$\leq +$0.5),
we use the value at the nearest point on the grid.
We include an additional 0.1~dex statistical uncertainty
in our error estimates to account for uncertainties
in the corrections.

Observational challenges have limited studies
of potassium in metal-poor stars.
Only two K~\textsc{i} lines are routinely detectable
in late-type metal-poor stars, the 
7664 and 7698~\AA\ K~\textsc{i} resonance doublet.
These stellar lines are often contaminated with atmospheric 
O$_{2}$ lines.
Our spectra extend redward enough to observe these lines.
As in the case of the Na~\textsc{i} resonance doublet
discussed in Section~\ref{abundances}, 
we only measure EWs of these lines
when one or both appears well separated from 
the model telluric absorption spectrum.
These resonance lines are likely formed out of LTE,
and we adopt corrections for the 7698~\AA\ line
from the grid of \citet{takeda02}.
Corrections for the 7664~\AA\ line are made from
an analogous grid kindly sent by 
Y.\ Takeda (2007, private communication).
The corrections are almost always negative, in the sense that
LTE underestimates the line strength and overestimates the abundance.
For stars with parameters outside the grid of their calculations 
(4500~$\leq$~\teff~$\leq$~6500~K, 
1.0~$\leq$~\logg~$\leq$~5.0,
1.0~$\leq$~\vt~$\leq$~3.0 \kmsec,
$-$3.0~$\leq$~[Fe/H]~$\leq$~0),
we use the value at the nearest point on the grid.
We report detections of the K~\textsc{i} 7664 and 7698~\AA\
lines in 41~stars and 72~stars, respectively,
for a total of 98~stars with potassium abundance derivations.
Both lines are detected in 15~stars, and 
the corrected abundances agree well in these stars:\ 
$\Delta = -$0.019~$\pm$~0.024 ($\sigma =$~0.093).
We include an additional 0.1~dex statistical uncertainty
in our error estimates to account for uncertainties
in the corrections.

\subsection{Line-by-line Abundance Offsets and Corrections}
\label{offsets}

One challenge in producing a homogeneous abundance dataset
for stars spanning several dex in metallicity
is that the set of useful lines for analysis changes
from metal-poor to metal-rich stars.
If systematic line-to-line differences in the derived abundances persist,
they will masquerade as subtle changes in the abundance trends.
While observers are generally aware of this effect
(e.g., \citealt{cohen04} discuss it in detail for the case of
Mg~\textsc{i} lines\footnote{
The line-by-line differences for five Mg~\textsc{i} lines
listed in Table~12 of \citet{cohen04} agree in sign but not magnitude
with ours in Table~\ref{offsettab}
after correcting for the different \loggf\ values 
and number of lines examined.
Their corrections would range from
$+$0.16~dex for the 4057~\AA\ line
to $-$0.23~dex for the 5183~\AA\ line
if computed on the same \loggf\ scale and in the same manner as ours.
We cannot trace the source of this difference.
}),
limitations in the size of the stellar sample often
preclude attempts to characterize it reliably
for large numbers of elements.
This is of little consolation to those wishing to make use
of abundance tables to constrain chemical evolution models.
We attempt to account for these effects by leveraging
our large dataset to identify and correct for lines that are systematically
discrepant.
In this section we discuss the process we use to identify those lines
and the empirical corrections that we apply to the abundances.

These differences may originate from 
inaccurate \loggf\ values or damping constants,
departures from LTE in the 
line-forming layers, poor estimates of the
continuous opacity, misidentification of the continuum,
or unidentified blends.
We can estimate which effects may dominate.
The line density is highest in the blue region of the spectrum.
If unidentified blends are the dominant source,
the abundances we derive from blue lines
should be higher on average than
abundances derived from red lines, so
the corrections would preferentially skew negative for blue lines.
The corrections are both positive and negative
in the blue region of the spectrum.
This indicates that unidentified blends are not the 
main source of systematic uncertainty here.
The magnitude of the corrections is higher
in the blue than in the red, however.
This suggests that 
higher line density and lower S/N ratios
are rendering the continuum placement more uncertain.
Simply adopting different analysis techniques---for example, using
spectrum synthesis in place of 
a traditional EW analysis---does not offer a panacea.
The dominant effects are likely to be 
uncertain continuum placement in the blue region of the spectrum and 
the deficiencies of using LTE to model the line formation.

For most species, we determine the corrections using the
following process.
We separate the stars into groups of different evolutionary 
classes (the RG, SG, HB, MS, and BS categories 
presented in Section~\ref{modelmethod}).
For each star in each group,
using only stars with large numbers of lines
of a given species measured,
we compute the difference between each line's resulting abundance 
and the mean abundance in that star.
Within each evolutionary group, 
we then compute the mean and standard deviation
of these differences.
Finally, we correct all abundances from a given line
in all stars of a given evolutionary group by 
subtracting these mean differences.
These corrections are listed in Table~\ref{offsettab}.
For example, we find that the Mg~\textsc{i} 5183~\AA\ line
in the SG class yields
abundances higher than the mean by 0.09~dex ($\sigma =$~0.08, 
N~$=$~49~stars).
Therefore
we reduce the abundance derived from the 5183~\AA\ line
by 0.09~dex for all stars in the SG class.
As a penalty incurred for making this statistical correction,
we add in quadrature the standard deviation, in this case 0.08~dex,
with the statistical uncertainty. 
Typically these uncertainties are
not the dominant component of the statistical error budget.
We apply all corrections, even if some are
not statistically meaningful, to preserve the overall
mean abundance of the sample.
Some evolutionary groups do not contain enough measurements
of a particular line to make a reliable assessment of
the mean offset, in which case no correction is made.
These cases are denoted by blanks in Table~\ref{offsettab}.
A few species present unique challenges that 
require minor modifications to this process, as discussed below.

We use four sodium abundance indicators in our analysis,
the Na~\textsc{i} resonance doublet at 5889 and 5895~\AA\ and the
higher excitation doublet at 5682 and 5688~\AA.
After correcting for non-LTE effects,
the two lines within each doublet yield
abundances in excellent agreement with each other on average:\
$\Delta_{\rm 5889 - 5895} = -$0.02~$\pm$~0.02 ($\sigma =$~0.06) and
$\Delta_{\rm 5688 - 5682} = -$0.01~$\pm$~0.03 ($\sigma =$~0.13).
There are only two stars where three or four of these lines 
could be measured reliably, 
so we are unable to assess whether they differ systematically.
Furthermore, there are insufficient data to assess
systematic differences among stars of different evolutionary state,
so we adopt a single set of corrections for all stars.

Previous studies 
(e.g., \citealt{preston06})
have revealed relationships between
[Si/Fe] and \teff\ when
the silicon abundance is derived from the 
Si~\textsc{i} 3905~\AA\ line.
When possible, we avoid using this line as an abundance indicator
and instead prefer the high excitation (4.90~$\leq$~EP~$\leq$~5.06~eV)
Si~\textsc{i} lines at
5665, 5701, 5708, and 5772~\AA.
Our analysis reveals that
the low excitation Si~\textsc{i} 3905 and 4102~\AA\ lines
(EP~$=$~1.91~eV)
give consistent results in the 
three stars where both can be reliably measured
($\Delta =$~0.00~$\pm$~0.04, $\sigma =$~0.07).
In the nine stars with at least one high excitation line
and at least one low excitation line used as abundance indicators,
the high excitation lines yield abundances higher by
0.14~$\pm$~0.05~dex ($\sigma =$~0.15) on average.
Here, we adopt the convention to correct the abundances 
of the low excitation lines in all stars
by $+$0.14~dex to match the average abundance of the high excitation lines.
We do not include the low excitation lines in
the reported mean silicon abundance for the 12~stars where
at least one high excitation line is also used.
This correction accounts for the higher silicon abundances
reported for the more metal-poor stars in our sample
when compared with other recent investigations.

Previous studies of Mn~\textsc{i} lines in late-type stars
have demonstrated that the 
three Mn~\textsc{i} resonance lines at 4030, 4033, and 4034~\AA\
yield abundances in LTE that are several tenths of a dex lower
than the high-excitation neutral or singly-ionized lines
(e.g., \citealt{cayrel04,roederer10}).
We thus take the following approach to identifying line-by-line
systematic offsets in our dataset.
First, we identify any offsets among the high-excitation 
neutral lines from the mean of all high-excitation neutral lines
within a given star.
Then, we recompute the mean manganese abundance derived
from the corrected high-excitation lines for each star.
Finally, we identify any offsets among the neutral resonance lines
relative to the corrected mean.
This forces the resonance lines, on average, to have the
same mean abundance as the high-excitation lines.
Fortuitously, the mean [Mn/Fe] ratio derived from 
neutral lines shows only a small difference from the
mean [Mn/Fe] ratio derived from the ionized lines,
with a mean difference (ion minus neutral) of only 
$-$0.038~$\pm$~0.011 ($\sigma =$~0.14).
This level of agreement, not enforced by our method, is encouraging.

For Cu~\textsc{i}, Tb~\textsc{ii}, Tm~\textsc{ii},
Ir~\textsc{i}, and Th~\textsc{ii}, there
are not enough stars with two or more lines measured
to assess systematic offsets.
For Co~\textsc{i}, Y~\textsc{ii}, Ba~\textsc{ii}, and La~\textsc{ii},
there are not enough stars with two or more lines measured
to assess systematic offsets in the MS class;
in these cases, we adopt the corrections from stars
in the SG class for the stars in the MS class.
For Co~\textsc{i}, a similar situation exists for stars
in the HB class, and we also adopt the corrections from
the stars in the SG class for stars in the HB class.
For K~\textsc{i}, Ce~\textsc{ii}, Pr~\textsc{ii}, Nd~\textsc{ii},
Sm~\textsc{ii}, Eu~\textsc{ii}, Gd~\textsc{ii}, Dy~\textsc{ii}, 
and Er~\textsc{ii} we adopt a single correction
for stars in all evolutionary states.
There are never enough stars in the BS class to 
define a separate set of corrections,
so the abundances in these stars are corrected by
adopting the offsets found for the stars in the SG class.

\subsection{Comparison with Previous Studies}
\label{comparestars}

\subsubsection{Differences in the [X/Fe] Ratios}

We compare our derived [X/Fe] ratios, where X is a given element, 
with those derived by previous studies.
In particular, we focus on the 18~red giants in common
with the First Stars analysis.
To keep the comparisons manageable when comparing with
other studies, we limit ourselves to two other studies
that also examined large numbers of stars in common with the
First Stars analysis, those of \citet{mcwilliam95b} and \citet{yong13}.
Note that
\citeauthor{yong13}\ rederived abundances of these stars
from published EW values.
We also compare with the \citet{bonifacio09} sample of dwarfs.
We perform a straight comparison of the results without accounting 
for differences in, e.g., the set of lines used or the
transition probabilities.

\begin{deluxetable*}{ccccc}
\tablecaption{Comparison of [X/Fe] Ratios with Previous Studies
\label{xfetab}}
\tablewidth{0pt}
\tabletypesize{\scriptsize}
\tablehead{
\colhead{Ratio} & 
\colhead{\citet{mcwilliam95b}\tablenotemark{a}} &
\colhead{\citet{cayrel04}\tablenotemark{b}} &
\colhead{\citet{yong13}} &
\colhead{\citet{bonifacio09}} }
\startdata   
~[C/Fe]              & $-$0.05 (0.18, 12) & $-$0.14 (0.20, 17) & \nodata            & $+$0.22 (0.15, 5)  \\
~[N/Fe]              & \nodata            & $-$0.12 (0.23, 13) & \nodata            & \nodata            \\
~[Na/Fe]             & $+$0.15 (0.13, 5)  & $+$0.08 (0.19, 6)  & $-$0.07 (0.21, 6)  & \nodata            \\
~[Mg/Fe]             & $+$0.23 (0.18, 14) & $+$0.26 (0.14, 18) & $+$0.28 (0.13, 18) & $+$0.17 (0.08, 9)  \\
~[Al/Fe]             & $+$0.01 (0.25, 14) & $+$0.16 (0.14, 18) & $+$0.08 (0.15, 18) & \nodata            \\
~[Si/Fe]             & $+$0.33 (0.29, 11) & $+$0.48 (0.20, 15) & \nodata            & $+$0.26 (0.12, 9)  \\
~[Ca/Fe]             & $+$0.04 (0.10, 14) & $+$0.15 (0.09, 18) & $+$0.16 (0.07, 18) & $+$0.14 (0.05, 9)  \\
~[Sc/Fe]             & $-$0.28 (0.25, 14) & $-$0.19 (0.08, 18) & $-$0.30 (0.11, 18) & $-$0.25 (0.20, 9)  \\
~[Ti~\textsc{i}/Fe]  & $-$0.11 (0.21, 12) & $-$0.05 (0.07, 18) & $-$0.10 (0.08, 18) & \nodata            \\
~[Ti~\textsc{ii}/Fe] & $-$0.16 (0.13, 14) & $-$0.07 (0.10, 18) & $-$0.11 (0.18, 18) & $-$0.23 (0.18, 9)  \\
~[Cr~\textsc{i}/Fe]  & $+$0.12 (0.15, 14) & $+$0.11 (0.08, 18) & $+$0.11 (0.10, 18) & $+$0.01 (0.10, 9)  \\
~[Mn~\textsc{i}/Fe]  & $+$0.22 (0.30, 13) & $-$0.08 (0.12, 18) & $+$0.05 (0.21, 8)  & $+$0.22 (0.11, 8)  \\
~[Fe~\textsc{i}/H]   & $-$0.41 (0.14, 14) & $-$0.31 (0.12, 18) & $-$0.31 (0.13, 18) & $-$0.27 (0.15, 9)  \\
~[Fe~\textsc{ii}/H]  & $-$0.27 (0.14, 14) & $-$0.20 (0.13, 18) & $-$0.21 (0.13, 18) & $-$0.10 (0.13, 9)  \\
~[Co/Fe]             & $-$0.28 (0.11, 14) & $-$0.29 (0.13, 18) & $-$0.16 (0.09, 18) & $-$0.15 (0.12, 7)  \\
~[Ni/Fe]             & $-$0.23 (0.31, 13) & $+$0.12 (0.10, 18) & $+$0.07 (0.11, 18) & $+$0.05 (0.11, 9)  \\
~[Zn/Fe]             & \nodata            & $+$0.21 (0.12, 16) & \nodata            & \nodata            \\
~[Sr/Fe]             & $-$0.06 (0.25, 14) & $-$0.03 (0.20, 17) & $+$0.07 (0.45, 7)  & $-$0.32 (0.45, 9)  \\
~[Ba/Fe]             & $+$0.04 (0.25, 11) & $-$0.24 (0.18, 16) & \nodata            & $+$0.01 (0.52, 3)  \\
\enddata
\tablecomments{
Differences are in the sense of this study 
\textit{minus} other.
Each entry represents the mean difference, 
standard deviation, and number of stars:\ 
$\langle\Delta\rangle$ ($\sigma$, $N$).
}
\tablenotetext{a}{Includes [Ba/Fe] from \citet{mcwilliam98}}
\tablenotetext{b}{Includes [C/Fe] and [N/Fe] from \citet{spite05} and
[Sr/Fe] and [Ba/Fe] from \citet{francois07}}
\end{deluxetable*}

The results of this comparison are listed in Table~\ref{xfetab}.
The [X/Fe] ratios, many derived from transitions in neutral atoms,
are generally overabundant in our analysis.
This can easily be traced to the fact that our 
iron abundances derived from Fe~\textsc{i} lines are 
$\sim$~0.1~dex lower than the iron abundance derived from
Fe~\textsc{ii} (Figure~\ref{delfeplot}), 
which are themselves lower than published [Fe/H] values
by 0.27~dex, on average, for red giants (Table~\ref{comparetab}).
A lower [Fe/H] value increases [X/Fe]
if element X is not also derived from lines of similar 
excitation potential and strength.
Ratios of neutral elements with the largest positive discrepancies, 
[Mg/Fe] and [Si/Fe], are often derived from (only) strong
Mg~\textsc{i} and Si~\textsc{i} 
lines that are sensitive to the
microturbulent velocity, and our derived \vt\ values
are lower than those derived by
previous studies (Table~\ref{comparetab}).
This drives the [Mg/Fe] and [Si/Fe] ratios to even higher values.
Our line-by-line corrections (Table~\ref{offsettab})
further increase [Si/Fe] when the
3905~\AA\ Si~\textsc{i} line is considered.
The underabundances of [Sc/Fe] and [Co/Fe]
with respect to \citet{cayrel04} can be accounted for
by our inclusion of the hyperfine structure for 
Co~\textsc{i} and Sc~\textsc{ii} lines, which desaturate 
the lines and lower the derived abundances.
In principle, additional discrepancies may arise
from differences in the transition probabilities,
although we have not checked this explicitly
since doing so would require examining
which sets of lines of each species
were employed in each star by each analysis.

\subsubsection{Dispersion in the [X/Fe] Ratios}

\begin{deluxetable*}{ccccc}
\tablecaption{Data Quality and Dispersions in Abundance Ratios 
for 18 Stars in Common among Four Large Surveys
\label{dispersiontab}}
\tablewidth{0pt}
\tabletypesize{\scriptsize}
\tablehead{
\colhead{} &
\colhead{This study} &
\colhead{First Stars} &
\colhead{\citet{yong13}} &
\colhead{\citet{mcwilliam95a,mcwilliam95b}}
}
\startdata
Resolution $\equiv \lambda/\Delta\lambda$ & 41,000 & 47,000 & \ldots & 22,000 \\
$\lambda$ of S/N estimate & 3950~\AA\ & 4000~\AA\ & \ldots & 4800~\AA\ \\
median S/N pix$^{-1}$     & 72        & 150       & \ldots & 35        \\
median S/N RE$^{-1}$      & 112       & 335       & \ldots & 67        \\
median photons \AA$^{-1}$    & 1.3$\times$10$^{5}$ & 1.3$\times$10$^{6}$ & \ldots & 2.0$\times$10$^{4}$ \\
\hline
  &
 $\sigma$ (N) &
 $\sigma$ (N) &
 $\sigma$ (N) &
 $\sigma$ (N) \\
\hline
~[C/Fe]              & 0.54 (17) & 0.49 (17) & \ldots    & 0.48 (12) \\
~[N/Fe]              & 0.78 (12) & 0.70 (15) & \ldots    & \ldots    \\
~[Na/Fe]             & 0.15  (6) & 0.44 (14) & 0.55 (15) & 0.64 (13) \\
~[Mg/Fe]             & 0.28 (18) & 0.33 (18) & 0.31 (18) & 0.31 (14) \\
~[Al/Fe]             & 0.27 (18) & 0.21 (18) & 0.25 (18) & 0.38 (14) \\
~[Si/Fe]             & 0.25 (15) & 0.17 (18) & \ldots    & 0.28 (14) \\
~[Ca/Fe]             & 0.13 (18) & 0.12 (18) & 0.12 (18) & 0.18 (14) \\
~[Sc/Fe]             & 0.12 (18) & 0.12 (18) & 0.16 (18) & 0.26 (14) \\
~[Ti~\textsc{i}/Fe]  & 0.17 (18) & 0.11 (18) & 0.14 (18) & 0.26 (12) \\
~[Ti~\textsc{ii}/Fe] & 0.15 (18) & 0.13 (18) & 0.21 (18) & 0.14 (14) \\
~[Cr/Fe]             & 0.08 (18) & 0.06 (18) & 0.06 (18) & 0.17 (14) \\
~[Mn/Fe]             & 0.20 (18) & 0.13 (18) & 0.24  (8) & 0.24 (13) \\
~[Co/Fe]             & 0.21 (18) & 0.16 (18) & 0.21 (18) & 0.19 (14) \\
~[Ni/Fe]             & 0.13 (18) & 0.14 (18) & 0.15 (18) & 0.34 (13) \\
~[Zn/Fe]             & 0.24 (16) & 0.19 (18) & \ldots    & \ldots    \\
~[Sr/Fe]             & 0.83 (18) & 0.72 (17) & 0.47  (7) & 0.80 (14) \\
~[Ba/Fe]\tablenotemark{a}
                     & 0.85 (17) & 0.81 (17) & \ldots    & 0.71 (11) \\
\enddata
\tablenotetext{a}{Includes [Ba/Fe] from \citet{mcwilliam98}}
\end{deluxetable*}

We can also compare the standard deviation of the [X/Fe] ratios
for stars in common with the First Stars, \citet{mcwilliam95b},
and \citet{yong13} samples.
In the absence of cosmic dispersion, this comparison
offers an independent quantitative measure of the
precision of the derived abundances.
In practice, many of these abundance ratios do exhibit 
at least a small amount of cosmic dispersion, so
the standard deviations are useful in a comparative sense.
Table~\ref{dispersiontab} lists the standard deviation in each
ratio and the number of stars used to compute it
for each of the four surveys.

The First Stars survey used higher quality spectroscopic
observations than our survey or that of \citet{mcwilliam95a}.
The typical spectral resolution and S/N ratios
are also listed in Table~\ref{dispersiontab} for the stars in common 
among these three surveys.
The First Stars observations correspond to an increase of a 
factor of $\approx$~10 in total photons \AA$^{-1}$
over our data and a factor of
$>$~70 over the data of \citeauthor{mcwilliam95a} 

\begin{figure}
\begin{center}
\includegraphics[angle=0,width=3.35in]{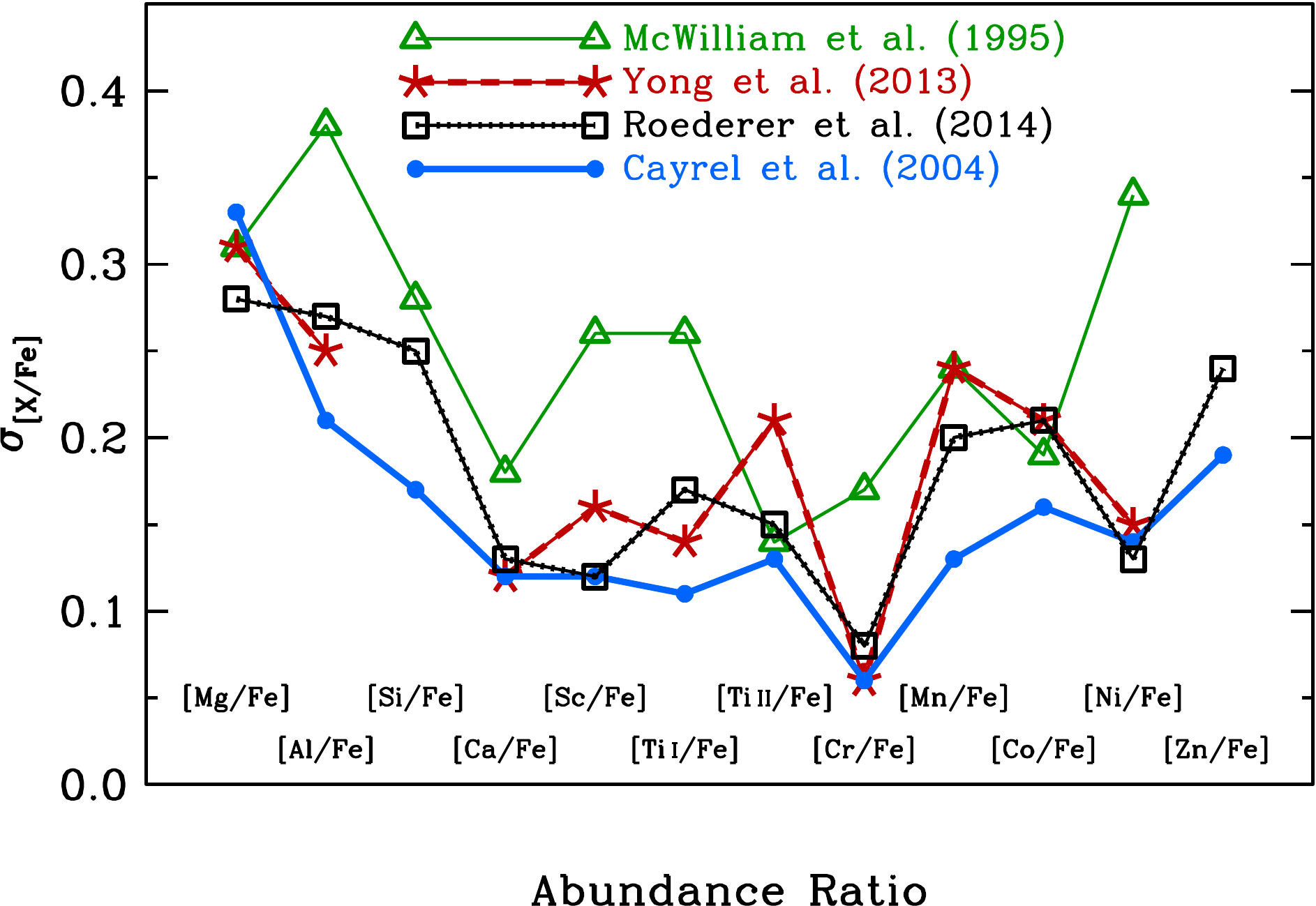}
\end{center}
\caption{
\label{dispersionplot}
Dispersion in the derived abundance ratios of stars
in common between our survey
(``Roederer et al.\ 2014''), \citet{mcwilliam95b}, 
\citet{yong13}, and \citet{cayrel04}.
The number of stars used to compute each dispersion is
listed in Table~\ref{dispersiontab}.
}
\end{figure}

Figure~\ref{dispersionplot} illustrates these results 
for the [Mg/Fe] through [Zn/Fe] ratios,
which are measured in most stars in each of these four surveys.
The na\"{i}ve expectation would be that the abundance precision
should roughly correlate with the data quality, and
the dispersions generally support this expectation.
The dispersions are generally smallest for the 
First Stars sample,
followed by the \citet{yong13} rederivation
and our sample, 
followed by the \citet{mcwilliam95b} sample.
We do not achieve the same internal precision 
as the First Stars survey does for the giants; however,
we have achieved reasonable precision
on a photon budget roughly 10~times lower per star.

\subsubsection{Detailed Comparisons of [X/Fe] for Individual Stars}

\begin{figure}
\begin{center}
\includegraphics[angle=0,width=3.35in]{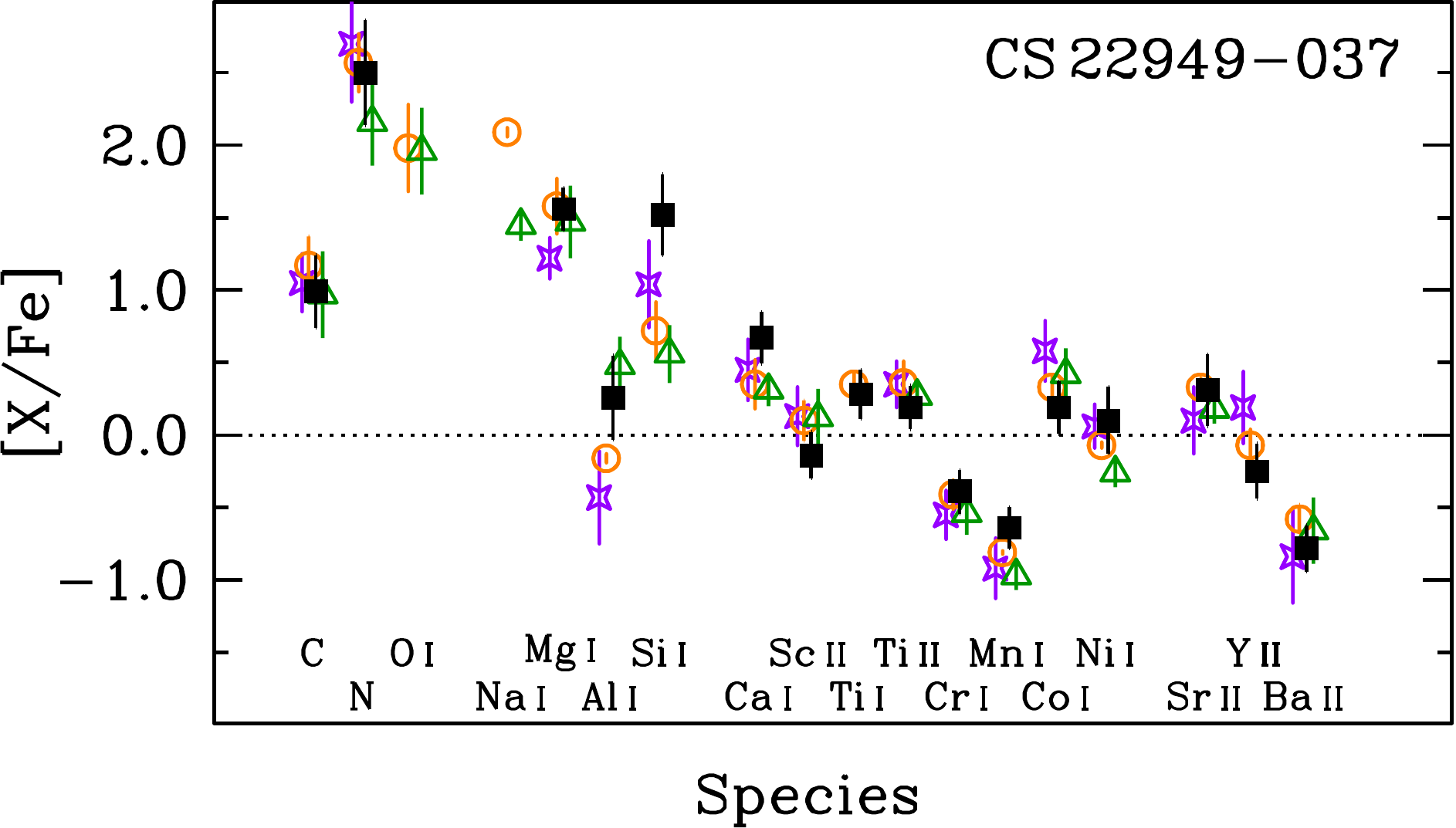} \\
\vspace*{0.1in}
\includegraphics[angle=0,width=3.35in]{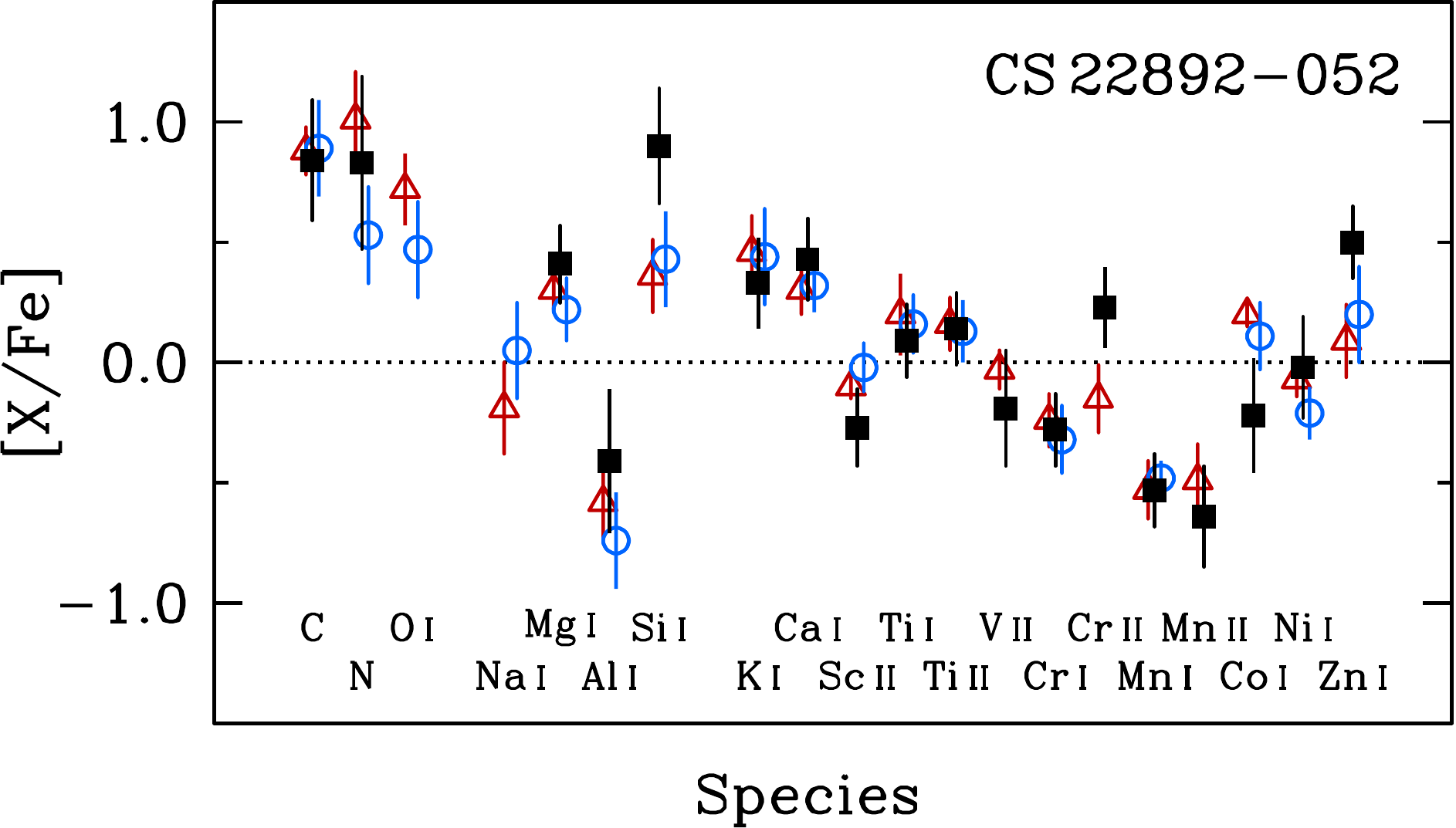} \\
\vspace*{0.1in}
\includegraphics[angle=0,width=3.35in]{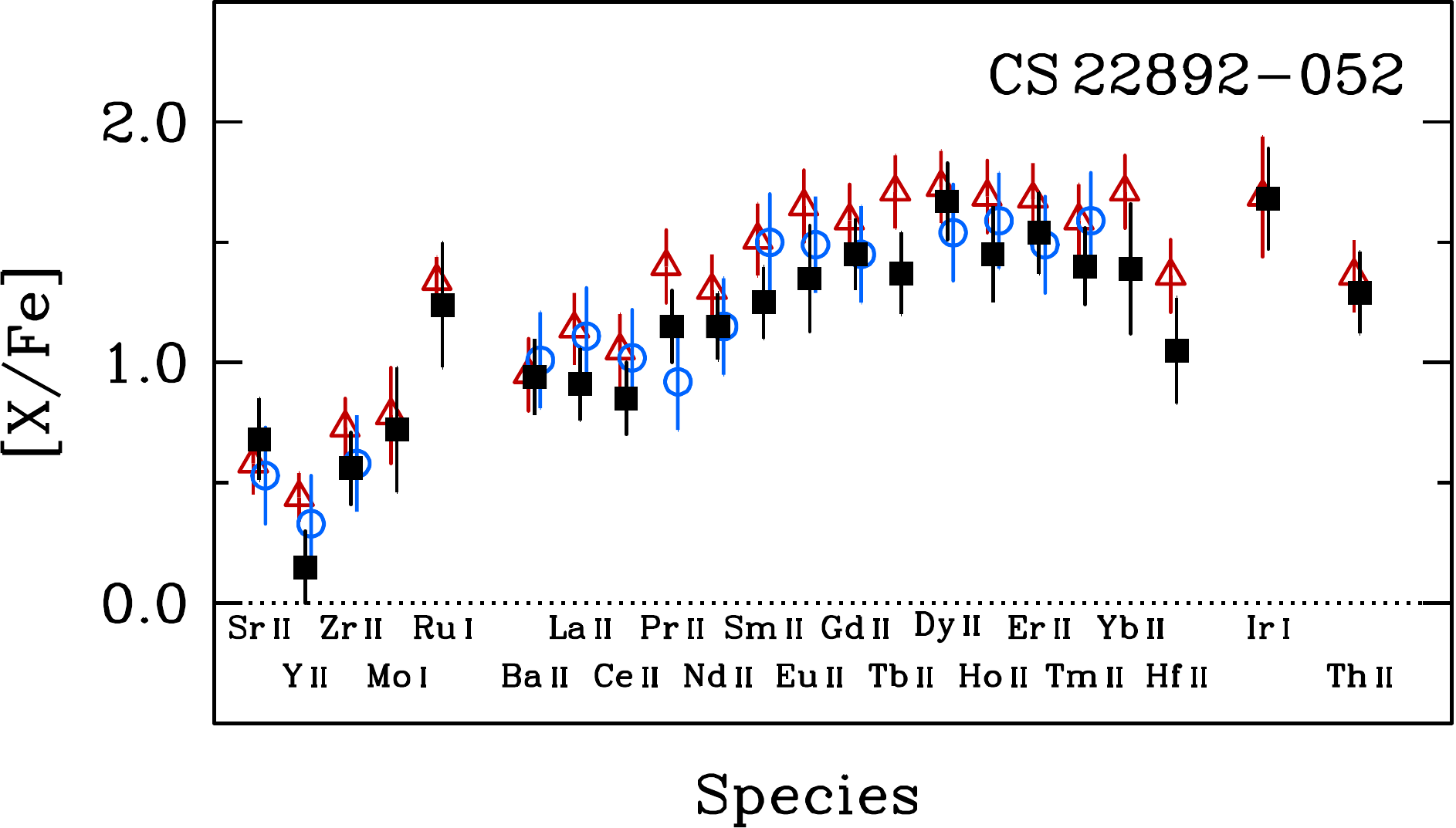} \\
\end{center}
\caption{
\label{compstar1}
Comparison of our derived [X/Fe] ratios 
with previous work
for two well-studied stars,
\mbox{CS~22949--037} and \mbox{CS~22892--052}.
In the top panel, our results (black squares)
are compared with those of
\citet{norris01} (purple stars),
\citet{depagne02} (orange circles), and 
\citet{cohen08} (green triangles).
In the middle and bottom panels, our results
are compared with those of
\citet{sneden03a,sneden09} and \citet{cowan05} (red triangles)
as well as
\citet{cayrel04} and \citet{francois07} (blue circles).
}
\end{figure}

We also compare the detailed abundance patterns 
of two well-studied stars.
\object[BPS CS 22892-052]{CS~22892--052}
is a well-studied giant with
a high level of \rpro\ enhancement.
\object[BPS CS 22949-037]{CS~22949--037}
is another well-studied giant with
substantial enhancement of carbon, nitrogen, and other light elements.
The abundance pattern of each of these stars has been subject to
close scrutiny by investigators
over the last two decades, and we
limit our comparison to a few
extensive studies of each.
Figure~\ref{compstar1} 
illustrates these comparisons, and
references are given in the figure caption.

In general the agreement is superb, but a couple of features stand out.
Our [Si/Fe] ratios are higher than found by previous work,
as discussed above.
The other prominent difference is that the [X/Fe]
ratios, where X represents one of the 
neutron-capture elements detected in its ionized state,
are lower in \object[BPS CS 22892-052]{CS~22892--052}
when compared with previous results
\citep{sneden94,sneden96,sneden03a,sneden09,mcwilliam95b,
honda04b,barklem05,francois07}.
For example, our derived [Eu/Fe] ratios are lower by
0.20~dex ($\sigma =$~0.07) on average.
Figure~\ref{compstar1} illustrates that
our derived ratios among the neutron-capture elements in
\object[BPS CS 22892-052]{CS~22892--052}
are in very good agreement with the previous studies.
In other words, there is generally a constant offset between our
[X/Fe] ratios and those of \citeauthor{sneden09}\ and 
\citeauthor{francois07} 
(The [X/Fe] ratios derived by \citeauthor{francois07}\ 
are also systematically lower than those derived by \citeauthor{sneden09}\
by about half as much as ours.)
Our results for 
\object[BPS CS 31082-001]{CS~31082--001},
another giant with a high level of \rpro\ enrichment,
exhibit a similar offset
with respect to \citet{hill02}, \citet{sneden09}, and others.
Our derived abundance ratios for well-studied giants
with sub-solar [Eu/Fe] ratios, like
\object[HD 122563]{HD~122563} and \object[HD 128279]{HD~128279},
are similarly low
(cf., e.g., \citealt{johnson02}).

The explanation for this offset is the cumulative
effect of several differences between our study
and previous ones, and we demonstrate this
using the \citet{sneden03a} analysis of
\object[BPS CS 22892-052]{CS~22892--052}.
For this test,
we adopt the model preferred by \citeauthor{sneden03a},
their Fe~\textsc{ii} EW measurements, 
their \loggf\ values,
their solar abundances,
and we only consider the four Eu~\textsc{ii} lines in common with our study.
Using the appropriate earlier version of MOOG,
we derive [Fe/H]~$= -$3.09
from Fe~\textsc{ii} lines
(identical to \citeauthor{sneden03a})\
and [Eu/Fe]~$= +$1.57 from our spectrum.
This is 0.22~dex higher than our result, and it is
in much better agreement with the value derived by
\citeauthor{sneden03a}, $+$1.64.
The remaining difference must be attributed 
to their higher-quality spectrum of 
\object[BPS CS 22892-052]{CS~22892--052}.

In summary, the abundance differences
can be attributed to different, but reasonable, choices
made during the course of the each analysis.

\subsection{Abundance Trends with Effective Temperature}
\label{tefftrends}

Figures~\ref{liteffplot} through \ref{ybteffplot} 
illustrate the relationship between the derived abundances
and \teff\ for most species examined.
Each of these figures is subdivided into four panels,
one each for the four classes of stars in our study.
Similar comparisons for Ga~\textsc{i}, Rb~\textsc{i}, 
Nb~\textsc{ii}, Mo~\textsc{i}, Tc~\textsc{i},
Ru~\textsc{i}, Sn~\textsc{i}, Ir~\textsc{i}, and Pb~\textsc{i}
are generally comprised of uninteresting upper limits.
Similar comparisons for 
most of the rare earth elements, Hf~\textsc{ii}, and Th~\textsc{ii}
closely resemble the La~\textsc{ii} and Eu~\textsc{ii}
abundances shown in Figures~\ref{lateffplot} and \ref{euteffplot}.

Many of these species show no [X/Fe] trends with \teff\
(where X stands for the element of interest).
The [Ca/Fe] ratios shown in Figure~\ref{cateffplot},
for example, demonstrate this scenario.
Other species show a false trend in that
lines of species of low abundance 
can only be detected in the coolest stars.
The [Eu/Fe] ratios shown in Figure~\ref{euteffplot},
for example, demonstrate this scenario.
Other species show a real trend 
that relates to internal mixing during the
course of normal stellar evolution.
The lithium abundances and [C/Fe] ratios
shown in Figures~\ref{liteffplot} and \ref{cteffplot},
for example, demonstrate this scenario.

Other species show genuine abundance trends with \teff,
including 
Si~\textsc{i}, Ti~\textsc{i}, Cr~\textsc{i},
Co~\textsc{i},
and---to a lesser degree---O~\textsc{i}, 
Sc~\textsc{ii}, V~\textsc{i}, V~\textsc{ii},
Mn~\textsc{i}, and Mn~\textsc{ii}.
Several of these trends are apparent 
in stars along the subgiant and red giant branches
as well as the horizontal branch,
suggesting it is not related to internal mixing
or processing.
\citet{preston06} identified such an effect for
Si~\textsc{i}.
\citet{lai08} also identified these trends in
Si~\textsc{i}, Ti~\textsc{i}, and Cr~\textsc{i},
although their study also found that
Ti~\textsc{i} and Ti~\textsc{ii} both show
a trend with \teff, while our
Ti~\textsc{ii} abundances show no such trend.

The silicon abundance trend is characterized as
showing higher abundances at lower temperatures.
The silicon abundance is derived mostly
from a single Si~\textsc{i} line at 3905~\AA,
so an unidentified blend that grows stronger 
in cooler stars could, in principle, 
explain this effect.
\citet{preston06} examined whether known CH molecular
features could account for this extra absorption
and concluded that this could
bias the silicon abundance by a few percent at most,
which is far insufficient to explain the observations.
The non-LTE calculations of \citet{shi09} suggest
that the use of LTE could account for part of the discrepancy
when using the Si~\textsc{i} 3905~\AA\ line,
but this matter is not fully resolved at present.

The oxygen trend with \teff\ goes in the same direction, but 
the oxygen abundance is usually derived from the
O~\textsc{i} triplet at 7771, 7774, and 7775~\AA.
At these wavelengths, line contamination is unlikely.
We have corrected our O~\textsc{i} triplet abundances
for departures from LTE according to the 
prescriptions of \citet{fabbian09}.
These corrections, while certainly better than
a pure LTE analysis, may still be imperfect,
and in a few cases our stars span a wider parameter range
than their grid
(4500~$\leq$~\teff~$\leq$~6500~K, 
2.0~$\leq$~\logg~$\leq$~5.0,
$-$3.0~$\leq$~[Fe/H]~$\leq$~0).
Given that the trend is mostly seen for warm subgiants
that do fall within the grid, this
explanation alone hardly seems adequate.

The Sc~\textsc{ii},
Ti~\textsc{i}, V~\textsc{i}, V~\textsc{ii}, Cr~\textsc{i},
Mn~\textsc{i}, Mn~\textsc{ii}, and
Co~\textsc{i} trends all run in the opposite sense
from the Si~\textsc{i} and O~\textsc{i} trends.
Most of these species are derived from many lines:\
the median number of lines used in the analysis
for Ti~\textsc{i}, Cr~\textsc{i},
Mn~\textsc{i}, Mn~\textsc{ii}, and Co~\textsc{i}
is 
eight, 
five,
three,
three,
and three, respectively.
V~\textsc{i} and V~\textsc{ii} are each derived from one or two lines.
For a trend of decreasing average abundance with
decreasing \teff, however, contamination by a molecular 
feature can be excluded.
Thus it seems that the cause of the correlation
in these cases is not unidentified blends.

We have included hfs components in our syntheses 
for all odd-$Z$ iron group species
except V~\textsc{ii}, 
for which we are unable to locate 
published values for the hyperfine A and B constants
for the levels of interest.
In principle, this could lead to an overestimate 
of the V~\textsc{ii} abundance, especially for 
stars with stronger absorption lines.
We have simulated possible ranges of broadening
for the V~\textsc{ii} lines up to 0.03~\AA,
which is larger than the broadening 
found for Sc~\textsc{ii} or Mn~\textsc{ii}.
In a cool red giant star with 
EWs in the 80$^{\rm th}$ percentile of strongest
V~\textsc{ii} lines for our sample,
where neglecting the hfs might affect the abundance most significantly,
we could potentially
underestimate the vanadium abundance by $<$~0.1~dex.
Regardless, this correction goes in the wrong direction.
Cooler stars with stronger absorption lines should
yield higher abundances when neglecting hfs.
Therefore neglecting the hfs for $^{51}$V is 
not the source of the V~\textsc{ii} trend with \teff.

In Figures~\ref{ti2ti1teffplot} through \ref{mn2mn1teffplot}
we show the 
[Ti~\textsc{ii}/Ti~\textsc{i}],
[V~\textsc{ii}/V~\textsc{i}],
[Cr~\textsc{ii}/Cr~\textsc{i}], and
[Mn~\textsc{ii}/Mn~\textsc{i}] ratios
plotted as functions of \teff.
We remind readers that these ratios denote the average
total abundance of each element as derived from the ionized or
neutral species after ionization corrections assuming LTE have been applied.
None shows any trend with \teff.

[Ti~\textsc{ii}/Ti~\textsc{i}] and
[V~\textsc{ii}/V~\textsc{i}] show a significant
trend when plotted as a function of metallicity,
as shown in Figures~\ref{ti2ti1feplot} and \ref{v2v1feplot}.
These ratios increase with increasing metallicity.
This trend is present in stars in each of the different
evolutionary classes, though the magnitude of the slope
differs.
One possible explanation is that
Saha equilibrium is an inadequate description of the
ionization distribution for stars in our sample.
[Cr~\textsc{ii}/Cr~\textsc{i}] shows no correlation 
with metallicity (Figure~\ref{cr2cr1feplot}), and 
[Mn~\textsc{ii}/Mn~\textsc{i}] shows, at most, a weak trend
among the SG class only (Figure~\ref{mn2mn1feplot}).

In conclusion, we cannot offer explanations for
all of the non-zero abundance trends with \teff.
We urge those who wish to make use of our abundances
to be careful with the species discussed in this section.
One approach to mitigate the influence of these effects
is to consider abundances of stars in only a limited
range of \teff, [Fe/H], and evolutionary state.
Our sample of 313~stars is large enough that 
sampling narrow ranges of parameter space still
provides satisfactory numbers for statistical comparison
in most cases.

\section{Summary}
\label{summary}

This paper presents the technical details of our analysis to 
measure equivalent widths, radial velocities, 
derive model atmosphere parameters, and derive chemical abundances
from hundreds of individual high resolution spectroscopic observations
of metal-poor halo stars.
Abundances or upper limits are reported for 
53~species of 48~elements in 313~metal-poor stars.
Our analysis finds 19~stars with metallicities [Fe/H]~$\leq -$3.5, 
84~stars with [Fe/H]~$\leq -$3.0, and 
210~stars with [Fe/H]~$\leq -$2.5.
For the stars selected from the HK~Survey, the numbers
of stars below these three metallicity thresholds
are 15, 67, and 173, respectively.
In subsequent papers, we will discuss
the interpretation of these abundances regarding 
the chemical evolution of the Galactic halo
and stellar nucleosynthesis
in the early Universe.
We welcome other investigators to make use of these
results in their own work;
a few words of caution, however, are appropriate.

First, our analysis is performed assuming that LTE
holds in the line-forming layers of the photosphere.
We employ static, 1D, 
plane-parallel model atmospheres constructed 
assuming LTE for a fixed set of abundances.
To relax these assumptions would require substantial
increases in computing power and relevant atomic data, and
it is currently not practical to do so 
for a survey of this scale.
Our results will differ, of course, from those
computed using such techniques for individual stars.

Second, as discussed in detail in Section~\ref{compareprevious},
our metallicity scale is slightly lower than that 
found by previous investigations of stars in common.
We have derived most model atmosphere parameters by
spectroscopic methods, whereas recent abundance
studies of extremely metal-poor stars use 
a combination of photometric and spectroscopic methods
to derive these quantities.
A natural consequence of this approach is that our
derived metallicities are, on average, lower by 
$\approx$~0.25~dex for red giants and 
$\approx$~0.04~dex for subgiants in common with previous studies.
The mean metallicity differences are a function of the evolutionary state,
as reported in Table~\ref{comparetab}.

Third, this is a biased sample,
and the biases are not easily quantified.
We have drawn our targets from a variety of sources,
and even those selected from the HK~Survey
are a heterogeneous sample 
where stars we deem to be chemically interesting
(based on previous studies)
are overrepresented.
Efforts to reconstruct the metallicity distribution function
or estimate the frequency of carbon-enhanced stars, for example,
using these data alone are not advised.
Our data can be used, however, to calibrate other
samples whose biases are well quantified.

Finally, some elemental ratios show a dependence on
the stellar evolutionary state.
For this reason, we strongly urge users 
to avoid plotting abundances of all 313~stars on the
same diagram when detailed comparisons are intended.
Instead, the size of this sample may be exploited
to minimize systematic errors arising from
the analysis techniques.
For example, it is possible to
select stars spanning a small range of
effective temperature, surface gravity, and metallicity
and still obtain statistically meaningful samples.
We intend to employ this strategy in our own analyses.

\acknowledgments

It is a pleasure to thank so many of our friends and colleagues for
encouragement and advice throughout the course of this project, 
especially T.\ Beers and A.\ McWilliam.
We appreciate the patience of the Magellan and McDonald TAC members.
I.U.R.\ also thanks D.\ Fabbian, K.\ Lind, A.\ McWilliam, and Y.\ Takeda for 
providing their interpolation codes,
J.\ Lawler and M.\ Wood for sending results in advance of publication,
J.\ Sobeck for assistance with abundance comparisons
and MOOG installations,
the referees for their diligent readings of the manuscript, and
T.\ Beers for commenting on portions of 
an earlier version of the manuscript.
We also thank G.\ Schwarz for generating the 
machine-readable tables included with this arXiv submission.

This research has made use of NASA's 
Astrophysics Data System Bibliographic Services, 
the arXiv preprint server operated by Cornell University, 
the SIMBAD and VizieR databases hosted by the
Strasbourg Astronomical Data Center, and 
the Atomic Spectra Database hosted by
the National Institute of Standards and Technology. 
Our appreciation for 
the reliability of these data archives cannot be understated.
IRAF is distributed by the National Optical Astronomy Observatories,
which are operated by the Association of Universities for Research
in Astronomy, Inc., under cooperative agreement with the National
Science Foundation.
This publication makes use of data products from the 
Two Micron All Sky Survey, which is a joint project of the 
University of Massachusetts and the Infrared Processing and 
Analysis Center/California Institute of Technology, funded by the 
National Aeronautics and Space Administration and the 
National Science Foundation.
The HET is a joint project of the University of 
Texas at Austin, the Pennsylvania State University, Stanford University, 
Ludwig-Maximilians-Universit\"{a}t M\"{u}nchen, and 
Georg-August-Universit\"{a}t G\"{o}ttingen. 
The HET is named in honor of its principal benefactors, William 
P.\ Hobby and Robert E.\ Eberly.

I.U.R.\ acknowledges support by the Barbara McClintock Fellowship
from the Carnegie Institution for Science.
C.S.\ acknowledges support by the U.S.\ National Science Foundation 
 (grants AST~06-07708, AST~09-08978, and AST~12-11585).

{\it Facilities:} 
\facility{HET (HRS)}, 
\facility{Magellan:Baade (MIKE)}, 
\facility{Magellan:Clay (MIKE)}, 
\facility{Smith (Tull)}

\appendix

\begin{figure*}
\begin{center}
\includegraphics[angle=00,width=4.5in]{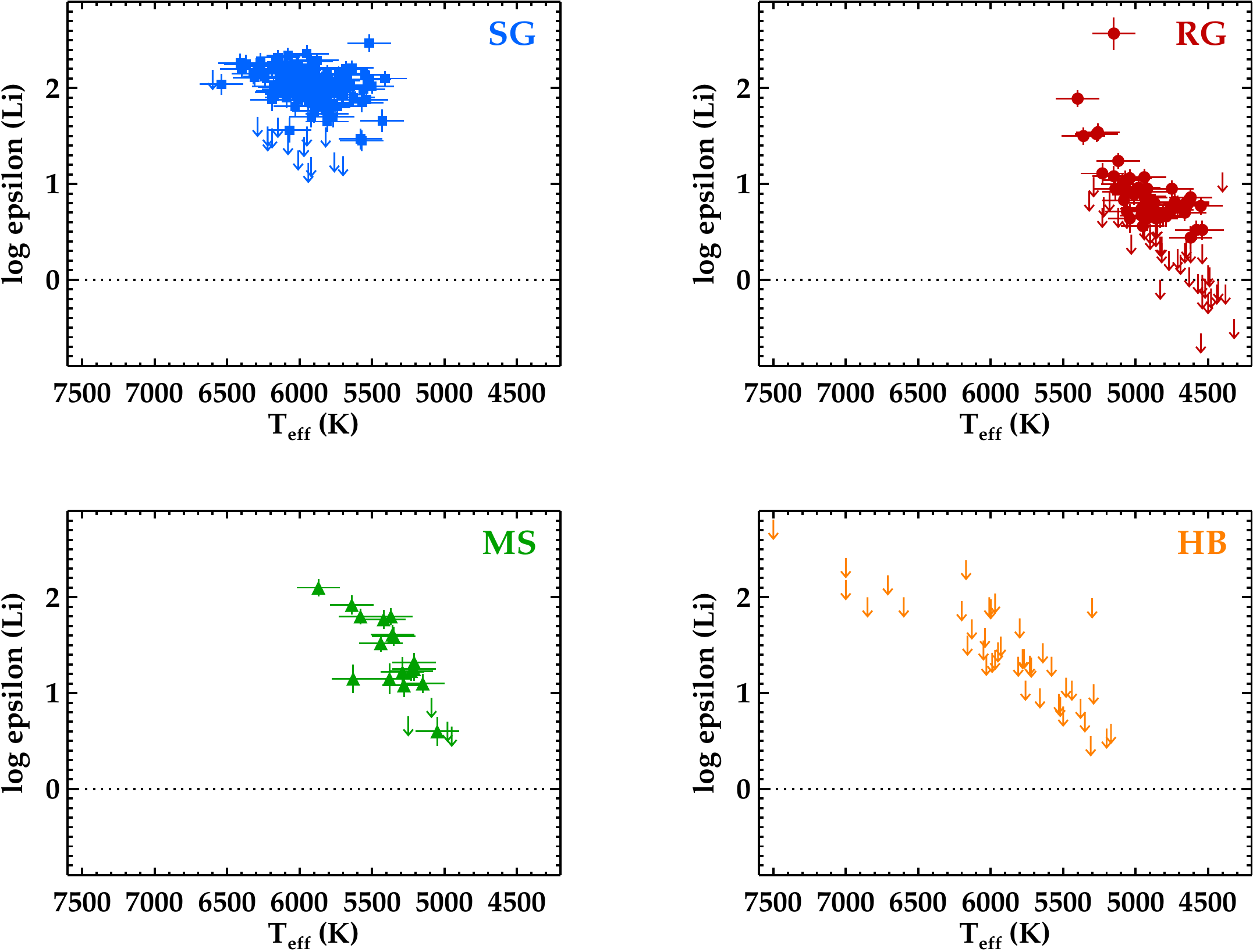}
\end{center}
\caption{
\label{liteffplot}
Derived lithium abundances as a function of \teff.
Each panel illustrates a different class of stars,
and the three stars in class ``BS'' are included in the
plot for stars in class ``HB.''
The 1~$\sigma$ uncertainties are indicated.
Downward arrows indicate 3~$\sigma$ upper limits, and
the actual value is marked by the end with no arrow head.
}
\end{figure*}

\clearpage
\begin{figure*}
\begin{center}
\includegraphics[angle=00,width=4.5in]{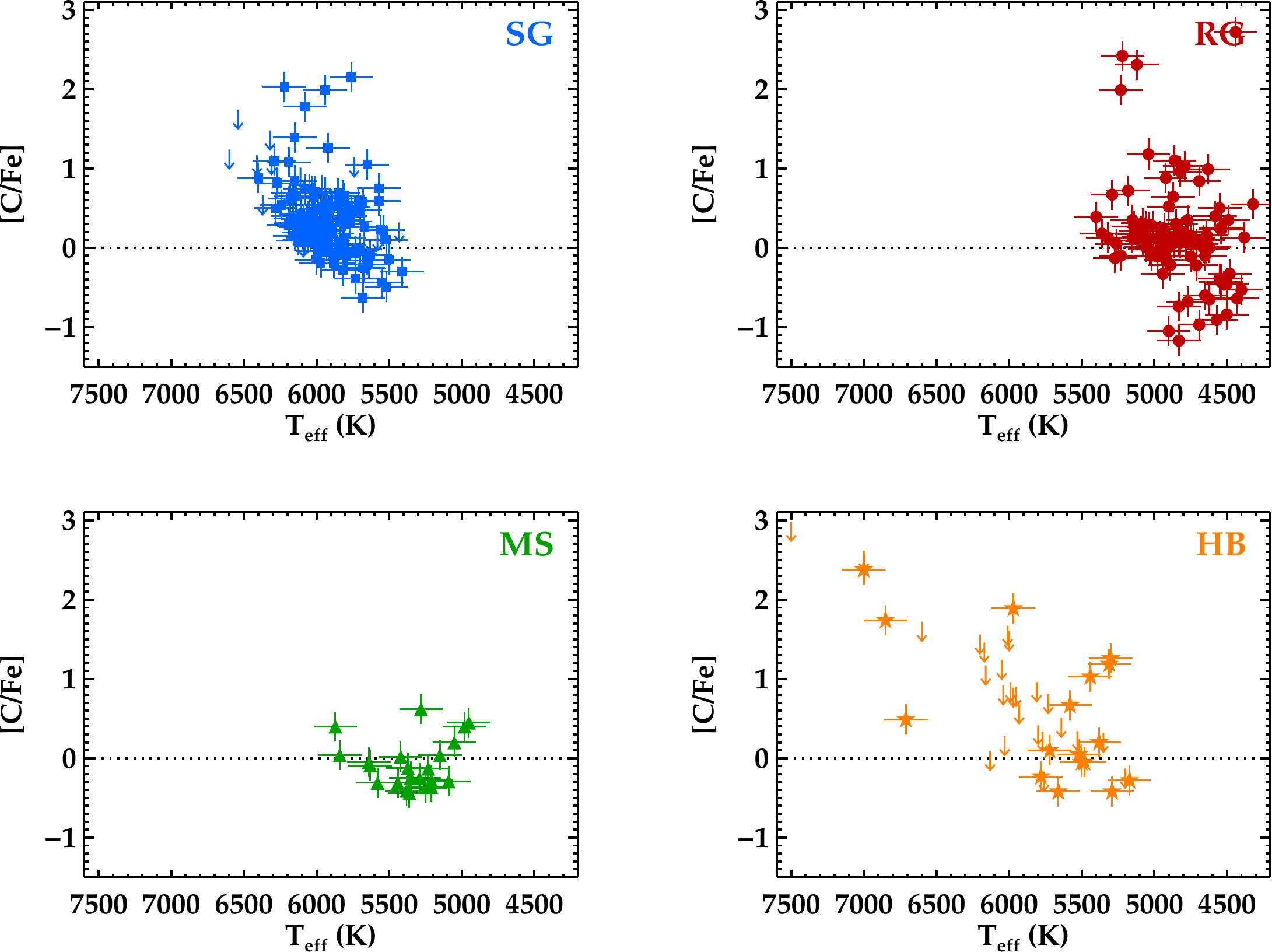}
\end{center}
\caption{
\label{cteffplot}
Derived [C/Fe] ratios as a function of \teff.
The carbon abundance is derived from the CH 
$A^2\Delta - X^2\Pi$ G band.
The dotted lines indicate the Solar ratio.
All other symbols are the same as in Figure~\ref{liteffplot}.
}
\end{figure*}

\begin{figure*}
\begin{center}
\includegraphics[angle=00,width=4.5in]{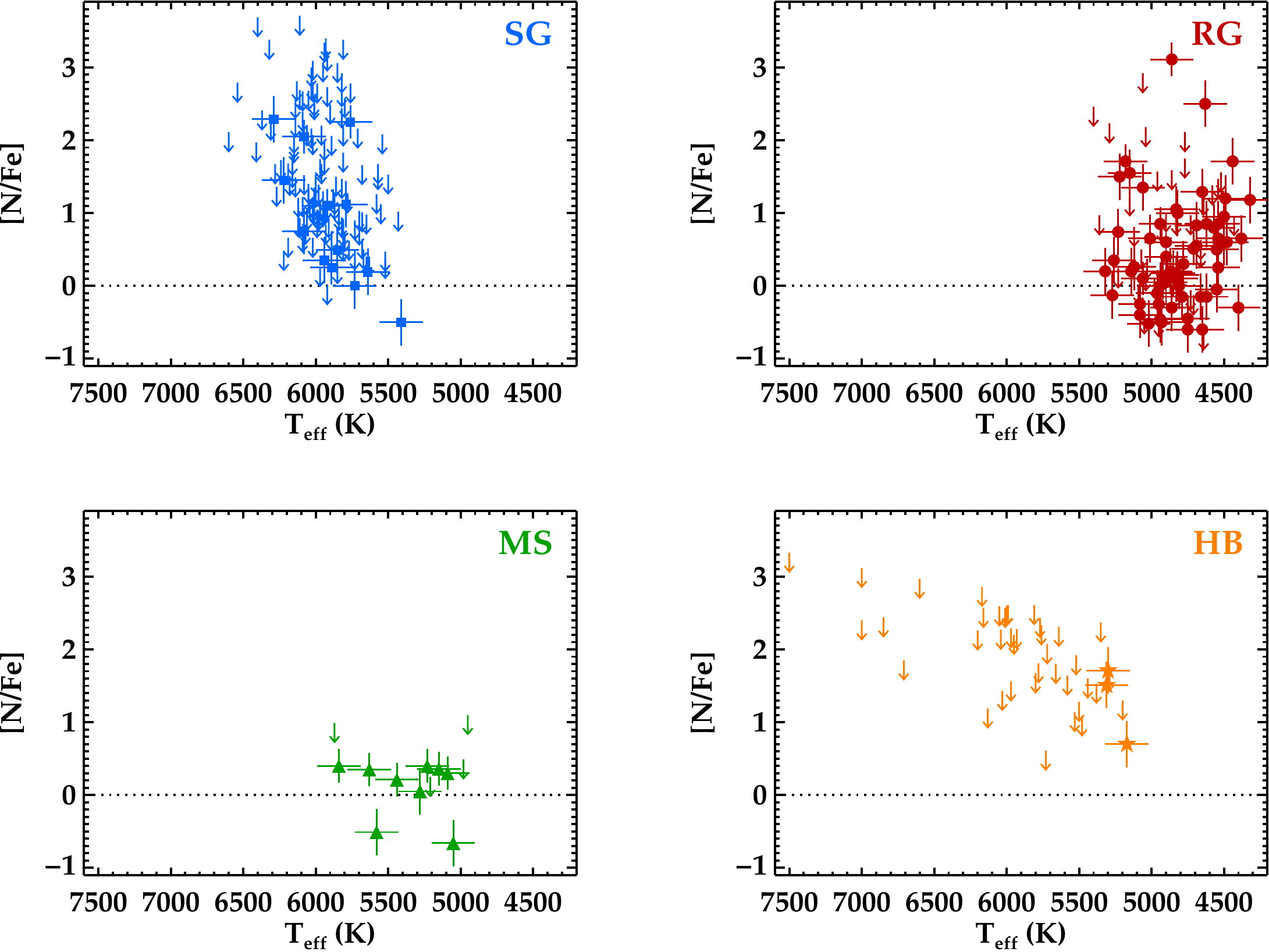}
\end{center}
\caption{
\label{nteffplot}
Derived [N/Fe] ratios as a function of \teff.
The nitrogen abundance is derived from the 
NH $A^3\Pi    - X^3\Sigma$ band or the
CN $B^2\Sigma - X^2\Sigma$ band.
Symbols are the same as in Figures~\ref{liteffplot} and \ref{cteffplot}.
}
\end{figure*}

\begin{figure*}
\begin{center}
\includegraphics[angle=00,width=4.5in]{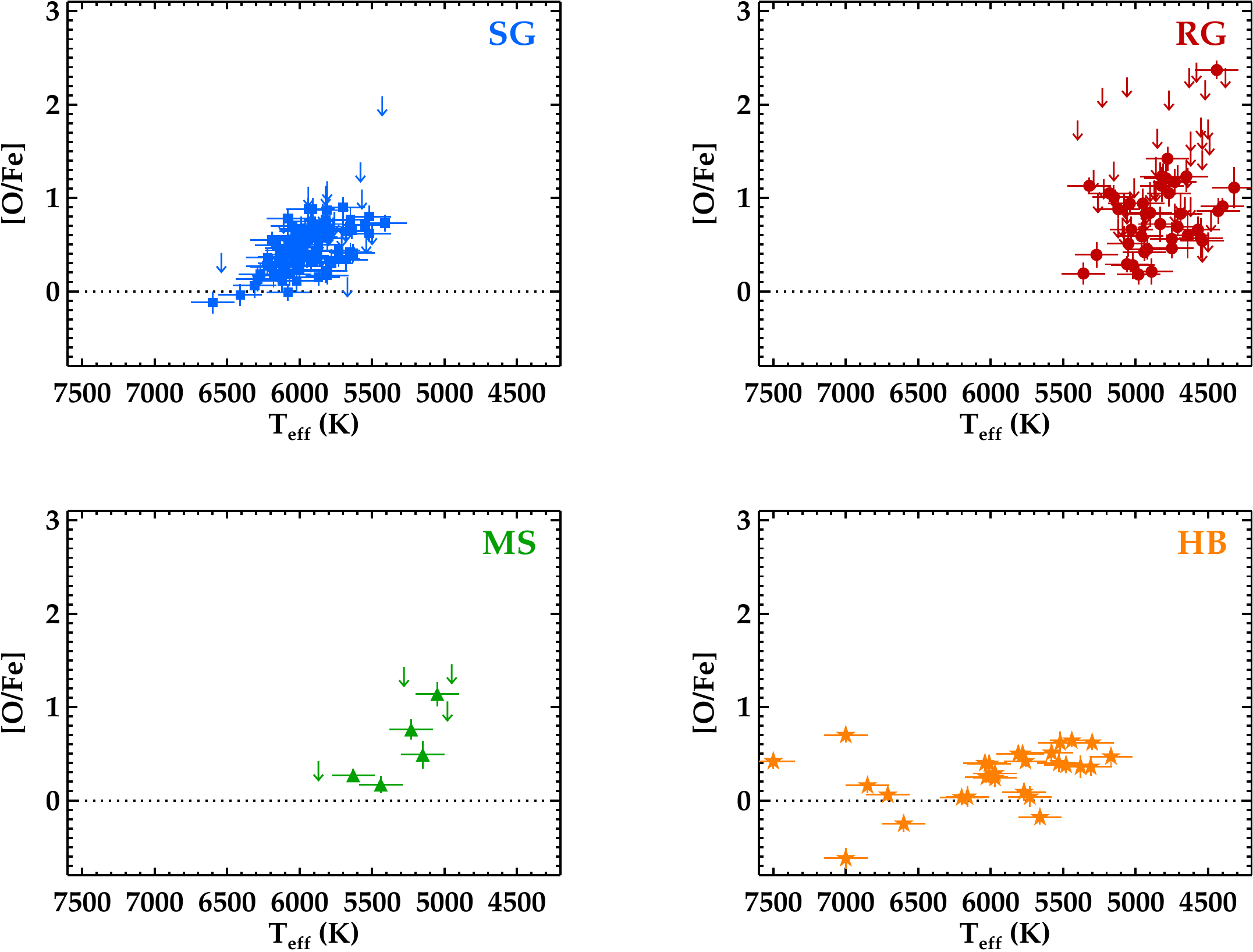}
\end{center}
\caption{
\label{oteffplot}
Derived [O/Fe] ratios as a function of \teff.
Symbols are the same as in Figures~\ref{liteffplot} and \ref{cteffplot}.
}
\end{figure*}

\begin{figure*}
\begin{center}
\includegraphics[angle=00,width=4.5in]{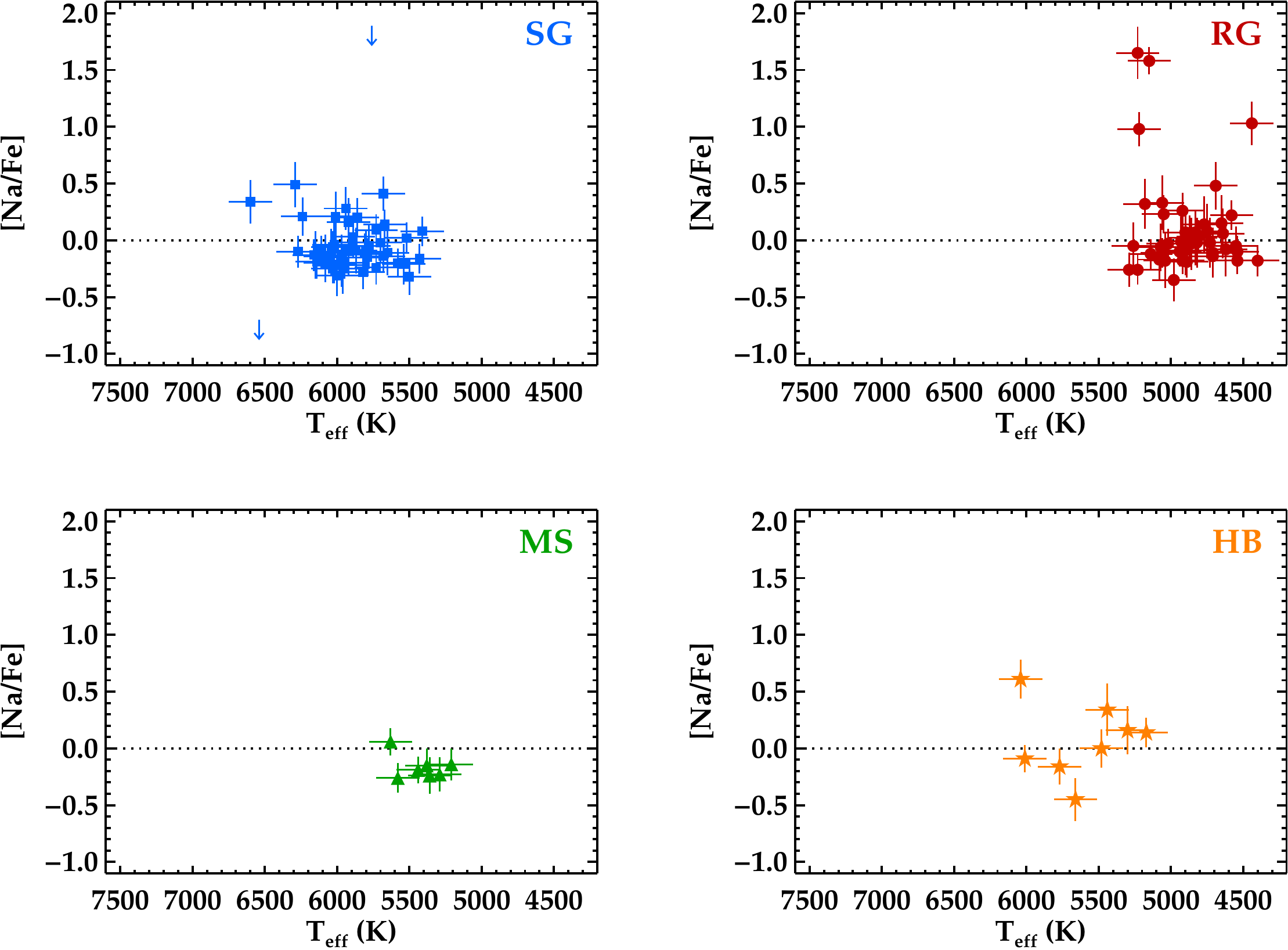}
\end{center}
\caption{
\label{nateffplot}
Derived [Na/Fe] ratios as a function of \teff.
Symbols are the same as in Figures~\ref{liteffplot} and \ref{cteffplot}.
}
\end{figure*}

\begin{figure*}
\begin{center}
\includegraphics[angle=00,width=4.5in]{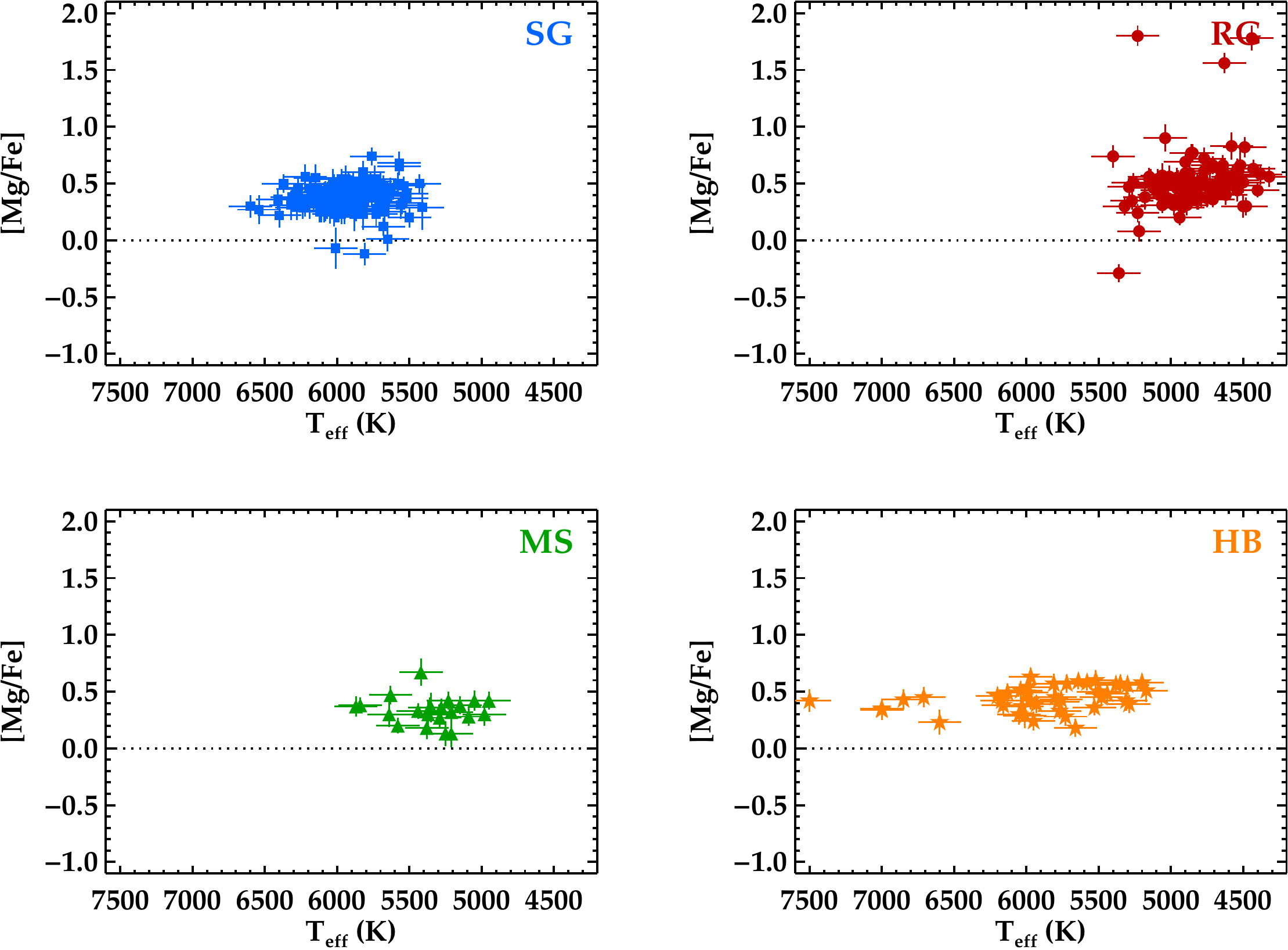}
\end{center}
\caption{
\label{mgteffplot}
Derived [Mg/Fe] ratios as a function of \teff.
Symbols are the same as in Figures~\ref{liteffplot} and \ref{cteffplot}.
}
\end{figure*}

\begin{figure*}
\begin{center}
\includegraphics[angle=00,width=4.5in]{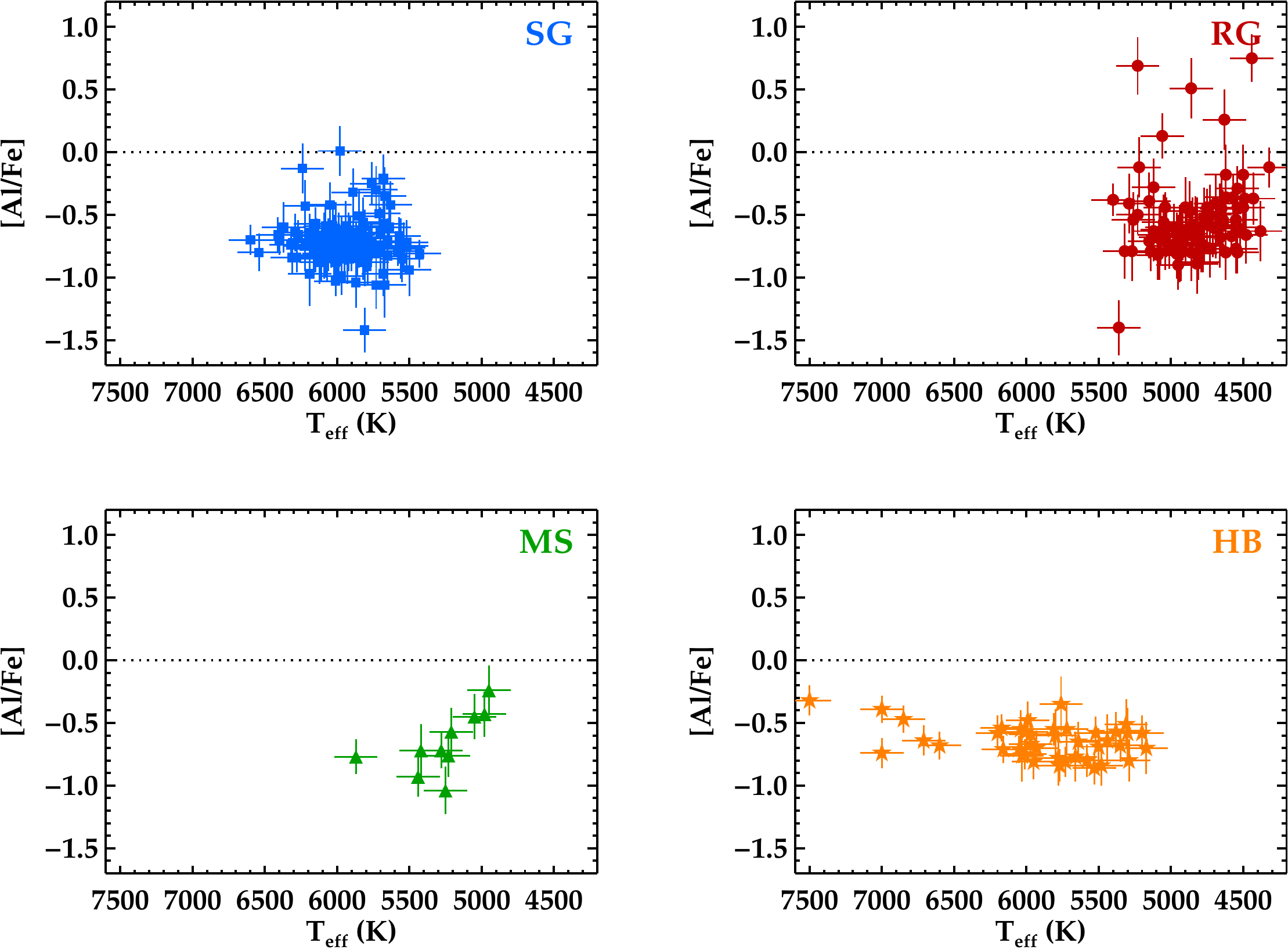}
\end{center}
\caption{
\label{alteffplot}
Derived [Al/Fe] ratios as a function of \teff.
Symbols are the same as in Figures~\ref{liteffplot} and \ref{cteffplot}.
}
\end{figure*}

\begin{figure*}
\begin{center}
\includegraphics[angle=00,width=4.5in]{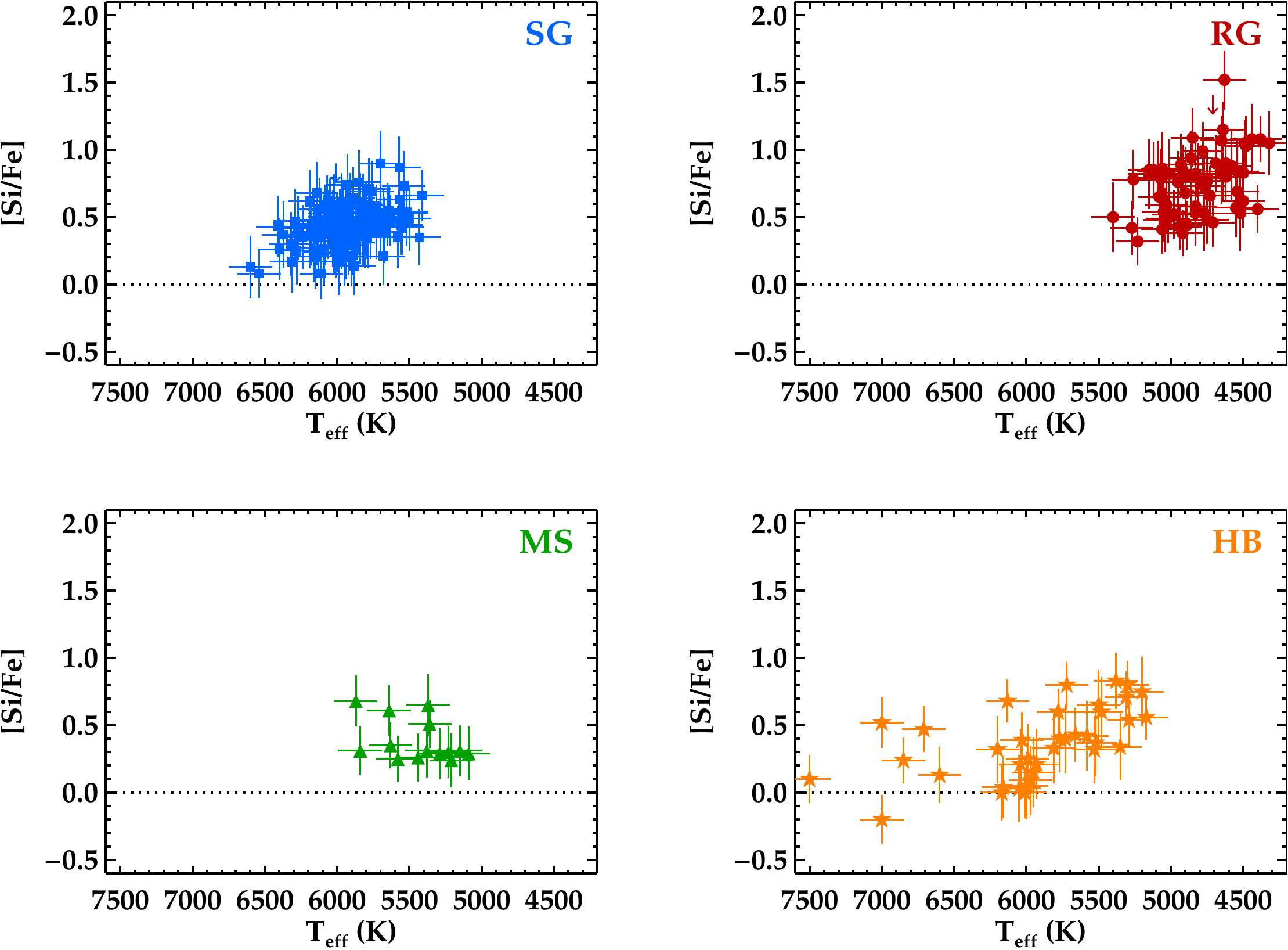}
\end{center}
\caption{
\label{siteffplot}
Derived [Si/Fe] ratios as a function of \teff.
Symbols are the same as in Figures~\ref{liteffplot} and \ref{cteffplot}.
}
\end{figure*}

\begin{figure*}
\begin{center}
\includegraphics[angle=00,width=4.5in]{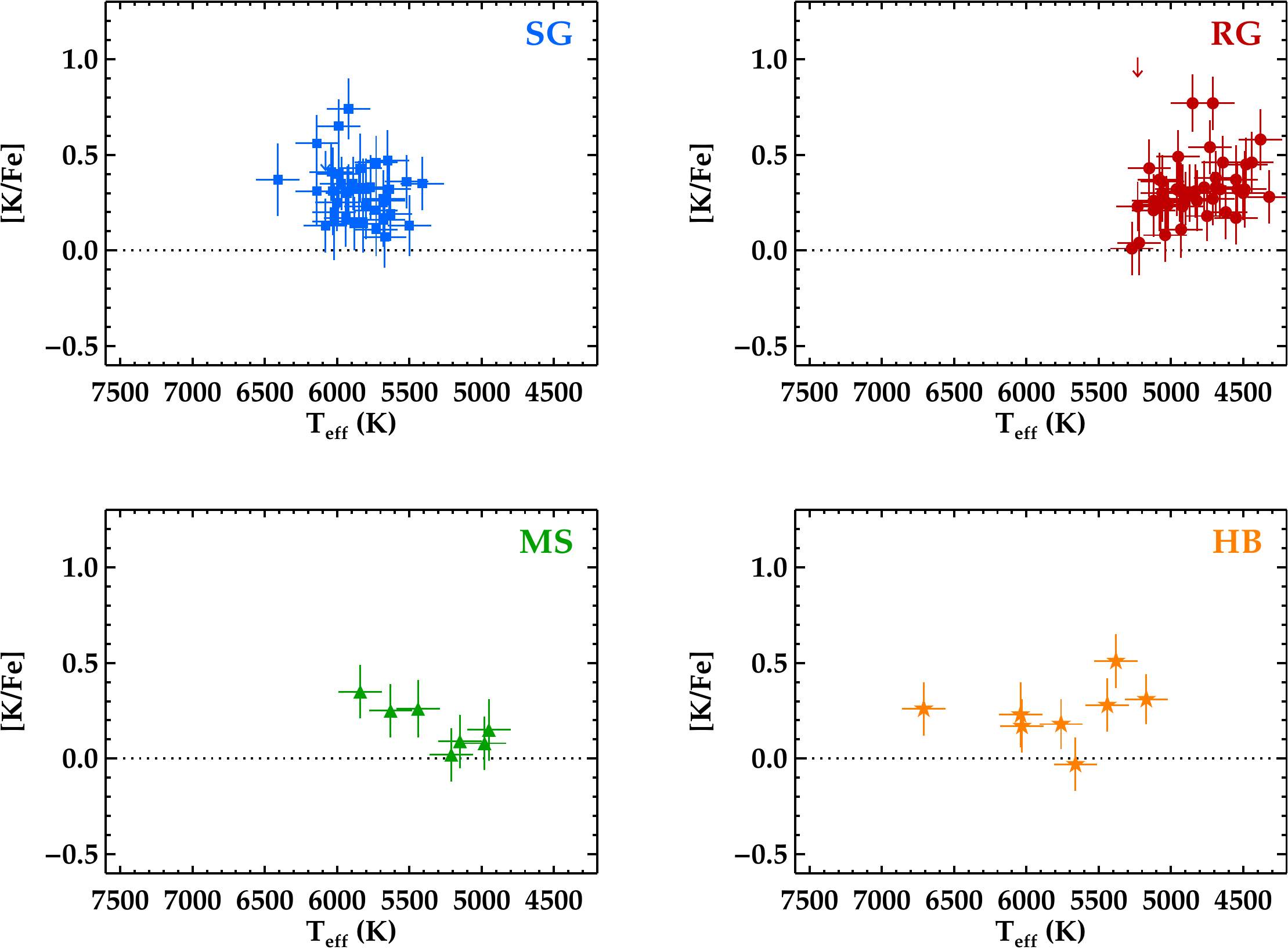}
\end{center}
\caption{
\label{kteffplot}
Derived [K/Fe] ratios as a function of \teff.
Symbols are the same as in Figures~\ref{liteffplot} and \ref{cteffplot}.
}
\end{figure*}

\begin{figure*}
\begin{center}
\includegraphics[angle=00,width=4.5in]{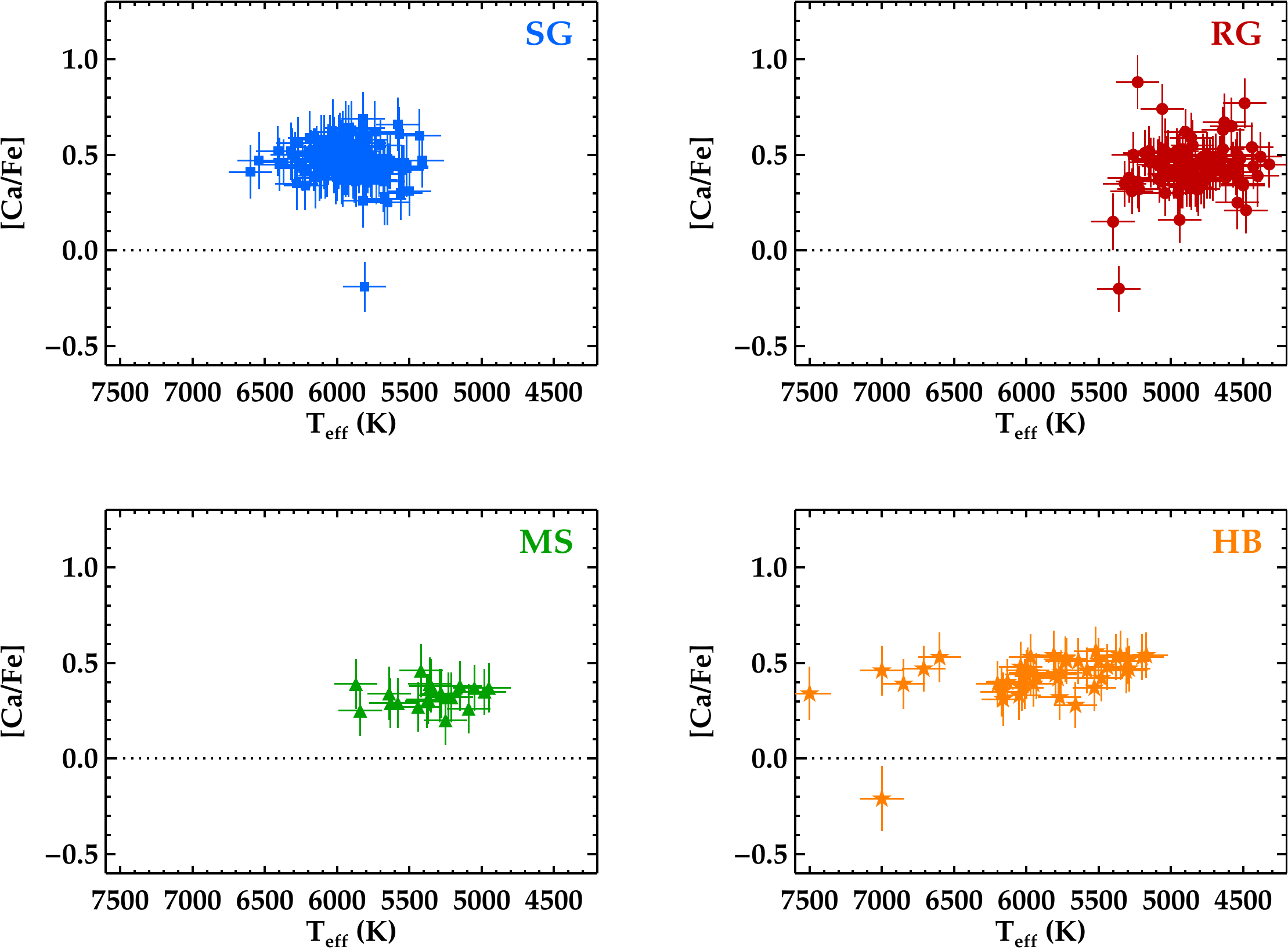}
\end{center}
\caption{
\label{cateffplot}
Derived [Ca/Fe] ratios as a function of \teff.
Symbols are the same as in Figures~\ref{liteffplot} and \ref{cteffplot}.
}
\end{figure*}

\begin{figure*}
\begin{center}
\includegraphics[angle=00,width=4.5in]{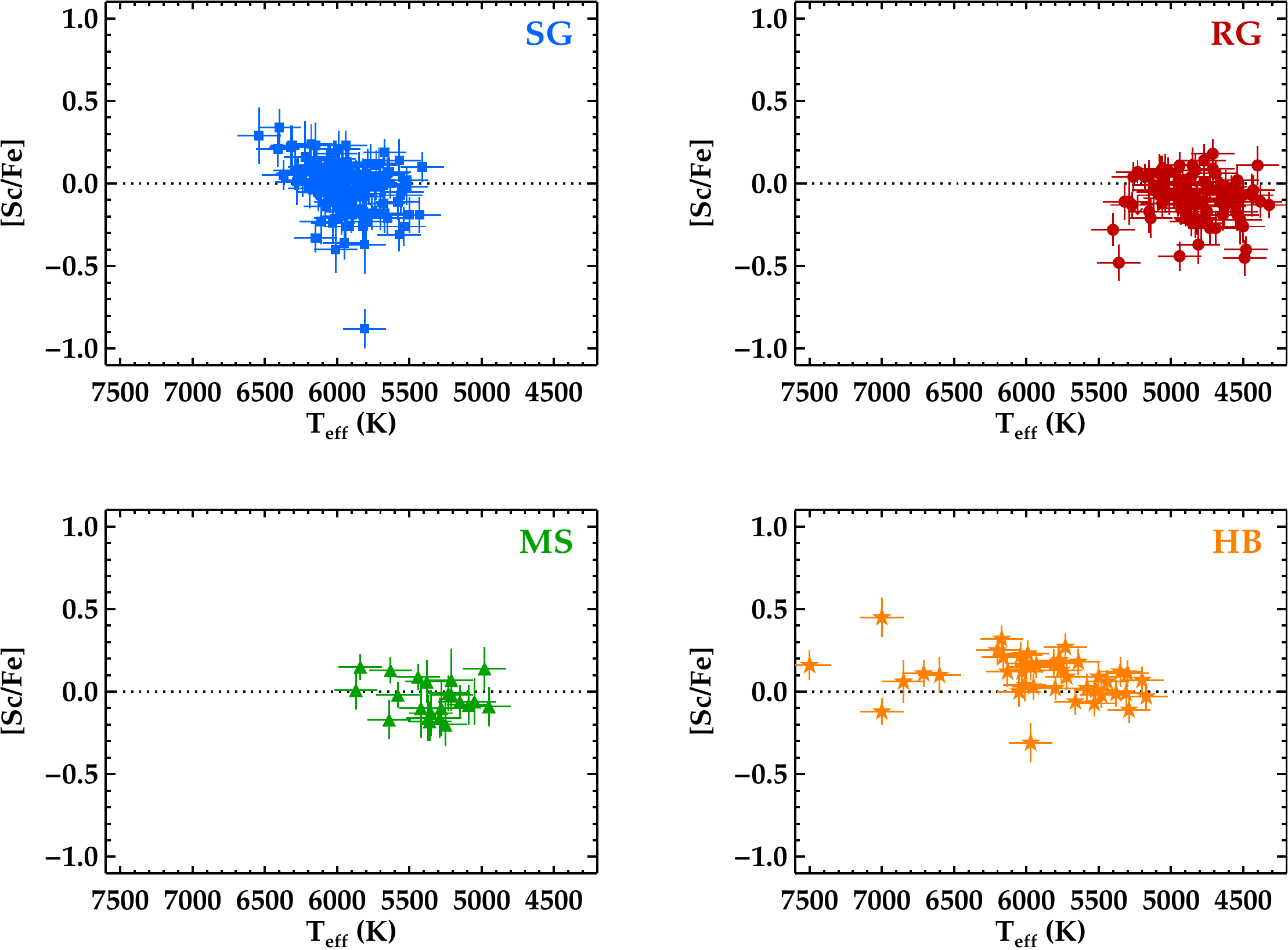}
\end{center}
\caption{
\label{scteffplot}
Derived [Sc/Fe] ratios as a function of \teff.
Symbols are the same as in Figures~\ref{liteffplot} and \ref{cteffplot}.
}
\end{figure*}

\begin{figure*}
\begin{center}
\includegraphics[angle=00,width=4.5in]{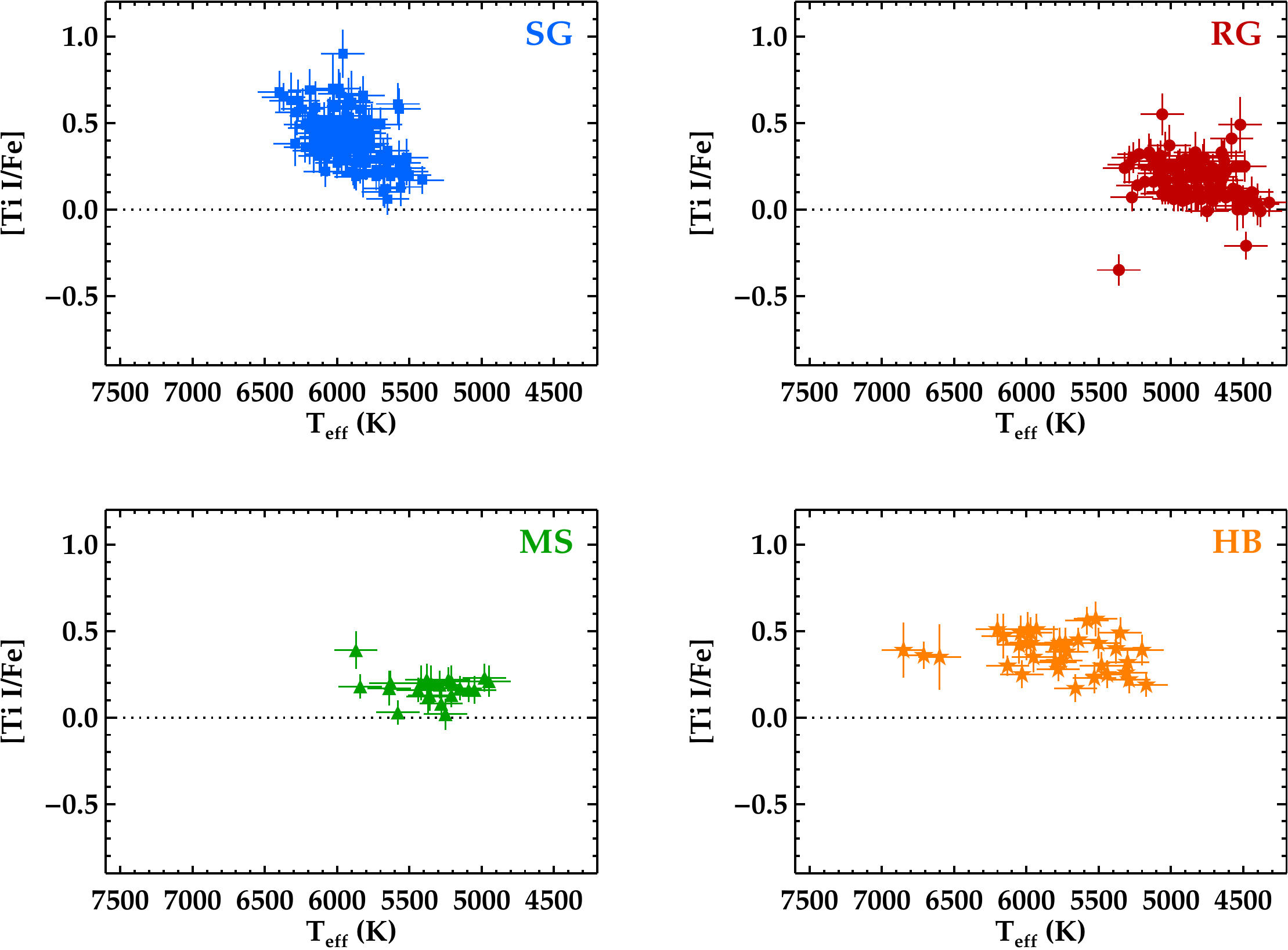}
\end{center}
\caption{
\label{ti1teffplot}
Derived [Ti/Fe] ratios for neutral lines as a function of \teff.
Symbols are the same as in Figures~\ref{liteffplot} and \ref{cteffplot}.
}
\end{figure*}

\begin{figure*}
\begin{center}
\includegraphics[angle=00,width=4.5in]{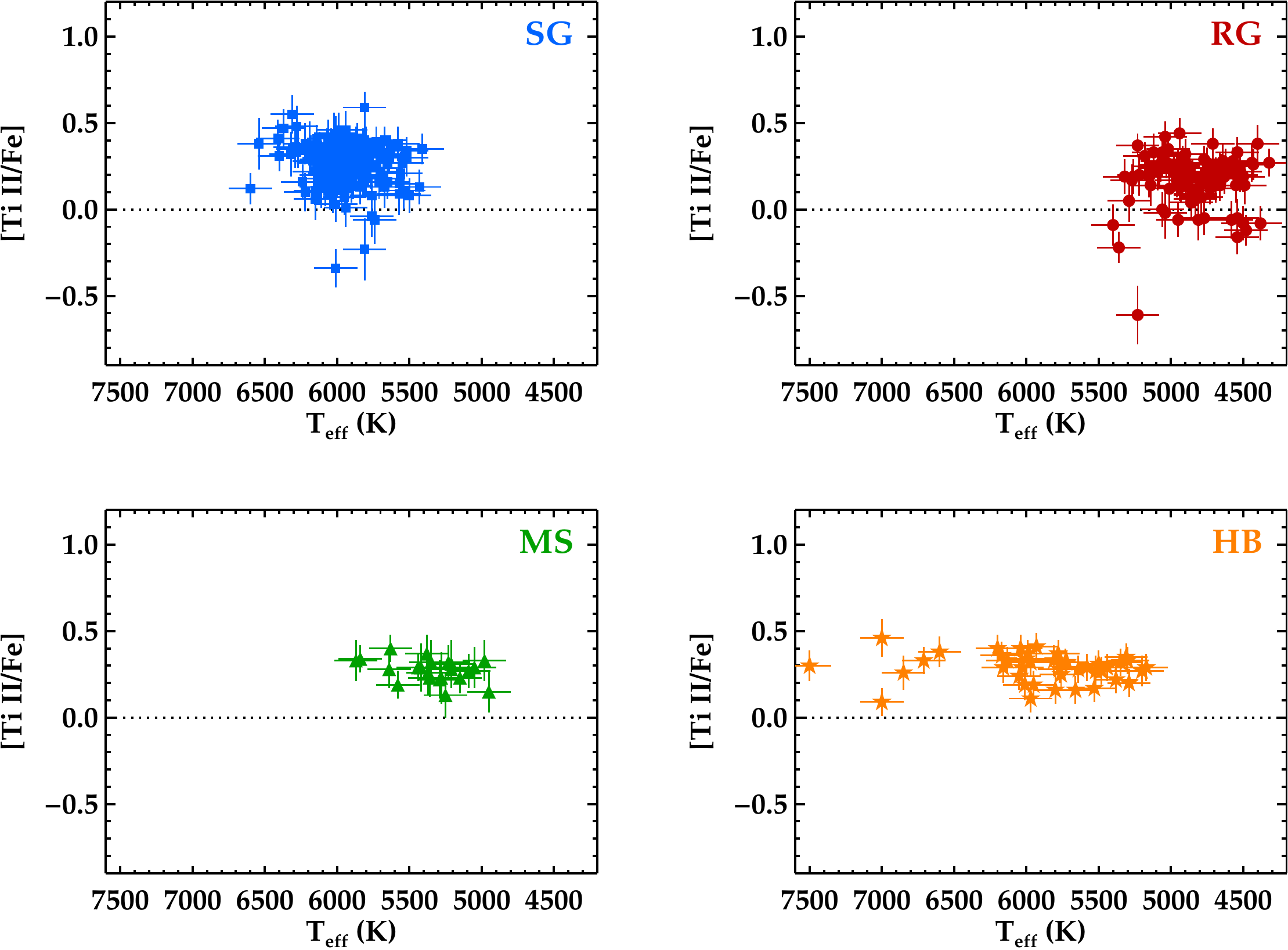}
\end{center}
\caption{
\label{ti2teffplot}
Derived [Ti/Fe] ratios for ionized lines as a function of \teff.
Symbols are the same as in Figures~\ref{liteffplot} and \ref{cteffplot}.
}
\end{figure*}

\begin{figure*}
\begin{center}
\includegraphics[angle=00,width=4.5in]{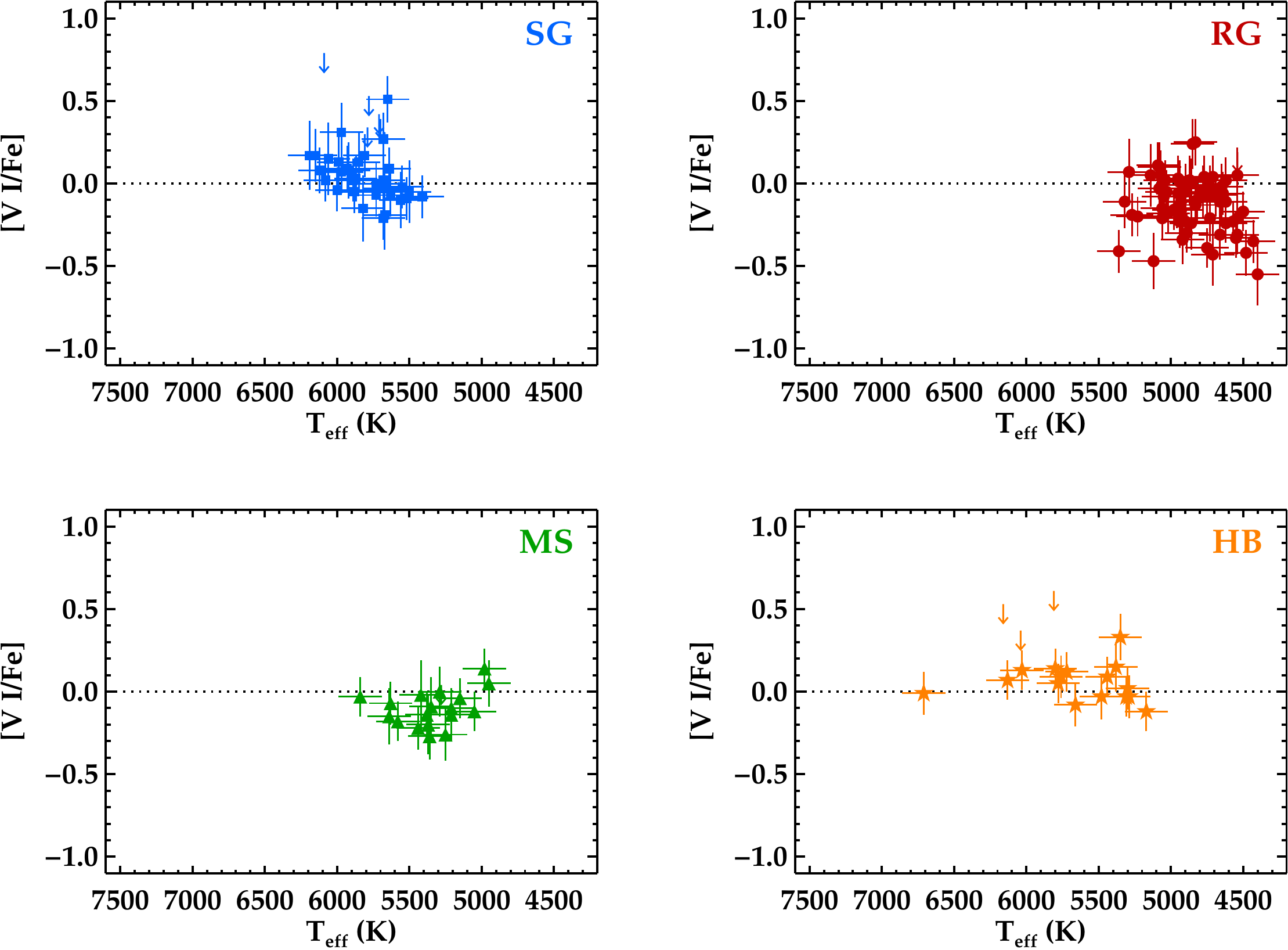}
\end{center}
\caption{
\label{v1teffplot}
Derived [V/Fe] ratios for neutral lines as a function of \teff.
Symbols are the same as in Figures~\ref{liteffplot} and \ref{cteffplot}.
}
\end{figure*}

\begin{figure*}
\begin{center}
\includegraphics[angle=00,width=4.5in]{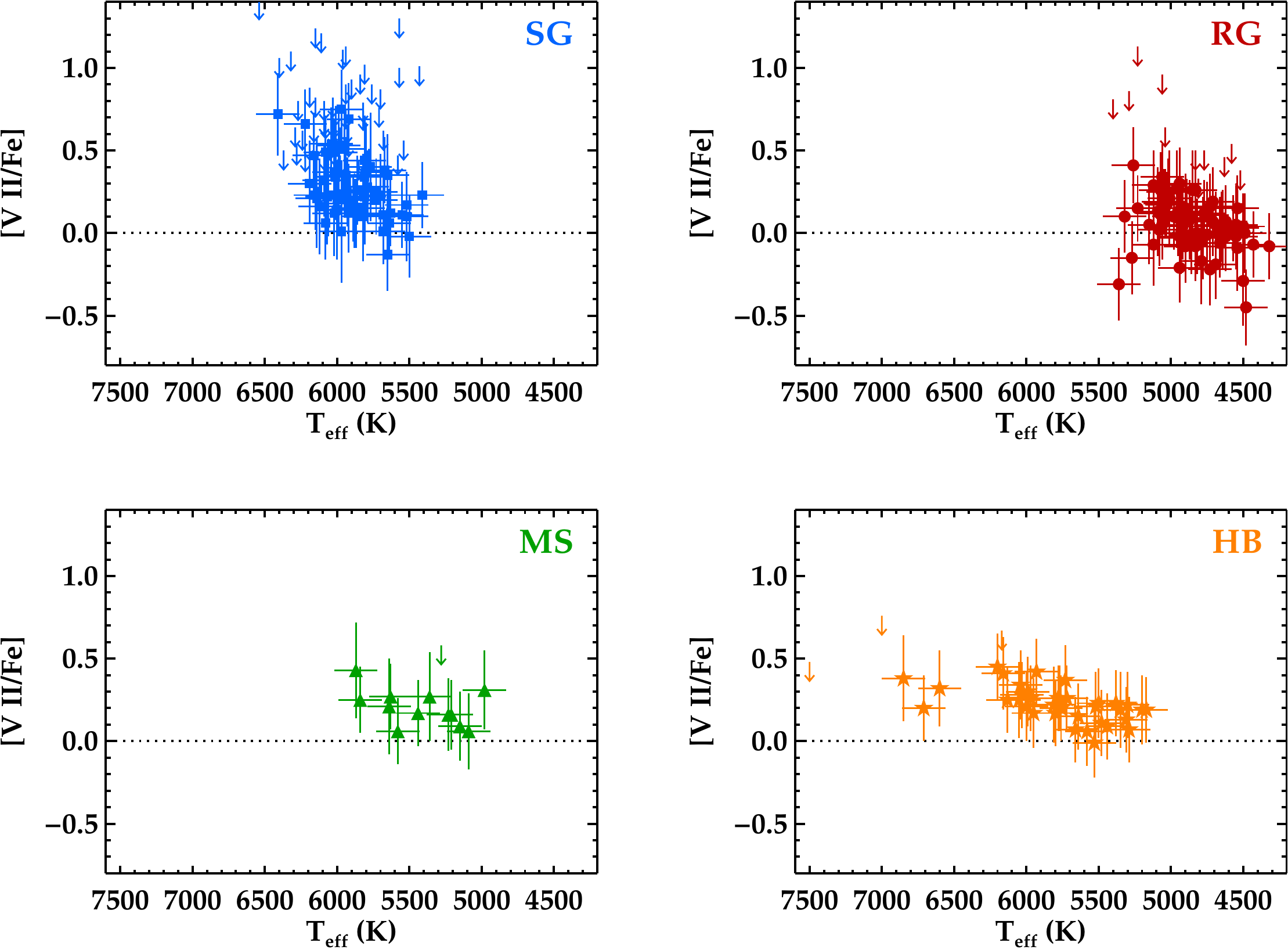}
\end{center}
\caption{
\label{v2teffplot}
Derived [V/Fe] ratios for ionized lines as a function of \teff.
Symbols are the same as in Figures~\ref{liteffplot} and \ref{cteffplot}.
}
\end{figure*}

\clearpage
\begin{figure*}
\begin{center}
\includegraphics[angle=00,width=4.5in]{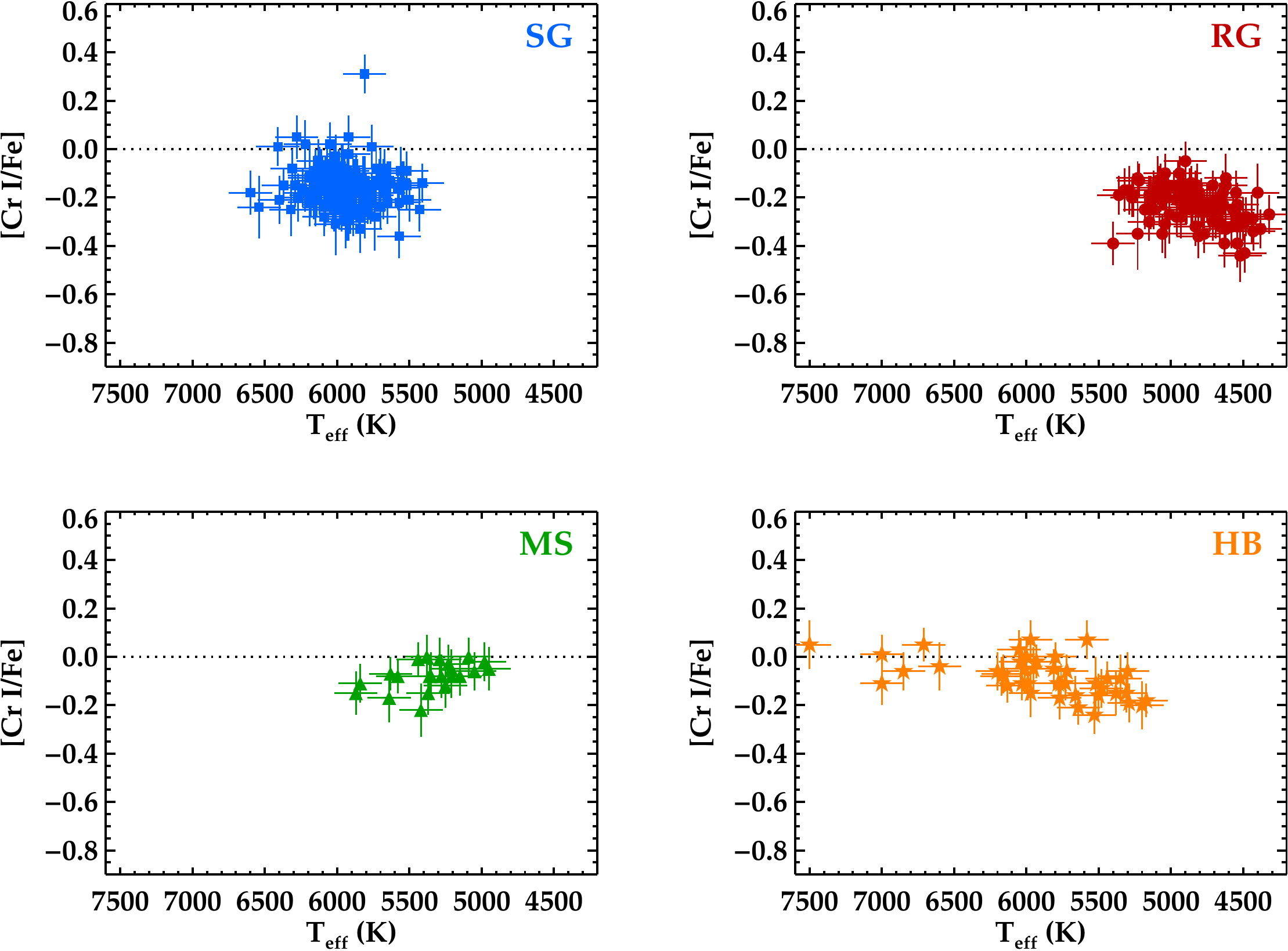}
\end{center}
\caption{
\label{cr1teffplot}
Derived [Cr/Fe] ratios for neutral lines as a function of \teff.
Symbols are the same as in Figures~\ref{liteffplot} and \ref{cteffplot}.
}
\end{figure*}

\begin{figure*}
\begin{center}
\includegraphics[angle=00,width=4.5in]{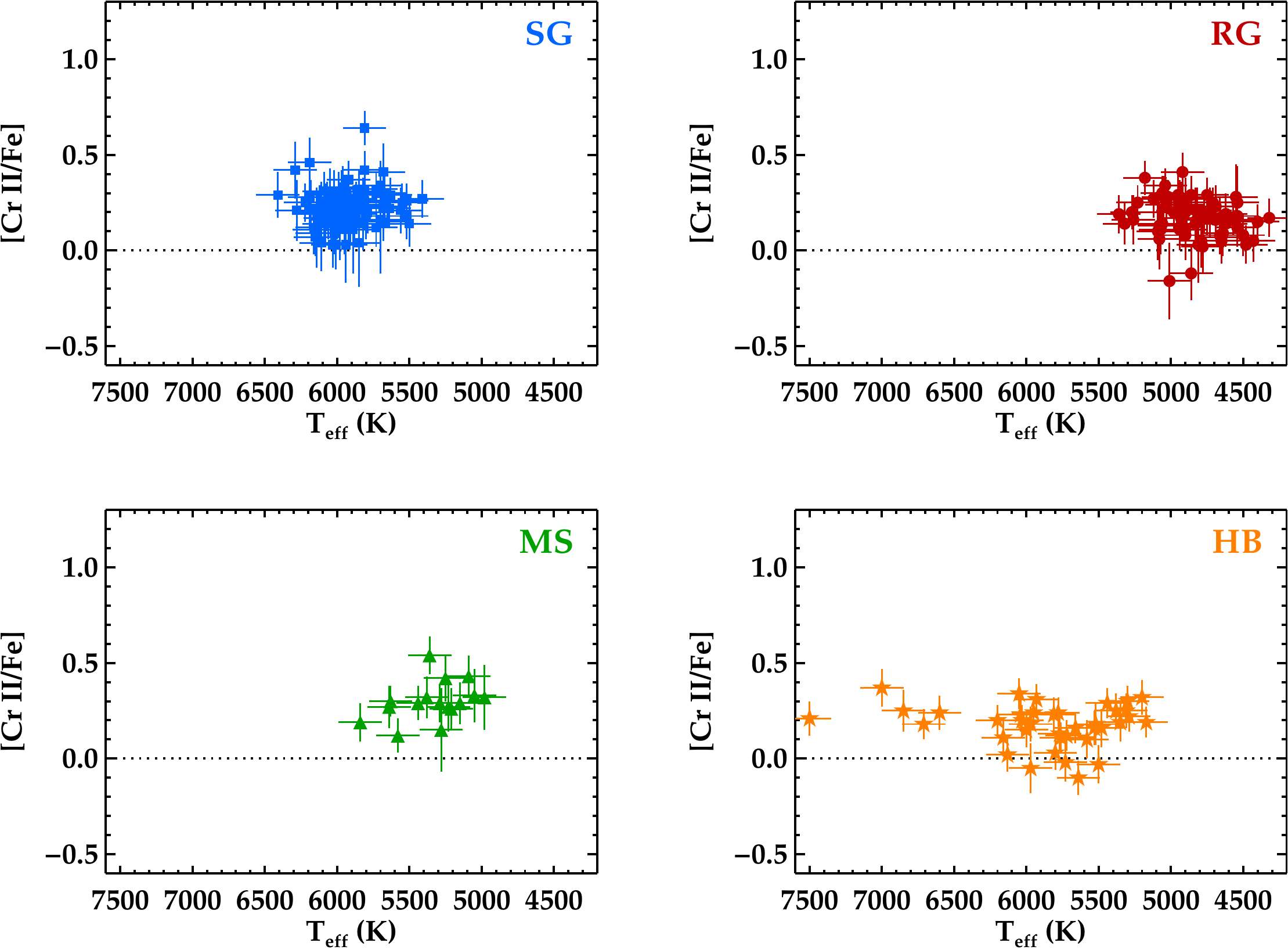}
\end{center}
\caption{
\label{cr2teffplot}
Derived [Cr/Fe] ratios for ionized lines as a function of \teff.
Symbols are the same as in Figures~\ref{liteffplot} and \ref{cteffplot}.
}
\end{figure*}

\begin{figure*}
\begin{center}
\includegraphics[angle=00,width=4.5in]{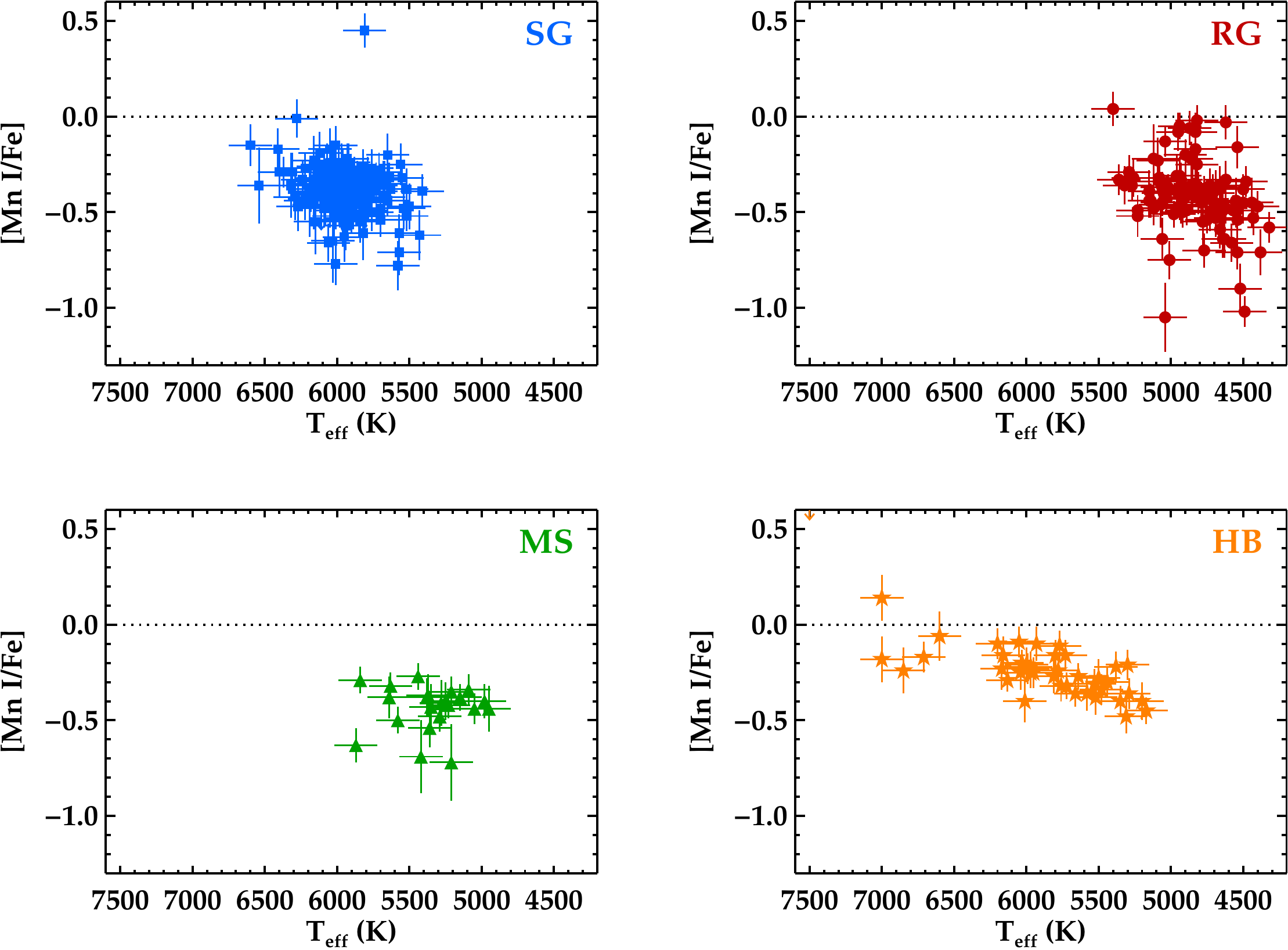}
\end{center}
\caption{
\label{mn1teffplot}
Derived [Mn/Fe] ratios for neutral lines as a function of \teff.
Symbols are the same as in Figures~\ref{liteffplot} and \ref{cteffplot}.
}
\end{figure*}

\begin{figure*}
\begin{center}
\includegraphics[angle=00,width=4.5in]{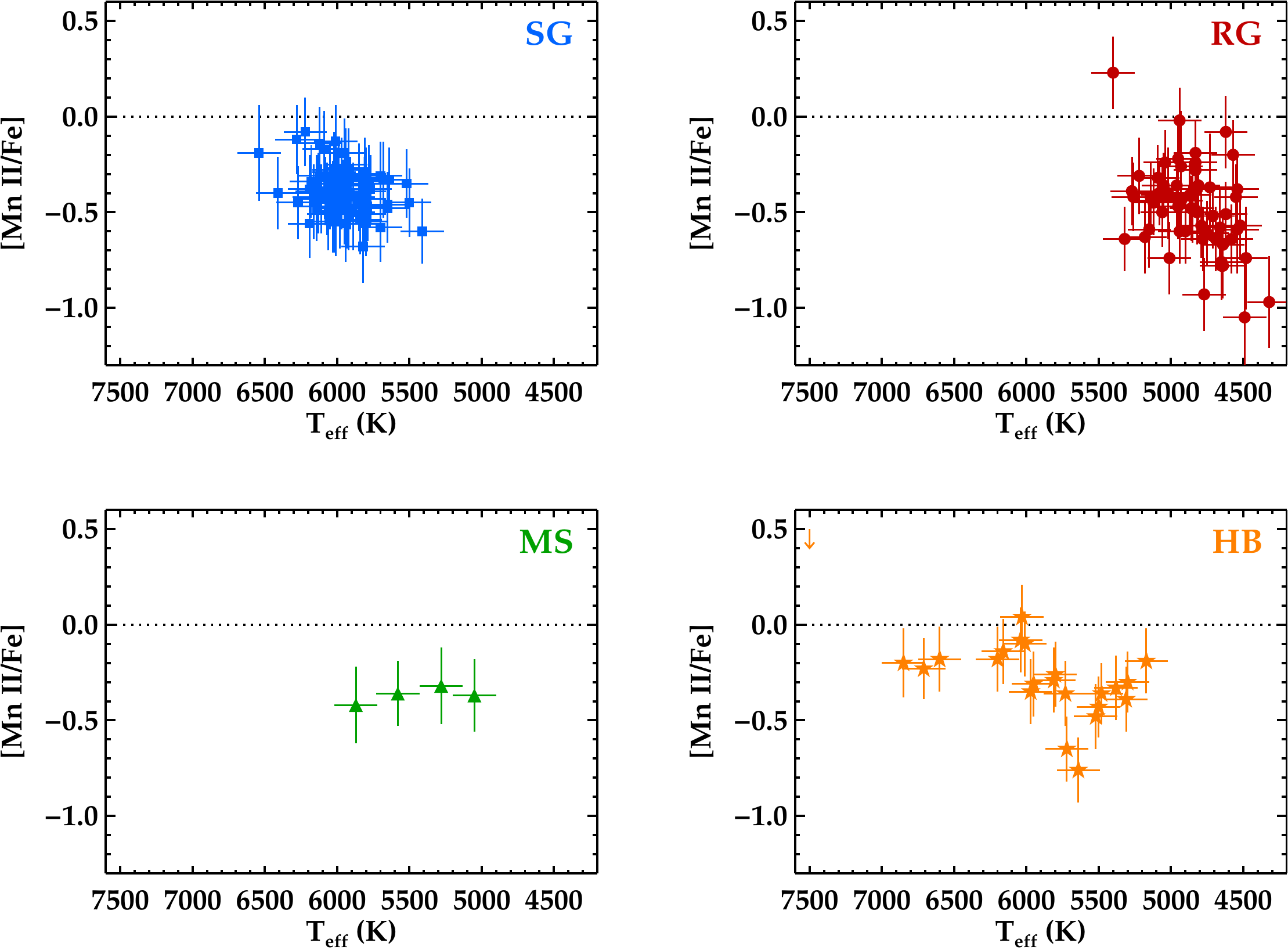}
\end{center}
\caption{
\label{mn2teffplot}
Derived [Mn/Fe] ratios for ionized lines as a function of \teff.
Symbols are the same as in Figures~\ref{liteffplot} and \ref{cteffplot}.
}
\end{figure*}

\begin{figure*}
\begin{center}
\includegraphics[angle=00,width=4.5in]{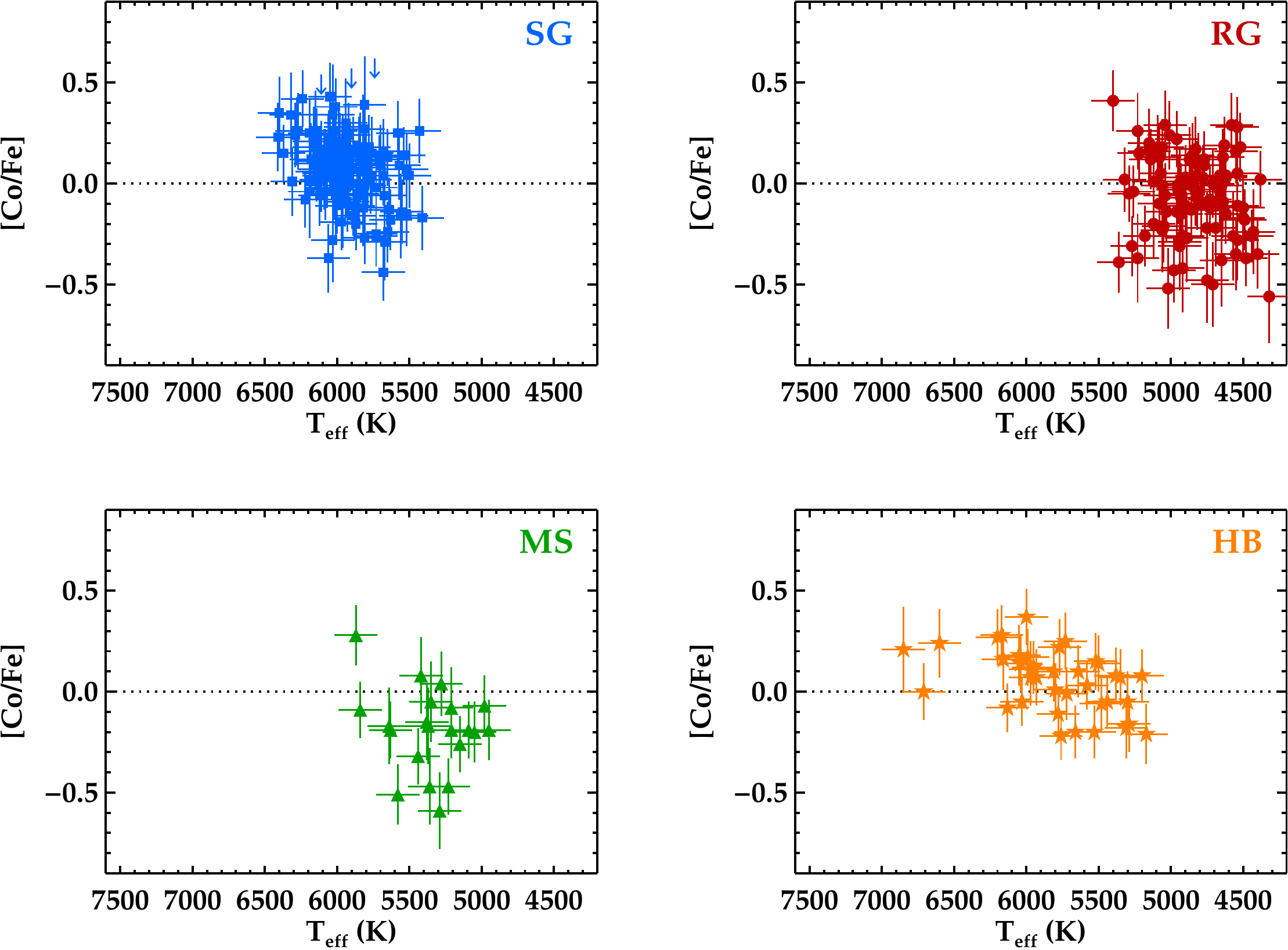}
\end{center}
\caption{
\label{coteffplot}
Derived [Co/Fe] ratios as a function of \teff.
Symbols are the same as in Figures~\ref{liteffplot} and \ref{cteffplot}.
}
\end{figure*}

\begin{figure*}
\begin{center}
\includegraphics[angle=00,width=4.5in]{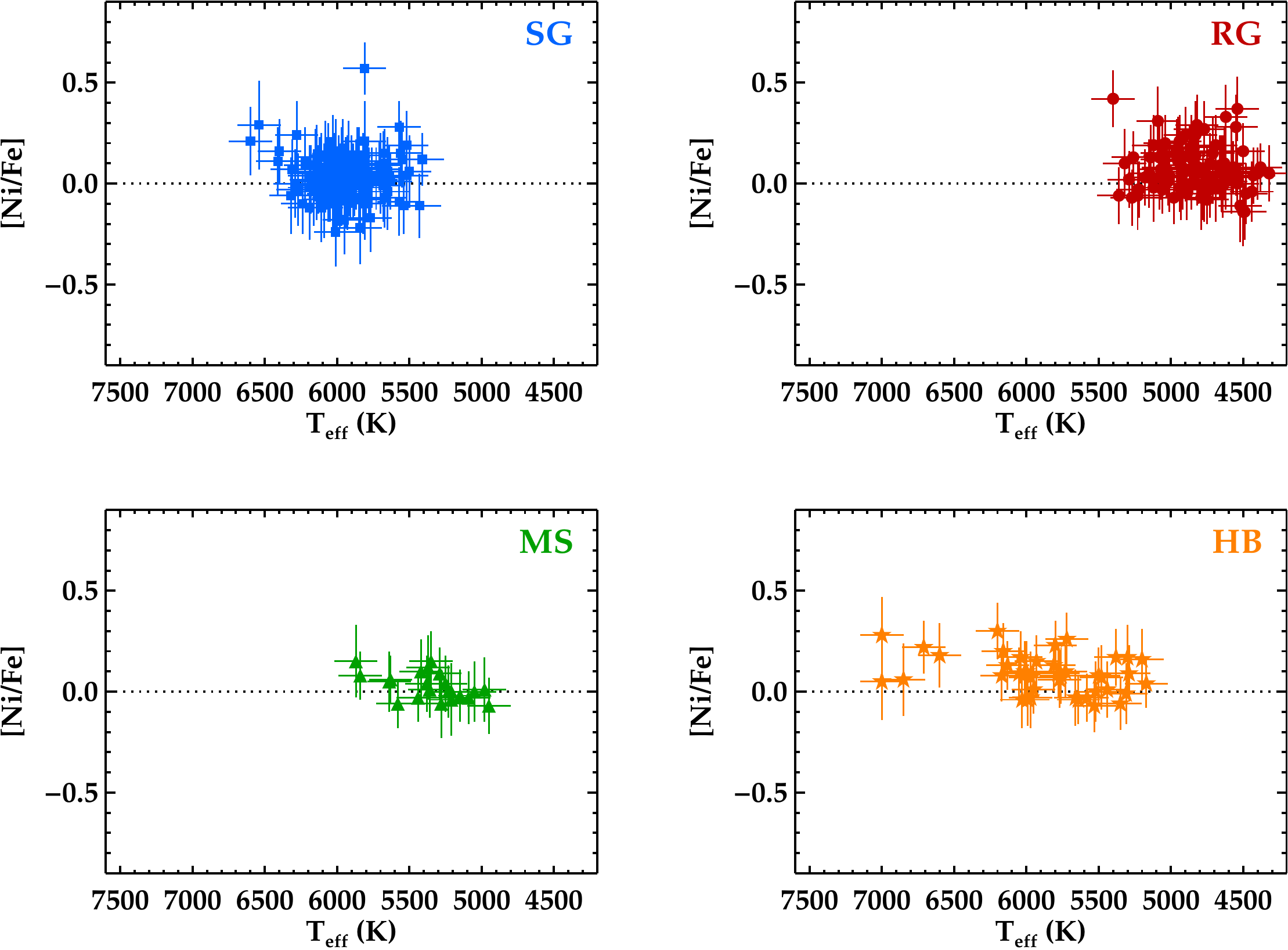}
\end{center}
\caption{
\label{niteffplot}
Derived [Ni/Fe] ratios as a function of \teff.
Symbols are the same as in Figures~\ref{liteffplot} and \ref{cteffplot}.
}
\end{figure*}

\begin{figure*}
\begin{center}
\includegraphics[angle=00,width=4.5in]{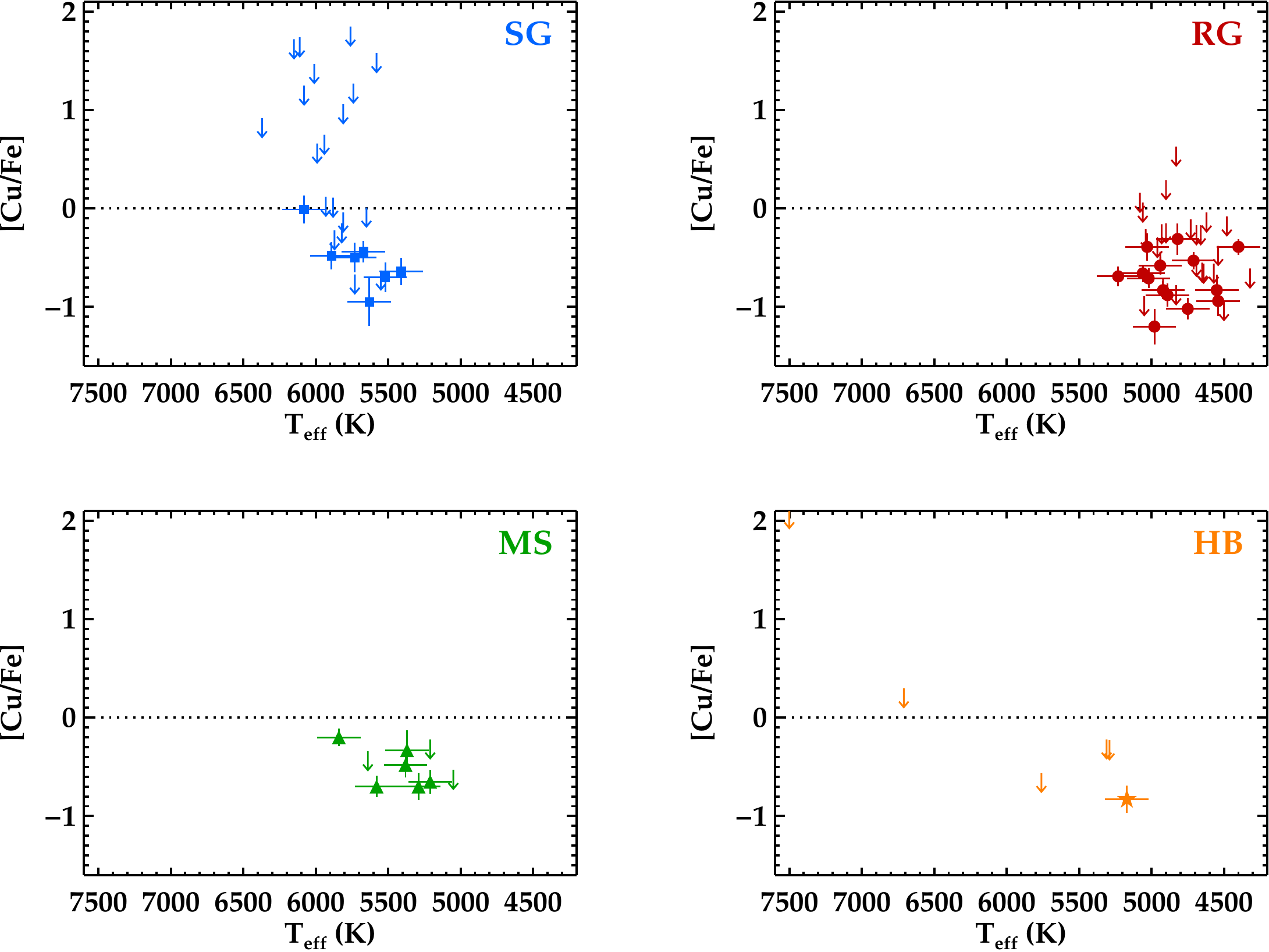}
\end{center}
\caption{
\label{cuteffplot}
Derived [Cu/Fe] ratios as a function of \teff.
Symbols are the same as in Figures~\ref{liteffplot} and \ref{cteffplot}.
}
\end{figure*}

\begin{figure*}
\begin{center}
\includegraphics[angle=00,width=4.5in]{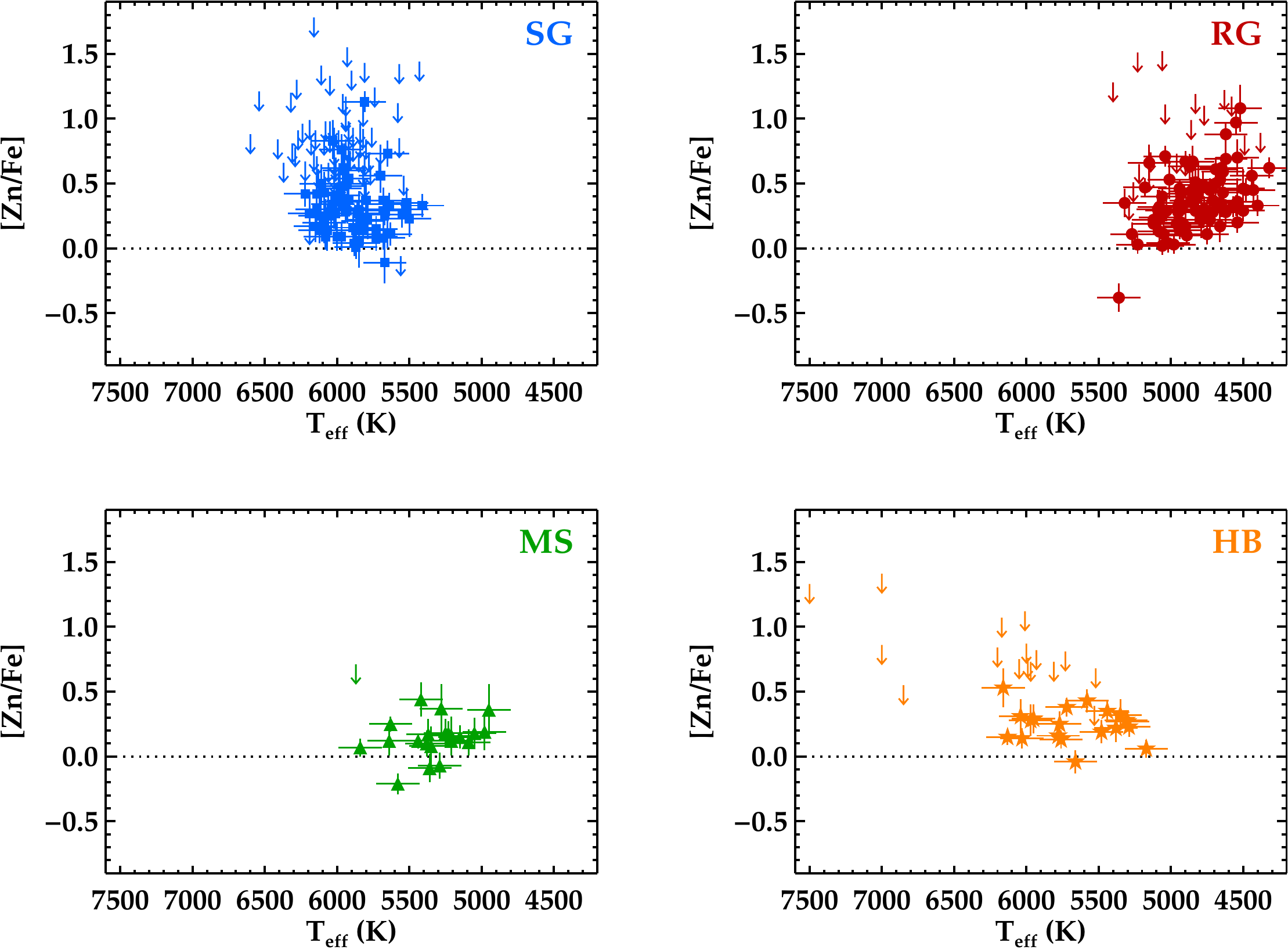}
\end{center}
\caption{
\label{znteffplot}
Derived [Zn/Fe] ratios as a function of \teff.
Symbols are the same as in Figures~\ref{liteffplot} and \ref{cteffplot}.
}
\end{figure*}

\begin{figure*}
\begin{center}
\includegraphics[angle=00,width=4.5in]{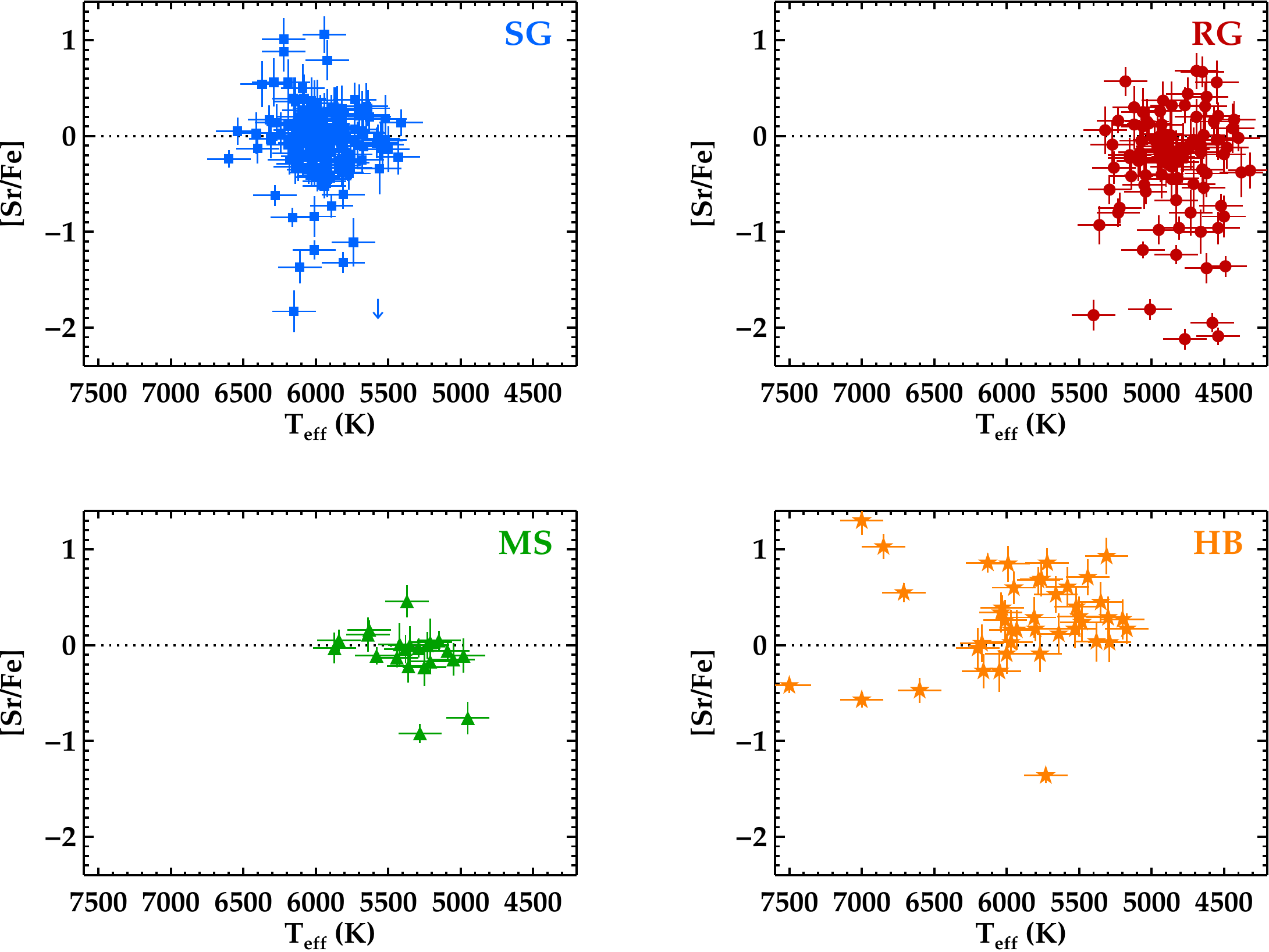}
\end{center}
\caption{
\label{srteffplot}
Derived [Sr/Fe] ratios as a function of \teff.
Symbols are the same as in Figures~\ref{liteffplot} and \ref{cteffplot}.
}
\end{figure*}

\begin{figure*}
\begin{center}
\includegraphics[angle=00,width=4.5in]{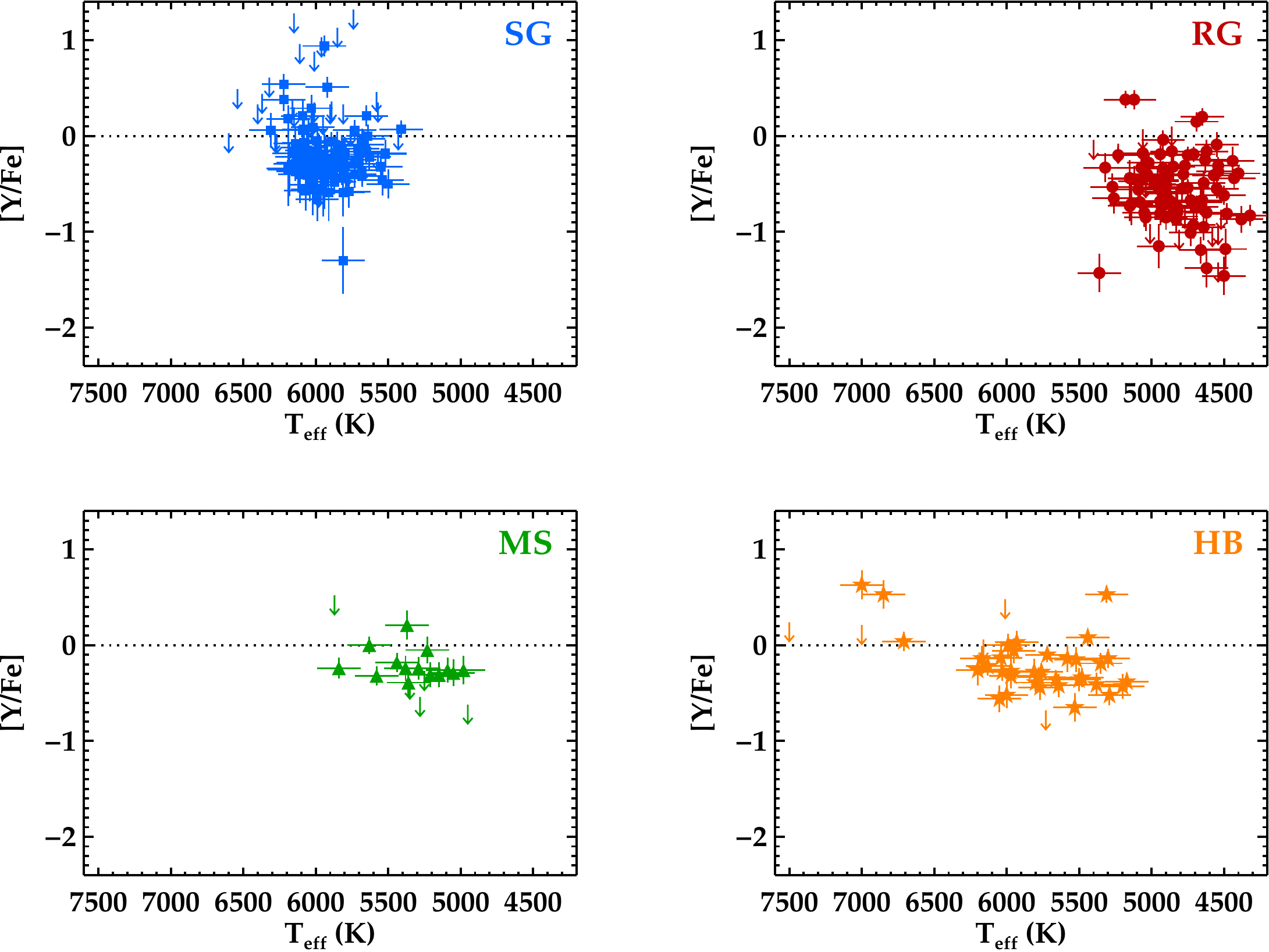}
\end{center}
\caption{
\label{yteffplot}
Derived [Y/Fe] ratios as a function of \teff.
Symbols are the same as in Figures~\ref{liteffplot} and \ref{cteffplot}.
}
\end{figure*}

\begin{figure*}
\begin{center}
\includegraphics[angle=00,width=4.5in]{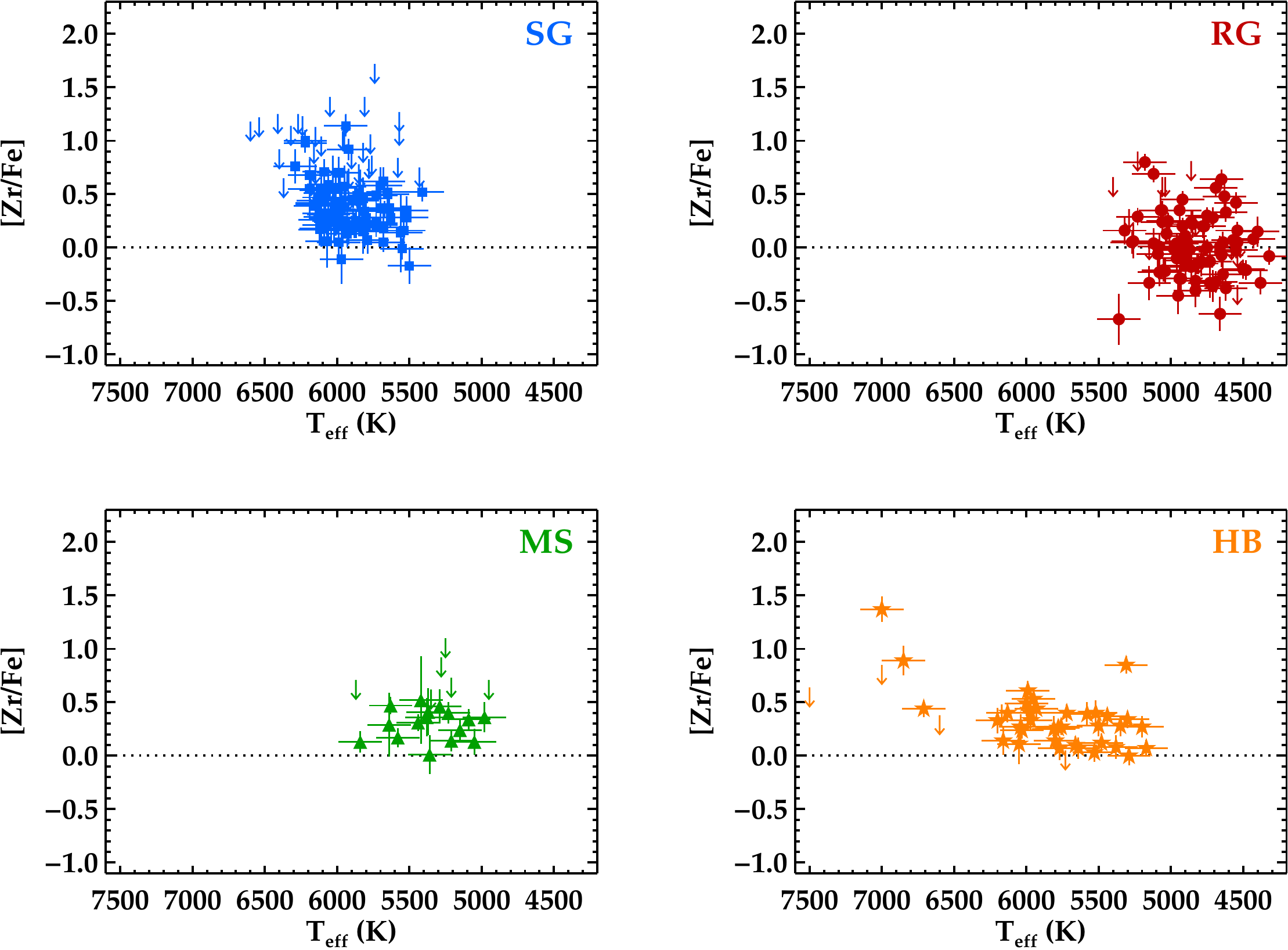}
\end{center}
\caption{
\label{zrteffplot}
Derived [Zr/Fe] ratios as a function of \teff.
Symbols are the same as in Figures~\ref{liteffplot} and \ref{cteffplot}.
}
\end{figure*}

\begin{figure*}
\begin{center}
\includegraphics[angle=00,width=4.5in]{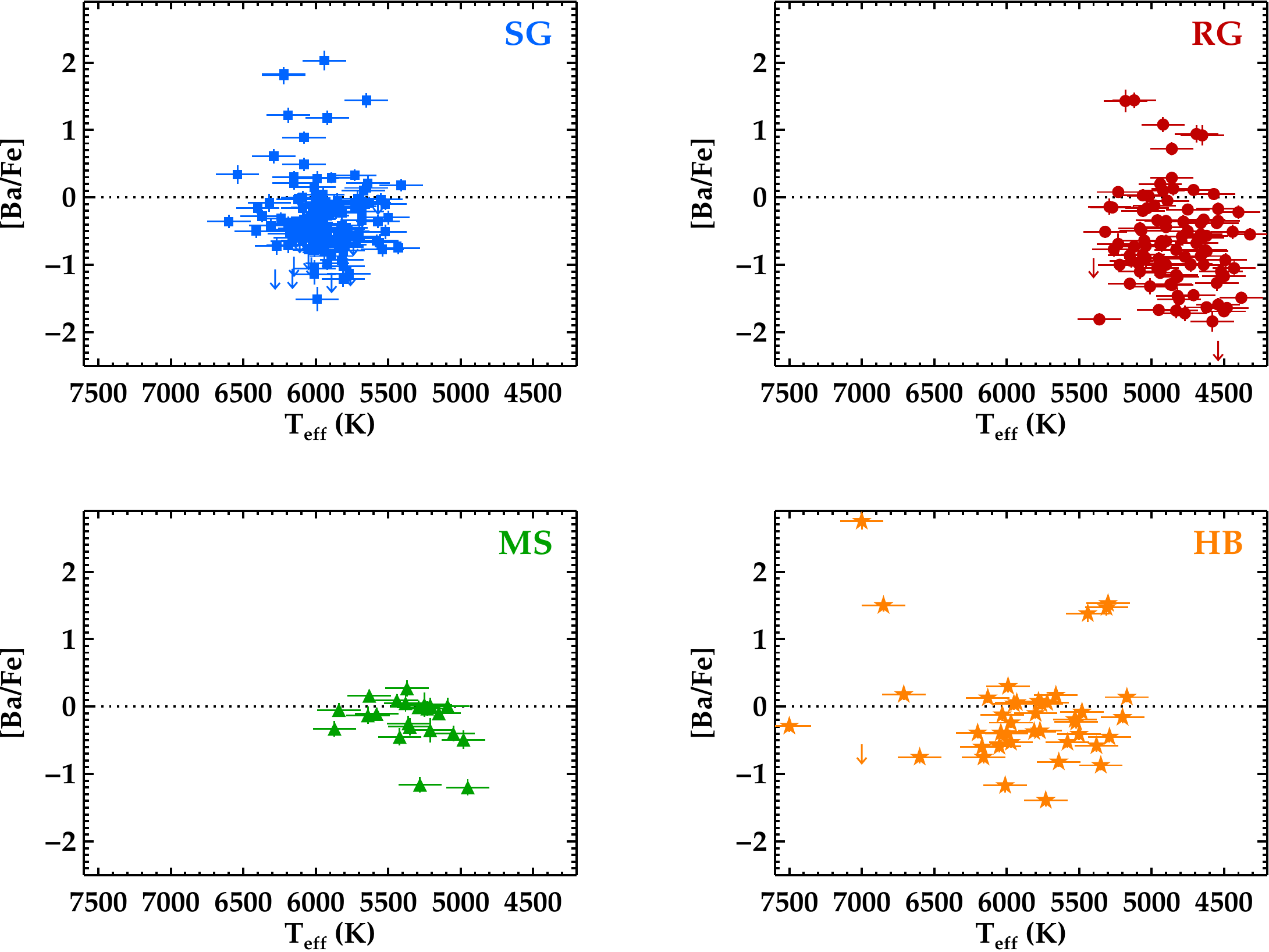}
\end{center}
\caption{
\label{bateffplot}
Derived [Ba/Fe] ratios as a function of \teff.
Symbols are the same as in Figures~\ref{liteffplot} and \ref{cteffplot}.
}
\end{figure*}

\begin{figure*}
\begin{center}
\includegraphics[angle=00,width=4.5in]{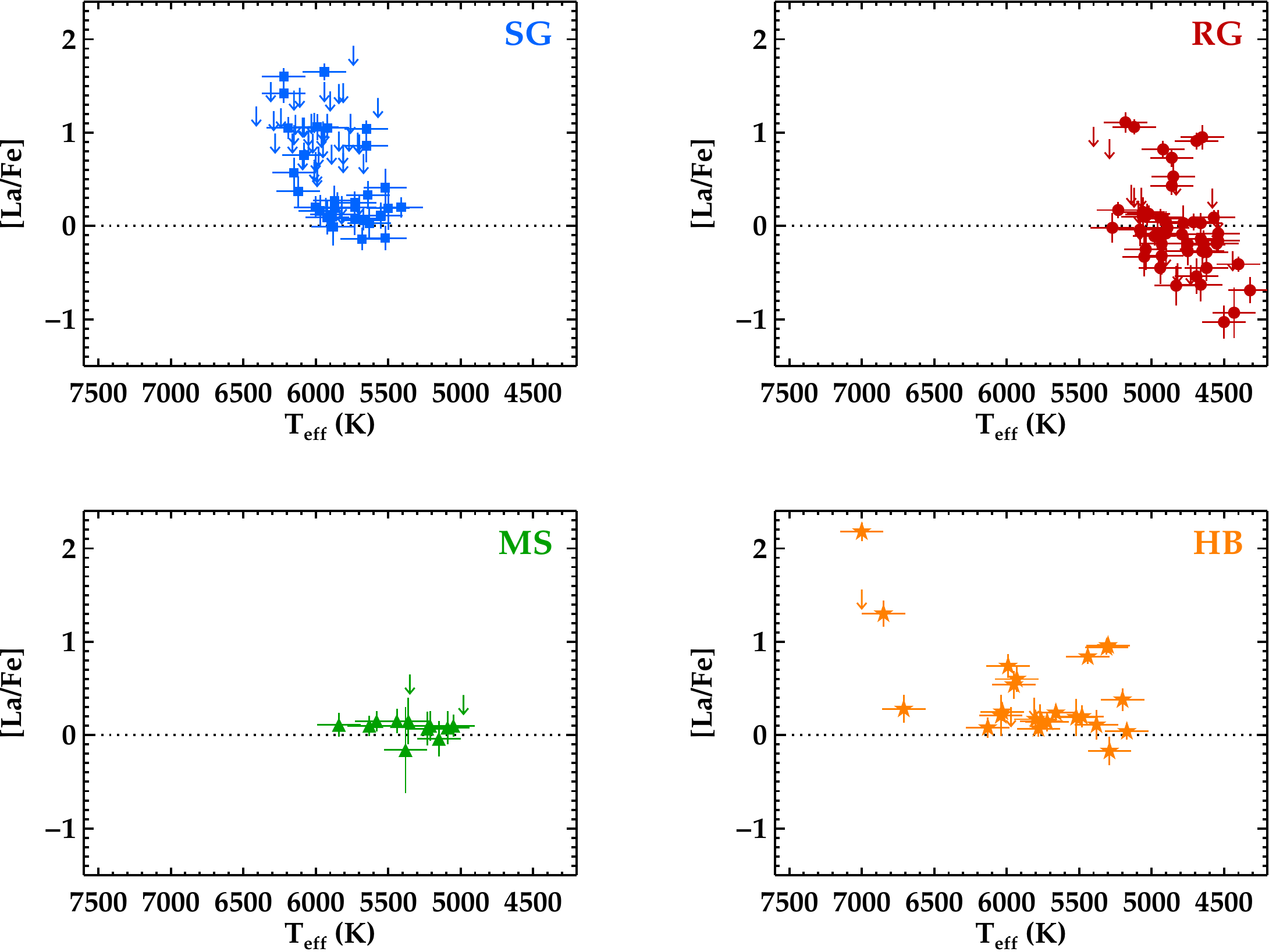}
\end{center}
\caption{
\label{lateffplot}
Derived [La/Fe] ratios as a function of \teff.
Symbols are the same as in Figures~\ref{liteffplot} and \ref{cteffplot}.
}
\end{figure*}

\begin{figure*}
\begin{center}
\includegraphics[angle=00,width=4.5in]{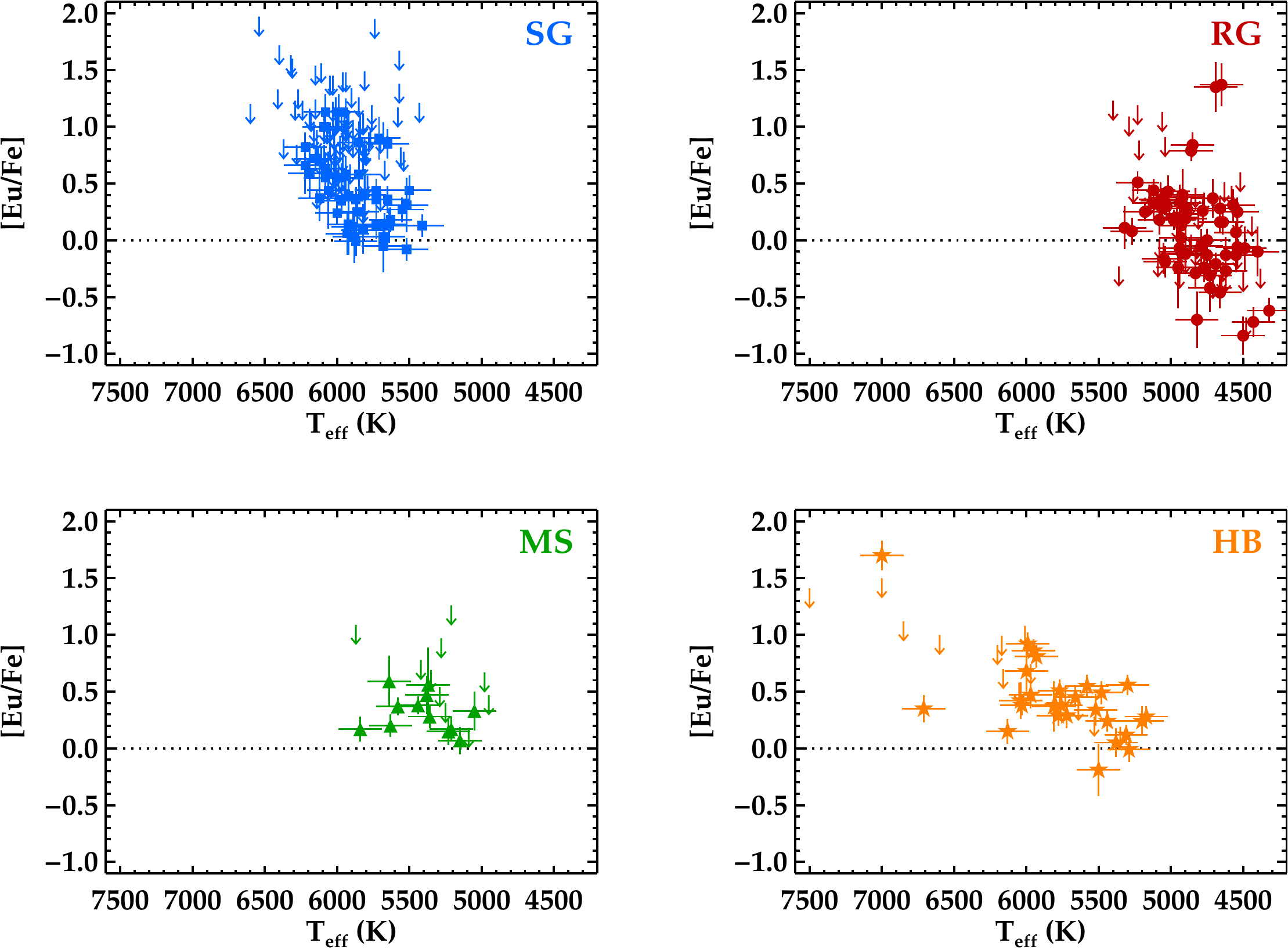}
\end{center}
\caption{
\label{euteffplot}
Derived [Eu/Fe] ratios as a function of \teff.
Symbols are the same as in Figures~\ref{liteffplot} and \ref{cteffplot}.
}
\end{figure*}

\begin{figure*}
\begin{center}
\includegraphics[angle=00,width=4.5in]{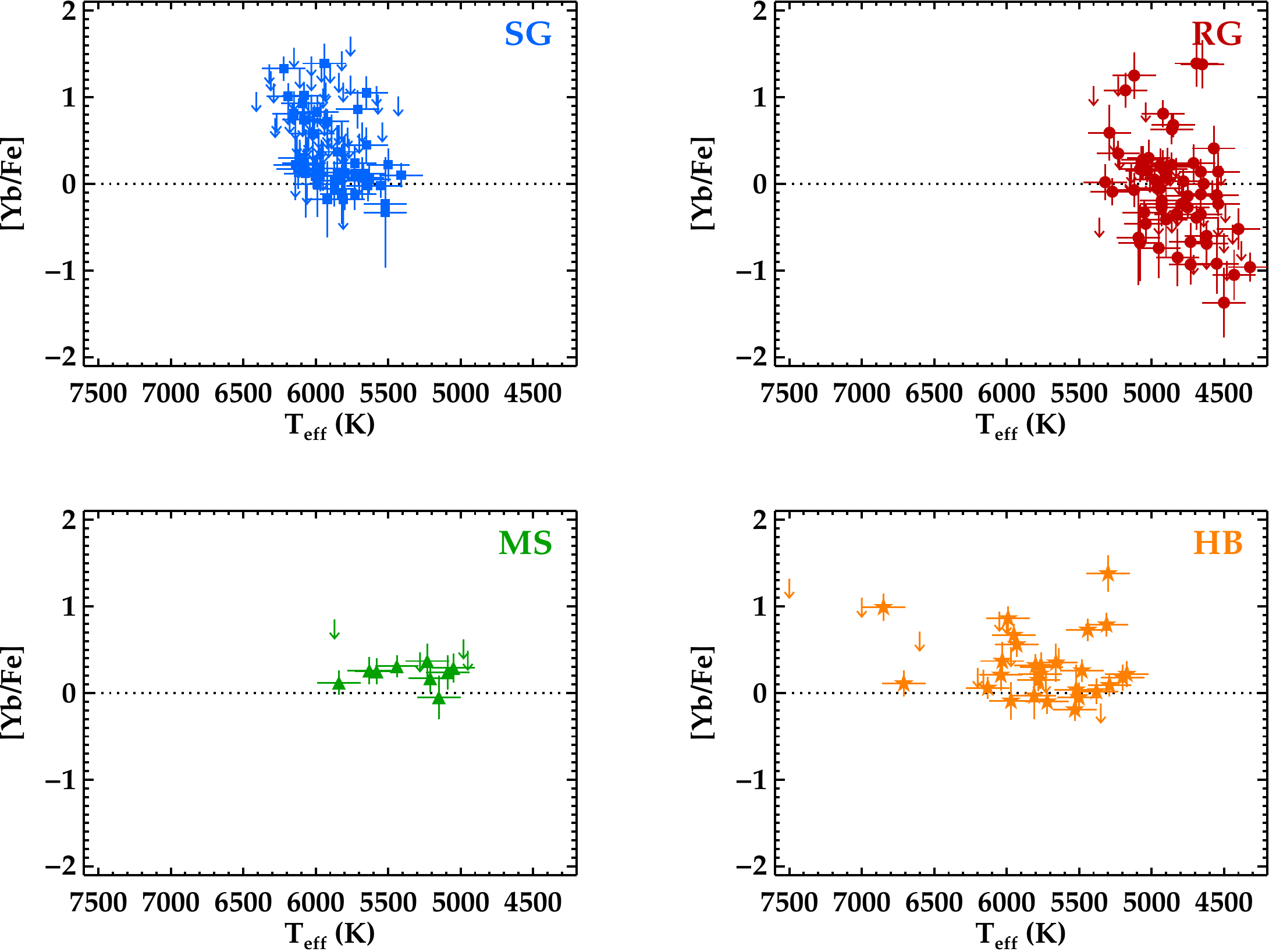}
\end{center}
\caption{
\label{ybteffplot}
Derived [Yb/Fe] ratios as a function of \teff.
Symbols are the same as in Figures~\ref{liteffplot} and \ref{cteffplot}.
}
\end{figure*}

\clearpage
\begin{figure*}
\begin{center}
\includegraphics[angle=00,width=4.5in]{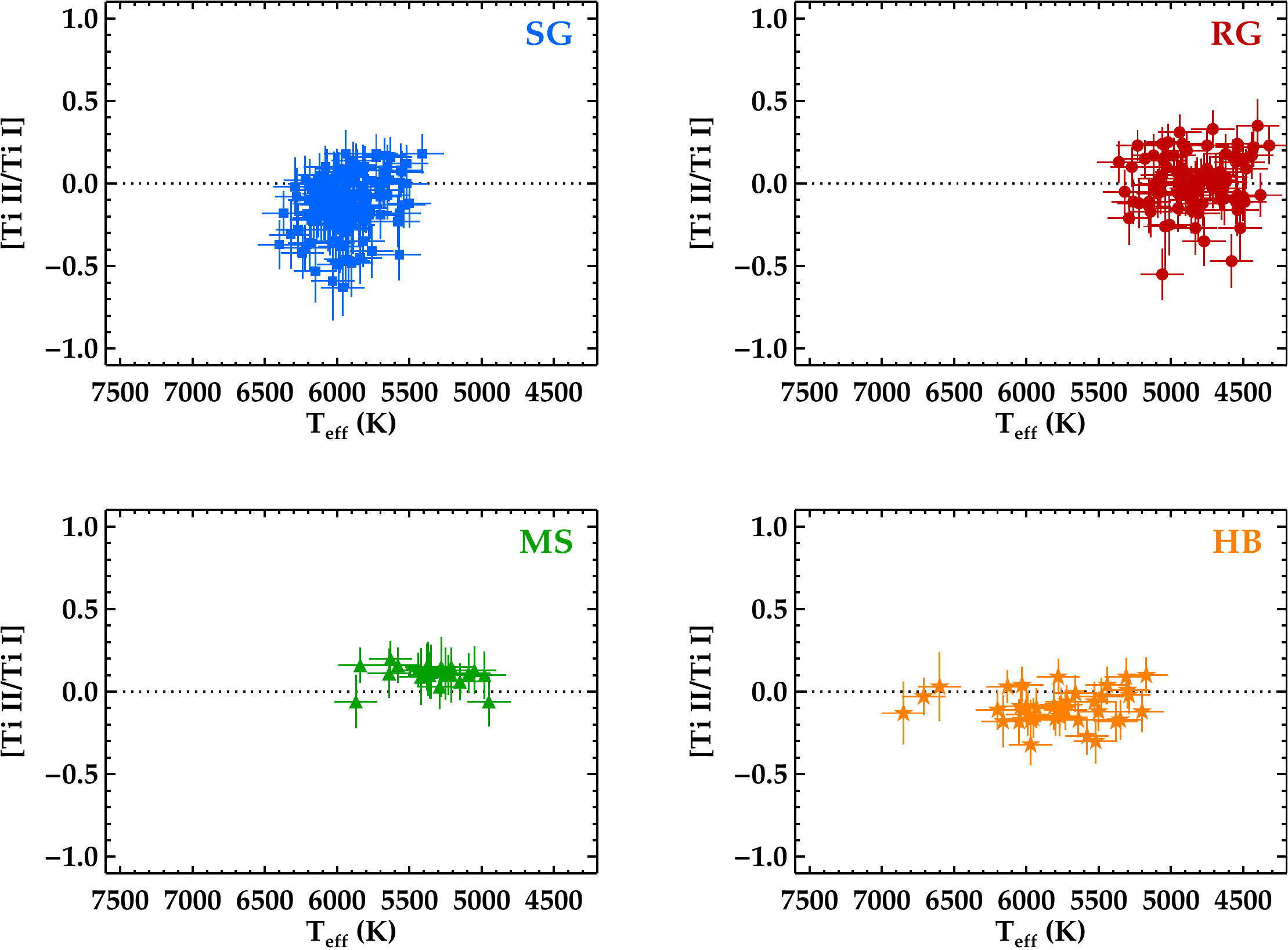}
\end{center}
\caption{
\label{ti2ti1teffplot}
Ratios of the total titanium abundance derived 
from each of the ionized and neutral species
as a function of \teff.
Each star is displayed only if both species have been detected.
Symbols are the same as in Figures~\ref{liteffplot} and \ref{cteffplot}.
}
\end{figure*}

\begin{figure*}
\begin{center}
\includegraphics[angle=00,width=4.5in]{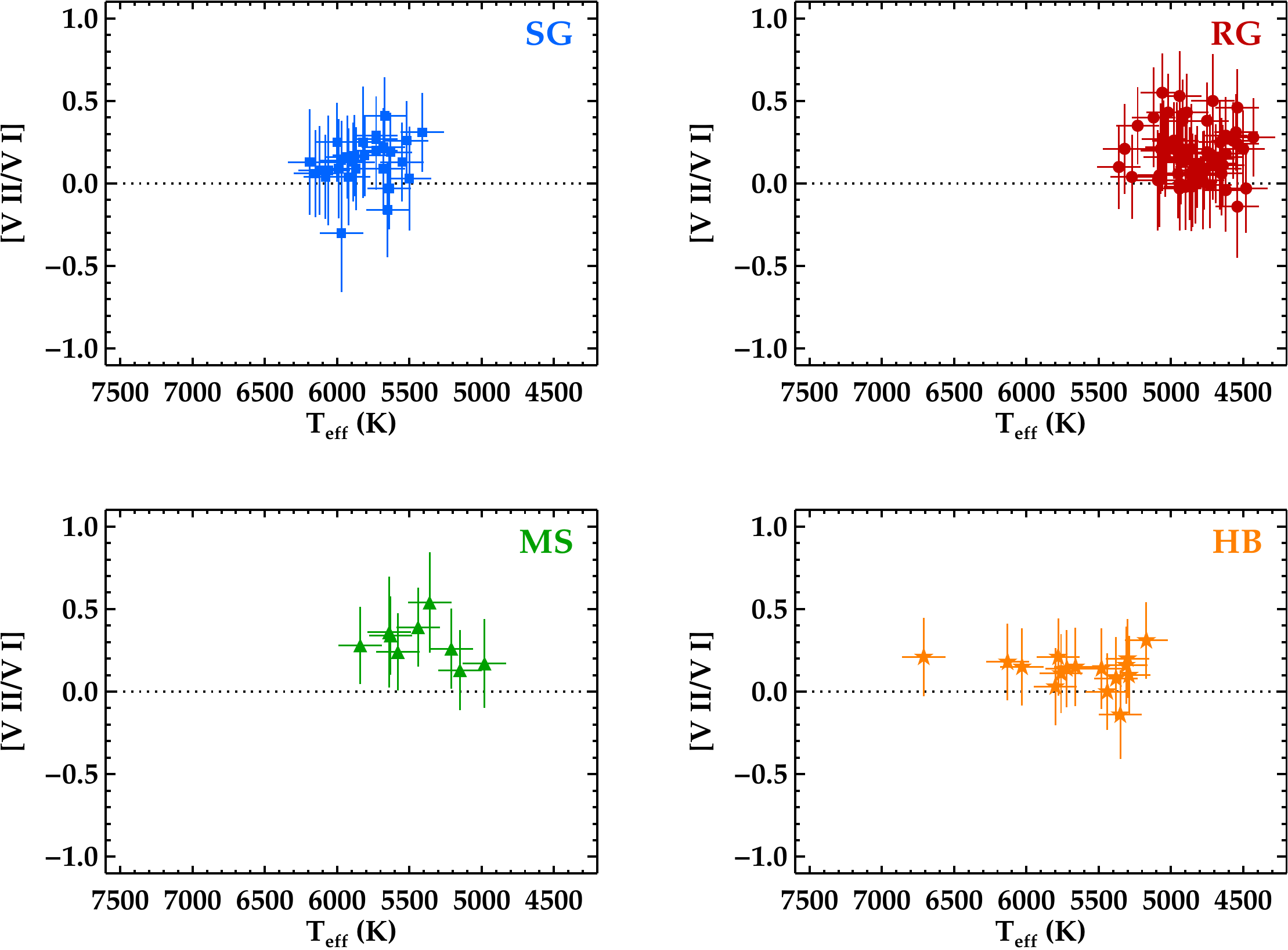}
\end{center}
\caption{
\label{v2v1teffplot}
Ratios of the total vanadium abundance derived 
from each of the ionized and neutral species
as a function of \teff.
Each star is displayed only if both species have been detected.
Symbols are the same as in Figures~\ref{liteffplot} and \ref{cteffplot}.
}
\end{figure*}

\begin{figure*}
\begin{center}
\includegraphics[angle=00,width=4.5in]{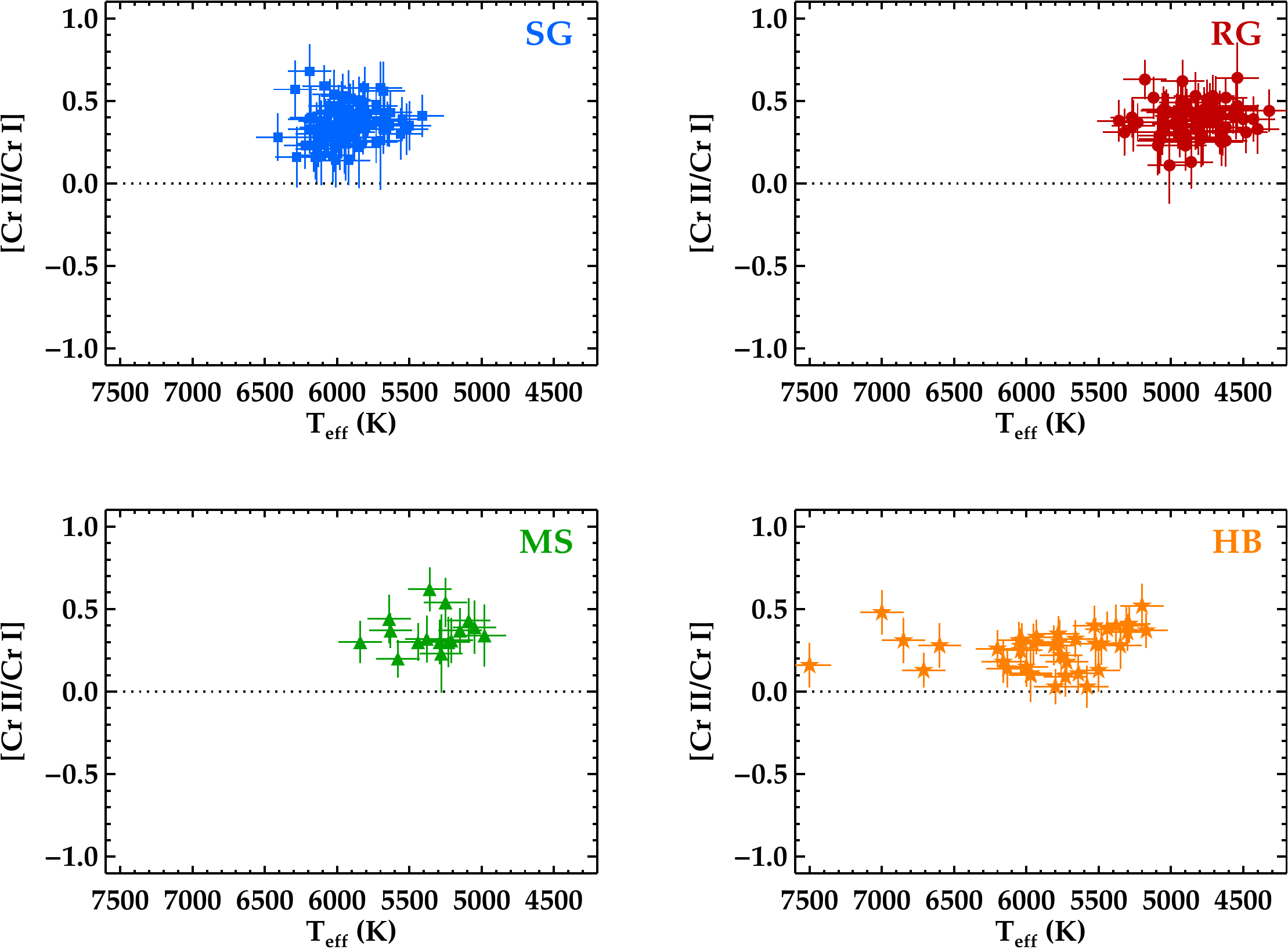}
\end{center}
\caption{
\label{cr2cr1teffplot}
Ratios of the total chromium abundance derived 
from each of the ionized and neutral species
as a function of \teff.
Each star is displayed only if both species have been detected.
Symbols are the same as in Figures~\ref{liteffplot} and \ref{cteffplot}.
}
\end{figure*}

\begin{figure*}
\begin{center}
\includegraphics[angle=00,width=4.5in]{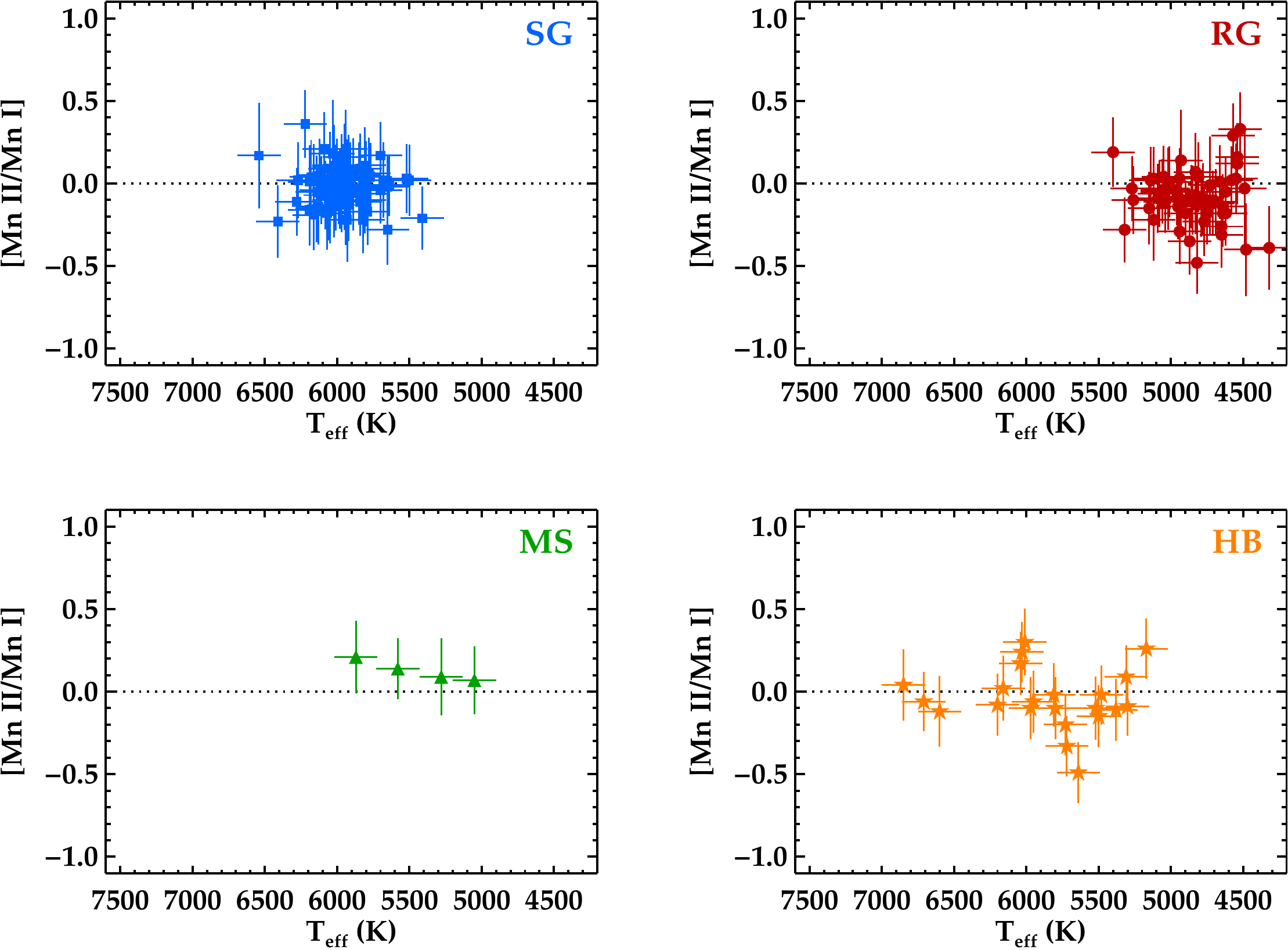}
\end{center}
\caption{
\label{mn2mn1teffplot}
Ratios of the total manganese abundance derived 
from each of the ionized and neutral species
as a function of \teff.
Each star is displayed only if both species have been detected.
Symbols are the same as in Figures~\ref{liteffplot} and \ref{cteffplot}.
}
\end{figure*}

\begin{figure*}
\begin{center}
\includegraphics[angle=00,width=4.5in]{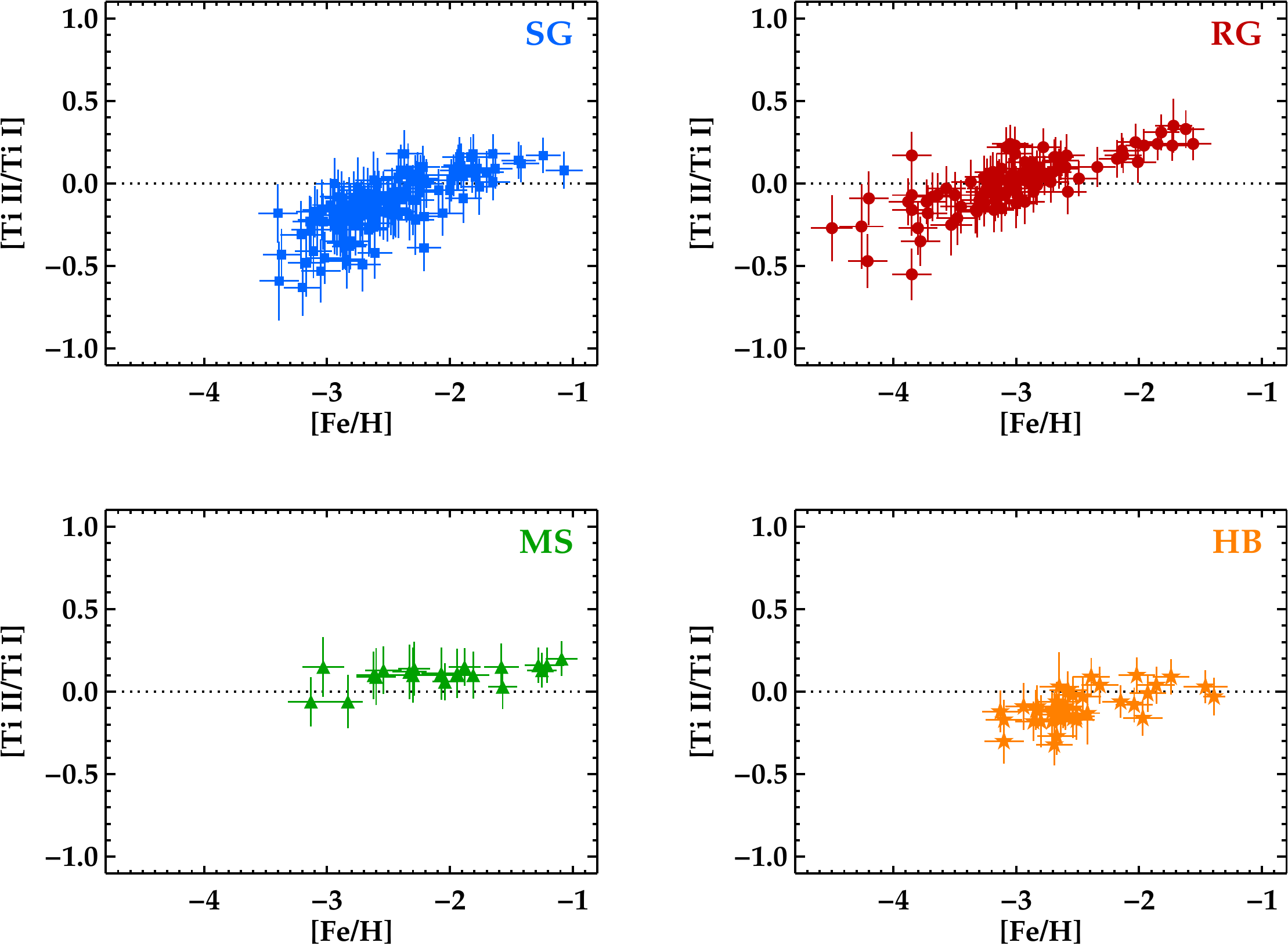}
\end{center}
\caption{
\label{ti2ti1feplot}
Ratios of the total titanium abundance derived 
from each of the ionized and neutral species
as a function of [Fe/H].
Each star is displayed only if both species have been detected.
Symbols are the same as in Figures~\ref{liteffplot} and \ref{cteffplot}.
}
\end{figure*}

\begin{figure*}
\begin{center}
\includegraphics[angle=00,width=4.5in]{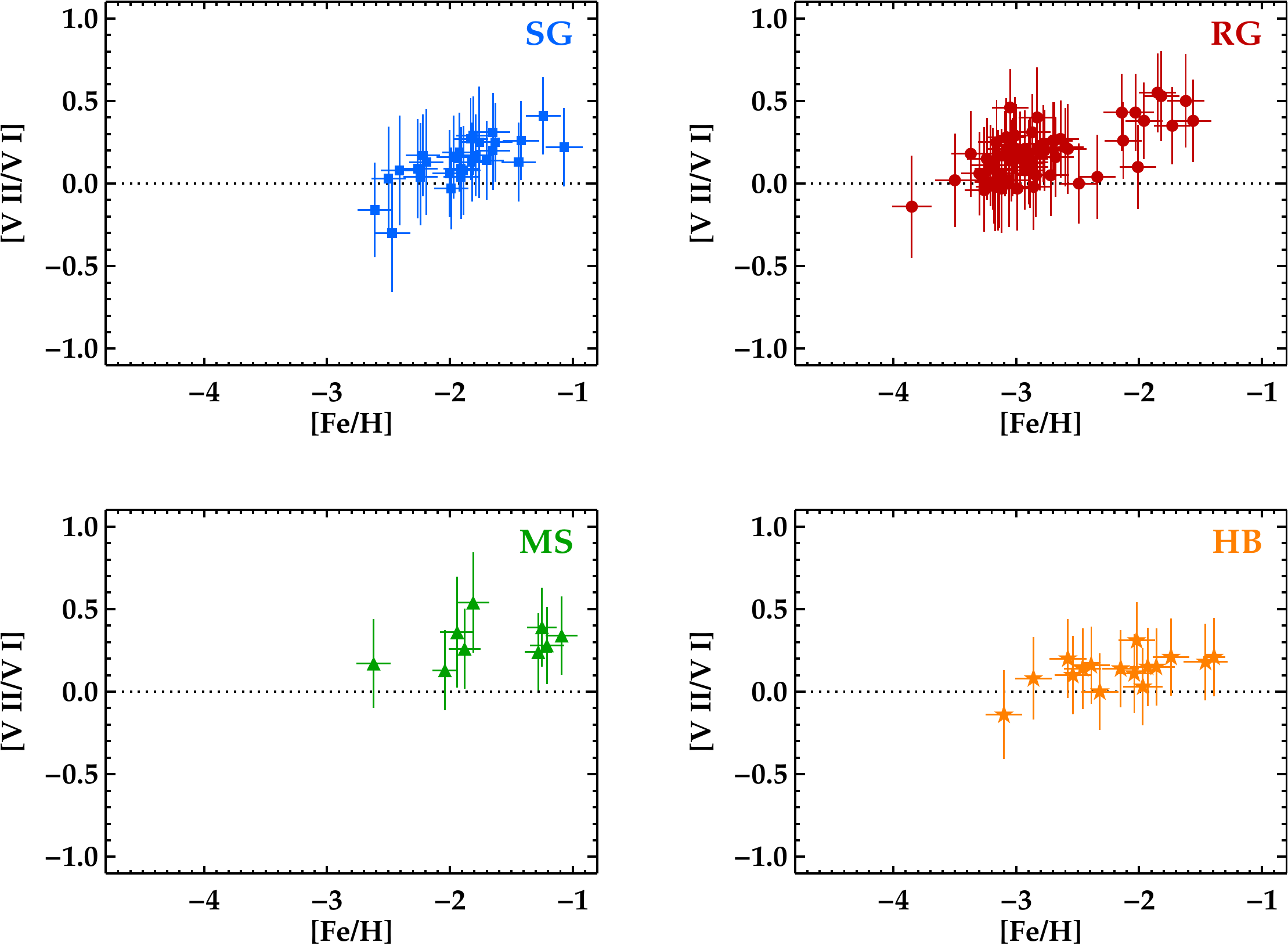}
\end{center}
\caption{
\label{v2v1feplot}
Ratios of the total vanadium abundance derived 
from each of the ionized and neutral species
as a function of [Fe/H].
Each star is displayed only if both species have been detected.
Symbols are the same as in Figures~\ref{liteffplot} and \ref{cteffplot}.
}
\end{figure*}

\begin{figure*}
\begin{center}
\includegraphics[angle=00,width=4.5in]{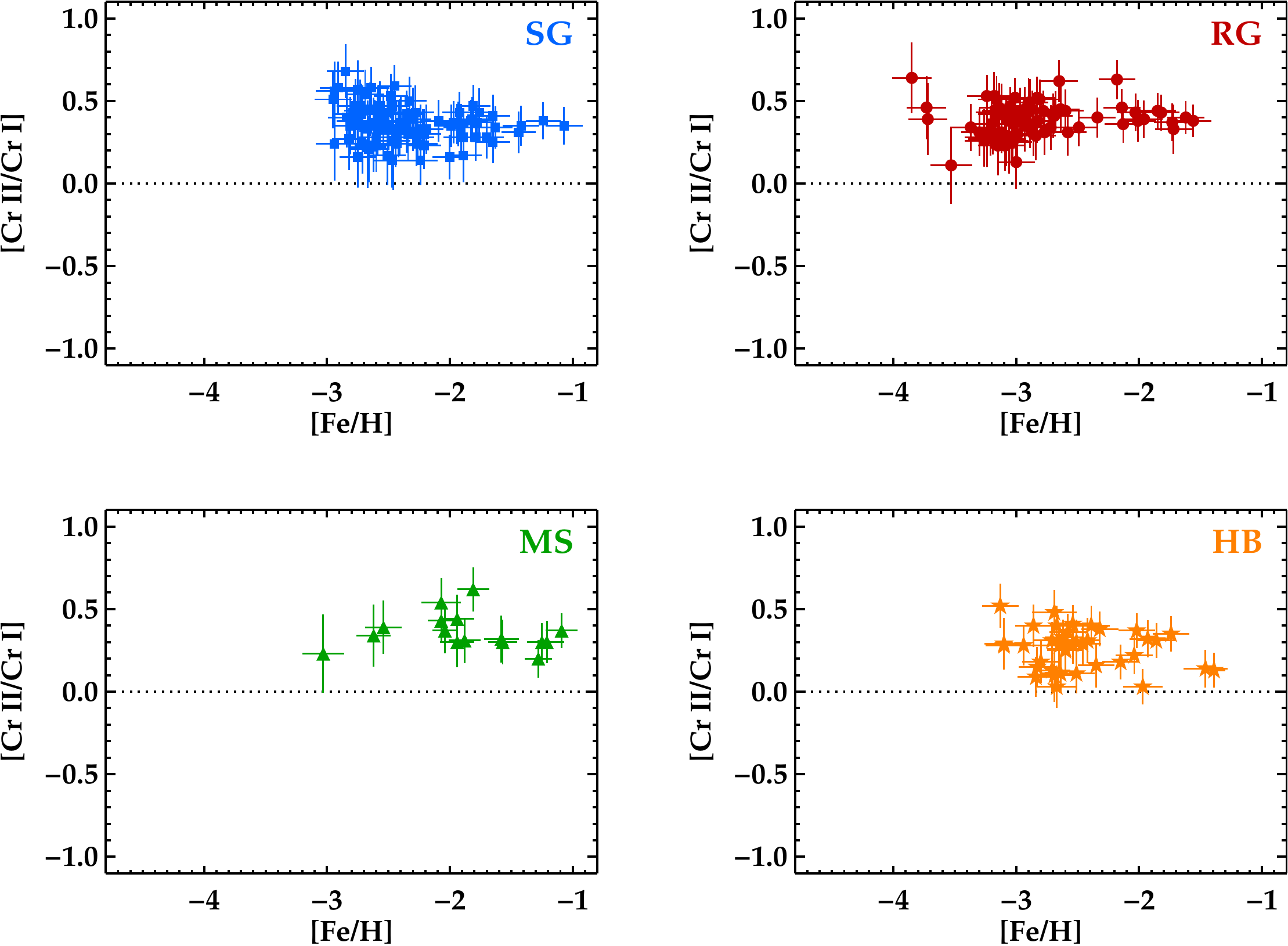}
\end{center}
\caption{
\label{cr2cr1feplot}
Ratios of the total chromium abundance derived 
from each of the ionized and neutral species
as a function of [Fe/H].
Each star is displayed only if both species have been detected.
Symbols are the same as in Figures~\ref{liteffplot} and \ref{cteffplot}.
}
\end{figure*}

\begin{figure*}
\begin{center}
\includegraphics[angle=00,width=4.5in]{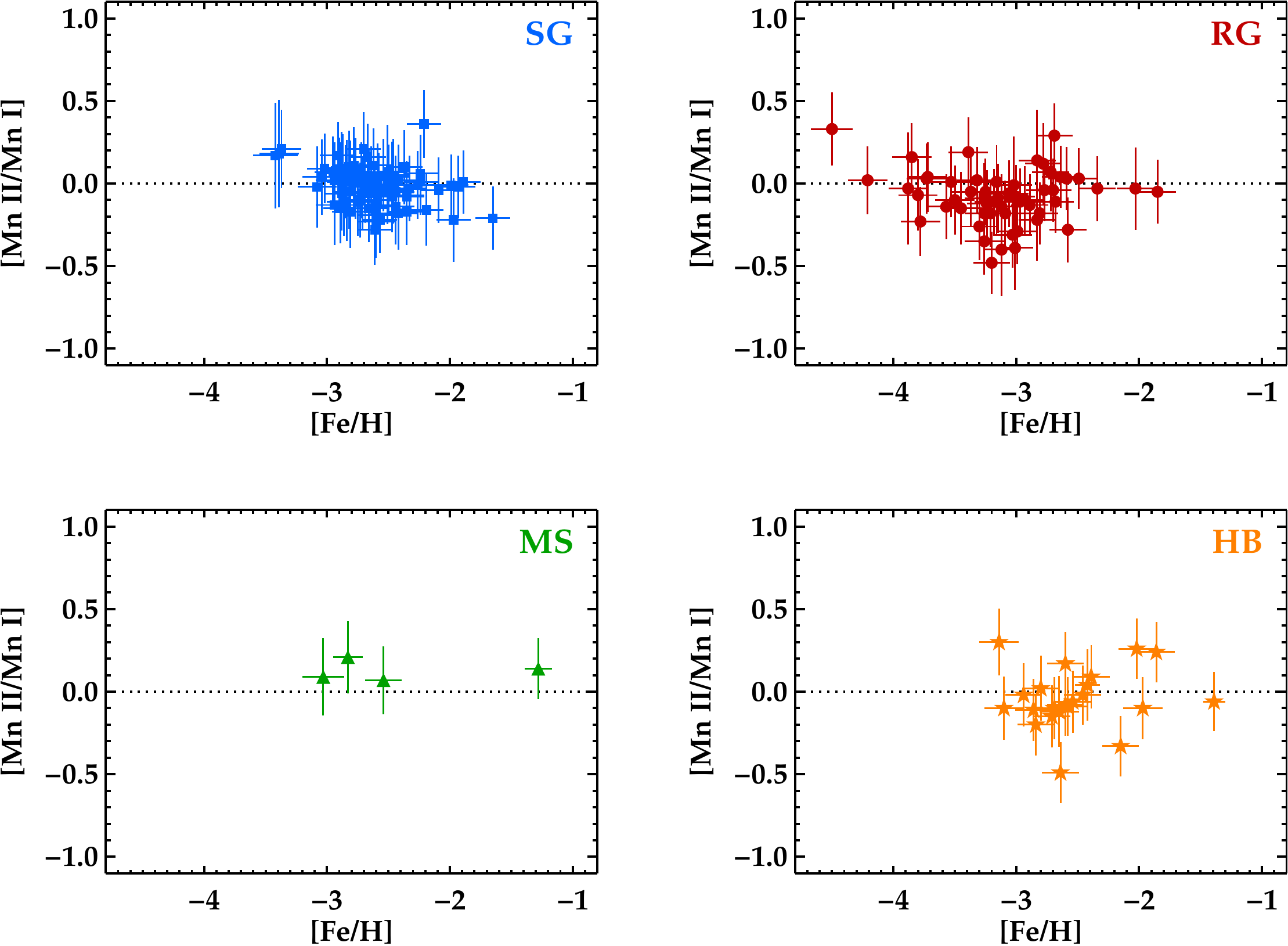}
\end{center}
\caption{
\label{mn2mn1feplot}
Ratios of the total manganese abundance derived 
from each of the ionized and neutral species
as a function of [Fe/H].
Each star is displayed only if both species have been detected.
Symbols are the same as in Figures~\ref{liteffplot} and \ref{cteffplot}.
}
\end{figure*}

\clearpage

\LongTables
\begin{deluxetable*}{clcccccccccccc}
\tablefontsize{\footnotesize}
\tablecaption{Magnitudes, Reddenings, Colors, Atmospheric Parameters, and Random Uncertainties
\label{atmtab}}
\tablewidth{0pt}
\tabletypesize{\scriptsize}
\tablehead{
\colhead{Class\tablenotemark{a}} &
\colhead{Star} &
\colhead{$V$} &
\colhead{$E(B-V)$} &
\colhead{$(B-V)_{0}$} &
\colhead{$(V-K)_{0}$} &
\colhead{\teff} &
\colhead{\logg} &
\colhead{\vt} &
\colhead{[M/H]} &
\colhead{$\sigma_{\teff}$} &
\colhead{$\sigma_{\logg}$} &
\colhead{$\sigma_{\vt}$} &
\colhead{$\sigma_{\rm [M/H]}$\tablenotemark{b}} \\
\colhead{} &
\colhead{} &
\colhead{} &
\colhead{} &
\colhead{} &
\colhead{} &
\colhead{(K)} &
\colhead{} &
\colhead{(\kmsec)} &
\colhead{} &
\colhead{(K)} &
\colhead{} &
\colhead{(\kmsec)} &
\colhead{} }
\startdata                                                                                                           
 RG & CS~22166--016                       & 12.75 & 0.025 &   0.63 &   1.92  & 4900 & 1.75 & 1.50 & $-$3.09 & 37 & 0.14 & 0.06 & 0.06  \\  
 SG & CS~22169--008                       & 14.99 & 0.057 &   0.37 &   1.27  & 5810 & 3.60 & 1.30 & $-$2.59 & 54 & 0.16 & 0.06 & 0.06  \\  
 RG & CS~22169--035                       & 12.88 & 0.043 &   0.89 &   2.41  & 4480 & 0.50 & 1.70 & $-$3.12 & 34 & 0.14 & 0.06 & 0.06  \\  
 SG & CS~22171--031                       & 13.93 & 0.023 &   0.37 &   1.22  & 6010 & 3.70 & 1.35 & $-$2.42 & 53 & 0.21 & 0.06 & 0.06  \\  
 SG & CS~22171--037                       & 14.93 & 0.020 &   0.33 &   1.11  & 6110 & 3.65 & 1.60 & $-$3.35 & 40 & 0.15 & 0.06 & 0.09  \\ 
 RG & CS~22172--029                       & 14.36 & 0.056 &\nodata &   1.81  & 4960 & 1.90 & 1.30 & $-$2.70 & 37 & 0.16 & 0.06 & 0.06  \\  
 SG & CS~22174--020                       & 15.06 & 0.040 &   0.36 &   1.18  & 5840 & 3.55 & 1.40 & $-$3.02 & 58 & 0.16 & 0.06 & 0.06  \\  
 SG & CS~22177--009                       & 14.27 & 0.044 &   0.36 &   1.20  & 5940 & 3.55 & 1.25 & $-$3.37 & 38 & 0.16 & 0.06 & 0.08  \\
\enddata
\tablenotetext{a}{
Classification scheme:
BS~$=$~blue straggler-like stars warmer than the main sequence turnoff;
HB~$=$~stars on the horizontal branch;
MS~$=$~stars on the main sequence;
RG~$=$~stars on the red giant branch;
SG~$=$~stars on the subgiant branch.
}
\tablenotetext{b}{
$\sigma_{\rm [M/H]} = \sigma_{\rm Fe\,II}$
}
\tablenotetext{*}{
Star with reliable parallax and $E(B-V) <$~0.12
}
\end{deluxetable*}

\clearpage

\LongTables
\begin{deluxetable*}{ccccccccccccc}
\tablecaption{Line Abundance Corrections and Uncertainties
\label{offsettab}}
\tablewidth{0pt}
\tabletypesize{\scriptsize}
\tablehead{
\colhead{Species} &
\colhead{Wavelength} &
\multicolumn{2}{c}{RG} &
\colhead{} &
\multicolumn{2}{c}{SG} &
\colhead{} &
\multicolumn{2}{c}{HB} &
\colhead{} &
\multicolumn{2}{c}{MS} \\
\cline{3-4} \cline{6-7} \cline{9-10} \cline{12-13}
\colhead{} &
\colhead{(\AA)} &
\colhead{Correction} & 
\colhead{$\sigma$} &
\colhead{} &
\colhead{Correction} & 
\colhead{$\sigma$} &
\colhead{} &
\colhead{Correction} & 
\colhead{$\sigma$} &
\colhead{} &
\colhead{Correction} & 
\colhead{$\sigma$} }
\startdata   
O~\textsc{i}   & 7771.94 & $-$0.50  &  0.25 & & $-$0.50  &  0.25 & & $-$0.50  &  0.25 & & $-$0.50  &  0.25 \\
O~\textsc{i}   & 7774.17 & $-$0.50  &  0.25 & & $-$0.50  &  0.25 & & $-$0.50  &  0.25 & & $-$0.50  &  0.25 \\
O~\textsc{i}   & 7775.39 & $-$0.50  &  0.25 & & $-$0.50  &  0.25 & & $-$0.50  &  0.25 & & $-$0.50  &  0.25 \\
Na~\textsc{i}  & 5682.63 & $+$0.00  &  0.13 & & $+$0.00  &  0.13 & & $+$0.00  &  0.13 & & $+$0.00  &  0.13 \\
Na~\textsc{i}  & 5688.20 & $+$0.00  &  0.13 & & $+$0.00  &  0.13 & & $+$0.00  &  0.13 & & $+$0.00  &  0.13 \\
Na~\textsc{i}  & 5889.95 & $+$0.01  &  0.06 & & $+$0.01  &  0.06 & & $+$0.01  &  0.06 & & $+$0.01  &  0.06 \\
Na~\textsc{i}  & 5895.92 & $-$0.01  &  0.06 & & $-$0.01  &  0.06 & & $-$0.01  &  0.06 & & $-$0.01  &  0.06 \\
Mg~\textsc{i}  & 3829.36 & $+$0.04  &  0.08 & & $+$0.19  &  0.07 & & $+$0.04  &  0.08 & & \nodata  &\nodata\\
\enddata
\end{deluxetable*}

\end{document}